\documentclass{aa} 
\usepackage{graphicx}
\usepackage{txfonts}
\usepackage{tabularx}
\usepackage{appendix}
\usepackage{threeparttable}  
\usepackage{dcolumn}
\usepackage{amsmath}

\newcommand{\hi}{H{\sc\,i}}
\newcommand{\hii}{H{\sc\,ii}}
\newcommand{\uchii}{UC\,H{\sc\,ii}}
\newcommand{\hchii}{HC\,H{\sc\,ii}}
\newcommand{\kms}{km\,s$^{-1}$}

\newcolumntype{d}[1]{D{.}{\cdot}{#1}}

\newcolumntype{.}{D{.}{.}{-1}}

\usepackage[normalem]{ulem}

\usepackage[]{xcolor}

\definecolor{tbc}{cmyk}{0.05002,1,0.9,0}

\definecolor{mygreen}{cmyk}{0.85002,0.3,1,0}

\definecolor{myyellow}{cmyk}{0.00,0.24,0.86,0.20}

\definecolor{mypink}{cmyk}{0.00,0.80,0.00,0.00}

\definecolor{grey}{cmyk}{0.30,0.30,0.30,0.30}

\usepackage{url}
\begin{document} 
\newcolumntype{L}[1]{>{\raggedright\arraybackslash}p{#1}}
\newcolumntype{C}[1]{>{\centering\arraybackslash}p{#1}}
\newcolumntype{R}[1]{>{\raggedleft\arraybackslash}p{#1}}
 \title{A population of hypercompact \hii\  regions identified from young \hii\ regions}
   \author{A. Y. Yang
        \inst{1}, 
        J. S. Urquhart\inst{2},
        M. A. Thompson\inst{3},
        K. M. Menten \inst{1},
        F. Wyrowski\inst{1},
        A. Brunthaler \inst{1},
        W. W. Tian\inst{4,5}, 
        M. Rugel\inst{1},
        X. L. Yang \inst{6},
        S.  Yao\inst{6},
        M. Mutale\inst{3}
        }
\institute{Max Planck Institute for Radio Astronomy, Auf dem H\"ugel 69, 53121, Bonn, Germany\\
\email{ayyang@mpifr-bonn.mpg.de}
\and
    Centre for Astrophysics and Planetary Science, University of Kent, Canterbury, CT2 7NH, UK\\
\and 
    Centre for Astrophysics Research, School of Physics Astronomy $\&$ Mathematics, University of Hertfordshire, College Lane, Hatfield, AL10 9AB, UK\\     %
\and 
    CAS Key Laboratory of FAST, National Astronomical Observatories, Chinese Academy of Sciences, Beijing, 100012\\
\and 
    University of Chinese Academy of Sciences, 19A Yuquan Road, Shijingshan District, Beijing 100049, China\\
\and
    Kavli Institute for Astronomy and Astrophysics, Peking University, Beijing 100871, China\\
    }
   \date{Received  ,  ; accepted    ,  }
%
  \abstract
  {The derived physical parameters for young \hii\ regions are normally determined assuming the emission region to be optically thin. 
   However, this assumption is unlikely to hold for young \hii\ regions such as hyper-compact \hii\ (\hchii) and ultra-compact \hii\ (\uchii) regions and leads to underestimation of their properties. 
   This can be overcome by fitting the SEDs over a wide range of radio frequencies.}
  {The two primary goals of this study are (1)  
   to determine the physical properties of young \hii\ regions  from radio SEDs in the search for potential \hchii\ regions, and (2) to use these physical properties to investigate their evolution. }
     { We used the Karl G. Jansky Very Large Array (VLA) to observe the X-band and K-band with angular resolutions of $\rm \sim1.7\arcsec$ and $\rm \sim0.7\arcsec$, respectively, toward 114 \hii\ regions with rising-spectra ($\alpha^{\rm 5\,GHz}_{\rm 1.4\,GHz} > 0$). We complement our observations with VLA archival data and construct SEDs in the range of 1$-$26 GHz and model them assuming an ionization-bounded \hii\ region with uniform density. }
      { Our sample has a mean electron density of $n_{\rm e}=\rm  1.6\times10^{4}\,cm^{-3}$, diameter $diam \rm =  0.14\,pc$, and emission measure $\rm EM =  1.9\times10^{7}\,pc\,cm^{-6}$. We identify 16 \hchii\ region candidates and 8  intermediate objects between the classes of \hchii\ and \uchii\, regions. The $n_{\rm e}$, $diam$, and EM change, as expected, but the Lyman continuum flux is relatively constant over time. We find that about 67\% of Lyman-continuum photons are absorbed by dust within these \hii\ regions and the dust absorption fraction tends to be more significant for more compact and younger \hii\ regions.} 
   { Young \hii\ regions are commonly located in dusty clumps; \hchii\ regions and intermediate objects are often associated with various masers, outflows, broad radio recombination lines, and extended green objects, and the accretion at the two stages tends to be quickly reduced or halted. }
   
   \keywords{ISM: \hii\ regions --ISM: evolution--radio continuum: stars--stars: massive--stars: formation } 
  
\titlerunning{hypercompact \hii\  regions identified from young \hii\ regions}
\authorrunning{A.Y. Yang, J.S. Urquhart, M.A. Thompson}
\maketitle
%
\section{Introduction}

One key question regarding massive star formation in the youngest \hii\ region relates to how accretion proceeds against the outward pressure therein \citep[e.g.,][]{Keto2006ApJ850K}, as massive stars reach the main sequence while still accreting \citep[e.g.,][]{Zinnecker2007ARAA45,Motte2018ARAA41M}. 
However, many details of the earliest stages of \hii\ regions are unclear.  
 Simple analytic models suggest that the \hii\ region can be created by either the  inner, ionized part of the inflowing material \citep{Keto2002ApJ980K,Keto2003ApJ} or the ionized photoevaporative outflow \citep{Hollenbach1994ApJ654H} fed by accretion \citep{Keto2007ApJK}. 
 The onset time for the development of a \hii\ region is found to be early in the \citet{McKee2003ApJM} and \citet{Peters2010ApJ711} turbulent core and ionization feedback models, but  the models of \citet{Hosokawa2009ApJ691} and \citet{Hosokawa2010ApJ721} for a bloated protostar suggest that this onset is later on.  
 After the birth of \hii\ regions, the subsequent expansion has been modeled as uniform spherical bubbles \citep{Spitzer1978ppimbookS}, or asymmetrical flows into outflow-driven cavities \citep{Peters2010ApJ711}, and the expansion rates predicted by different models could also be different \citep[e.g.,][]{Bisbas2015MNRAS4531324B}.  
Detailed observations toward the youngest \hii\ regions are crucial to investigate their initial development and constrain theoretical models \citep{thompson2015,Thompson2016mksconfE15T}.

The two youngest \hii\ region stages are commonly known as hyper-compact \hii\ (\hchii) and ultra-compact \hii\ (\uchii) regions \citep[e.g.,][]{Kurtz2005IAUS}. 
The youngest is the \hchii\  region with a typical physical size ($diam$) of $diam \rm \lesssim 0.05\,pc$, an electron density ($ n_{\rm e}$) of $ n_{\rm e} \gtrsim \rm 10^{5}\,cm^{-3}$, an emission measure (EM) of $\rm EM\gtrsim 10^{8}\,pc\,cm^{-6}$, and a radio recombination line (RRL) with  a {\bf line} width of   $\Delta V \gtrsim \rm 40\,km\,s^{-1}$ \citep{Kurtz2000prplconfK, Sewilo2004ApJ,Hoare2007prplconfH,Murphy2010MNRASa2}. 
The \uchii\  region is thought to be the next evolutionary stage after the \hchii\ region, with $diam\rm \lesssim 0.1\,pc$, $ n_{\rm e} \gtrsim \rm 10^{4}\,cm^{-3}$, $\rm EM\gtrsim 10^{7}\,pc\,cm^{-6}$, and $\Delta V\sim \rm 25-30\,km\,s^{-1}$ \citep[e.g.,][]{Wood1989ApJS, Afflerbach1996ApJS106,Hoare2007prplconfH}. 
The defining characteristics of these two stages (i.e., $diam$, $n_{\rm e}$, and EM) are somewhat arbitrary, as the evolution from \hchii\ regions to \uchii\ regions is thought to be continuous  \citep[e.g.,][]{Garay1999PASP1049G, Yang2019MNRAS4822681Y}. Compared to the hitherto discovered $\sim$ 600 \uchii\ regions \citep{Urquhart2007AA461, Urquhart2009AA501, Lumsden2013ApJS208, Urquhart2013MNRAS435, Cesaroni2015AA579A, Kalcheva2018AA615A103K,Djordjevic2019MNRAS4871057D}, only 16 \hchii\ regions have been identified in previous studies (summarized by \citealt{Yang2019MNRAS4822681Y} and references therein). It is not yet clear at what stage and how an \hchii\ region evolves into an \uchii\ region. Given the fact that the observed sizes of young \hii\ regions are found to vary with observing frequency   \citep{Panagia1975AA391P, Avalos2006ApJ641406A}, it has been suggested that the classical quantitative criteria for identifying \hchii\ regions should be modified   in order to consider the variations \citep{Yang2019MNRAS4822681Y}, which could lead to a better understanding of the intermediate object between an \hchii\ region and an \uchii\ region. 
However, to understand the relation between the two classes, and eventually to understand the early stages of newly formed massive stars, reliable properties toward a large sample of \hchii\ regions and \uchii\ regions are needed to be determined. 

Although young \hii\ regions around massive stars are heavily obscured by a thick cocoon of molecular gas, they can nevertheless be studied at radio wavelengths thanks to the ability of radio radiation to penetrate the dense molecular gas. 
Therefore, most studies of young \hii\ regions are based on radio continuum observations  \citep[e.g.,][]{Wood1989ApJS,Kurtz1994ApJS659K,vandertak2005AA947V, Gibb2007MNRAS246G}. 
The radio continuum spectrum of an \hii\ region with spectral index $\alpha$ ($S_{\nu}\propto\nu^{\alpha}$) varies from  $ +2$ (optically thick) at low frequency to $-0.1$ (optically thin) at high frequency. 
The turnover frequency between the optically thick and thin regimes for thermal bremsstrahlung is essentially a linear function of the electron density \citep{Mezger1967ApJ}. 
A younger \hii\ region with higher density will remain optically thick at higher frequencies. For instance, \uchii\ regions have a typical turnover frequency of $ \nu_{\rm t} \sim$ 5\,GHz, while \hchii\ regions have $ \nu_{\rm t}= 10  \rm \,to\, 100\rm\,GHz$ \citep[e.g.,][]{Beltran2007AA,Hoare2007prplconfH,Keto2008ApJ672,Zhang2014ApJ}. Therefore, young \hii\ regions  with spectra still rising in a higher frequency are potentially young and dense, which might correspond to an early stage of \uchii\ region or a stage connecting \uchii\ and \hchii\ regions.  

The physical properties of young \hii\ regions have been measured in several previous studies \citep[e.g.,][]{Wood1989ApJS, Murphy2010MNRASa2,Urquhart2013MNRAS435, Kalcheva2018AA615A103K,Medina2019AA627A175M}, either by a targeted multi-band observation on small samples of \uchii\ regions \citep[e.g.,][]{Murphy2010MNRASa2} or using single-band surveys assuming that the gas is optically thin to free-free emission \citep[e.g.,][]{Urquhart2013MNRAS435, Kalcheva2018AA615A103K}. The assumption that \hii\ regions are optically thin would give unreliable physical properties if the \hii\ region is actually optically thick at the observed frequency.  
Therefore, multi-band data taken over a large range of frequencies are crucial in order to  reliably determine the physical properties of  young \hii\ regions. 
  
 In this work, we present the results of multi-band observations with the Karl G. Jansky Very Large Array\footnote{The Karl G.~Jansky Very Large Array of the National Radio Astronomy Observatory: \url{https://science.nrao.edu/facilities/vla} \label{vla_footnote}} (VLA) in X-band (8--12\,GHz) and K-band (18--26\,GHz) of a sample of 114 young \hii\ regions. 
These sources were selected from a sample of \hii\ regions with rising spectra between  1.4\,GHz and 5\,GHz, that is, $\rm \alpha^{\rm 5\,GHz}_{\rm 1.4\,GHz}>0$ \citep{Yang2019MNRAS4822681Y}. Together with archival VLA data at 1.4\,GHz and 5\,GHz (see Sect.\,\ref{sect:sample_section} for details), we measure the spectral energy distribution (SED) between 1 and 26 GHz for each source in the sample, which covers both optically thick and thin portions of their radio spectra. We model every SED to find the best estimates for the physical properties. 

This paper is organized as follows: Section\,\ref{sect_obs} describes the details of the sample, observation, and data reduction. Section\,\ref{sec_results_analysis} presents and discusses the observational results, the modeled SEDs, and  the radio properties  of the
sources and their distributions. 
In Section\,\ref{sect_optically_thick_hii}, 
we discuss \hchii\ region candidates, plus a small sample of objects considered to be in an intermediate phase between \hchii\ and \uchii\ regions. 
We use our observations to derive the physical properties ($ n_{\rm e}$, $ diam$, EM, Lyman continuum flux) of \hii\ regions and compare our multi-band results to those estimated using the optically thin assumption. 
In Section\,\ref{sect:discussion} we discuss the relations and distribution of all of the \uchii\ and \hchii\ regions. We present a summary of this work and highlight our conclusions in Section\,\ref{sec_conclusion}.

\begin{table}
\caption[]{
Observed 114 rising spectra \hii\ regions. Columns, in order, show source name, flux density at 1.4\,GHz and 5\,GHz respectively (see \citealt{Yang2019MNRAS4822681Y} for details), heliocentric distance, and  bolometric luminosity, and the reference these values are drawn from. Uncertainties on the fluxes and distances are estimated to be 10\%, and those on  luminosity, 20\%. 
\label{tab_obser_list}
}
\centering
\footnotesize
\setlength{\tabcolsep}{2pt}
\begin{tabular}{lccccc}
\hline
\hline
Name  & $S_{1.4\,GHz}$ & $\rm S_{5\,GHz}$ & Dist & $L_{\rm bol}$ & [Ref.] \\
    &  ($\rm mJy$) & ($\rm mJy $) & (kpc) & ($L_{\odot}$) &    \\
\hline
& (2) & (3) & (4) & (5) & (6) \\
\hline
G010.3009$-$00.1477 & 426.2 & 631.4 & 3.5 & 5.2 & [1] \\
G010.4724$+$00.0275 & 31.3 & 38.4 & 8.5 & 5.7 & [1] \\
G010.6223$-$00.3788$\dagger$ & 327.6 & 483.3 & 2.4 &  5.7 &  [1] \\
G010.6234$-$00.3837 & 571.3 & 1952.2 & 5.0 & 5.7 & [1] \\ 
G010.9584$+$00.0221 & 47.9 & 196.0 & 2.9 & 4.0 & [1] \\
G027.9782$+$00.0789 & 89.3 & 124.0 & 4.8 & 4.2 & [2] \\
G028.2003$-$00.0494 &  $-$  & 161.0 & 6.1 & 5.1 & [1] \\
G028.2879$-$00.3641 & 410.9 & 552.8 & 11.6 & 5.9 & [1] \\
G028.6082$+$00.0185 & 168.2 & 210.2 & 7.4 & 5.0 & [1] \\
G029.9559$-$00.0168 & 1610.8 & 3116.2 & 5.2 & 5.7 & [1] \\
G049.3704$-$00.3012 & 252.9 & 414.4 & 5.4 & 5.1 & [3] \\
G048.6099$+$00.0270 & 56.5 & 131.2 & 9.8 & 5.1 & [4] \\
\hline

\end{tabular}
\tablefoot{
Only a small portion of the data is provided here, the full table is presented in Table \ref{tab_obser_list_appendix} and will be available in electronic form at the CDS. 
Source names appended with a $\dagger$ refers to the sources observed that could not be imaged. 
References: [1] \citet{Urquhart2018MNRAS4731059U}, [2] \citet{Cesaroni2015AA579A}, [3] \citet{Urquhart2013MNRAS435}, [4] \citet{Kalcheva2018AA615A103K}. 
}
\end{table}


\section{Observation}
\label{sect_obs}

\setlength{\tabcolsep}{2pt}
\begin{table}
\centering 
\caption{Summary of VLA observation parameters.}\label{tab_obsparms}
\begin{tabular}{lc}\hline \hline
 
Parameter              & \\
\hline
Project              & VLA18B-065 \\  
Frequency (GHz)      & X-band (8 -- 12) $\&$ K-band(18--26)   \\ 
Array configuration  &  C \\  
Observing mode       & continuum \\ 
Bandwidth per channel      & 128 MHz \\ 
No. Channels         &  30 $\&$ 60\\ 
Primary beam         & $\sim$ 4.2$\arcmin$ $\&$ $\sim$ 1.9$\arcmin$  \\ 
Synthesized beam     & $\sim$ {$2.0\arcsec$} $\times$ {$1.4\arcsec$} $\&$  $\sim$ {0.7\arcsec} $\times$ {$0.6\arcsec$}  \\
Observing dates      & 2019 Feb 07 $\&$ 2019 Feb 26 \\ 
Time on-source per source & $\sim$ 1\,min \\ 
No. Targets &  114  \\  
Total observing time & 2h $\&$ 2.5h   \\  
Flux density calibrator (Jy) & 3C286~(4.5\,Jy) \\ 
Phase calibrators (Jy)  & J1832-1035~(1.28\,Jy)\\
& J1851-0035~(1.10\,Jy)\\ 
&J1922+1530~(1.0\,Jy) \\ 
\hline
\end{tabular}
\end{table}

\subsection{Sample selection}
\label{sect:sample_section}

In \citet{Yang2019MNRAS4822681Y}, 
we constructed a parent sample of 534 objects with rising radio spectral indexes between 1.4\,GHz and 5\,GHz using three JVLA surveys, THOR \citep[The HI, OH, Recombination line survey of the Milky Way,][]{Bihr2016AA,Beuther2016AA595A32B}, MAGPIS \citep[The Multi-Array Galactic Plane Imaging Survey,][]{White2005AJ,Helfand2006AJ}, and CORNISH \citep[Coordinated Radio ``N'' Infrared Survey for High-mass star formation,][]{Hoare2012PASP,Purcell2013ApJS}. 
From an analysis of the combined radio, infrared, and submillimeter emission properties \citep{Yang2019MNRAS4822681Y}, we identified 120 young \hii\ regions from the parent sample. 
This sample not only recovers previously known \hchii\ regions, but also includes broad RRL objects with line widths of $\rm \Delta V > 40\,km\,s^{-1}$ and a number of \uchii\ regions with positive spectra  \citep{Yang2019MNRAS4822681Y}. 
 We observed 114 young \hii\ regions in X- and K-band data taken with the VLA. We use the data from archives and the literature for the four sources in the initial sample that have not been observed in the project, marked with a star in Tables \ref{tab_obser_list} and \ref{tab_classify_hii}. 
The final sample includes 118 young \hii\ regions.

The flux densities and angular diameters of the 118 observed sources are given in Table\,\ref{tab_obser_list}. The 1.4 and 5 GHz flux densities are taken from \citet{Yang2019MNRAS4822681Y} and references therein. 
The distances and bolometric luminosities are mainly drawn from the results reported in \citet{Urquhart2018MNRAS4731059U} \footnote{ This study by \citet{Urquhart2018MNRAS4731059U} is based on the ATLASGAL compact source catalog, which consists of $\sim$10\,000 clumps showing submillimeter wavelength emission from dust \citep{Contreras2013AA549A45C,Urquhart2014AA568A41U}.}, which includes 105 objects of the sample.
For the remaining 13 sources with no measurements in \citet{Urquhart2018MNRAS4731059U}, 
their distances and bolometric luminosities are taken from three studies, namely \citet{Cesaroni2015AA579A}, \citet{Urquhart2013MNRAS435}, and \citet{Kalcheva2018AA615A103K}. 
The kinematic distances were computed by fitting the radial velocity of each source to the Galactic rotation curve. The kinematic distances near/far ambiguity (KDA) for sources within the solar circle was resolved by CO emission line data and \hi\ absorption \citep[e.g.,][]{Urquhart2013MNRAS435, Cesaroni2015AA579A, Yang2016ApJS2236Y, Kalcheva2018AA615A103K} or using a combination of \hi\ analysis, maser parallax, and spectroscopic measurements \citep{Urquhart2018MNRAS4731059U}. 
The bolometric luminosity of the sample was taken from the same reference as the distance and was determined by integrating the SED from near-infrared to submillimeter wavelengths  \citep[e.g.,][]{Konig2017AA599A139K}.

\subsection{Observations and data reduction}

Observations of 114 young \hii\ regions were carried out using the VLA in C configuration. Instrument parameters used are shown in Table\,\ref{tab_obsparms}. The observations were made at X-band (8--12\,GHz) and K-band (18--26\,GHz), split into two subbands with 30 channels at X-band, and four subbands with 60 channels at K-band, each channel with a bandwidth of 128\,MHz, full stokes. The synthesized beams in C configuration at X-band and K-band are $\rm \sim 1.8\arcsec $ and $\rm \sim 0.7\arcsec $, and the FWHM primary beams sizes are $\rm \sim 4.2\arcmin $ and $\rm \sim 2\arcmin $, respectively. The typical on-source time for each target is about one minute and the total observation time is 4.5 hours. The phase calibrators (J1832-1035, J1851-0035, and J1922+1530) were observed every half hour at X-band and every 12 minutes at K-band to correct the amplitude and phase of the interferometer data by atmospheric and instrumental effects. The pointing corrections at the high-frequency K-band were determined by observing the nearby phase calibrators in interferometric pointing mode. The absolute flux density scale at X-band and K-band was calibrated by comparing the observations of the standard flux density scale calibrator J1331+305 (3C286) with its models provided by the NRAO.

Standard calibration and data reduction were performed using the {Common Astronomy Software Applications package}  \citep[CASA,][]{McMullin2007ASPC376127M}. 
Raw VLA data were calibrated and reduced by running the CASA pipeline. We discarded the first 3 s of data of every scan for calibrators to exclude the antenna settling time. 
Flux and phase calibrator data were carefully examined to ensure high-quality data.
A calibration table was produced and applied to all targeted data. Each target was inspected by eye to flag  bad data such as phase scatters, errant amplitudes, system-temperature spikes, which resulted in a mean on-source integration time of $\sim$ 50\,s for each source.   

Images were constructed using the default Briggs robust parameter of zero, which provides a good trade-off between the low thermal noise of natural weighting and the high resolution of uniform weighting. 
Because of short on-source time ($\sim$50 s), we adopted to widest possible frequency ranges  for each image to do the {\tt clean} task in CASA. 
In order to measure flux density at different frequencies, we produced multi-band images at X-band and K-band. At X-band, three images were produced at central frequencies of 9\,GHz (8--10\,GHz), 10\,GHz (8--12\,GHz), and 11\,GHz (10--12\,GHz). Also, at K-band, three images were produced at central frequencies of 20\,GHz (18--22\,GHz),  22\,GHz (18--26\,GHz), and 24\,GHz (22--26\,GHz). 
The final beam size of images at the central frequency of X-band, namely 10 GHz, and at the central frequency of K-band, that is 22 GHz, are $\sim 2.1\arcsec \times 1.4\arcsec$ and $\sim \rm 0.7\arcsec \times 0.6\arcsec$, respectively. Sources with $\theta < 1.8\arcsec$ (X-band) and $\theta < 0.8\arcsec$ (K-band) are considered to be unresolved. Sources with angular size $\theta > 1.8\arcsec$ (X-band) and $\theta > 0.8\arcsec$ (K-band) are considered to be resolved and the deconvolved sizes are given in Table\,\ref{summary_hii_obsparam}.

\section{Results and analysis}
\label{sec_results_analysis}

  \begin{figure*}
 \centering
    \includegraphics[width = 0.3\textwidth, trim= 20 10 30 30]{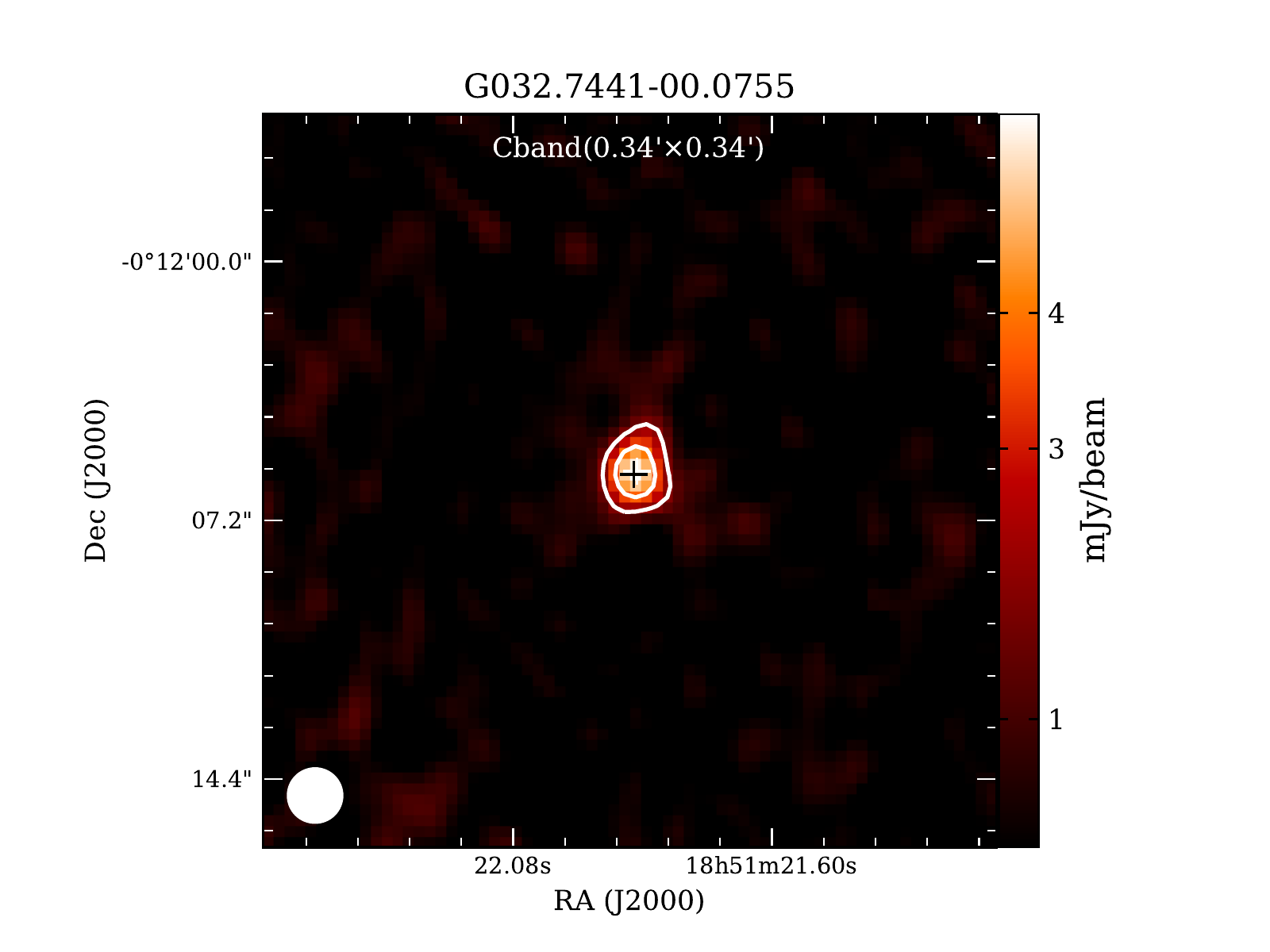}
    \includegraphics[width = 0.3\textwidth, trim= 20 10 30 30]{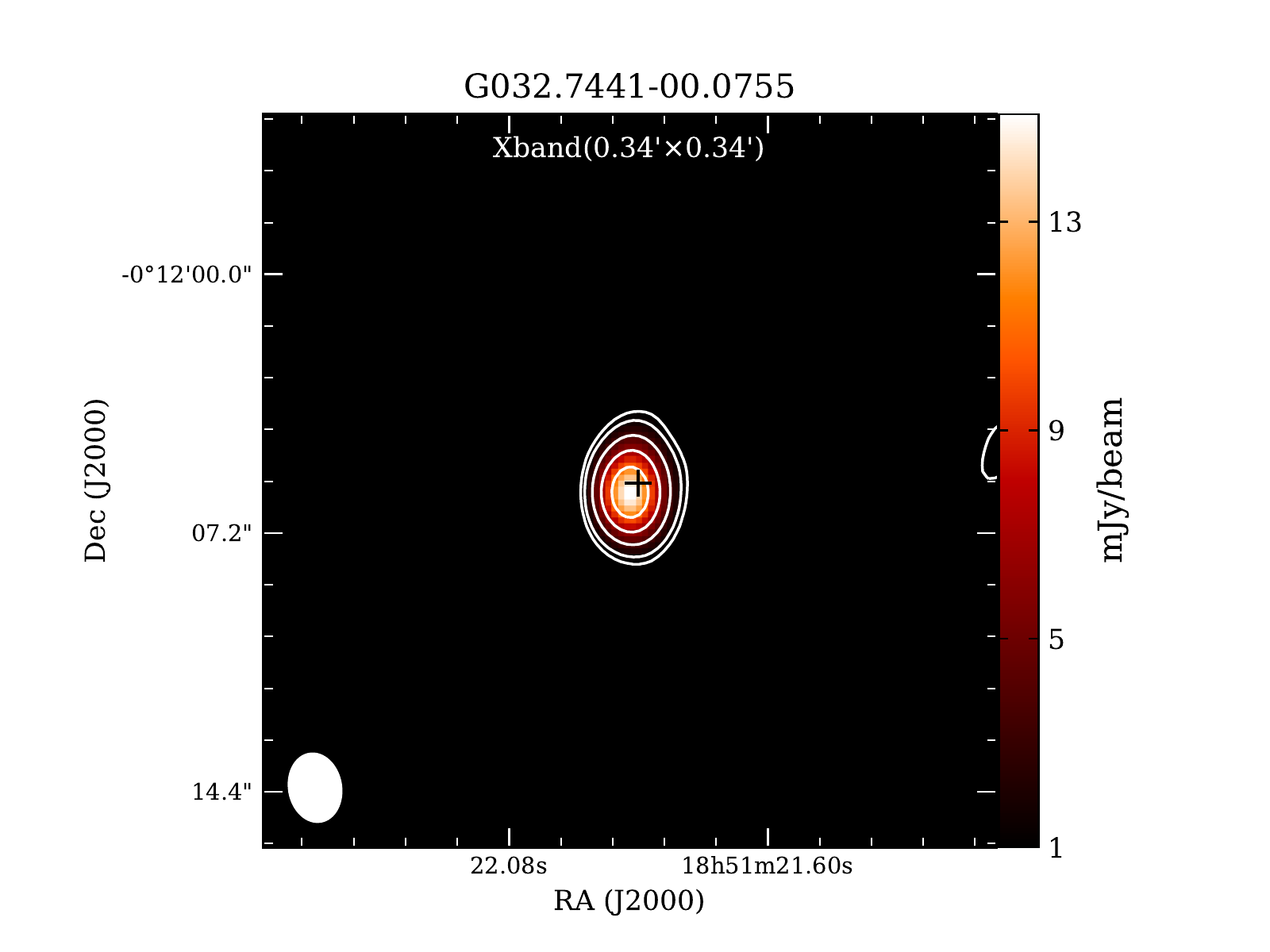}
    \includegraphics[width = 0.3\textwidth, trim= 20 10 30 30]{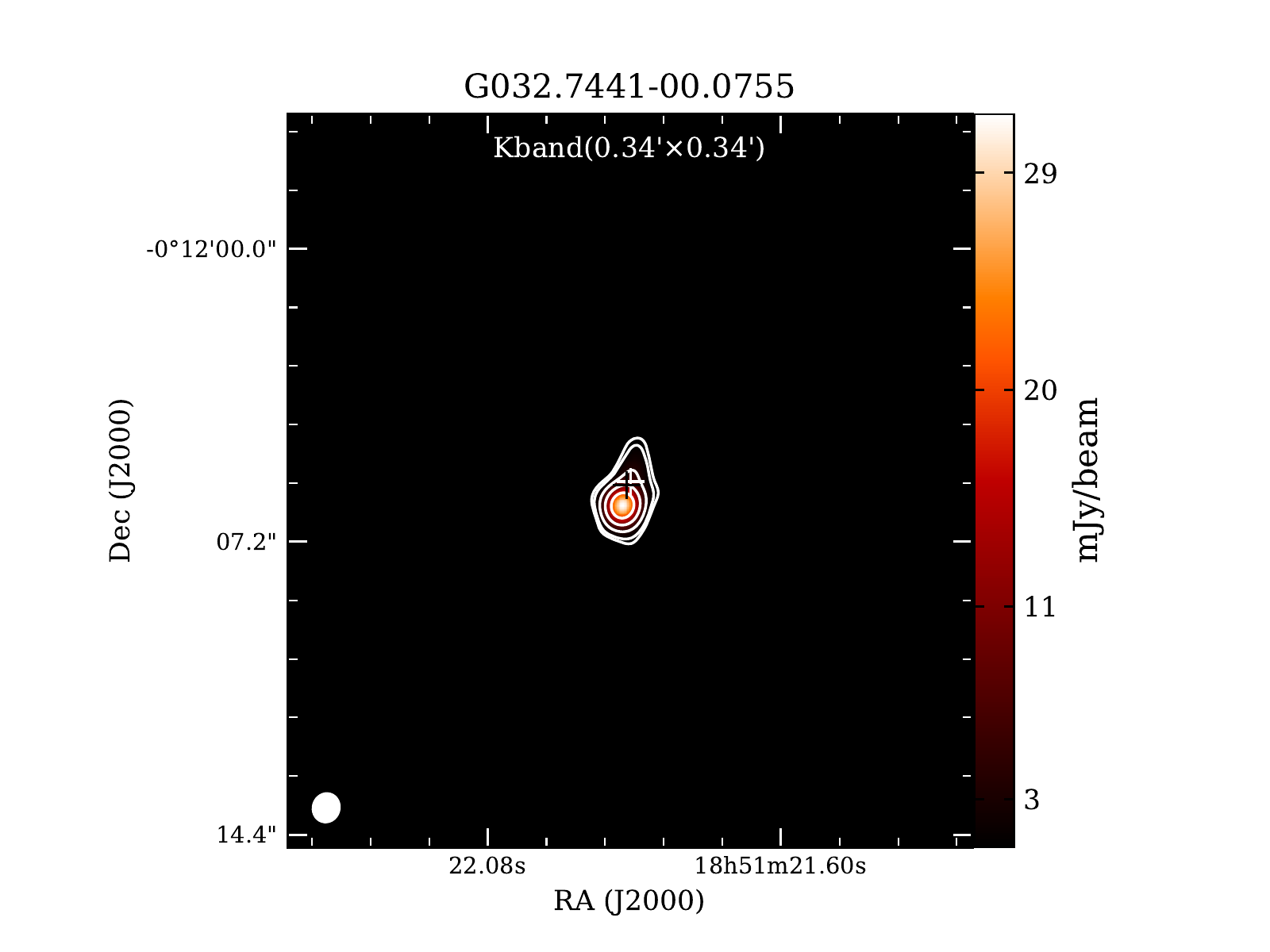}  \\
  \includegraphics[width = 0.3\textwidth, trim= 20 10 30 30]{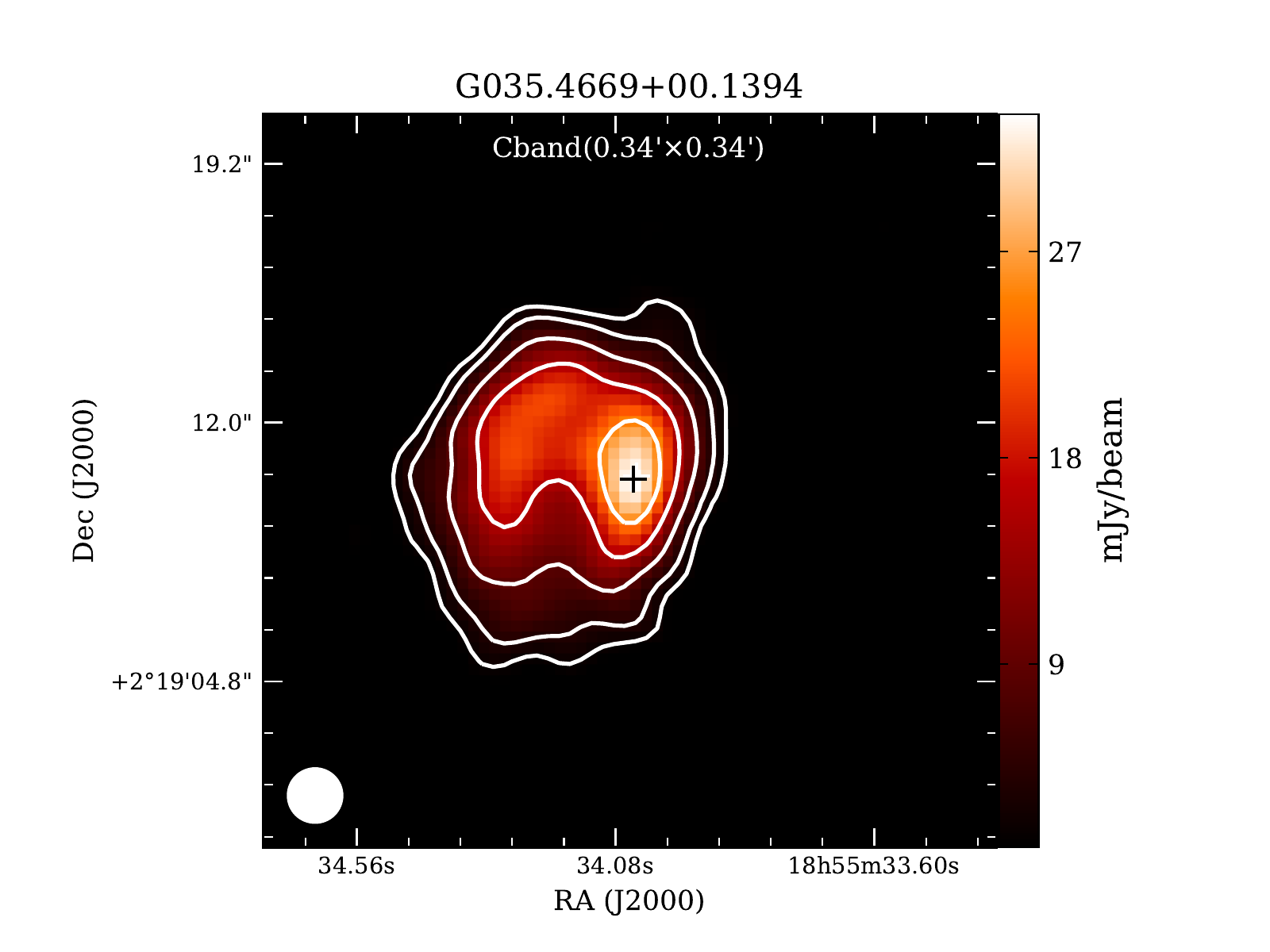} 
    \includegraphics[width = 0.3\textwidth, trim= 20 10 30 30]{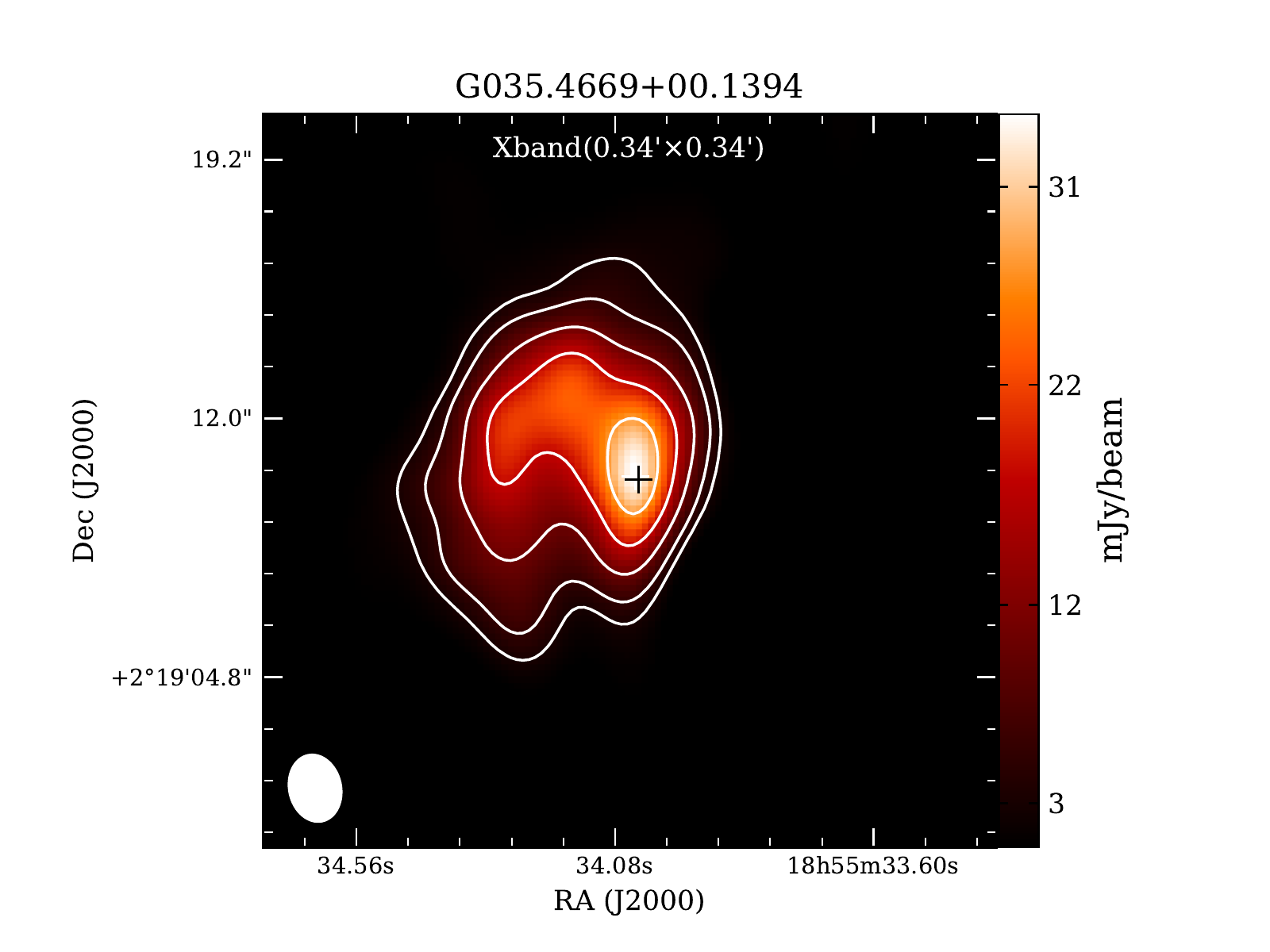}
    \includegraphics[width = 0.3\textwidth, trim= 20 10 30 30]{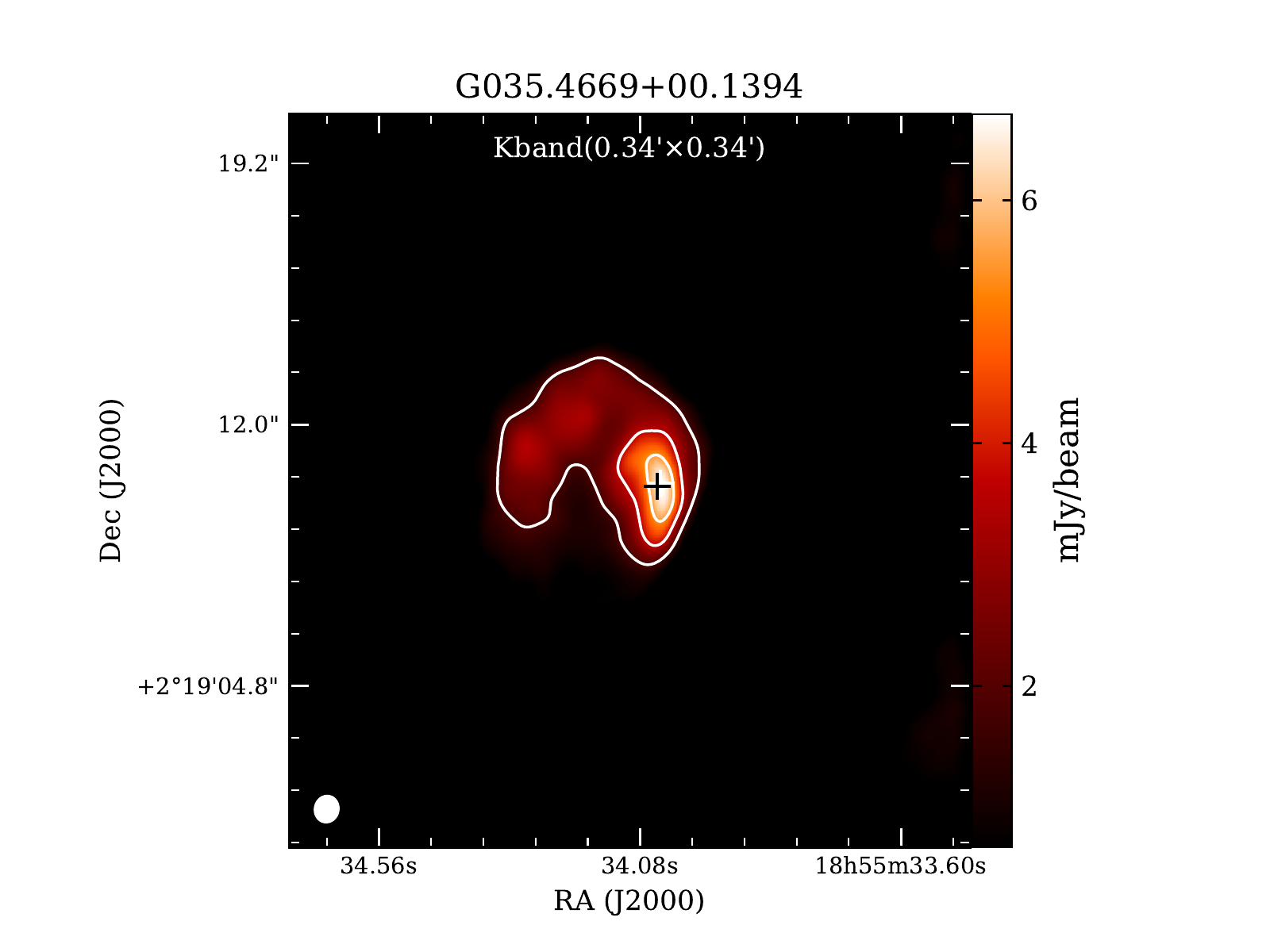}  \\
  \includegraphics[width = 0.3\textwidth, trim= 20 10 30 30]{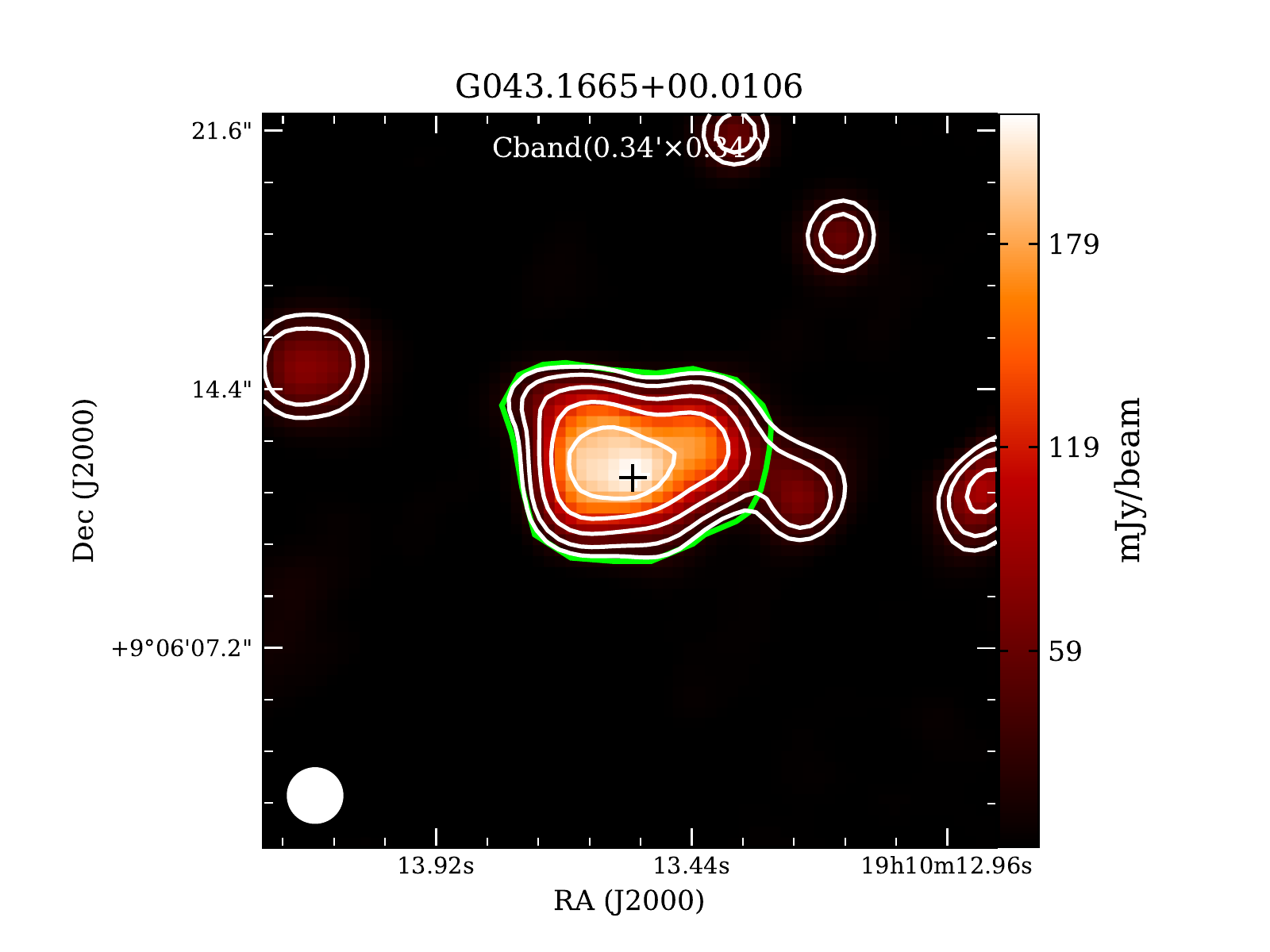} 
    \includegraphics[width = 0.3\textwidth, trim= 20 10 30 30]{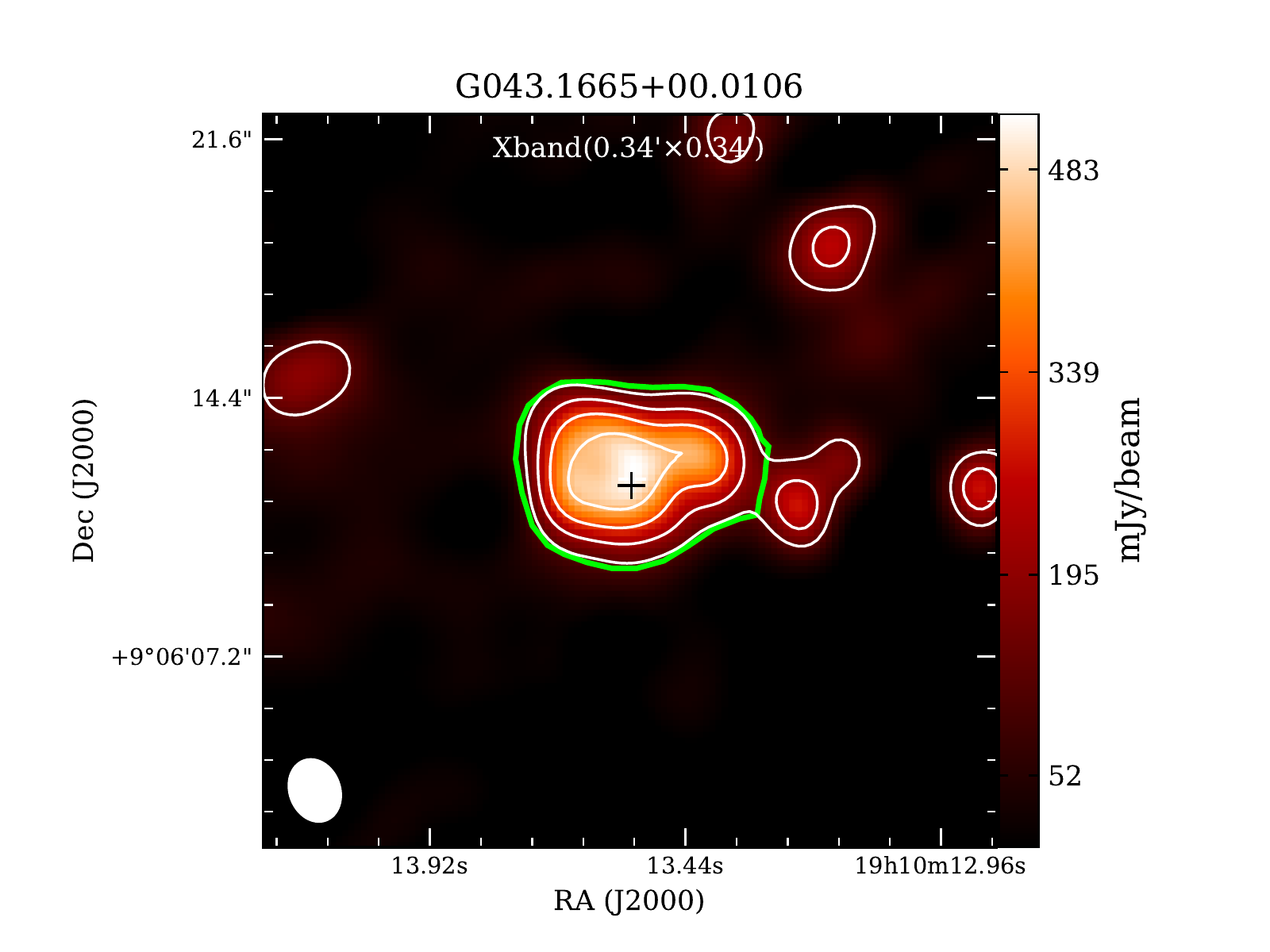}
    \includegraphics[width = 0.3\textwidth, trim= 20 10 30 30]{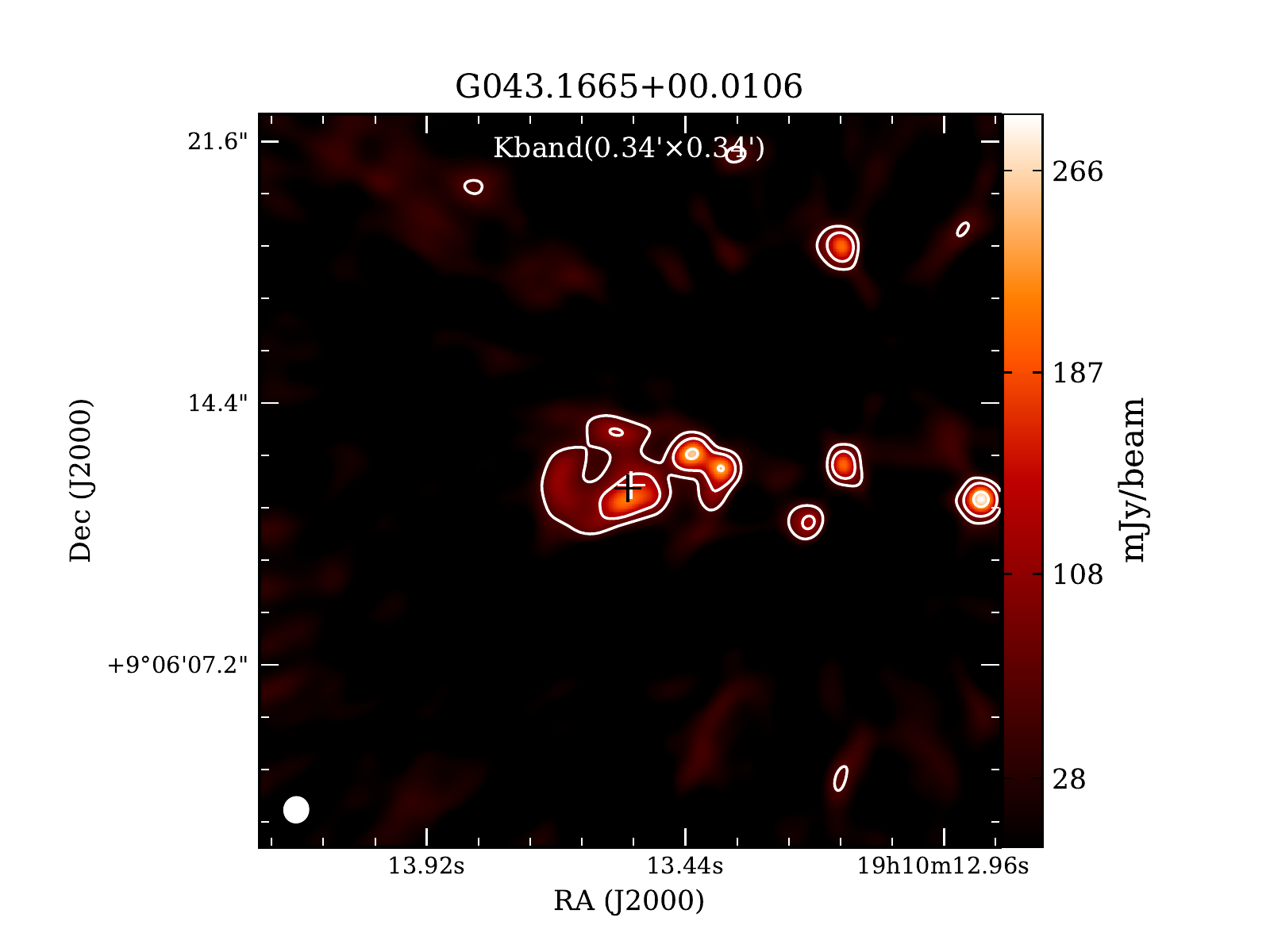}  \\    
  \caption{Example images of three radio sources at C-band (left-column), X-band (middle-column), and K-band (right-column). The position of the \hii\ region is marked with a plus. In the upper, middle, and lower rows, we show the maps for the compact \hii\ region G032.7441$-$00.0755, the extended \hii\ region G035.4669$+$00.1394, and the \hii\ region G043.1665$+$00.0106 located in a cluster (see Sect.\,\ref{sect_obs_results}), respectively. 
 C-band images are from the CORNISH survey. The white contour levels of each image are equally spaced by 5$\sigma$ and start at a level of 5$\sigma$. The green outline shown in the lower row shows the polygon that was manually drawn around the \hii\ region located in a cluster. 
  The image size and beam size are shown in the upper-middle and lower-left of each image. 
  The C-band, X-band, and K-band images for the whole sample are shown in Appendix Figure\,\ref{summary_sed_multiband_images}.
  }
 \label{example_sources}
 \end{figure*}
 
\subsection{Observational results}
 \label{sect_obs_results}
\begin{table*}
\setlength{\tabcolsep}{0.6pt}
\caption[]{ 
Observational results of 112 young \hii\ regions at X-band (8--12\,GHz) and K-band (18--26\,GHz). Columns: (1) Source name; (2) and (3) peak flux density and local RMS at X-band; (4-6) flux density at 9\,GHz, 10\,GHz and 11\,GHz, respectively; (7) deconvolved source size at X-band; (8) and (9) peak flux density and RMS at K-band;
(10-12) flux density at 20\,GHz, 22\,GHz, and 24\,GHz, respectively; (13) deconvolved source size at K-band. The uncertainties in the flux measurements are estimated to be $10\%$.
\label{summary_hii_obsparam}}
\centering
\footnotesize
\begin{tabular}{l.....c.....c}
\hline
\hline
Source name & \multicolumn{1}{c}{$ S_{\rm Peak}(X)$} & \multicolumn{1}{c}{$\sigma$(X)} & \multicolumn{1}{c}{$S_{\rm 9\,GHz}$} &\multicolumn{1}{c}{$S_{\rm 10\,GHz}$} & \multicolumn{1}{c}{$ S_{\rm 11\,GHz}$} & \multicolumn{1}{c}{$ {\theta}_{\rm s}$(X)} & \multicolumn{1}{c}{$S_{\rm Peak}$(K)}  & \multicolumn{1}{c}{$\sigma$(K)} & \multicolumn{1}{c}{$ S_{\rm 20\,GHz}$} & \multicolumn{1}{c}{$S_{\rm 22\,GHz}$} & \multicolumn{1}{c}{$ S_{\rm 24\,GHz}$}    & \multicolumn{1}{c}{${\theta}_{\rm s}$(K)}  \\ 
  & \multicolumn{1}{c}{$\rm (mJy/beam)$}& \multicolumn{1}{c}{$\rm (mJy)$}  & \multicolumn{1}{c}{$\rm (mJy)$} & \multicolumn{1}{c}{$\rm (mJy)$} & \multicolumn{1}{c}{$\rm (mJy)$}   
 & \multicolumn{1}{c}{$\rm ({\arcsec\times \arcsec})$}  
 & \multicolumn{1}{c}{$\rm (mJy/beam)$} & \multicolumn{1}{c}{$\rm (mJy)$} & \multicolumn{1}{c}{$\rm (mJy)$} & \multicolumn{1}{c}{$\rm (mJy)$} & \multicolumn{1}{c}{$\rm (mJy)$} 
 & \multicolumn{1}{c}{$\rm {(\arcsec\times \arcsec)}$}  \\ 

\multicolumn{1}{c}{(1)}  & \multicolumn{1}{c}{(2)}  &  \multicolumn{1}{c}{(3)} &
\multicolumn{1}{c}{(4)} & \multicolumn{1}{c}{(5)} & \multicolumn{1}{c}{(6)}  & \multicolumn{1}{c}{(7)} & \multicolumn{1}{c}{(8)} & \multicolumn{1}{c}{(9)} & \multicolumn{1}{c}{(10)} & \multicolumn{1}{c}{(11)} & \multicolumn{1}{c}{(12)} & \multicolumn{1}{c}{(13)} \\ 
\hline
G010.3009$-$00.1477$\oplus$ &    88.1   &    3.1  &    700.3   &    686.7   &     661.4   &      6.8$\times$6.6  &     15.9  &     0.7   &    433.4  &     419.6  & 392.6  &      6.5$\times$6.4   \\ 
 G010.4724$+$00.0275          &    82.5   &    1.6  &    100.7   &    105.9   &     115.4   &      1.4$\times$0.4  &     66.9  &     0.8   &     156.8  &     159.9  &     172.0  &      1.6$\times$0.5   \\ 
  G010.6234$-$00.3837$\oplus$ &    1099.9 &    7.9  &    3071.4  &    3072.1  &     3314.8  &      4.2$\times$3.8  &     572.3 &     14.6  &     2884.8 &     2857.2 &     2851.5 &      3.1$\times$3.0   \\ 
  G010.9584$+$00.0221          &    186.3  &    1.5  &    258.3   &    256.2   &     265.0   &      1.2$\times$0.9  &     91.3  &     1.2   &     210.7  &     202.6  &     200.6  &      1.0$\times$0.7   \\ 
   G011.0328$+$00.0274  &    3.9    &    0.2  &    4.8     &    4.3     &     4.1     &      1.3$\times$0.9  &     1.7   &     0.1   &     3.4    &     2.9    &     2.8    &      0.9$\times$0.4   \\ 
  G011.1104$-$00.3985$\oplus$ &    70.5   &    0.7  &    350.4   &    334.8   &     327.7   &      9.5$\times$9.4  &     15.6  &     0.4   &     136.1  &     123.3  &     126.1  &      2.2$\times$1.7   \\ 
  G011.1712$-$00.0662$\oplus$          &    4.1    &    0.1  &    95.1    &    92.7    &     100.1   &      11.9$\times$8.6  &     0.6   &     0.1   &     $-$    &     $-$    &     $-$    &     $-$              \\ 
\multicolumn{1}{c}{$\cdots$}  &    \multicolumn{1}{c}{$\cdots$}  &    \multicolumn{1}{c}{$\cdots$}  &   \multicolumn{1}{c}{$\cdots$}  \\
    G011.9368$-$00.6158$\oplus$ &    306.7  &    2.0  &    1116.4  &    1083.5  &     1098.3  &      3.4$\times$3.2  &     76.4  &     2.0   &    656.1  &     652.4  &    629.6 &      2.8$\times$1.8   \\ 
    G011.9446$-$00.0369$\oplus$ &    85.7   &    2.0  &   709.6   &    691.4   &    724.6   &      6.3$\times$4.7  &     20.0  &     0.6   & 307.2  &     291.6  &     289.9  & 4.3$\times$2.1   \\ 
   G012.1988$-$00.0345  &    29.6   &    0.4  &    66.0    &    64.8    &     63.9    &      2.0$\times$1.9  &     6.5   &     0.2   &     59.6   &     54.7   &     55.9   &      2.0$\times$2.0   \\ 
  G012.2081$-$00.1019          &    88.0   &    0.8  &    212.5   &    209.3   &     206.5   &      2.3$\times$1.9  &     27.2  &     0.6   &     159.0  &     142.1  &     140.8  &      2.0$\times$1.2   \\ 
\hline
\hline
\end{tabular}
\tablefoot{
 The  `-' symbol means no measurement is available. 
$\oplus$ indicates the sources are extended and their K-band flux densities should be considered to be lower limits. 
Only a small portion of the data is provided here, the full table is shown in Table \ref{summary_hii_obsparam_appendix} and will be available in electronic form at the CDS. 
$\dagger$ refers to the added 5 \uchii\ regions with rising spectra between C and X band, see Sect\,\ref{sect_obs_results}.
}
\end{table*}

  \begin{figure*}
 \centering
 \begin{tabular}{ccc}
    \includegraphics[width = 0.3\textwidth]{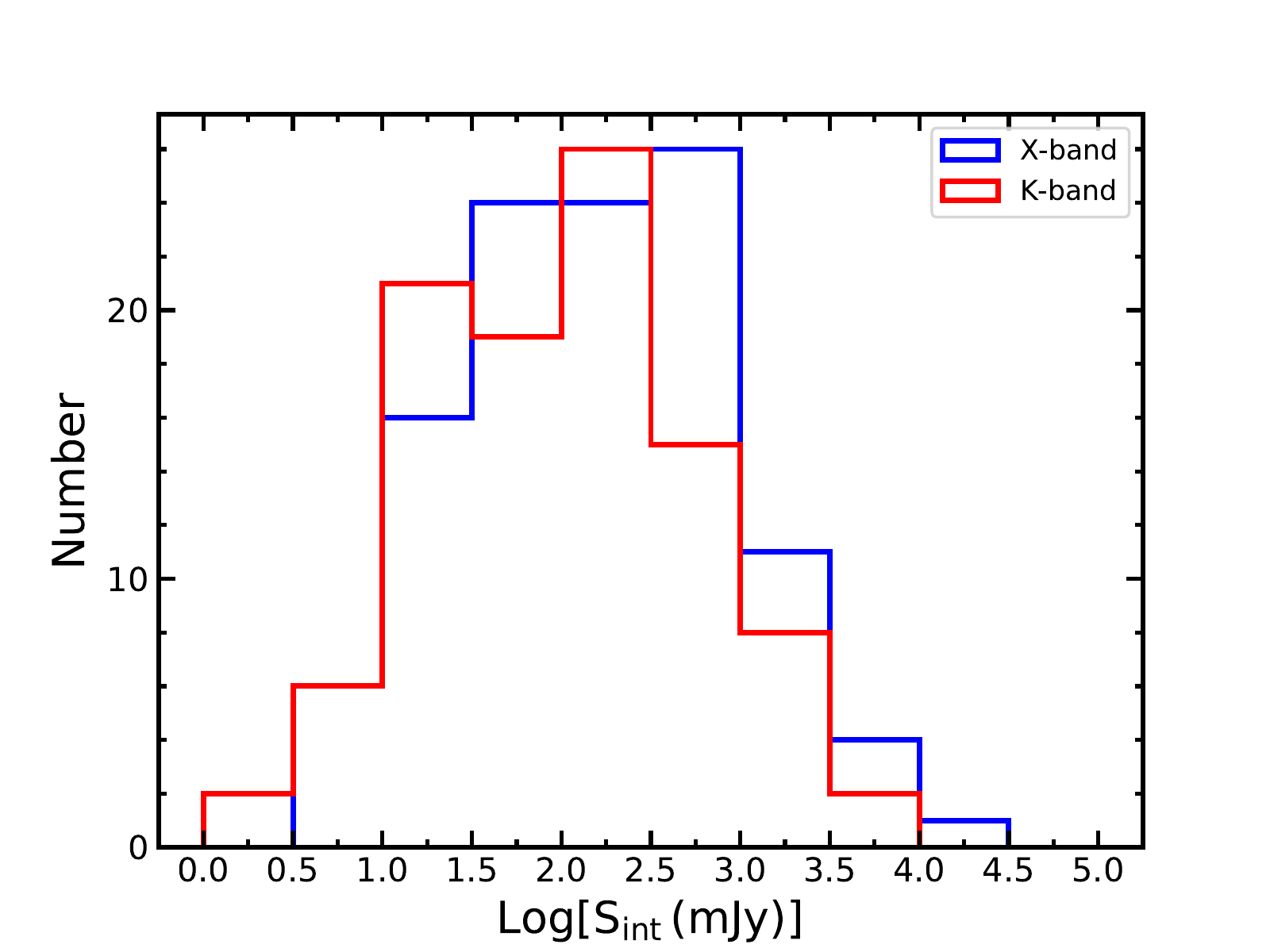} & 
    \includegraphics[width = 0.3\textwidth]{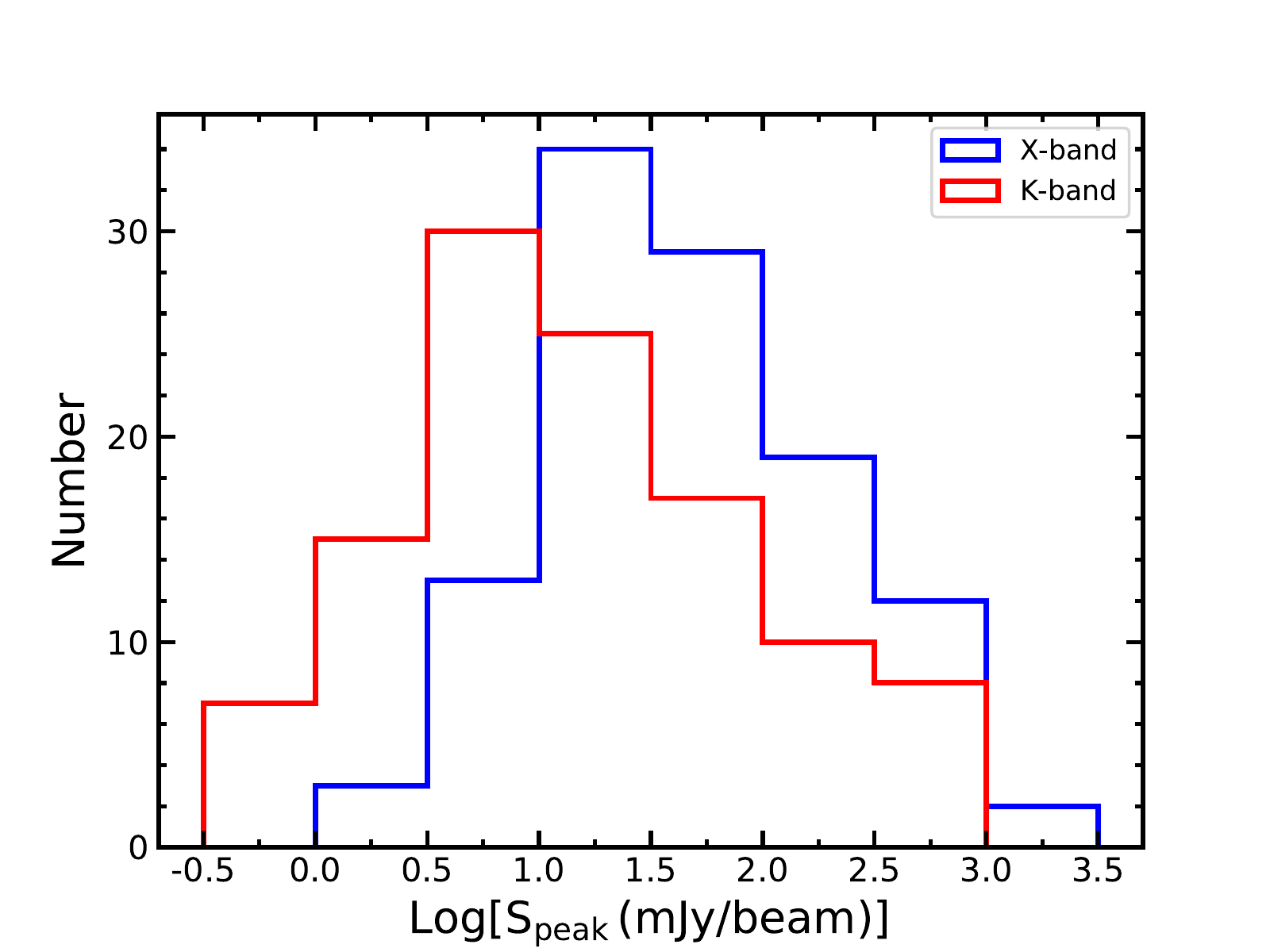} &
  \includegraphics[width = 0.3\textwidth]{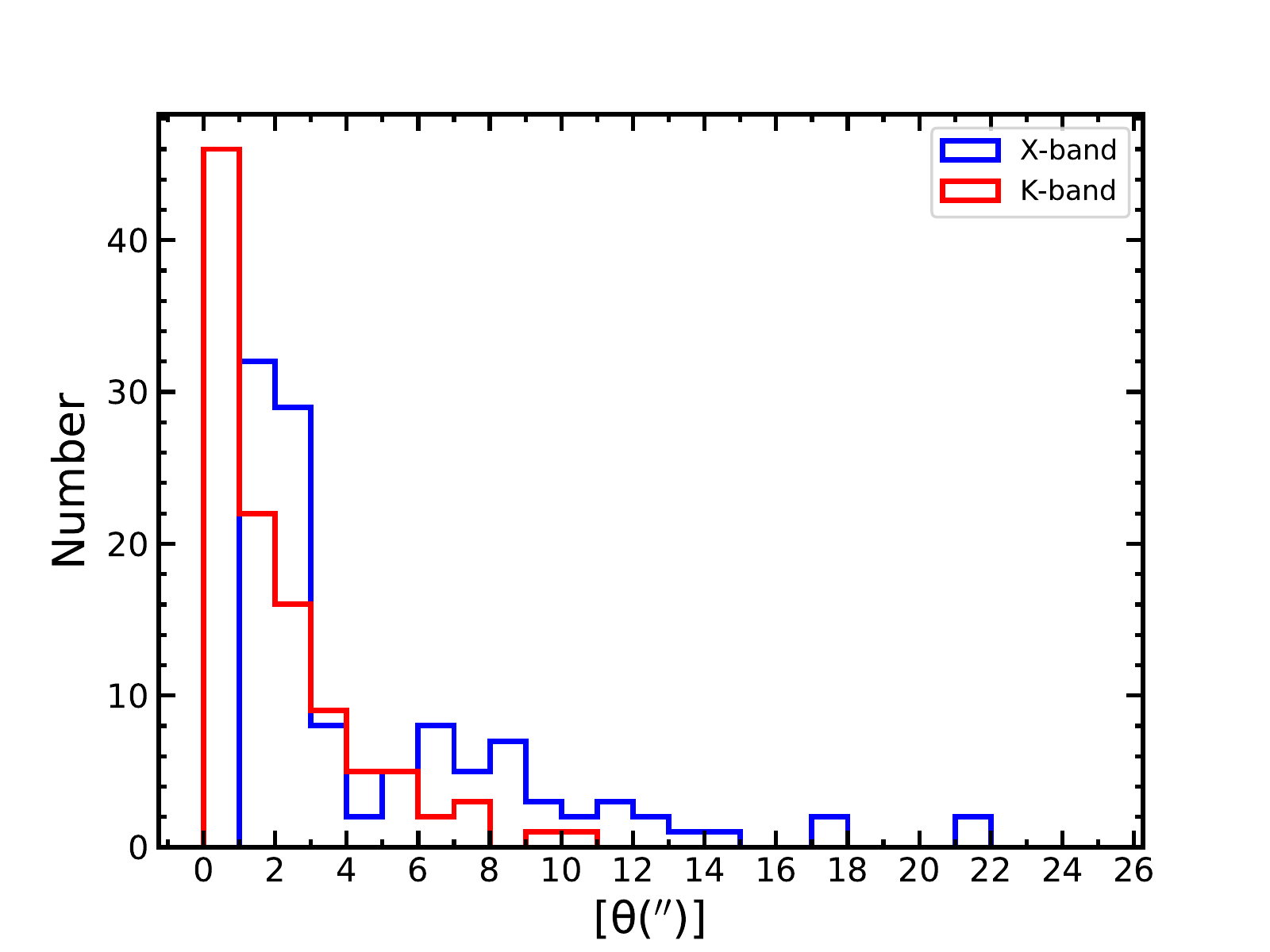} \\
 \end{tabular}
  \caption{Distributions of observation results such as integrated flux density $ S_{\rm int}$, peak flux density $ S_{\rm peak}$, and angular size $\rm \theta$, of 116 young \hii\ regions at X-band (blue solid line) and K-band (red solid line). The bin sizes are 0.5\,dex, 0.5\,dex, and $1\arcsec$ for $ S_{\rm int}$, $ S_{\rm peak}$, and $\rm \theta$, respectively.}
 \label{fig:distr_obs_results}
 \end{figure*}

In Fig.\,\ref{example_sources}, we present images of three sources that show the typical  variation in emission structure observed in our sample. 
The contour levels shown in these images were determined using a dynamic range power-law fitting scheme to meaningfully represent both high and low dynamic range images \citep[e.g.,][]{Thompson2006AA,Urquhart2009AA501,Yang2018ApJS2353Y}. This has been  slightly altered from the scheme described  by \citet{Thompson2006AA} and can be described as the following relationship $D = 5\times N_{\rm i} + 5$, where $D$ is the dynamic range of the map (defined as the ratio between the peak brightness and the $1\sigma$ RMS noise), $N$ is the number of contour lines, and `i' is the contour power-law index. Here, the minimum power-law index was one, which resulted in linearly spaced contours starting at $5\sigma$ and increasing in steps of $5\sigma$. The starting contour level we adopted for each target is variable, ranging from $5\sigma$ to $7\sigma$ according to the RMS level of each image. The RMS noise level ($\sigma$) of each image was determined using the standard deviation ($\rm STDEV= 1.4826 \times MADFM$), where MADFM is the median absolute deviation from the median ($\rm MADFM = median (\left |X_{i} - median(X)  \right |)$, where X is one element in the data set), in order to reduce the effects of outliers on noise measurement \citep[e.g.,][]{Purcell2013ApJS}. 
The short on-source integration time of the target observation ($\rm \sim 50\,s$) could lead to a rather high RMS level on the observed field for some sources located in complex star formation regions. 
 
The compact sources in the sample are directly fitted by 2D Gaussian models using the {\tt imfit} task in CASA (see upper panels of Fig.\,\ref{example_sources}). The resolved  \uchii\ regions are classified into a variety of morphologies ranging from spherical to irregular \citep[e.g.][]{Wood1989ApJS, Urquhart2007AA461, Urquhart2009AA501, Urquhart2013MNRAS435,Purcell2013ApJS}. 
The properties of the extended sources (see the middle row of Fig.\,\ref{example_sources}) and \uchii\ regions located within a cluster (see the lower row of Fig.\,\ref{example_sources}) are determined from the flux enclosed within a polygon fitted around the emission profile of the source; this is determined by the noise level for an extended source manually fitted around the emission for a cluster source, which follows the same strategy used in the construction of the CORNISH survey catalog \citep{Purcell2013ApJS}. 
The observational results of the extended sources or cluster sources in the sample such as flux density (defined as the difference between the aperture summed flux and background flux density divided by the beam-area) and angular diameter (defined as intensity-weighted diameter), as well as their uncertainties, can be measured by aperture photometry (for details of the aperture photometry method that we used see Sect.\, 5.3.2 of \citealt{Purcell2013ApJS}).

Analysis of the poor-quality X-band and K-band data for seven young \hii\ regions (marked by $\dagger$ in Table\,\ref{tab_obser_list}) revealed that their images are too confused to obtain reliable results and so these have been excluded.  
We also add five sources identified as \uchii\ regions in CORNISH by \citet{Purcell2013ApJS} that are located within our fields and have rising spectra between C-band and X band in this work. 
Thus, the final observed sample consists of 112 \hii\ regions. 
In Table\,\ref{summary_hii_obsparam}, we give the observational results and the derived properties for all of these sources.

\setlength{\tabcolsep}{4pt}
\begin{table}
\caption{Summary of observational results and the derived physical parameters for 116 young \hii\ regions. In Columns (2-5) we give the minimum, maximum, mean$\pm$ standard deviation, and median values, respectively, of each parameter. 
}
\begin{tabular}{lcccc}
\hline
\hline
Parameter                                       &  $x_{min}$ & $x_{max}$   & $x_{mean}\pm x_{std}$ & $x_{med}$ \\
\hline
\multicolumn{5}{c}{observational properties at X-band}     \\
\hline
$\log[S_{\rm int}\,(\rm mJy)]$                &     0.58      & 4.20            &  2.20$\pm$0.76               & 2.20           \\
$\log[S_{\rm peak}\,(\rm mJy/beam)]$  &      0.40      & 3.04            &  1.71$\pm$0.66               & 1.68           \\
$\rm Angular\,size\,\theta \arcsec$                        &    1.68      & 21.25           & 4.54$\pm$3.76               &  2.4    \\
\hline
\multicolumn{5}{c}{observational properties at K-band}     \\
\hline
$\log[S_{\rm int}\,(\rm mJy)]$          &     0.33       & 3.80            & $1.99\pm0.74$               & 2.0           \\
$\log[S_{\rm peak}\,(\rm mJy/beam)]$  &     -0.40      & 2.88            &  1.23$\pm$0.79              & 1.68          \\
$\rm Angular\,size\, \theta \arcsec$ &    0.6       & 10.24           & $2.12\pm1.95$               & 1.34          \\
\hline
\multicolumn{5}{c}{Physical properties}     \\
\hline
${\log}[n_{\rm e}\rm(cm^{-3})]$                   &   3.14            &  5.65       &  $4.20\pm 0.05$               &      4.10            \\
$diam\,\rm [pc]$                              &   0.004          & 0.81        &  $0.14\pm 0.01$              &     0.08             \\
$\log[\rm EM\,(pc\,cm^{-6})]$           &   5.96            & 9.05        &  7.28$\pm$ 0.06              &     7.09             \\
${\log}[N_{\rm Ly}\rm\,(s^{-1})]$              &    45.37          & 49.82      &  47.81$\pm$ 0.09          &    47.95           \\
$ \nu_{\rm t}\,\rm [GHz]$            &   0.56              &  16.67     &   $3.29 \pm 0.31$             &   1.95              \\
Dust\,absorption\,fraction\,$f_{\rm d}$          &   0.14             & 0.99      &   0.67 $\pm$ 0.03             &   0.75            \\

\hline
\hline
\end{tabular}
\label{tab_summary_param}
\end{table} 
 
In Table\,\ref{tab_summary_param}, we provide a statistical summary of the observed and derived properties for each source at both the X-band and K-band. 
We estimate the uncertainties on the flux density and angular size at both frequencies to typically be $\sim$10\% by considering the calibration errors and errors of the measurement method  \citep[e.g.,][]{Murphy2010MNRASa2,Sanchez_Monge2013AA21S}. In Fig.\,\ref{fig:distr_obs_results}, we show the distributions of the derived parameters. The distributions of integrated flux $S_{\rm int}$ and peak flux density $S_{\rm peak}$ in the left and middle panels of Fig.\,\ref{fig:distr_obs_results} are similar at X-band and K-band, which suggests that the majority of sources are optically thin between these frequencies.  
The X-band shows a slightly higher peak value of $S_{\rm int}$ and $S_{\rm peak}$ than K-band, some of which may be due to the majority of sources having a turnover frequency below X-band and the fluxes start to decrease afterwards following the power-law of $S_{\nu}\propto \nu^{-0.1}$ at the optically thin regime of an \hii\ region. 
Some sources may be  due to the larger beam at X-band collecting more flux. The X-band has a larger field of view and is more sensitive to larger angular scales than K-band, which is why a larger proportion of the sources detected at X-band are more extended in the right panel of Fig.\,\ref{fig:distr_obs_results}.

\subsection{Radio properties from the SED models}
\label{sect:sed_fitting}

The physical characteristics of \hii\ regions (e.g., $\rm EM$, $ n_{\rm e}$, Lyman-continuum flux $N_{\rm Ly}$) can be estimated by the observed angular sizes and flux densities at a given frequency, 
assuming that the continuum emission comes from a homogeneous, optically thin ionized gas \citep[e.g.,][]{Urquhart2013MNRAS435,Kalcheva2018AA615A103K}. However, one should keep in mind that the physical properties of young \hii\ regions might be underestimated or overestimated by using a single frequency observation for two reasons: (i) The young \hii\ region might be optically thick at the observed frequency \citep[e.g.,][]{Cesaroni2015AA579A}; and (ii) the apparent angular size depends on the observing frequency \citep[e.g.,][]{Panagia1975AA391P,Avalos2006ApJ641406A,Yang2019MNRAS4822681Y}. Therefore, to determine the properties of young \hii\ regions, it is essential to know their spectral energy distribution (SED) over a wide frequency range that covers both optically thick and thin emission \citep[e.g.,][]{Murphy2010MNRASa2}.

  \begin{figure}
 \centering
 
    \includegraphics[width = 0.45\textwidth]{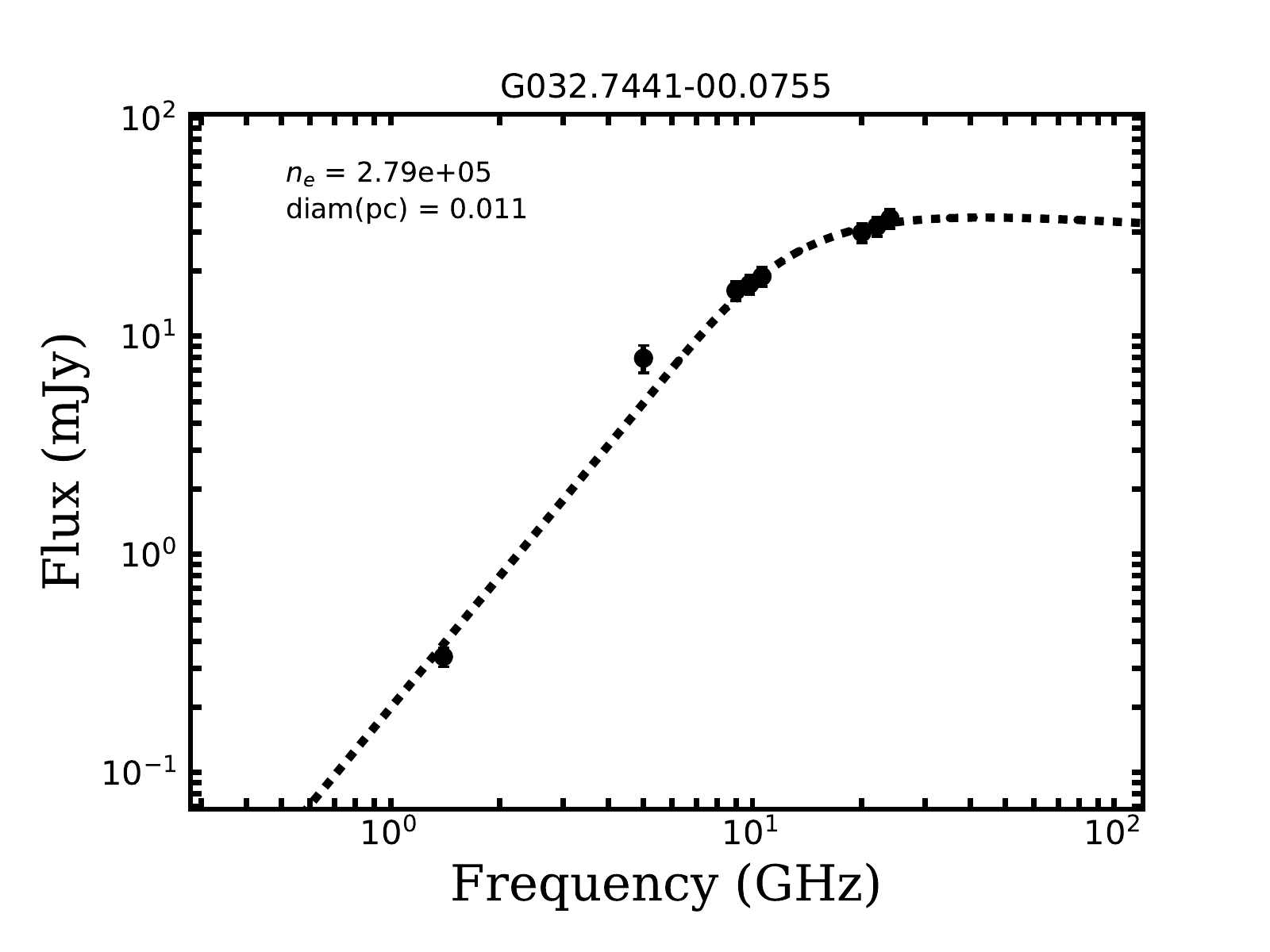}   
    \includegraphics[width = 0.45\textwidth]{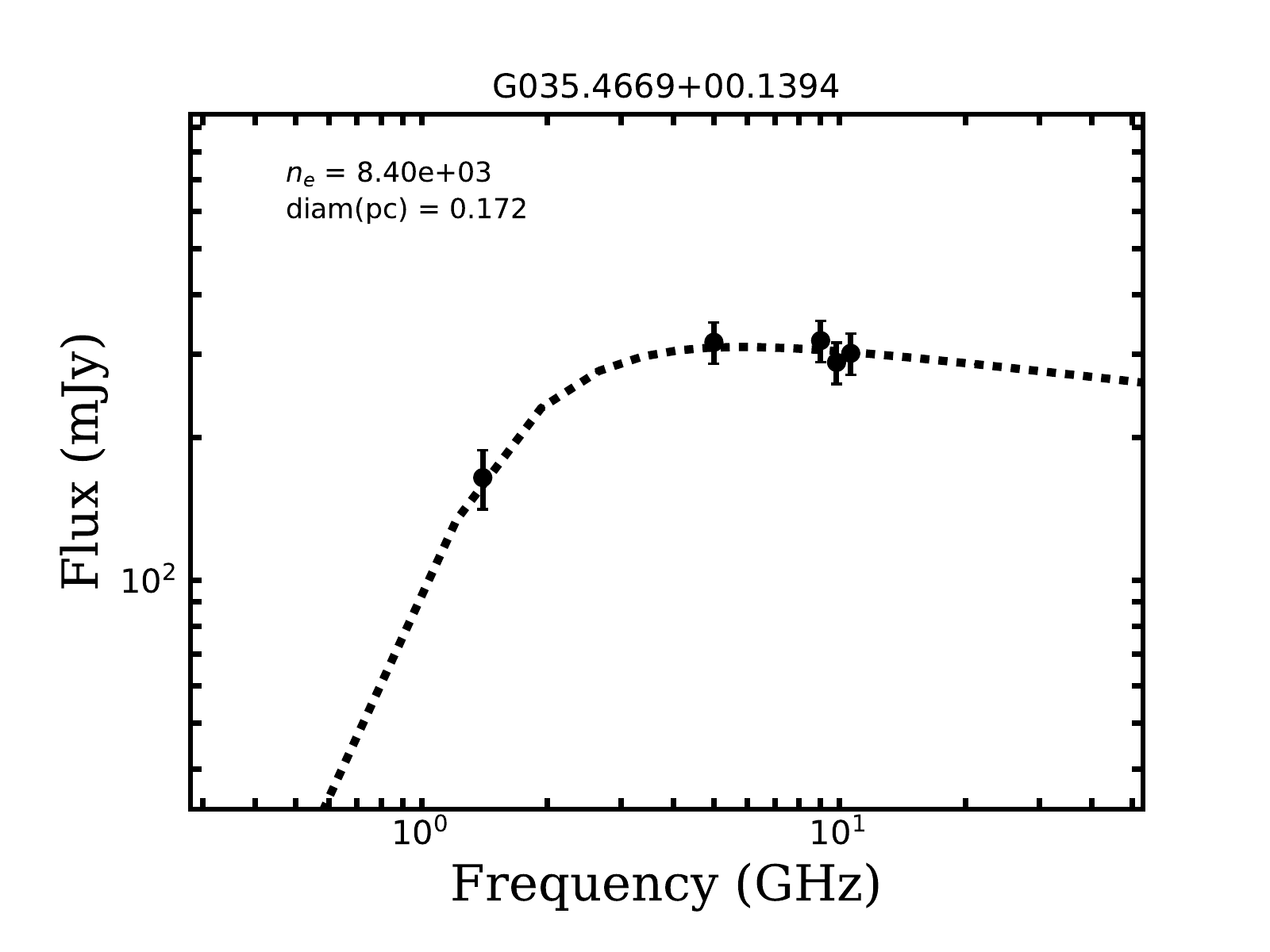}
  \caption{ The radio SED fitting to flux density points for for example compact and extended sources. 
  Upper panel: SED fitting to flux density points between 1 and 26 GHz for the compact example G032.7441$-$00.0755 (upper row of Fig.\,\ref{example_sources}). 
  Lower panel: SED fitting to flux density points between 1 and 11 GHz for extended example G035.4669$+$00.1394 (middle row of Fig.\,\ref{example_sources}) by excluding K-band flux measurements.
  The best-fitting results for the electron density $n_{\rm e}$ and physical linear diameter $ diam $ are shown in the upper-left corner of each plot.  The best-fitting SEDs for the whole sample are shown in Fig\,\ref{summary_sed_multiband_images}. }

 \label{fig:ffsed}
 \end{figure}

  \begin{figure*}
 \centering
 \begin{tabular}{ccc}
\includegraphics[width = 0.3\textwidth]{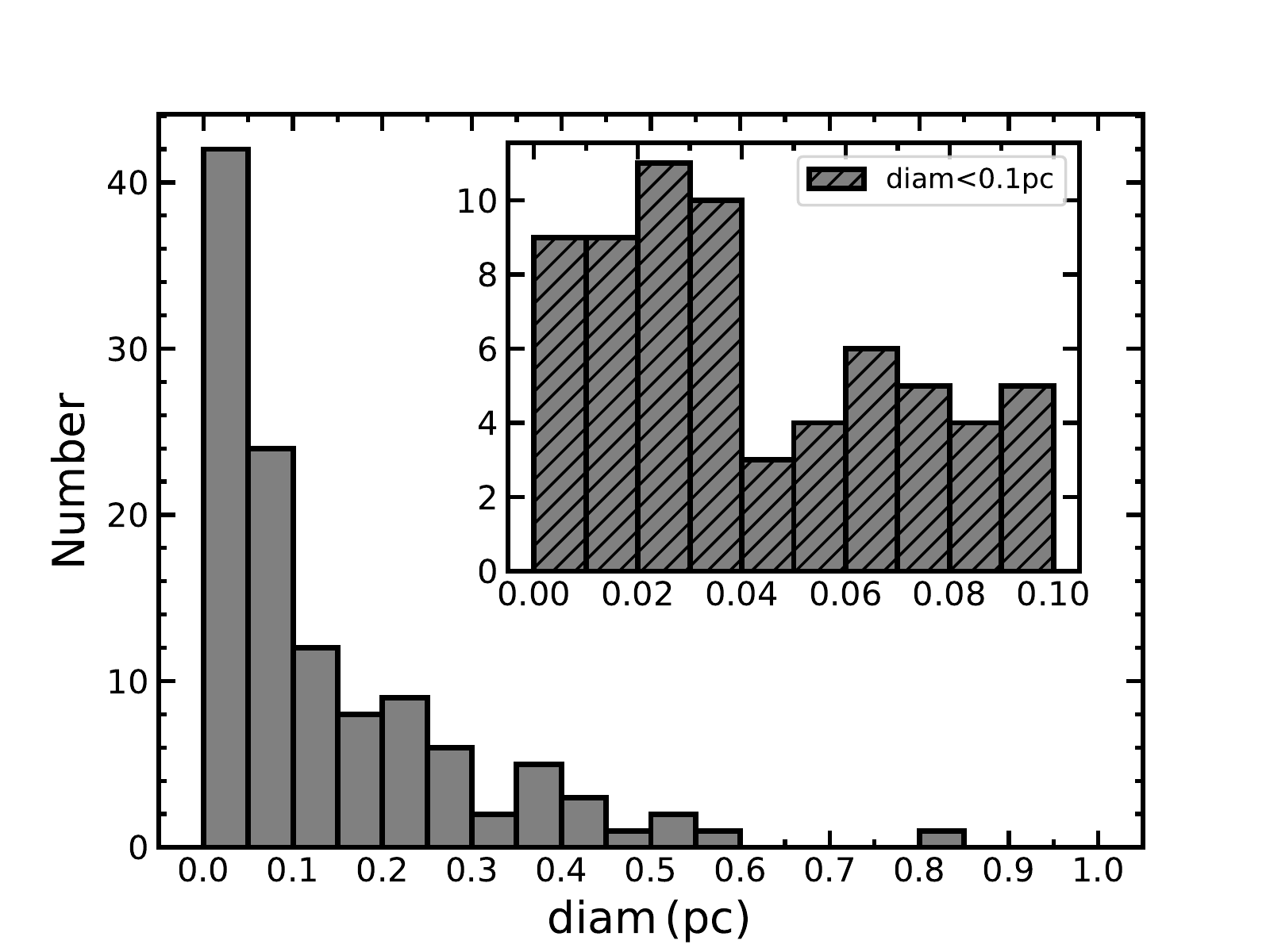} & 
\includegraphics[width = 0.3\textwidth]{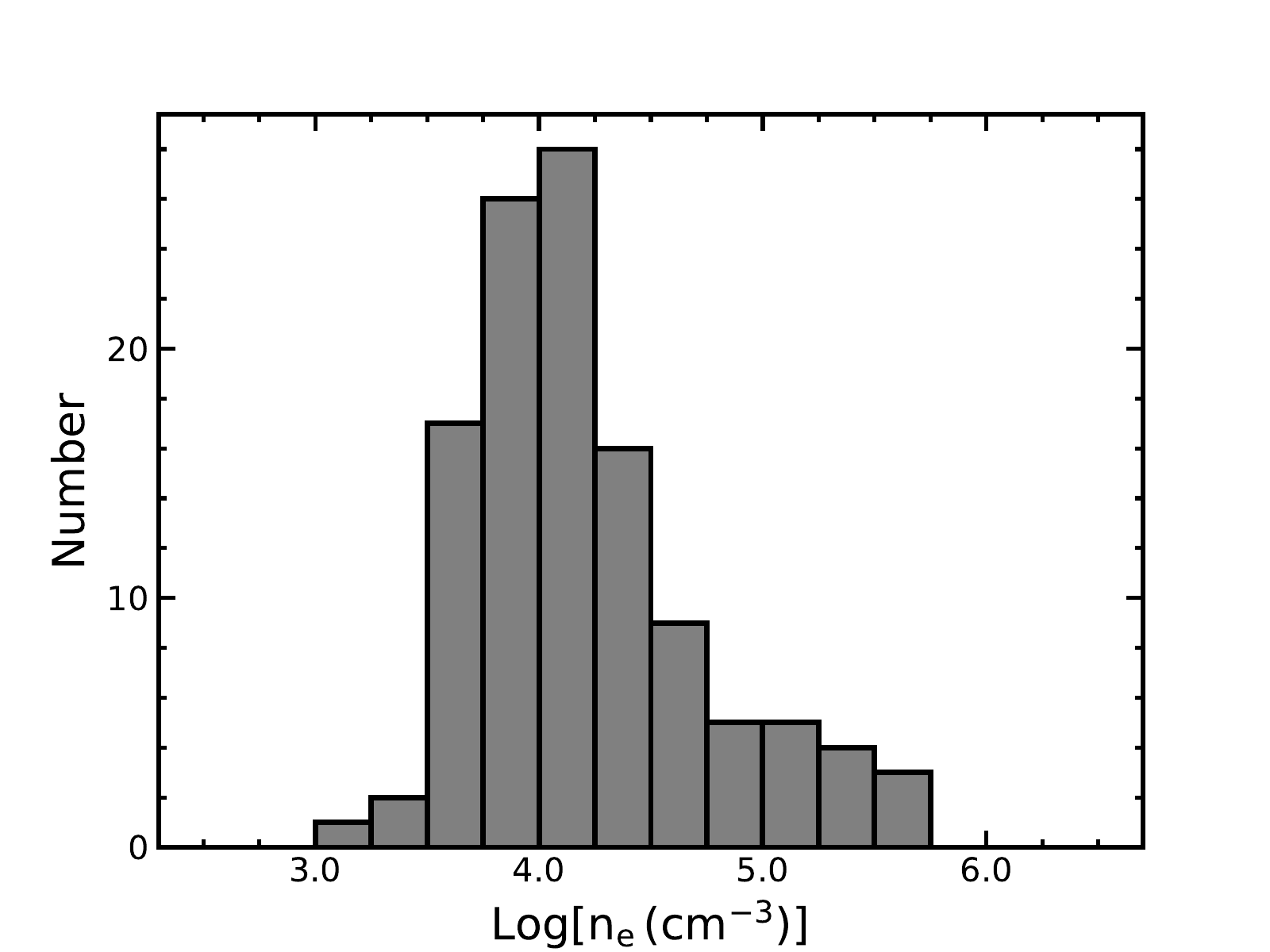} &
\includegraphics[width = 0.3\textwidth]{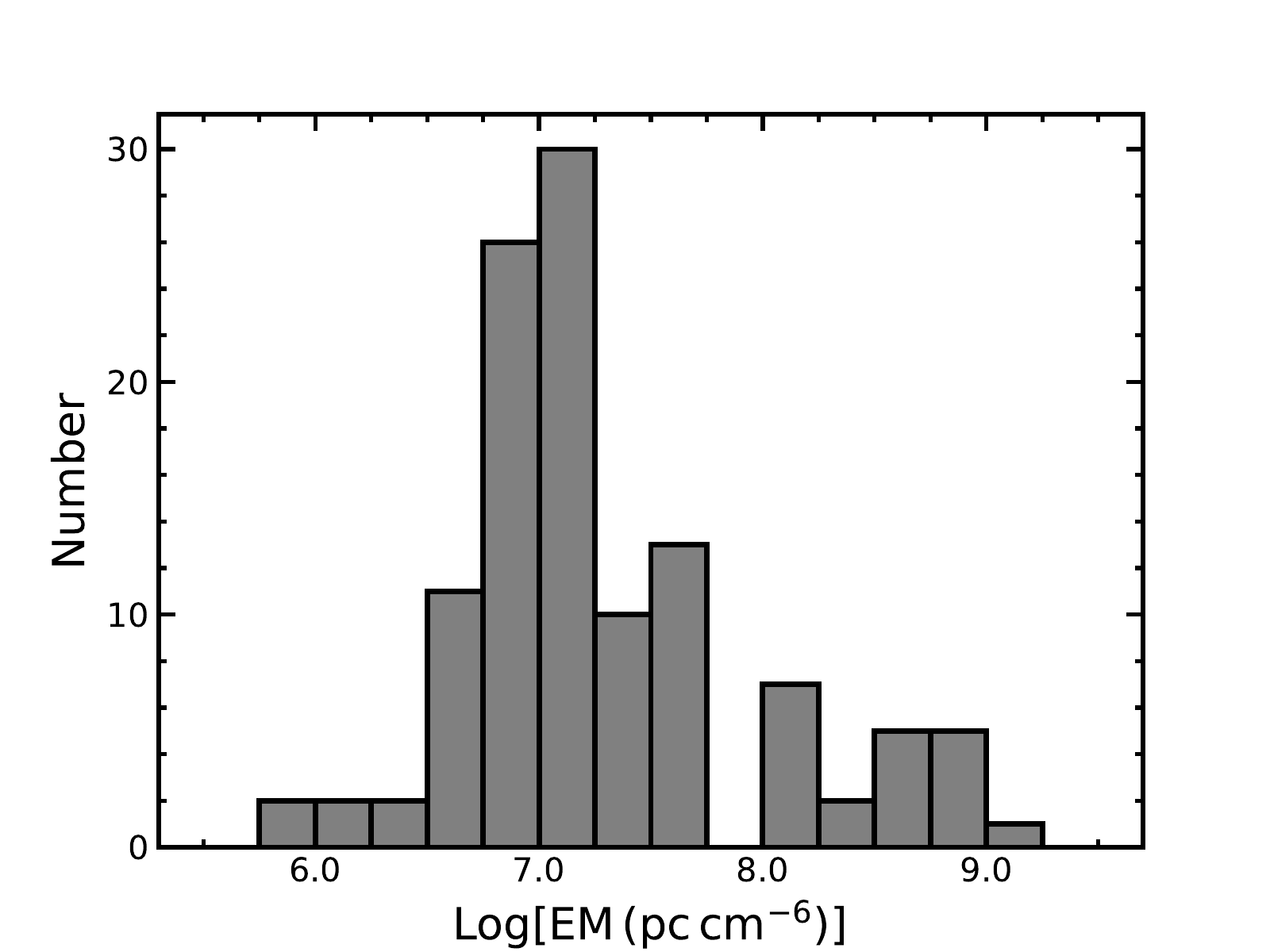} \\
    (a) & (b) & (c) \\
\includegraphics[width = 0.3\textwidth]{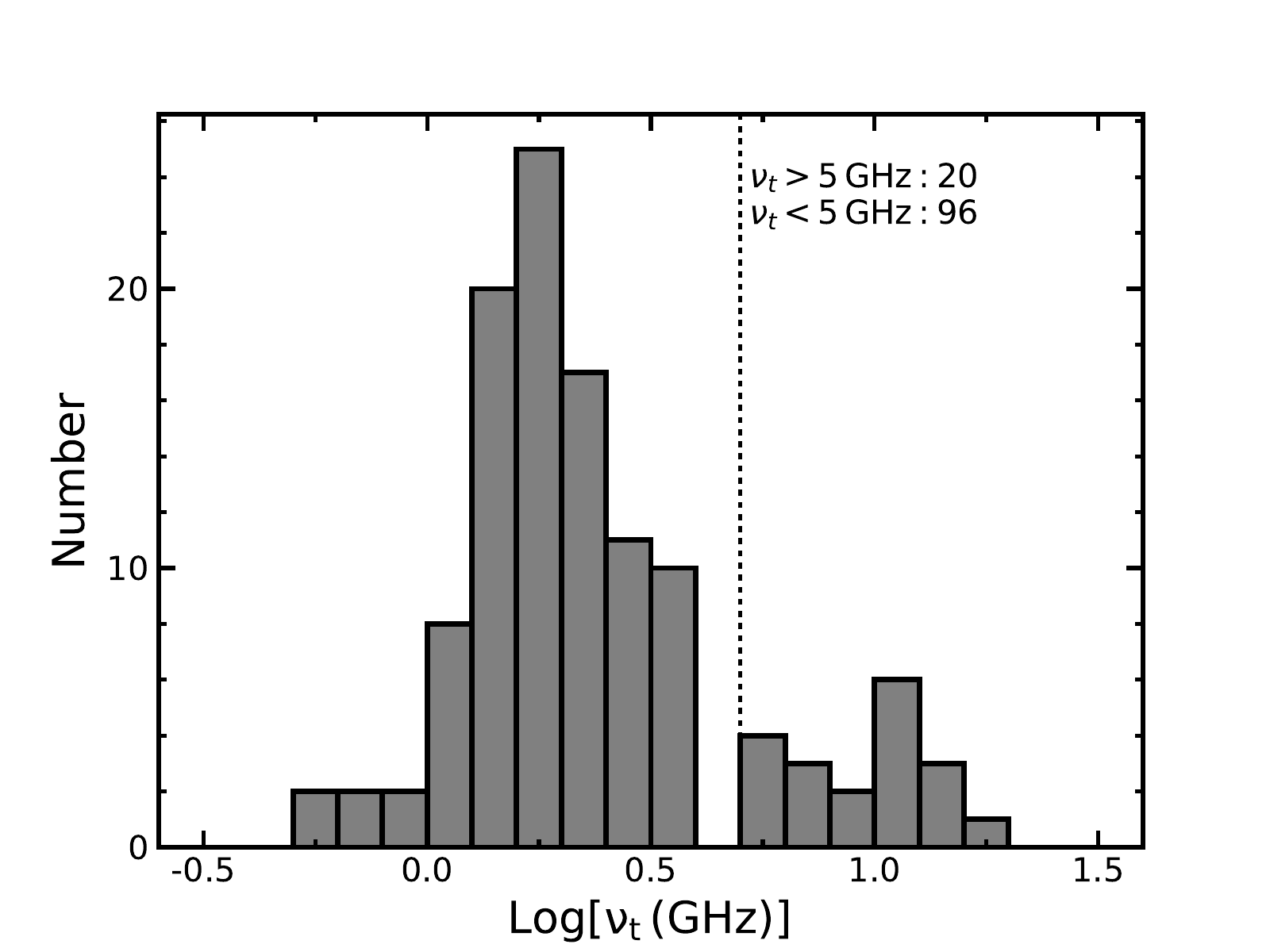}&
\includegraphics[width = 0.3\textwidth]{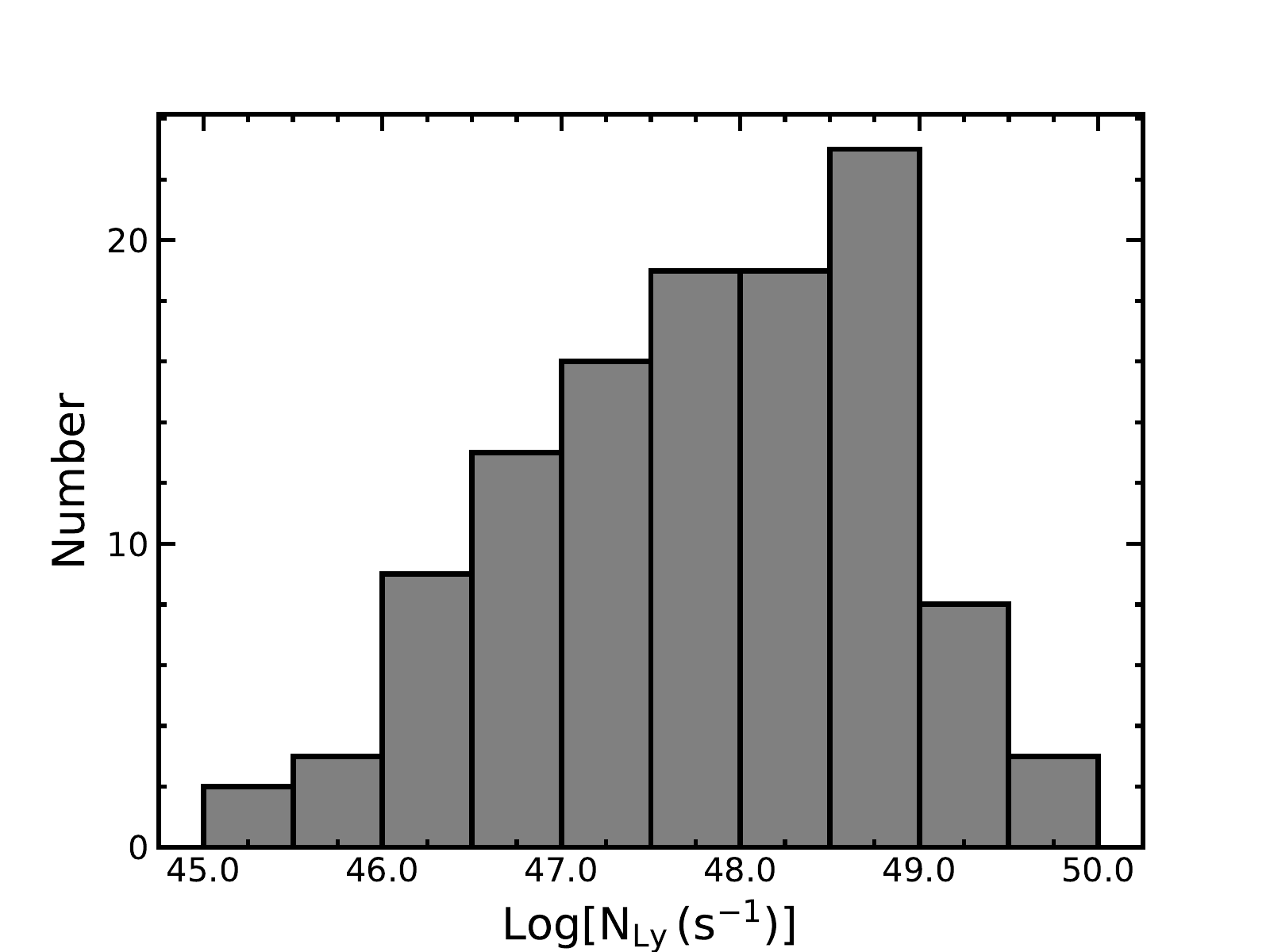} & 
\includegraphics[width = 0.3\textwidth]{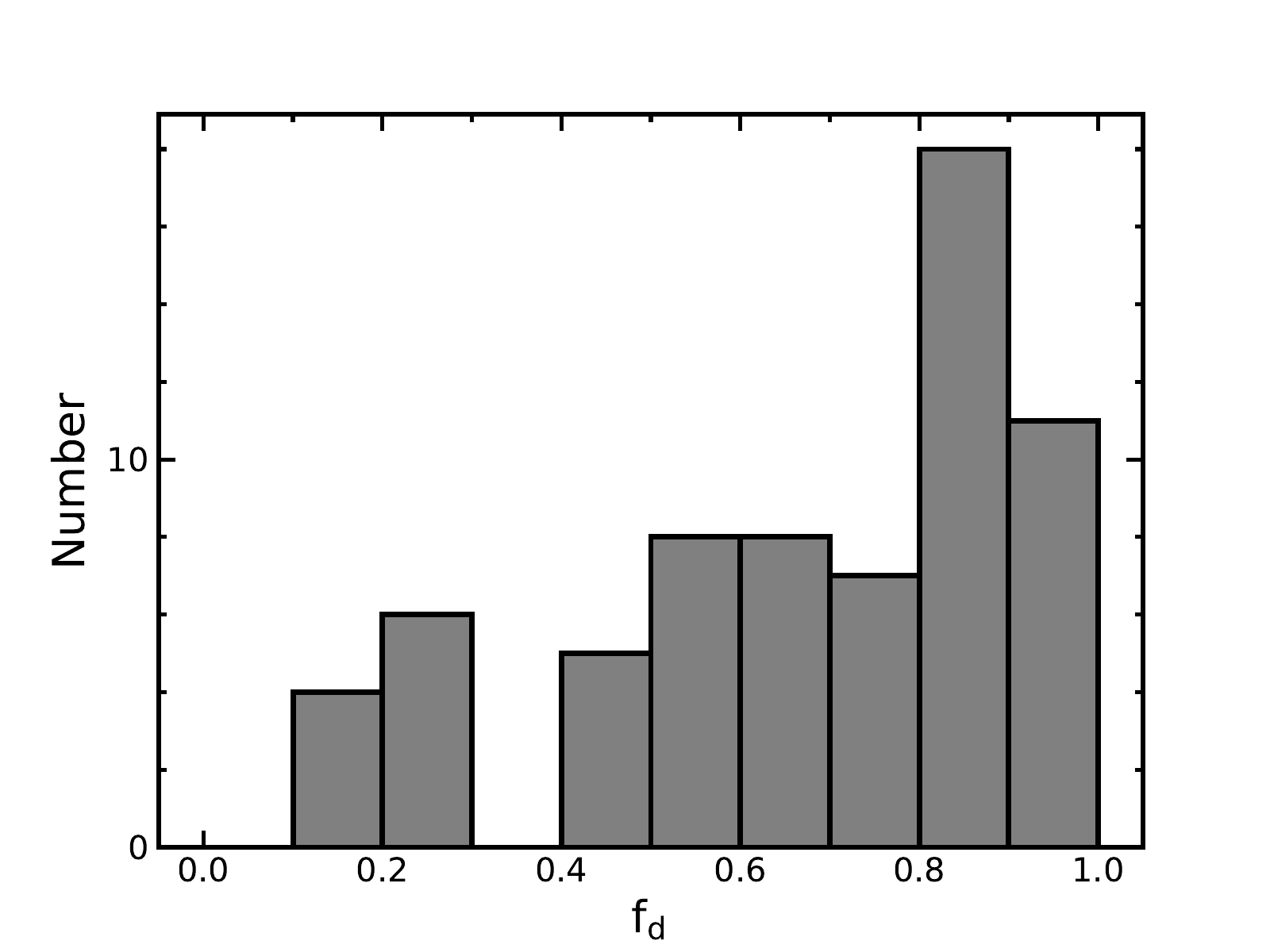}  \\
   (d)  &  (e) &  (f) \\
 \end{tabular}
  \caption{ Distributions of the derived physical properties of 116 young \hii\ regions. 
  Panels (a), (b), and (c):  
  Distributions of the physical linear diameter $diam$ (a), the electron density $ n_{\rm e}$ (b), and the $\rm EM$ (c). The bin sizes are 0.05\,pc, 0.25\,dex, and 0.25\,dex for $diam$,  $n_{\rm e}$, and $\rm EM$, respectively. 
  Panels (d), (e), and (f): Distributions of the turnover frequency $\nu_{\rm t}$ (d), the Lyman continuum flux $N_{\rm Ly}$ (e), and the dust absorption fraction $f_{\rm d}$ (f). The bin sizes are 0.1\,dex, 0.1\,dex, and 0.5\,dex for $ \nu_{\rm t}$, $f_{\rm d}$, and $N_{\rm Ly}$, respectively. 
  }
 \label{fig:distr_phys_param}
 \end{figure*}

We use our multi-wavelength VLA data to construct SEDs for the free-free emission in order to measure the radio properties of our sample of young \hii\ regions. 
We model each SED for an ionization-bounded \hii\ region using the standard uniform electron density model given by \citet{Mezger1967ApJ}. 
In this standard model, the integrated flux density at a given frequency $\nu$ is given by $S_{\rm \nu} = \frac{2k\nu^{2}\Omega\,T_{e}(1-e^{-\tau})}{c^{2}}$ using the Rayleigh Jeans approximation, where $\Omega$ is the solid angle related to the physical diameter $diam$ and distance $d$ of each source.  The optical depth $\tau$ of free-free radiation can also be represented as a function of frequency \citep{Mezger1967ApJ,Dyson1997pismbook}, $\tau \propto T_{\rm e}^{-1.35}\,\nu^{-2.1}\,n_{\rm e}^{2}\,diam$, where we assume an electron temperature $T_{\rm e} =\rm 10^4\,K$ \citep{Dyson1997pismbook}.  
Therefore, the radio SED of an \hii\ region from the standard model is expected to have a rising spectrum at low frequencies $s_{\rm \nu}\propto \nu^{+2}$ ($\tau\gg 1$)  and a flat spectrum at high frequencies $s_{\rm \nu}\propto \nu^{-0.1}$ ($\tau \ll 1$). 
Based on the distances $d$ in Table\,\ref{tab_obser_list} and the observed fluxes $s_{\rm \nu}$ in Table\,\ref{tab_obsparms}, the SED model of each source has two free parameters: 
the electron density $n_{\rm e}$ and the physical diameter $diam$.  
The best estimate for the two parameters can be obtained by fitting the radio-frequency continuum spectrum of each source. 
The uncertainties on flux measurements at these points are taken into account in the fitting process.
 For compact and spherical \hii\ regions in the sample, the derived density $n_{\rm e}$ and diameter $diam$ from SED fitting represent averaged properties over the ionized gas that are responsible for the free-free emission between 1 and 26 GHz. 
 For the \hii\ regions with non-spherical geometry, 
this spherical morphology model might introduce additional uncertainty into the determination of the geometry-dependent parameters such as the electron density and diameter. 
Ideally, the calculation should consider the three-dimensional structure of the volume responsible for the radio emission; however, 
we do not know the internal structure and any model of the source geometry would introduce additional unknown parameters. 
 Moreover, the morphologies of the nonspherical \hii\ regions are variable between X-band and K-band as shown in Fig.\,\ref{summary_sed_multiband_images}. 
To avoid the complication when calculating the geometry-dependent parameters, the peak physical properties averaged over the beam rather than the entire source are commonly used for these nonspherical and irregular \hii\ regions in previous studies \citep[e.g.,][]{Wood1989ApJS,Kurtz1994ApJS659K}. 
In this work, the uniform spherical model is sufficient to match the SEDs of the nonspherical \hii\ regions, and the SED of each source takes into account the multi-band radio emission of the entire source. 
Therefore, the fitted $n_{\rm e}$ and $diam$ represent averaged properties over the entire emission gas at multi-bands and {\bf can be used to shed light on} the physical condition of these nonspherical \hii\ regions as a whole.

Figure\,\ref{fig:ffsed} shows examples of the fitted SEDs for a compact source G032.7441$-$0.076 and an extended source G035.4669$+$00.1394. 
Owing to the lack of short baseline spacings, the K-band flux measurements have been excluded from the SED fitting of the extended sources in the sample. 
Including the four sources with data from archives and references (see Sect.\,\ref{sect_optically_thick_hii} and Table \ref{tab:10hchii_candidates}), 
the SEDs and best-fitting models of all 116 \hii\ regions are presented in Appendix Fig.\,\ref{summary_sed_multiband_images}. 
The EM of each \hii\ region is then calculated using ${\rm EM} = {n_{\rm e}}^2\times\,diam$. 
Considering a mean error of $\sim$10\% both in the flux density at each frequency and the distance measurement, 
this gives typical errors of $\sim$20\% in $n_{\rm e}$, $\sim$10\% in $diam,$ and $\sim$40\% in EM. 
The typical errors that we adopted refer to the uncertainty on measurements, as in previous studies \citep[e.g.,][]{Sanchez_Monge2013AA21S,Kalcheva2018AA615A103K}, 
and would be larger if  the uncertainty on the assumptions in the model were considered.

The fitted parameters from radio SEDs are given in Table\, \ref{112hii_result_phyparam} along with the physical parameters derived from the analysis presented in the following section. 
In panels (a), (b), and (c) of  Fig.\,\ref{fig:distr_phys_param}, we present the distributions of the fitted parameters. 
The physical sizes peak at 0.02\,pc in panel (a), and 57\% of the sources (66/116) have physical diameters of less than 0.1\,pc, as shown in the subplot of that panel. 
This is consistent with the majority of these being classified as \uchii\ regions or smaller. There are 9 sources with $diam<\rm 0.01\,pc$ and the mean diameter is $diam=$0.006\,pc, 
corresponding to $\rm \sim1000\,AU$. 
This physical scale implies that the sample could have coincidences with radio jets and jet candidates from massive young stellar objects \citep[MYSOs;][]{Purser2016MNRAS4601039P}.

Figure\,\ref{fig:distr_phys_param}\,(b) shows the distribution of $n_{\rm e}$, which peaks at $\rm 10^{4}\,cm^{-3}$. 
About 60\% (70/116) of the sources have high densities with $ n_{\rm e}>\rm 10^{4}\,cm^{-3}$. 
The 70 high-density \hii\ regions are compact with a mean diameter of $diam=$ 0.06\,pc, 
implying that there might exist small-scale and high-density objects in the sample such as \hchii\ regions \citep{Kurtz2005IAUS} and MYSO jets  \citep{Purser2016MNRAS4601039P}.

Figure\,\ref{fig:distr_phys_param}\,(c) shows the distribution of EM, which peaks at $\rm 10^{7}\,pc\,cm^{-6}$, 
and most sources have EM between $3.2\times\,\rm 10^{6}\,pc\,cm^{-6}$ and $1.0\times\,\rm 10^{8}\,pc\,cm^{-6}$. 
There are two groups in the distribution of EM: one with $\rm EM<10^{8}\,pc\,cm^{-6}$ and the other with $\rm EM>10^{8}\,pc\,cm^{-6}$, 
which indicates that there are sources in the sample connected to the very early stages of \hii\ regions.

The median values of diameter ($diam=$0.08\,pc), electron density ($n_{\rm e}=1.3\times10^{4}\,cm^{-3}$), and EM (EM = 1.9$\times10^{7}\,pc\,cm^{-6}$) of our sample are consistent with typical values for \uchii\ regions. About 10\% of the sources have $n_{\rm e}> \rm 10^{5}\,cm^{-3}$, 36\% of the sample show $ diam\rm < 0.05 \,pc$,  and 17\% of them have $\rm EM> 10^{8}\,pc\,cm^{-6}$, which fulfill the standard quantitative criteria of \hchii\ regions. 
We discuss the potential \hchii\ regions in the sample in Section\,\ref{sect_optically_thick_hii}.

\subsection{Derived physical characteristics}

\subsubsection{Turnover frequency $ \nu_{\rm t}$}
\label{sec_turnover_frequency}

As a dividing line between the optically thin and thick regimes of the radio spectrum of  \hii\ region, the turnover frequency $\nu_{\rm t}$ is defined as the frequency where $\tau=1$ \citep{Kurtz2005IAUS}. 
The flux density of \hii\ region peaks at $\nu > \nu_{t}$, and decreases as the square of frequency at $\nu < \nu_{t}$. Using the formula provided in \citet{Mezger1967ApJ} for a homogeneous \hii\ region, 
the optical depth can be expressed as a function of observing frequency $\nu$, electron temperature $ T_{\rm e}$, which is assumed to be $10^{4}\rm\,K$, and emission measure EM:

\begin{equation}
\label{eq:Snu}
 \tau = 0.082\times{\left[\frac{T_{\rm e}}{\rm K}\right]}^{-1.35}\times\,{\left[\frac{\nu}{\rm GHz}\right]} ^{-2.1}\times\,\left[\frac{\rm EM}{\rm pc\,cm^{-6}}\right],
\end{equation}
\noindent Setting $ \tau=1$, the turnover frequency can be expressed as  \citep{Kurtz2005IAUS}: 

\begin{equation}
\label{eq:nu_t}
 \left[\frac{\nu_{\rm t}}{\rm GHz}\right]  = 0.082\times{\left[\frac{T_{\rm e}}{\rm K}\right]}^{-1.35}\times \left[\frac{n_{\rm e}^{2}\times diam}{\rm cm^{-3}\,pc}\right]^{0.476}\,\\
\end{equation}

The typical error for the $\nu_{\rm t}$ is 30\% by considering the typical 20\% error in density estimation and 10\% in diameter measurements. Panel (d) of Fig.\,\ref{fig:distr_phys_param} presents the distribution of the turnover frequency $\nu_{\rm t}$ for this sample of young \hii\ regions  (i.e., young \uchii\ regions), which peaks at $\nu_{\rm t}\sim 2\,\rm GHz$ and has a mean value of $\nu_{\rm t}\sim 3.3\,\rm GHz$. Both of the peak and mean turnover frequencies of this sample of young \hii\ regions  are lower than the expected value of $\sim \rm 5\,GHz$ of \uchii\ regions in \citet{Kurtz2005IAUS} with typical $ n_{\rm e}\sim3\times10^{4}$\,cm$^{-3}$ and $diam\sim \rm 0.1\,pc$.
This lower turnover frequency found in the sample may be due to  a large fraction of detected emission from the optically thin low-density region surrounded by a \hii\ region, as suggested in \citet{Steggles2016PhDT374S} and \citet{Steggles2017MNRAS4664573S}.
Alternatively, many of these \hii\ regions are simply optically thin.
 
  The  Fig.\,\ref{fig:distr_phys_param}\,(d)  indicates two populations of \hii\ regions: one with $\nu_{\rm t}\rm <5\,GHz$ and the other with $\nu_{\rm t}\rm >5\,GHz$, 
 which are referred to as optically thin and optically thick \hii\ regions in this work, respectively. 
 The optically thick \hii\ regions are found to have higher density, higher emission measure, 
 and smaller physical linear size compared to optically thin \hii\ regions, as shown in Table\,\ref{tab_summary_thin_thick_hii}.

\subsubsection{Lyman continuum flux}
\label{sec_lyc}

For an optically thin \hii\ region in the photoionization equilibrium, the Lyman continuum ionizing flux $ N_{\rm Ly}$ emitted by the embedded massive star can be calculated from the radio continuum flux and heliocentric distance to the source \citep{Sanchez_Monge2013AA21S}, as

\begin{equation}
\label{eq:Ni}
\left[\frac{N_{\rm Ly}}{\mathrm{s}^{-1}}\right]\,=\,8.9\times10^{40}~\left[\frac{S_\nu}{\mathrm{Jy}}\right] \left[ \frac{\nu}{{\rm GHz}} \right]^{0.1} \left[\frac{T_\mathrm{e}}{10^4\mathrm{K}}\right]^{-0.45} \left[\frac{d}{\mathrm{pc}}\right]^2,
\end{equation}

\noindent where $S_{\nu}$ is the integrated flux density at frequency $\nu$, $T_{\rm e}$ is electron temperature assumed to be  $10^{4}\rm\,K,$ and $d$ is the distance to the source. 
For each source in the sample, we use the $S_{\nu}$ measured in the optically thin part of the radio SED to calculate the Lyman continuum flux. 
The distance for each source is taken from  the literature (as discussed in Sect.\,\ref{sect:sample_section}). 
The typical error of the derived Lyman continuum flux is $\sim40\%$ considering the error in both kinematic distance and the integrated flux measurement  \citep[e.g.,][]{Urquhart2013MNRAS435}.

The distribution of the derived Lyman continuum flux is shown in  Fig.\,\ref{fig:distr_phys_param}\,(e), which peaks at $\rm 10^{48}\,s^{-1}$ and ranges from $\rm 10^{45.4}\,s^{-1}$ to $\rm 10^{49.9}\,s^{-1}$.
The corresponding spectral types of the zero-age main sequence (ZAMS) stars are between B0 and O4 listed in Table\,\ref{112hii_result_phyparam}, assuming that a single star is responsible for the ionization and there is no dust in the ionization-bounded \hii\ region \citep[e.g.,][]{Garay1993ApJ418368G,Wood1989ApJS}. 
The derived spectral type of the ZAMS star would be earlier or later \citep[e.g.,][]{Wood1989ApJS}, 
if multiple stars are responsible for the ionization or if there is dust absorption within the \hii\ region \citep[e.g.,][]{Garay1993ApJ418368G}. 
For instance, the presence of dust may lower the flux by a factor of two or more as the dust absorption fraction ranges from $\sim$50\% to $\sim$90\% for \uchii\ regions \citep[e.g.,][]{Wood1989ApJS, Garay1993ApJ418368G, Kurtz1994ApJS659K}, but if the emission was from a cluster then the spectral type would be typically earlier by a subclass or two  \citep{Wood1989ApJS,Urquhart2013MNRAS435}. 
The effects of cluster and dust on determining the spectral type are probably comparable and counterbalance each other. Therefore, the values we estimated are reliable within a few subclasses.

\subsubsection{Dust within \hii\ regions}
\label{sect_dust}

Previous studies found that a significant fraction of the Lyman continuum photons are absorbed by the dust within \hii\ regions  \citep{Garay1993ApJ418368G,Wood1989ApJS,Kim2001ApJ549}. 
By assuming that a single star is responsible for the observed luminosity and the observed Lyman continuum flux of an \hii\ region, 
the fraction of UV photons absorbed by dust within \hii\ regions is defined as 
$f_{\rm d} = 1- N'_{\rm c}/N^{\star}_{\rm c}$ \citep[e.g.,][]{Wood1989ApJS},
\noindent where $N'_{\rm c}$ is the number of observed ionizing photons and $N^{\star}_{\rm c}$ the predicted Lyman continuum photons derived from spectral type based on the total infrared luminosity. 
As discussed in previous studies  \citep[e.g.,][]{Garay1993ApJ418368G, Wood1989ApJS}, $f_{\rm d}$ should be taken as an upper limit as it is very likely to be overestimated if the expected Lyman continuum photons are excited by clusters of young stars rather than by a single star. For instance, at a given total luminosity, the spectral type estimated assuming a cluster that provides the entire infrared luminosity is typically two or three subclasses later than the spectral type estimated assuming a single star \citep{Wood1989ApJS}, and thus leads to a lower expected Lyman continuum flux $N^{\rm \star}_{c}$ than derived assuming a single star. 
The observed $N'_{\rm c}$ would be dominated by the earliest spectral type in the clusters as the properties of O-type stars change so dramatically between two subclasses \citep[e.g.,][]{Panagia1973AJ78,Wood1989ApJS}, which has also been found by \citet{Urquhart2013MNRAS435} who suggested that the most massive stars within clumps dominate the observed properties.  
The upper limit of the fraction of Lyman continuum photons absorbed by dust within \hii\ regions can range from 50\% \citep{Garay1993ApJ418368G,Kim2001ApJ549} to 90\% \citep{Wood1989ApJS,Kurtz1994ApJS659K}.

There is evidence of dust existing in the \hii\ regions in our sample: all of them show bright 24$\mu m$ emission in the MIPSGAL survey \citep{Carey2009PASP76C} 
and strong 70$\mu m$ emission in the Hi-GAL survey \citep{Molinari2010PASP}, at a high angular resolution ($\sim$6\arcsec).   
After excluding $\sim$40\% of the sources with Lyman excess (see Sect.\,\ref{sec:disc_lym}), 
the upper limit of the mean fraction absorbed by dust within \hii\ regions for our sample is $f_{\rm d}=0.67\pm0.03$,  
which is consistent with previous results \citep[e.g.,][]{Garay1993ApJ418368G, Kim2001ApJ549,Wood1989ApJS}, as shown in panel (f) of Fig.\,\ref{fig:distr_phys_param}. 
Among the 67 \hii\ regions with dust absorption, 43\% (29/67) of the sources with physical diameters $diam<\rm 0.1\,pc$ have a mean of $f_{\rm d} =0.79\pm0.04$ , and 57\%  (38/67) of the sources with $diam>\rm 0.1\,pc$ have a mean of $f_{\rm d} =0.58\pm0.04$.  
This indicates that the dust absorption fraction tends to be more significant for the more compact and presumably younger \hii\ regions compared to the larger and more evolved \hii\ regions, 
which agrees with the model in \citet{Arthur2004ApJ608282A} who suggest that the fraction of ionizing photons in \hii\ regions absorbed by dust decreases with time.

\begin{table*}
\setlength{\tabcolsep}{3pt}
\caption[]{{ 
 Derived physical properties of 116 young \hii\ regions. 
}
\label{112hii_result_phyparam}}
\centering
\begin{tabular}{l.......c}
\hline
\hline
 Name & 
\multicolumn{1}{c}{$n_{\rm e}$} 
& \multicolumn{1}{c}{$diam$} 
& \multicolumn{1}{c}{EM} 
& \multicolumn{1}{c}{$\nu_{\rm t}$} 
&  \multicolumn{1}{c}{$\log N_{\rm Ly}$} 
& \multicolumn{1}{c}{Spectral} 
&  \multicolumn{1}{c}{$f_{\rm d}$}  
&   \\
 &  \multicolumn{1}{c}{($\rm 10^{5}\,cm^{-3}$)} 
 & \multicolumn{1}{c}{($\rm pc$)} 
 & \multicolumn{1}{c}{($\rm 10^{7}\,pc\,cm^{-6}$)} 
 & \multicolumn{1}{c}{($\rm GHz$)} 
 &  \multicolumn{1}{c}{(photons\,s$^{-1}$)} 
 &   \multicolumn{1}{c}{Type}  
 &  \\
 
\hline
 \multicolumn{1}{c}{(1)}  & 
 \multicolumn{1}{c}{(2)} & 
 \multicolumn{1}{c}{(3)} & 
 \multicolumn{1}{c}{(4)} & 
 \multicolumn{1}{c}{(5)} & 
 \multicolumn{1}{c}{(6)} & 
 \multicolumn{1}{c}{(7)} & 
 \multicolumn{1}{c}{(8)} \\ 
 
\hline
\hline
  G010.3009$-$00.1477 & 0.09 & 0.119 & 0.92   & 1.69  & 47.94   & O9.5          & 0.86           \\ 
  G010.4724$+$00.0275 & 1.43 & 0.022 & 45.2   & 10.77 & 48.11   & O9            & 0.94           \\ 
  G010.6234$-$00.3837 & 0.16 & 0.166 & 4.39   & 3.55  & 48.9    & O6.5          & 0.81           \\ 
  G010.9584$+$00.0221 & 0.36 & 0.029 & 3.78   & 3.31  & 47.35   & B0            & $-$            \\ 
  G011.0328$+$00.0274 & 0.13 & 0.014 & 0.24   & 0.89  & 45.57   & B1            & $-$            \\ 
  G011.1104$-$00.3985 & 0.07 & 0.145 & 0.62   & 1.4   & 47.94   & O9.5          & 0.27           \\ 
  G011.1712$-$00.0662 & 0.09 & 0.053 & 0.45   & 1.21  & 46.91   & B0            & $-$            \\ 
  G011.9368$-$00.6158 & 0.07 & 0.155 & 0.86   & 1.63  & 48.12   & O9            & 0.19           \\ 
  G011.9446$-$00.0369 & 0.17 & 0.075 & 2.2    & 2.56  & 47.84   & O9.5          & $-$            \\ 
  G012.1988$-$00.0345 & 0.07 & 0.148 & 0.65   & 1.43  & 47.98   & O9            & 0.77           \\

\hline
\hline
\end{tabular}
\tablefoot{
Only a small portion of the data is provided here, the full table is presented in Table \ref{112hii_result_phyparam_append} and will be available in electronic form at the CDS.  
}
\end{table*}

\setlength{\tabcolsep}{4pt}
\begin{table}
\caption {Summary of the derived physical  parameters for the 96 optically thin \hii\ regions ($\nu_{t}<\rm 5\,GHz$) and the 20 optically thick \hii\ regions ($\nu_{t}>\rm 5\,GHz$). Columns (2-5) provide the minimum, maximum, mean$\pm$ standard\,deviation, and median values, respectively, of each parameter.
}
\begin{tabular}{lcccc}
\hline
\hline
Parameter                                       &  $x_{min}$ & $x_{max}$   & $x_{mean}\pm x_{std}$ & $x_{med}$ \\
\hline
\multicolumn{5}{c}{The 20 optically-thick \hii\ regions sample}     \\
\hline
${\log}[n_{\rm e}\rm(cm^{-3})]$                   &   4.37            &  5.65       &  5.11$\pm$ 0.07               &      5.12            \\
$diam\,\rm [pc]$                              &   0.004          & 0.23        &  $0.035\pm 0.01$              &     0.023             \\
$\log[\rm EM\,(pc\,cm^{-6})]$           &   8.00            & 9.05        &  8.50$\pm$ 0.08              &    8.58             \\
${\log}[N_{\rm Ly}\rm\,(s^{-1})]$              &    46.21          & 49.55      &  47.77$\pm$ 0.20            &    47.80           \\
$ \nu_{\rm t}\,\rm [GHz]$ &   5.28              &  16.67     &   9.73 $\pm$ 0.80             &   9.94             \\
{ \footnotesize Dust\,absorption\,fraction\,$f_{\rm d}$}          &   0.16             & 0.99      &   0.81 $\pm$ 0.05             &   0.88            \\
\hline
\multicolumn{5}{c}{The 96 optically-thin \hii\ regions sample}     \\
\hline
${\log}[n_{\rm e}\rm(cm^{-3})]$                   &   3.15            &  4.69       &   4.02$\pm$ 0.03 &     4.01           \\
$diam\,\rm [pc]$                              &   0.006          & 0.81        &  $0.16\pm 0.02$              &     0.11             \\
$\log[\rm EM\,(pc\,cm^{-6})]$           &   5.96            & 7.73        &  $7.16\pm 0.04$              &     7.02             \\
${\log}[N_{\rm Ly}\rm\,(s^{-1})]$              &    45.37          & 49.83      &  47.82$\pm$ 0.1            &    47.97          \\
$ \nu_{\rm t}\,\rm [GHz]$            &   0.56              &  3.60     &   $1.91 \pm 0.08$             &   1.80              \\
{ \footnotesize Dust\,absorption\,fraction\,$f_{\rm d}$}          &   0.14             & 0.97      &   0.62 $\pm$ 0.03             &   0.66            \\
\hline
\hline
\end{tabular}
\label{tab_summary_thin_thick_hii}
\end{table} 
\setlength{\tabcolsep}{6pt}

\setlength{\tabcolsep}{4pt}
\begin{table}
\caption { Quantitative criteria for \hchii\ regions, \uchii\ regions and intermediate objects (\hchii\ $\rightarrow$ \uchii\,) between the two stages, summarized from the literature. 
}
\begin{tabular}{p{2.8cm}|L{1.0cm}|L{2.8cm}|L{1.0cm}}
\hline
\hline
Parameters                                       &  \hchii\ & {\hchii\ $\rightarrow$ \uchii\,}  & \uchii\  \\
\hline
Size\,(pc)  & $ \rm \lesssim 0.05$ &  $ \rm\sim[0.05-0.1]$ &  $ \rm\lesssim 0.1$\\
\hline
$n_{\rm e}$ ($\rm cm^{-3}$) & $ \rm\gtrsim 10^{5}$ & $ \rm\sim[10^{4}-10^{5}]$  & $ \rm\gtrsim 10^{4}$ \\
\hline
EM\,($\rm pc\,cm^{-6}$) & $ \rm\gtrsim 10^{8}$ &  $ \rm\sim[10^{7}- 10^{8}]$ &  $ \rm\gtrsim 10^{7}$ \\
\hline
RRL $\rm \Delta V$\,($\rm km\,s^{-1}$) &$\gtrsim 40$  & $\rm \sim[25-40]$ &  $< 40$ \\
\hline
\end{tabular}
\label{tab_classify_hii}
\end{table} 

\begin{table*}
\centering
\caption[]{Summary of the physical parameters and the classification of the 20 optical thick \hii\ regions identified in this work. The classifications given in col.\,6 are \hchii\ regions (Class: HC), \hchii\ region candidates (Class: HC?),  and intermediate objects (Class: HC-UC); these have been assigned based on their electron density $n_{\rm e}$, physical diameter $diam$ and emission measure $\rm EM$, derived from the SED fitting method. We also include four sources with $\nu_{\rm t} \sim \rm  3.5\,GHz$ , such as  G010.9584$+00.0221$ and G035.5781$-$00.0305 that have previously been identified as \hchii\ regions but not recovered by this work (see Sect.\,\ref{hchii_not_resolved}), as well as G030.8662+00.1143 and G030.7197-00.0829 that show broad RRL with $\Delta V \rm > 40\,km\,s^{-1}$. }
\begin{tabular}{p{3.1cm}C{1.0cm}C{1.0cm}C{1.5cm}C{1.9cm}C{1.1cm}C{0.95cm}C{0.95cm}C{1.1cm}C{1.0cm}}
\hline
\hline
Name & $n_{\rm e}$ & $diam$ & $\rm EM$  & $\Delta V(\rm RRL)[ref.]$ & Class &  $\log[ L_{\rm bol}$]   & $\log[N_{\rm Ly}]$ & Maser & clump  \\
   & ($\rm 10^{5}\,cm^{-3}$) & ($\rm pc$) & ($\rm 10^{8}\,pc\,cm^{-6}$) & ($\rm km\,s^{-1}$) &  & (L$_{\odot}$)    &  (s$^{-1} $)   & emission? & outflow?  \\
\hline
 \multicolumn{1}{c}{(1)}  & \multicolumn{1}{c}{(2)} & \multicolumn{1}{c}{(3)} & \multicolumn{1}{c}{(4)} & \multicolumn{1}{c}{(5)} & \multicolumn{1}{c}{(6)} & \multicolumn{1}{c}{(7)} & \multicolumn{1}{c}{(8)} & \multicolumn{1}{c}{(9)} & \multicolumn{1}{c}{(10)} \\ 
 \hline
 \hline
\multicolumn{10}{c}{\hchii\ and candidate \hchii\ identified in this work}\\
\hline
G010.4724+00.0275  & 1.43 & 0.022 & 4.52 &-  & HC?  & 5.7    & 48.11 & Yes  & Yes \\
G024.7898+00.0833$\dagger$  & 3.38 & 0.008 & 9.18 & 40 (H66$\alpha$)[1]& HC &  5.2    & 47.52  & Yes  & Yes \\
G028.2003-00.0494$\dagger$  & 1.41 & 0.027 & 5.35 & 74(H92$\alpha$)[2] & HC   &  5.1   &  48.37  & Yes &  Yes \\
G030.0096-00.2734  & 1.91 & 0.0043 & 1.56 & -& HC?  & 3.8   &  46.21 & Yes  & Yes  \\
G032.7441-00.0755  & 2.79 & 0.011 & 8.28 & 40.3(Hn$\alpha$)[3]& HC  &  5.0    & 47.69 & Yes  & Yes \\
G034.2573+00.1523  & 3.55 & 0.0046 & 5.82 &48.7(H42$\alpha$)[3]& HC   & 4.8   & 46.58   &   Yes & Yes \\
G034.2581+00.1533$\dagger$  & 3.01 & 0.0041 & 3.73 & 48.4(H76$\alpha$)[2]& HC   &  4.8    & 46.54  &  Yes & Yes   \\
G043.1657+00.0116$\dagger$  & 1.57 & 0.046 & 11.32 & 63.9(H66$\alpha$)[4]& HC    & 6.9   & 48.69 & Yes & Yes \\
G045.0712+00.1321$\dagger$  & 1.22 & 0.040 & 5.89 &40(H76$\alpha$)[5]& HC     &  5.7   & 48.73  & Yes  & Yes \\
G045.4656$+$00.0452  & 1.02 & 0.023 & 2.36 &47.8(H39$\alpha$)[3]& HC  & 5.0   &  47.88 &   Yes & Yes  \\
G061.4770$+$00.0892  & 4.45 & 0.0040 & 7.88 &-& HC?  & 5.1   &  46.81  & Yes  & Yes \\
G030.5887$-$00.0428$\star$   & 2.08 & 0.009 & 3.91 & 56.2(H40$\alpha$)[3] & HC  & 4.0   &  47.28  & Yes  & Yes \\ 
\hline
\multicolumn{10}{c}{Intermediate objects (\hchii\ $\rightarrow$ \uchii\ regions)}\\
\hline
G034.2572$+$00.1535   & 0.52 & 0.038 & 1.01 &22.8(H76$\alpha$)[2] & HC$-$UC   & 4.8   & 48.03   &   Yes & Yes \\
G045.0694$+$00.1323 & 0.74 & 0.026 & 1.40 &16.7(H92$\alpha$)[2]& HC$-$UC     &  5.7   & 47.76  & -  & Yes \\
G049.3666$-$00.3010 & 0.94 & 0.031 & 2.69 &34.5(H40$\alpha$)[3]& HC$-$UC     &  5.1   & 48.05  & Yes  & Yes \\
G051.6785$+$00.7193 & 0.64 & 0.026 & 1.07 &  - & HC$-$UC   &  5.0   & 47.68  & Yes  & Yes  \\
G060.8842$-$00.1286 & 0.93 & 0.012 & 1.08 &- & HC$-$UC     &  4.2   & 46.40  & Yes  & Yes  \\
G030.7197$-$00.0829$\star$   & 0.22 & 0.093 & 0.45 & 43.0(H40$\alpha$)[3] & HC$-$UC  & 4.7   &  48.44  & Yes  & Yes \\
G030.8662$+$00.1143$\star$   & 0.37 & 0.031 & 0.43 & 44.9(H39$\alpha$)[3] & HC$-$UC  & 4.1   &  47.46  & Yes  & Yes \\ 
G033.1328$-$00.0923$\star$   & 0.21 & 0.10 & 0.46 & 43.0(H39$\alpha$)[3] & HC$-$UC  & 5.0   &  48.54  & Yes  & Yes \\ 
\hline
\multicolumn{10}{c}{Previously identified \hchii\ regions not resolved in the current work}\\
\hline
G010.9584$+$00.0221$\ast$   & 0.36 & 0.029 & 0.38 &43.8(H92$\alpha$)[2]& HC &  4.0    & 47.35  & Yes  & Yes  \\
G035.5781$-$00.0305$\ast$   & 0.22 & 0.093 & 0.45 &50.0(H42$\alpha$)[3]& HC   & 5.3   &  48.36  &  Yes & Yes \\
G043.1652$+$00.0129$\ast$  & 0.88 &  0.053 & 4.15 &53.7(H66$\alpha$)[2]& HC    & 6.9   &  48.91 & Yes  &  Yes \\
G043.1665$+$00.0106$\ast$  & 0.24 & 0.22 &  1.22 &48.6(H66$\alpha$)[2]& HC  &  6.9   & 49.55  & Yes  & Yes \\
\hline
\end{tabular}
\begin{tablenotes}  
\item 
References: 1. \cite{Beltran2007AA}; 2, \cite{Sewilo2004ApJ,Sewilo2011ApJS}; 3, \cite{Kim2017AA602A}; 4, \cite{dePree1997ApJ}; 5, \cite{Keto2008ApJ672}; 
6, \citep{Zhang2014ApJ}. 
Notes: for G032.7441$-$00.0755, the RRL $\rm Hn\alpha$ indicates n=39,40,41,42.
Symbols $\dagger$ and $\ast$ indicate the known \hchii\ regions summarized in Table\,1 of \citet{Yang2019MNRAS4822681Y}. 
Symbol $\star$ represents the four \hii\ regions with data from the literature and archives. 
\end{tablenotes}
\label{tab:10hchii_candidates}
\end{table*}

\section{Classification and properties of the optically thick  \hii\ regions}
\label{sect_optically_thick_hii}


In Sect.\,\ref{sec_turnover_frequency} we identified 20 young optically thick \hii\ regions with turnover frequencies larger than 5\,GHz.  
As the turnover frequency of an \uchii\ region is $\sim$5\,GHz \citep{Kurtz2005IAUS}, the 20 optically thick \hii\ regions are very likely to be in the \hchii\ region stage or in an intermediate stage connecting the \hchii\ region and \uchii\ region stages. 
The quantitative criteria for \hchii\ regions, \uchii\ regions, and the intermediate objects between the two stages, as summarized from the literature \citep[e.g.,][]{Wood1989ApJS,Kurtz1994ApJS659K,Afflerbach1996ApJS106,Kurtz2005IAUS,Hoare2007prplconfH}, are presented
in Table \ref{tab_classify_hii}.

\begin{figure}[!htp]
 \centering
    \includegraphics[width = 0.45\textwidth]{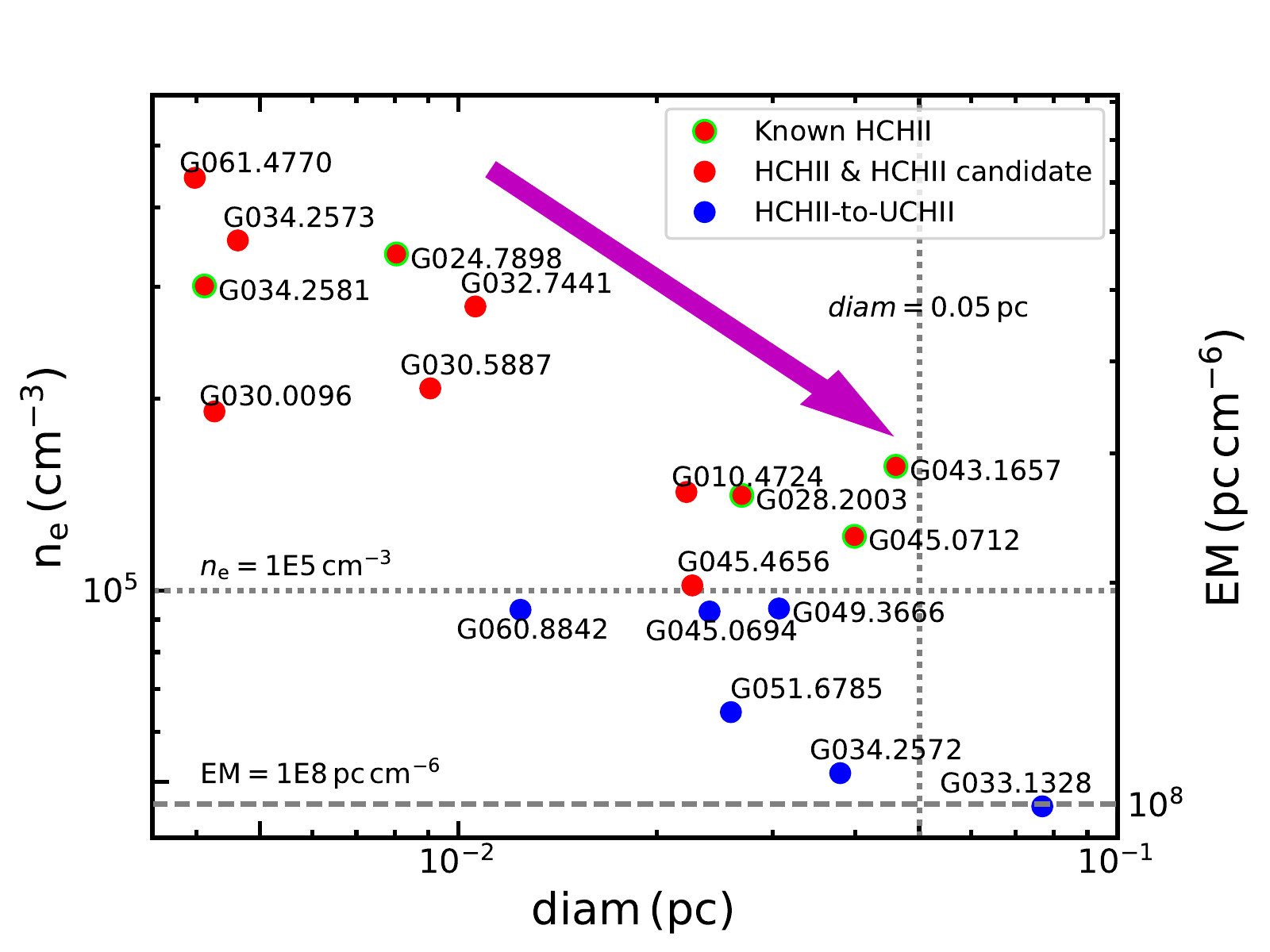}   
  \caption{Distribution of properties of 18 optically thick \hchii\ regions identified in Sect.\,\ref{sect_optically_thick_hii}. 
  The vertical and horizontal dotted and dashed lines indicate the standard quantitative criteria of an \hchii\ region. 
  The red filled circles show \hchii\ regions and \hchii\ region candidates identified in this work while the red filled circles with lime circles identify the previously known \hchii\ regions.  
  The filled blue circles show the intermediate objects between the \hchii\ and \uchii\ region stages.  
   The magenta arrow shows the evolutionary trend. 
  }
 \label{fig:classify_summary}
 \end{figure}

Among the 20 optically thick \hii\ regions, 7 sources are associated with previously identified \hchii\ regions that have been summarized in Table\,1 of  \citet{Yang2019MNRAS4822681Y}. 
In Fig.\,\ref{fig:classify_summary} we show the distribution of the $n_{\rm e}$, EM, and $diam$ of 18 optically thick \hii\ regions, as we excluded two objects (G043.1652 $\&$ G043.1665) in the optically thick sample that are associated with unrecovered \hchii\ regions listed in Table\,\ref{tab:10hchii_candidates} and  marked with an asterisk (see Sect. \ref{hchii_not_resolved}).
On this plot, we indicate the region of parameter space where \hchii\ regions are expected to reside (i.e., $n_{\rm e} > 10^5$\,cm$^{-3}$ and $diam < 0.05$\,pc), and we show the evolutionary trend from \hchii\ region to the stage between \hchii\ region and \uchii\ region in the physical parameter space. Of the optically thick \hii\ regions, 14 satisfy these criteria. 
The remaining sources all satisfy the size criterion for \hchii\ regions but their electron densities are too low and so these are considered to be intermediate between the \hchii\ and \uchii\ region stages.

\begin{figure*}
 \centering
  \begin{tabular}{cc}
    \includegraphics[width = 0.42\textwidth]{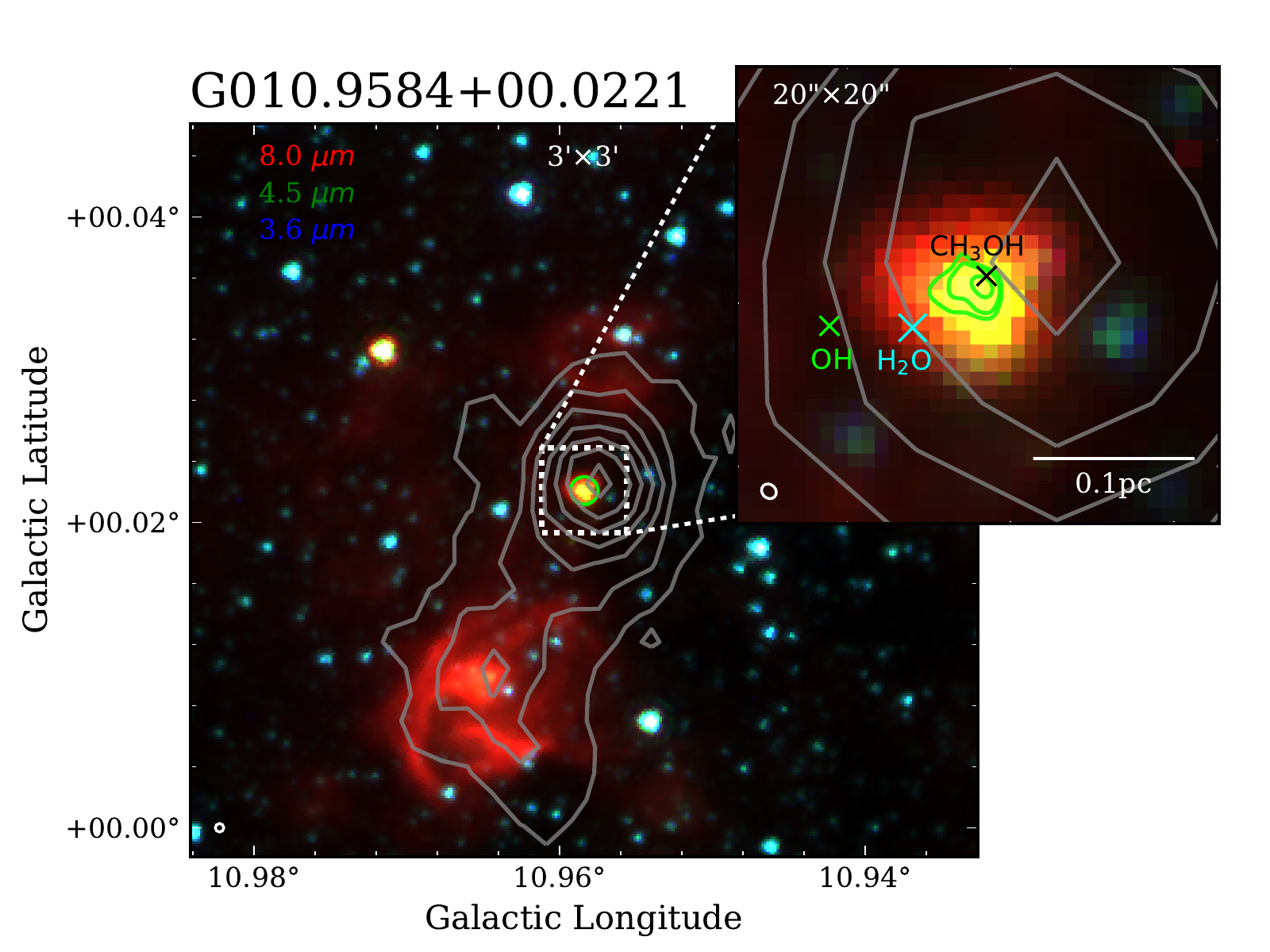}   &    \includegraphics[width = 0.42\textwidth]{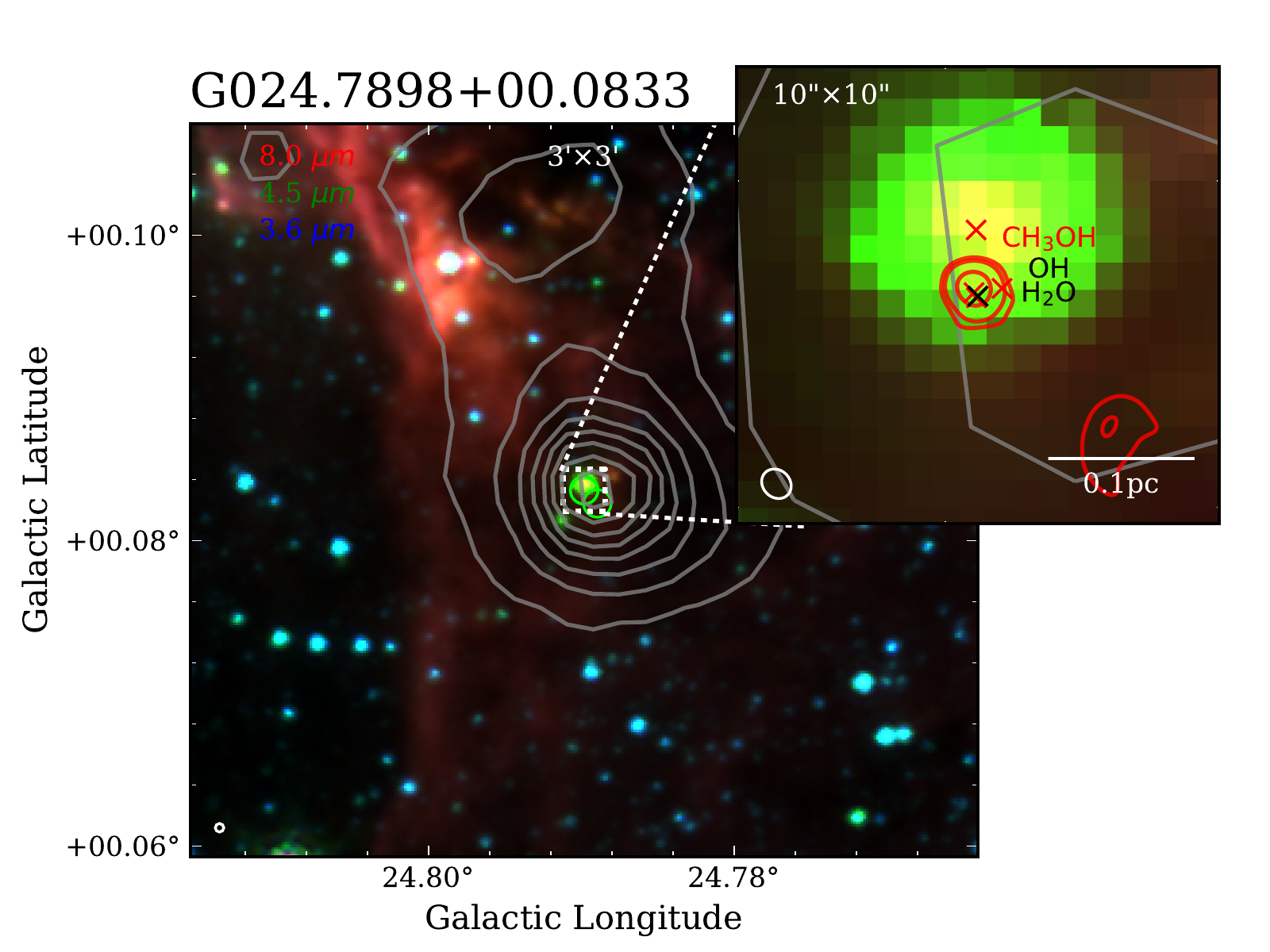}   \\
    \includegraphics[width = 0.42\textwidth]{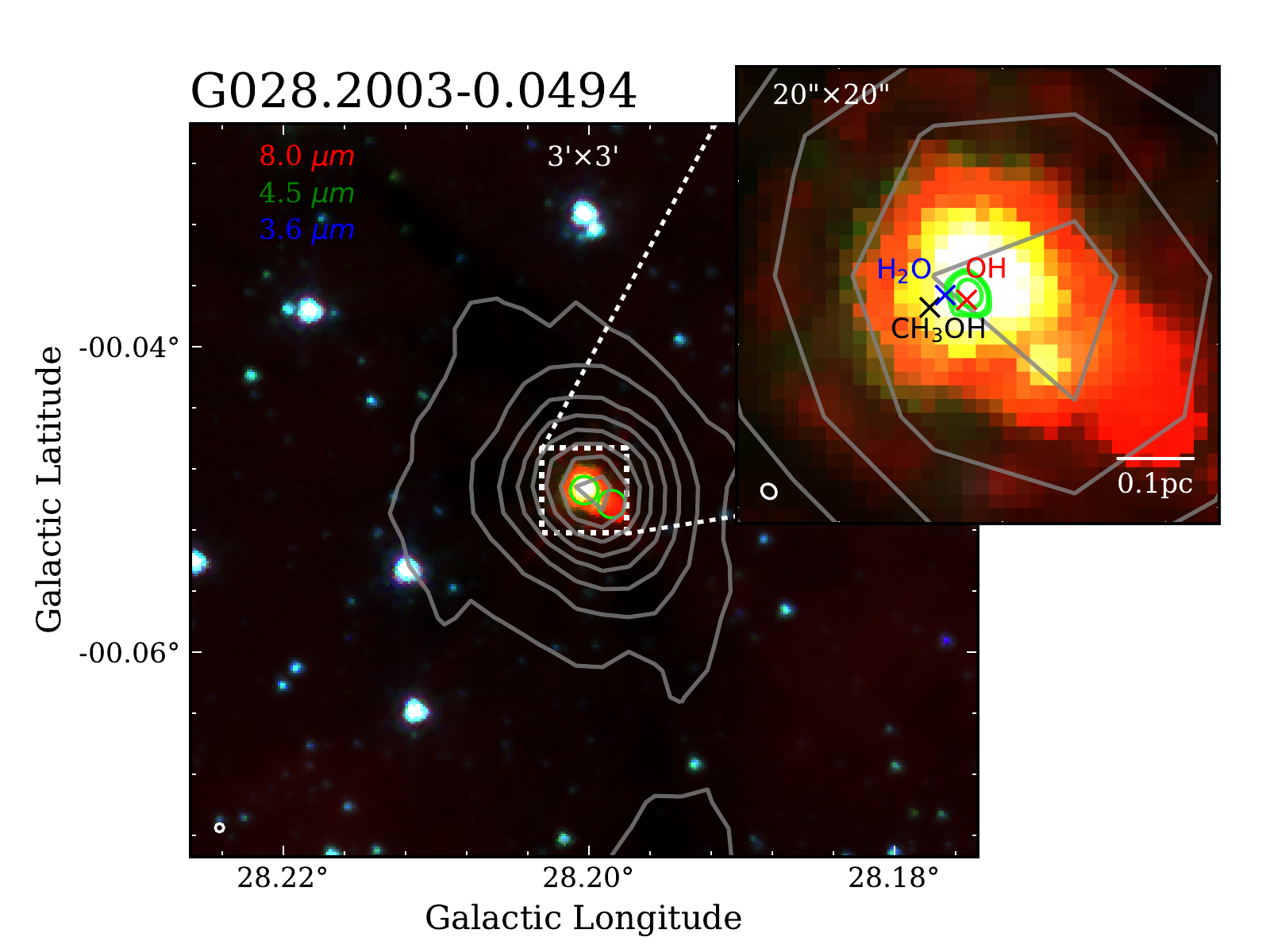}   &    \includegraphics[width = 0.42\textwidth]{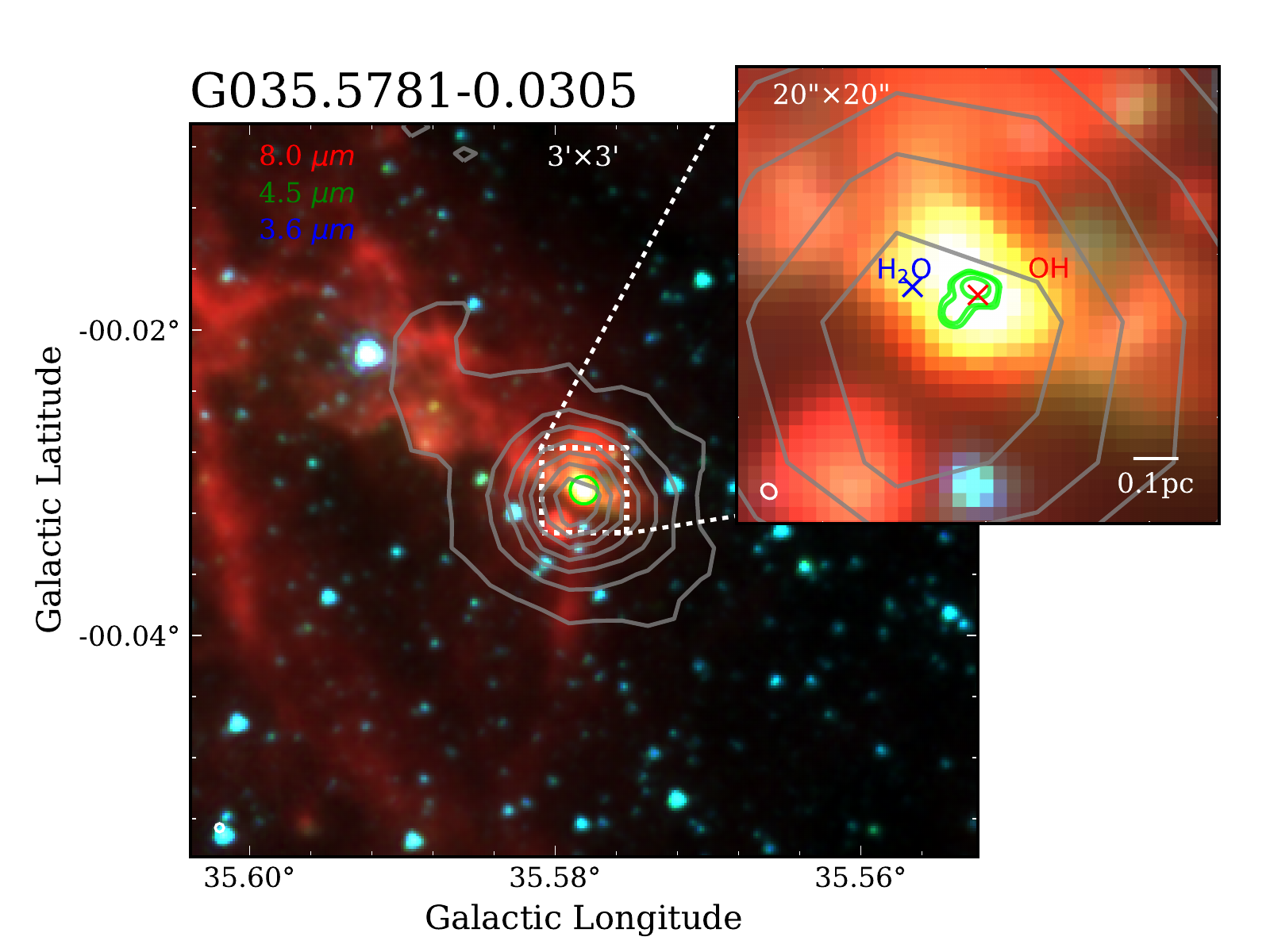}   \\
    \includegraphics[width = 0.42\textwidth]{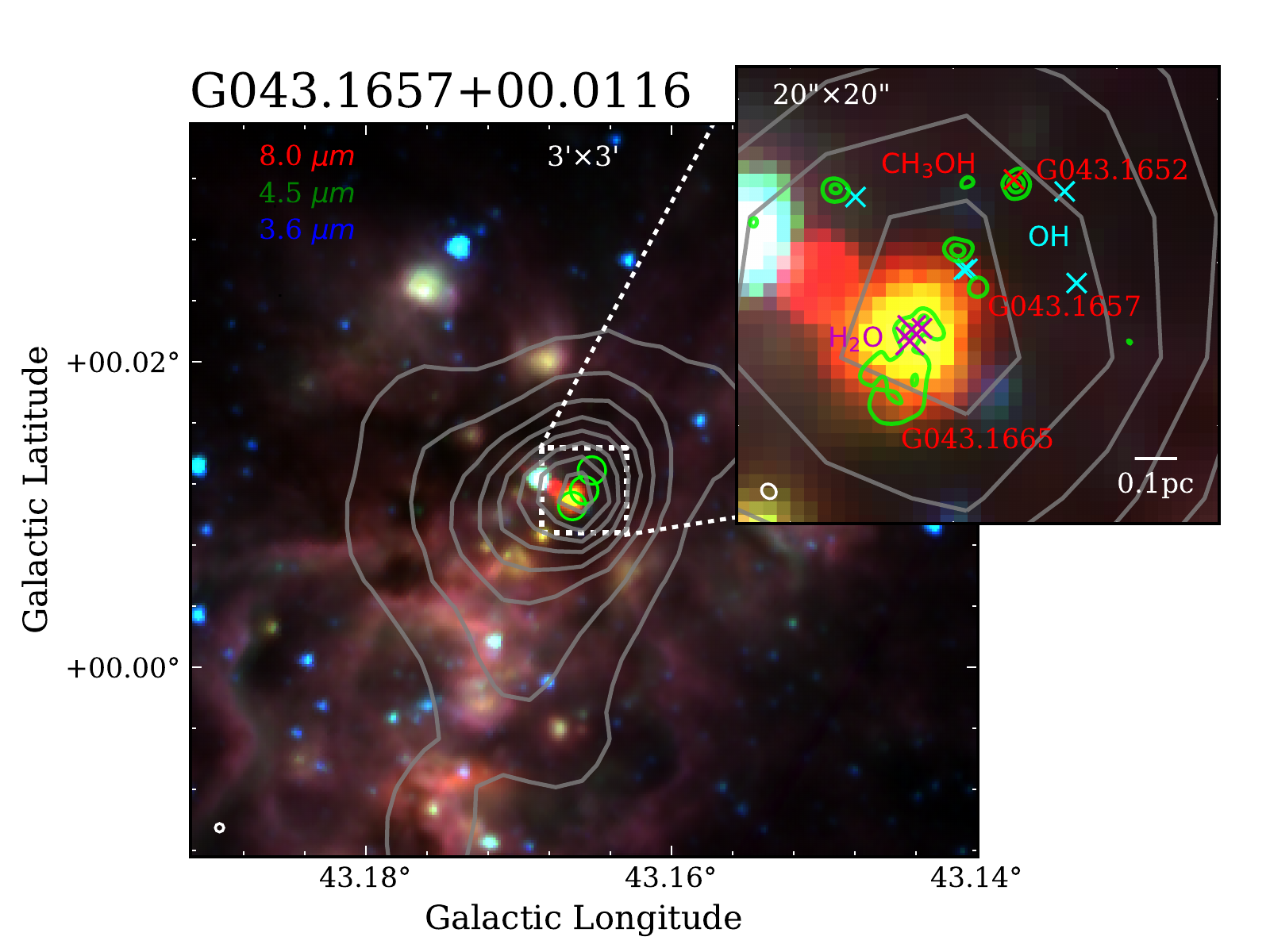}   &    \includegraphics[width = 0.42\textwidth]{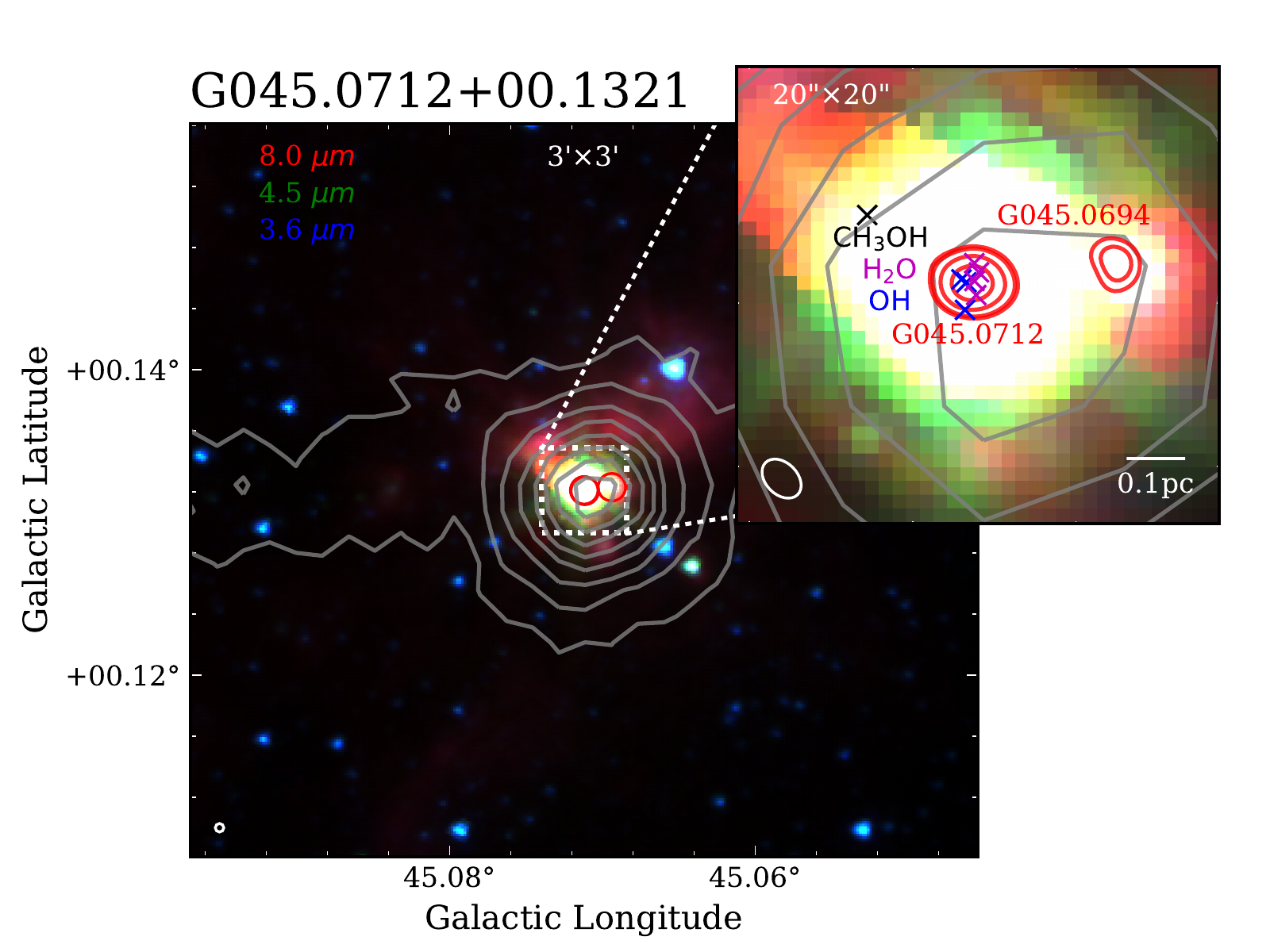}   \\
       \end{tabular}
  \caption{Three-color composition image (or RGB image) from $\emph{Spitzer}$ GLIMPSE $8\,\mu m$ (red), $4.5\,\mu m$ (green), and $3.6\,\mu m$ (blue)
bands \citep{Benjamin2003PASP,Churchwell2009PASP} for the \hchii\ regions discussed in Sect.\,\ref{sect_optically_thick_hii}. 
Lime or red circles show the radio sources in the field from the CORNISH survey. The upper-right zoomed-in images for each panel show the peak position of $\rm H_{2}O$ maser (magenta cross) and OH maser (black cross), and the linear scale-bar of 0.1\,pc in white. Gray contours in the image show 870\,$\mu m$ emission from ATLASGAL \citep{Schuller2009AA}, and the lime (or red) contours show K-band 22\,GHz emission presented in this work. 
 The red contours in the bottom-right panel show X-band 10 GHz emission as the K-band emission is missing for source G045.0694.  
The FWHM beam of GLIMPSE (2\arcsec) and K-band observations are indicated by the white circles shown in the lower-left corner of each image.
  }
 \label{fig:optically_thick_hiis_1}
 \end{figure*}

\begin{figure*}
 \centering
  \begin{tabular}{cc}
    \includegraphics[width = 0.42\textwidth]{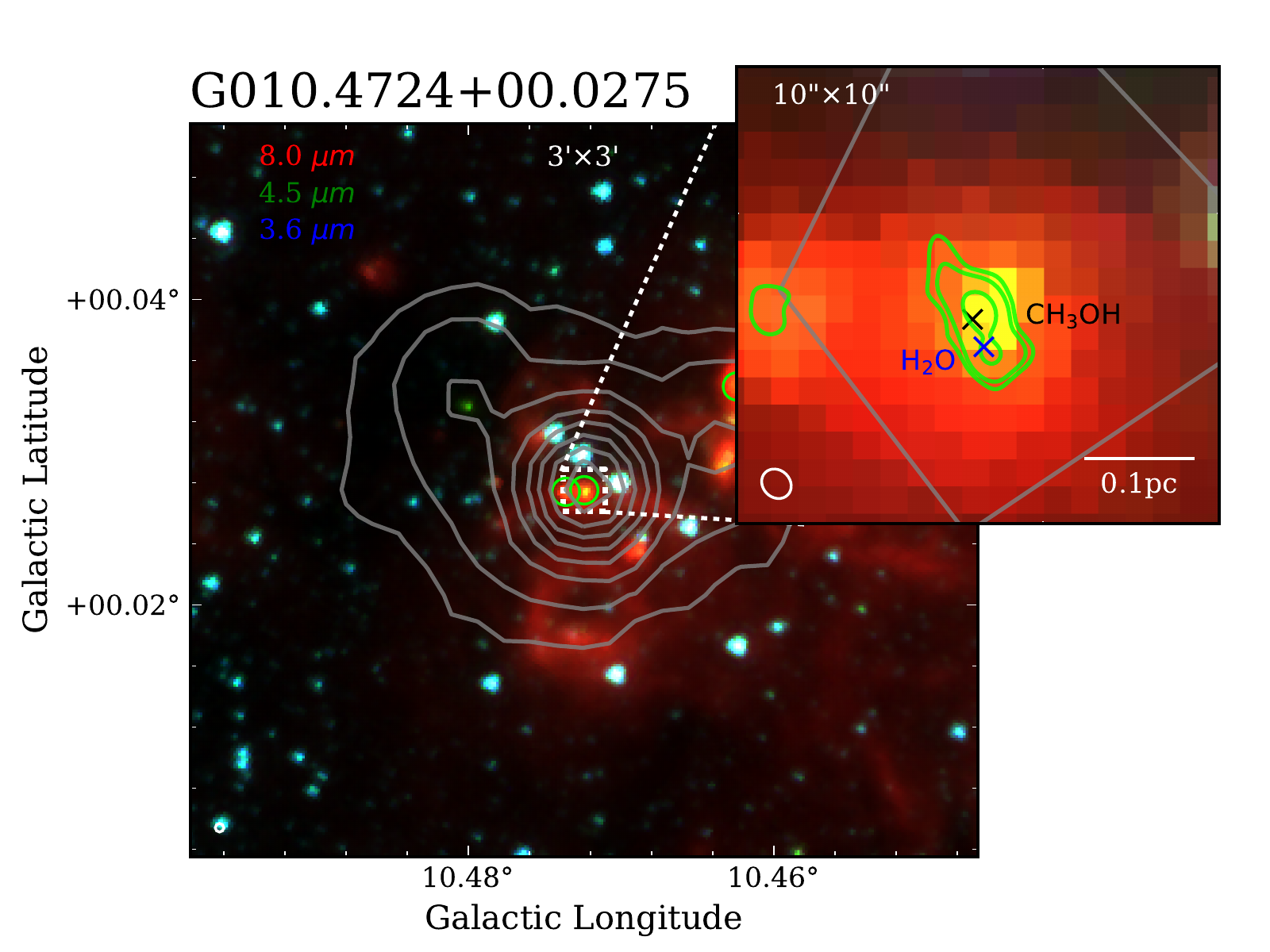}   &    
    \includegraphics[width = 0.42\textwidth]{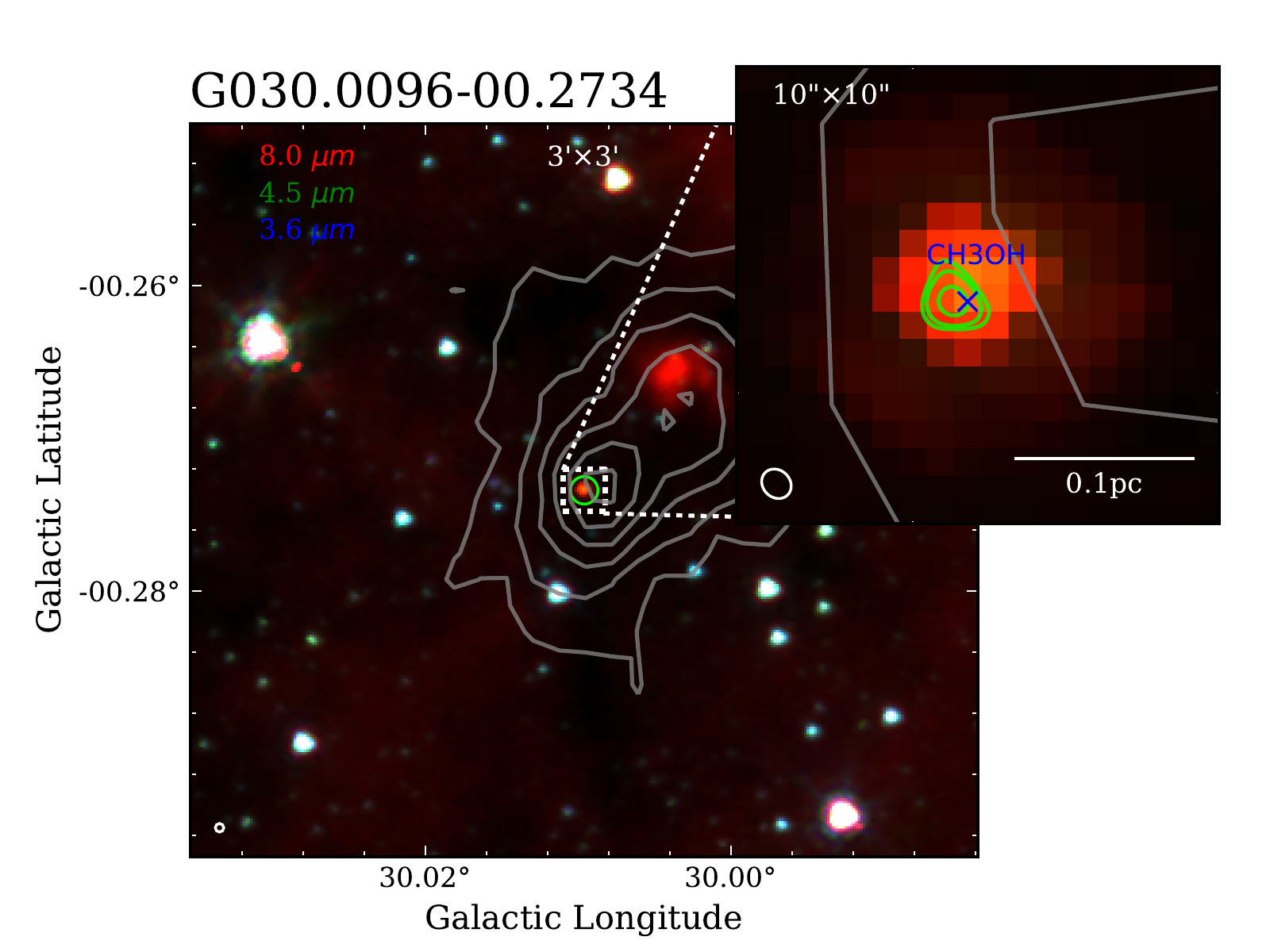}   \\
    \includegraphics[width = 0.42\textwidth]{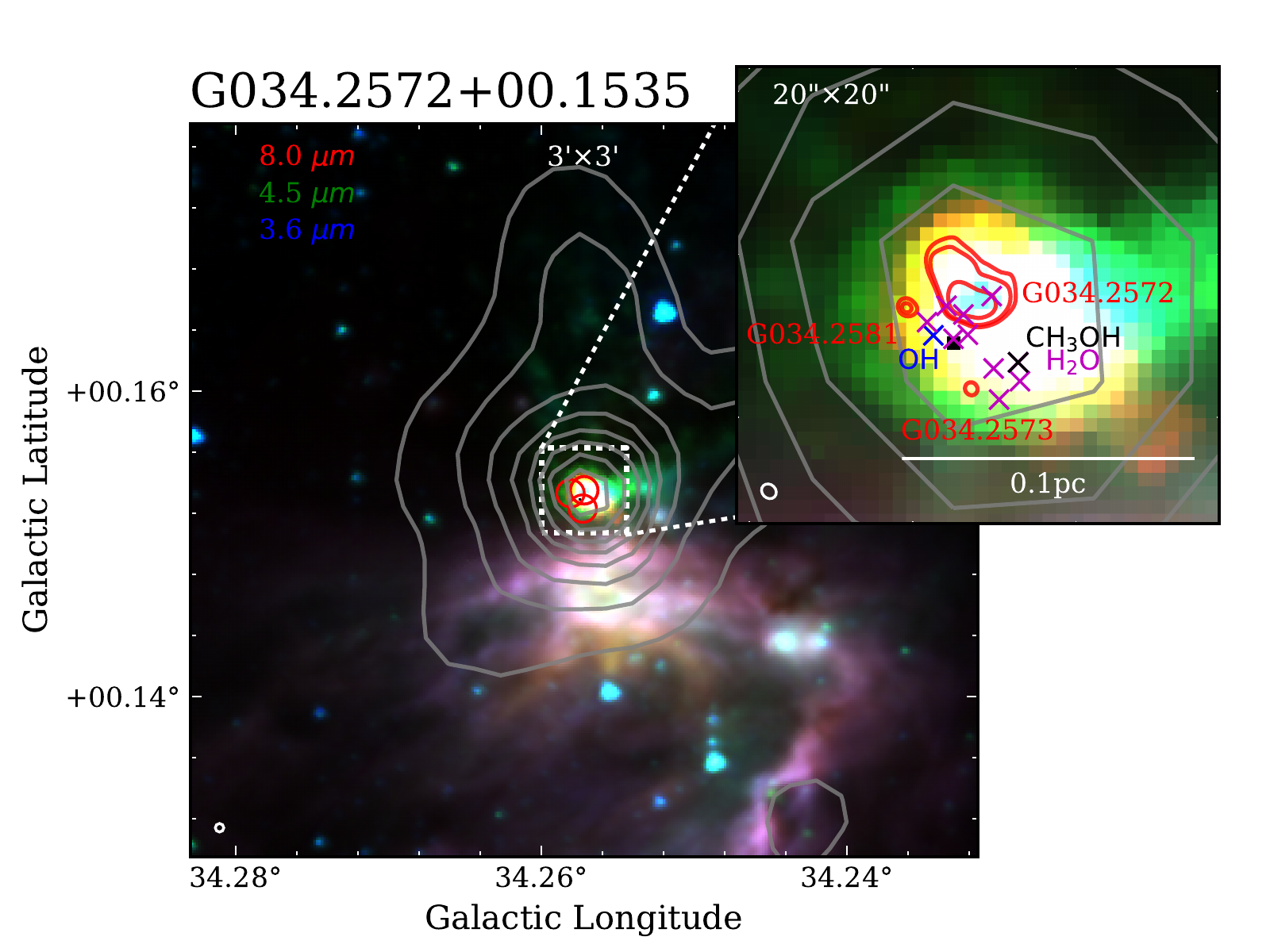}   &    
    \includegraphics[width = 0.42\textwidth]{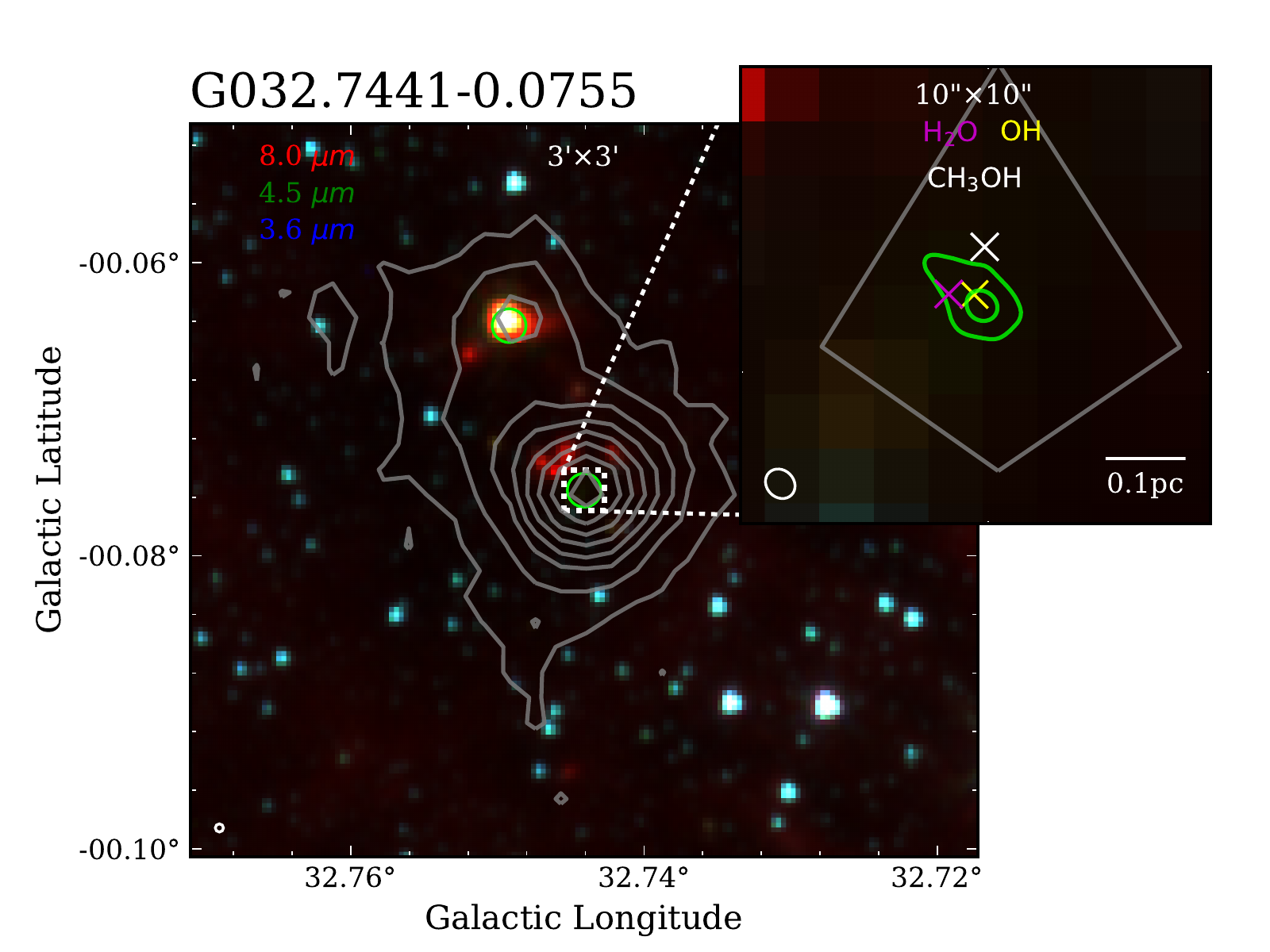}   \\
    \includegraphics[width = 0.42\textwidth]{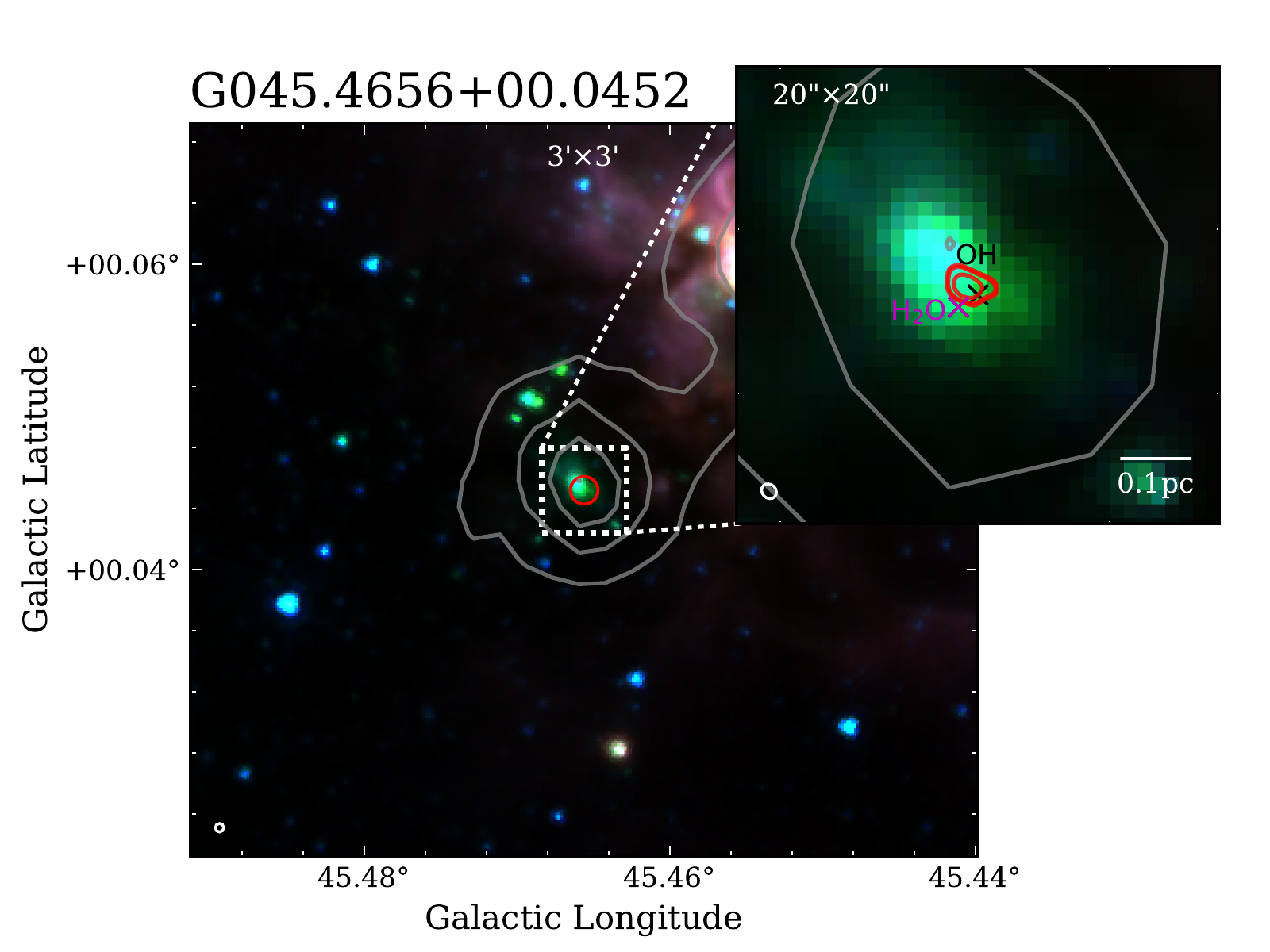}   &    
    \includegraphics[width = 0.42\textwidth]{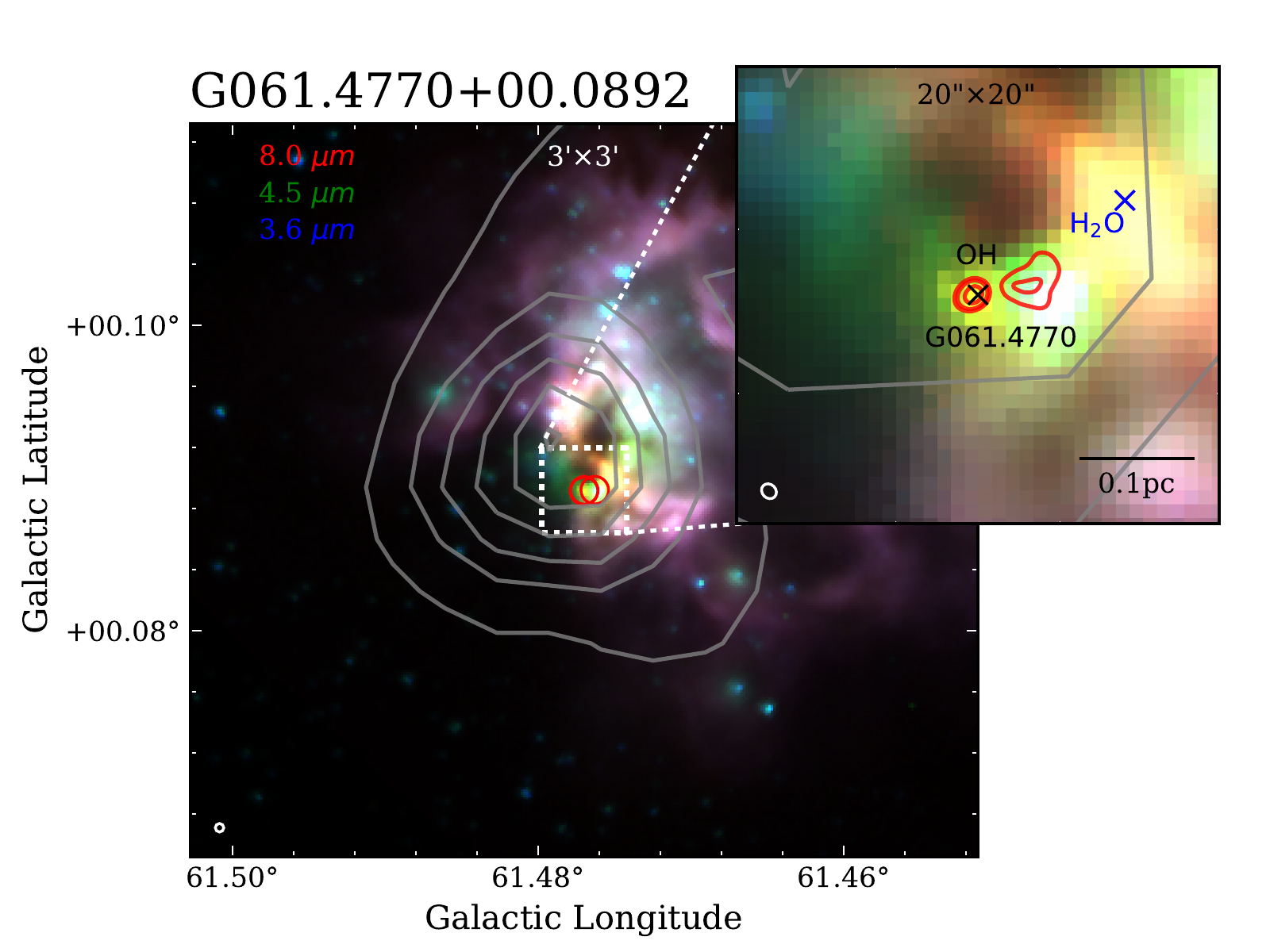}   \\
    \end{tabular}
  \caption{ As described in Fig.\,\ref{fig:optically_thick_hiis_1}, except in this case the sources include newly identified \hchii\ regions and intermediate objects. 
  The gray contours in the image of G061.4770$+$00.0892 show the 500$\mu $m emission from Hi$-$GAL.
  }

 \label{fig:optically_thick_hiis_2}
 \end{figure*}

In Figure\,\ref{fig:optically_thick_hiis_1}-\ref{fig:optically_thick_hiis_3}, we present three-color infrared maps of each \hii\ region. 
In these maps, we include contours of the dust and radio emission and any coincident masers so that we can investigate their environments and associations with other star-formation tracers. 
We individually discuss the properties of the optically thick \hii\ regions with respect to their environment, their association with dense gas,  and star-formation tracers in the following sections, and we follow the order that is presented in Table\,\ref{tab:10hchii_candidates}.

\subsection{\hchii\ regions and candidate \hchii\ regions identified in this work}

\paragraph{G010.4724$+$00.0275:}
\label{G010p4724}

This source is located in the G10.47$+$0.03 complex region that hosts three \uchii\ regions \citep{Wood1989ApJS}, water masers \citep{Hofner1996AAS120283H}, 6.7\,GHz methanol masers \citep{Pestalozzi2005AA432737P}, various complex molecules \citep{Hatchell1998AAS13329H}, and massive molecular outflows along the NE--SW direction \citep{Sepulcre2009AA499811L}. 
This object is resolved into two compact sources, G10.47$+$0.03A and G10.47$+$0.03B, in \citet{Wood1989ApJS} with a resolution of 0.4\arcsec, which is also seen in the K-band emission shown as contours in the upper-left panel of Fig.\,\ref{fig:optically_thick_hiis_2} with two blended compact components. The radio source is positionally coincident with methanol and water masers, a bright mid-infrared point source and is embedded in a dense molecular clump as traced by the ATLASGAL emission, and therefore clearly associated with star formation activity. 
Its physical properties such as $n_{\rm e} =\rm 1.43\times10^{5}\,cm^{-3}$, $diam=\rm 0.022\,pc$, $\rm EM=4.52\times10^{8}\,pc\,cm^{-6}$, and $\log N_{\rm Ly}\rm  = 48.11$, imply that it is likely an \hchii\ region. 
Its natal clump has a mass of  $2.57\times10^{4}\,M_{\odot}$ and a bolometric luminosity of 5.0 $\times10^{5}\,L_{\odot}$  \citep{Urquhart2018MNRAS4731059U}. 
Its spectral type of O5.5 derived from the bolometric luminosity is earlier than O9 derived from Lyman continuum flux, which supports the hypothesis that this source is located in a cluster, as reported in  \citet{Pascucci2004AA426523P}.


\paragraph{G024.7898$+$0.0833:} 
\label{G024p7898}

This source is an \hchii\ region identified by  \citet{Beltran2007AA}, which is found to be associated with many CH$_{3}$OH masers \citep{Surcis2015AA578A102S,Bartkiewicz2016AA587A104B} and OH masers \citep{Forster2000ApJ371F,Caswell2013MNRAS4311180C}, $\rm H_{2}O$ masers \citep{Caswell1983AuJPh36443C,Forster2000ApJ371F}, and outflows traced by CO \citep{Furuya2002AA390L1F,Beltr2011AA532A91B} and SiO \citep{Codella2013AA550A81C}. 
Its physical properties such as $n_{\rm e}$,  $diam$, and EM (Table \ref{tab:10hchii_candidates}) are consistent with previous results \citep{Beltran2007AA,Cesaroni2019AA624A100C}.
Its natal clump has a mass of  $7.64\times10^{3}\,M_{\odot}$ and a bolometric luminosity of $1.58\times10^{5}\,L_{\odot}$  \citep{Urquhart2018MNRAS4731059U}. 
The spectral type of this \hchii\ region O6.5 derived from the infrared luminosity (Table \ref{tab_obser_list}) is much earlier than O9.5 derived from the Lyman continuum flux which includes contributions from the nearby \uchii\ region G024.7889$+$00.0824 in the field. 
One possible explanation for the discrepancy of spectral type is that this source is located in a cluster and/or a significant amount of Lyman continuum photons are absorbed by the surrounding dust, with an upper limit on the dust absorption fraction of $f_{\rm d}$= 92\% (see Sects. \ref{sec_lyc} and \ref{sect_dust}). 
As this \hchii\ region shows extended 4.5 $\mu$m emission, it is associated with an extended green object as defined by \citet{Cyganowski2008AJ1362391C}.


\paragraph{G028.2003$-$0.0494:}
\label{G028p2003}

This source is a known \hchii\ region identified by \citet{Sewilo2004ApJ}, which is found to be associated with the 37.7 GHz $\rm CH_{3}OH$ maser \citep{Ellingsen2011ApJ109E}, OH masers \citep{Argon2000ApJS,Caswell2013MNRAS4311180C}, and $\rm H_{2}O$ masers \citep{Urquhart2011MNRAS4181689U}. 
Its physical properties such as $n_{\rm e}$,  $diam$, and EM listed in Table \ref{tab:10hchii_candidates} are consistent with previous results  \citep{Sewilo2011ApJS}.
 Its natal clump has a mass of $4.45\times10^{3}\,M_{\odot}$ and a bolometric luminosity of $1.30\times10^{5}\,L_{\odot}$  \citep{Urquhart2018MNRAS4731059U}, which is associated with molecular outflows \citep{Maud2015MNRAS453645M,Yang2018ApJS2353Y}. 
 Its spectral type O6.5 derived from the bolometric luminosity is earlier than O7.5 derived from the Lyman continuum flux that includes the contribution from its nearby \uchii\ region G028.1985$-$00.0503 with $N_{\rm Ly}=5.0\times10^{47}$.  
 This could be the result of this source being located in a cluster, as shown in the middle-left panel of Fig.\,\ref{fig:optically_thick_hiis_1}, or could be due to the fact that about 43\% of the Lyman continuum photons are absorbed by the surrounding dust.


\paragraph{G030.0096$-$00.2734: }
\label{G030p0096}

This compact radio source, located in the W43 star-forming complex  \citep[e.g.,][]{Blum1999AJ1171392B,Medina2019AA627A175M,Gao2019AA623A105G}, is the first of the sample that was found to be associated with an infrared dark cloud (G030.01$-$0.27; \citealt{Battersby2011AA535A128B}), 
which itself is associated with many molecular lines \citep{Schlingman2011ApJS19514S} as well as methanol masers \citep{Breen2015MNRAS4109B}. 
Its natal clump,  AGAL030.008$-$0.272, is associated with a molecular outflow identified by \citet{Yang2018ApJS2353Y}, which has a maximum outflow velocity of $\rm 4.5\,km\,s^{-1}$. 
It is the only radio source in its natal clump, and its spectral type B1 derived from the bolometric luminosity is consistent with B0.5 derived from the Lyman continuum photons, indicating a lack of dust within this \hii\ region. 
The radio emission is coincident with a compact mid-infrared point source confirming it is associated with an embedded protostellar object.
The physical properties of G030.0096$-$00.2734 are consistent with this source being an \hchii\ region at a very early evolutionary stage.

\paragraph{G030.5887$-$00.0428: }
\label{G030p5887}

\begin{figure*}
 \centering
  \begin{tabular}{cc}
    \includegraphics[width = 0.42\textwidth]{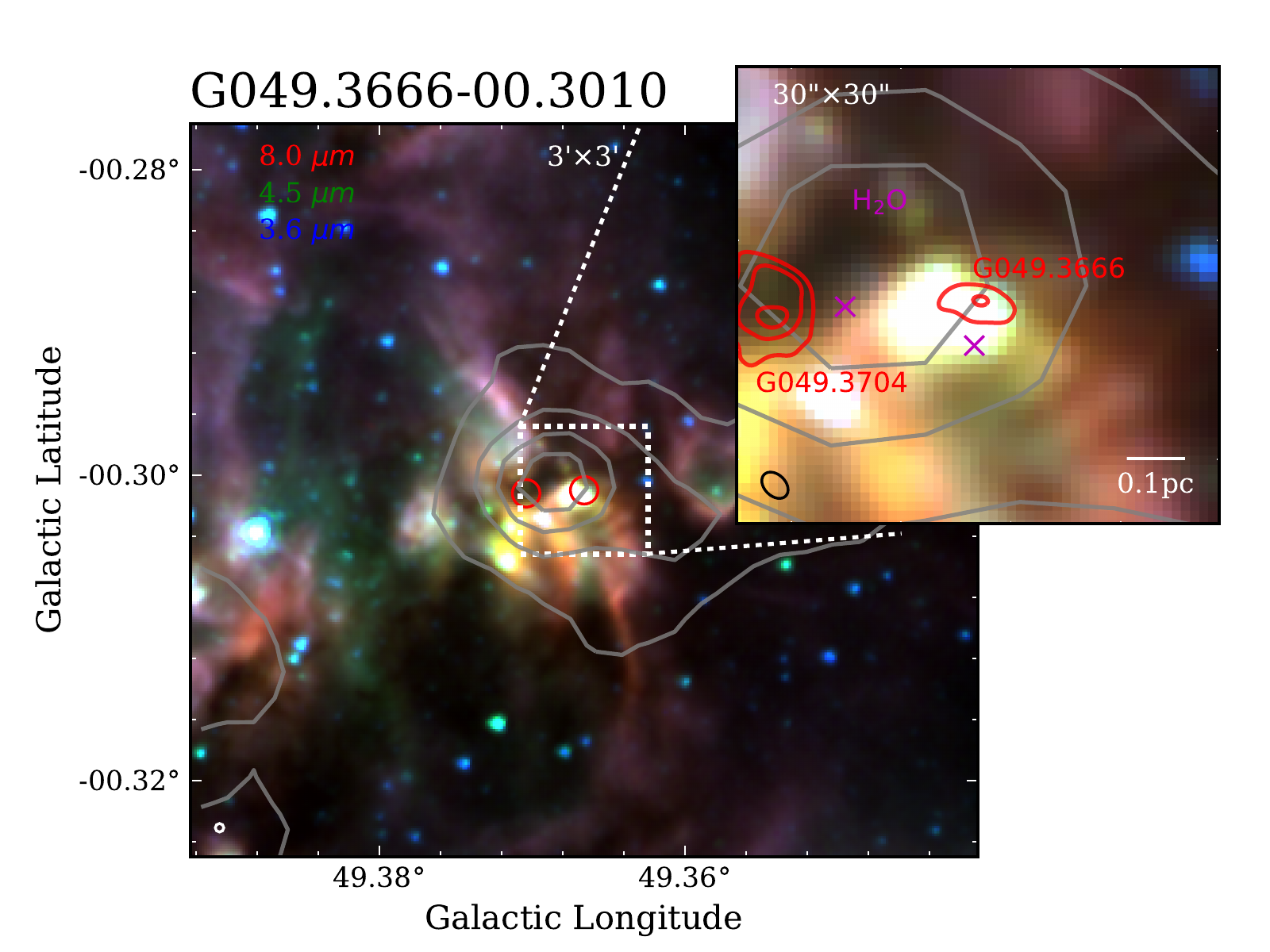}   &    
    \includegraphics[width = 0.42\textwidth]{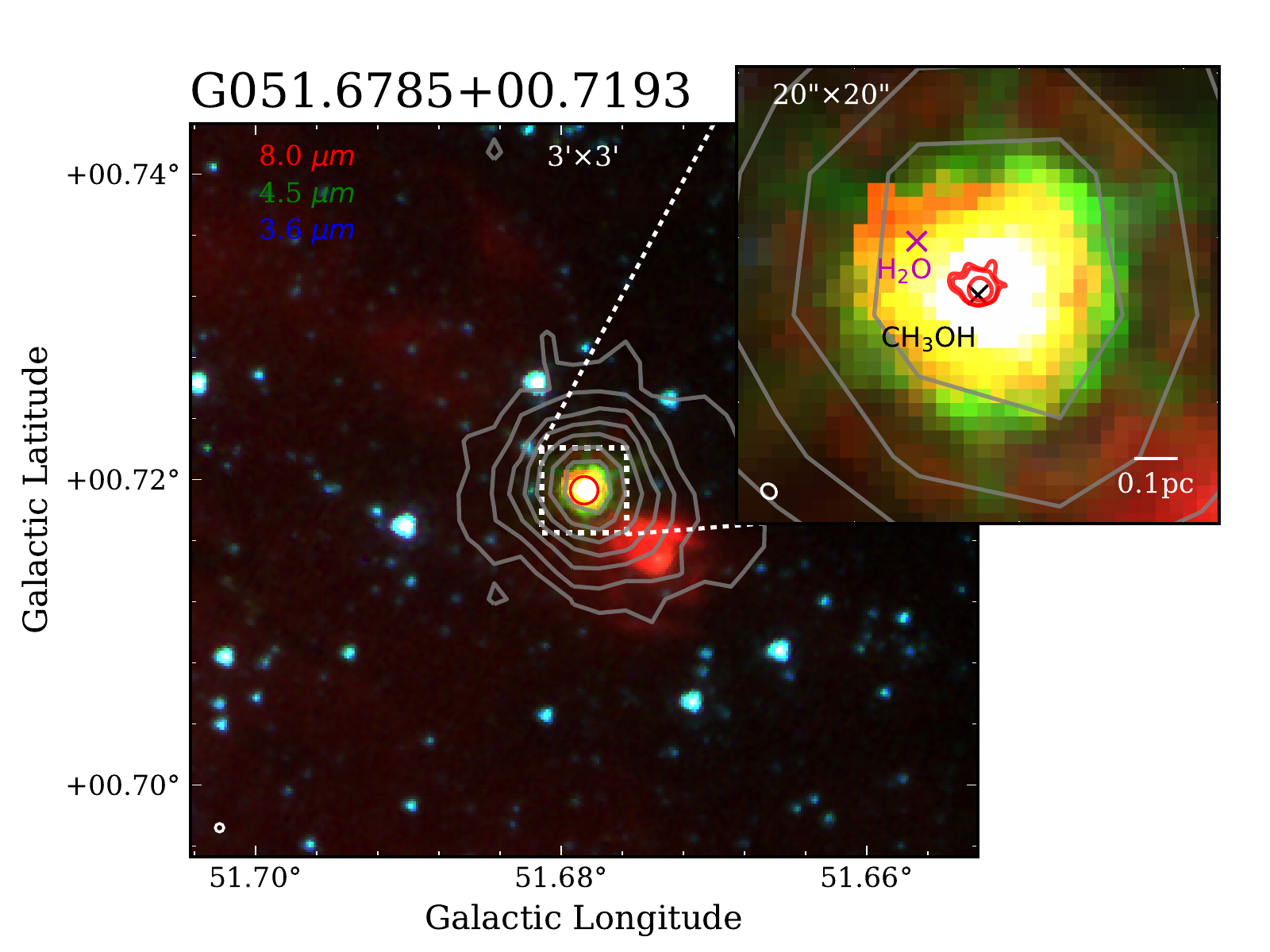}   \\
     \includegraphics[width = 0.42\textwidth]{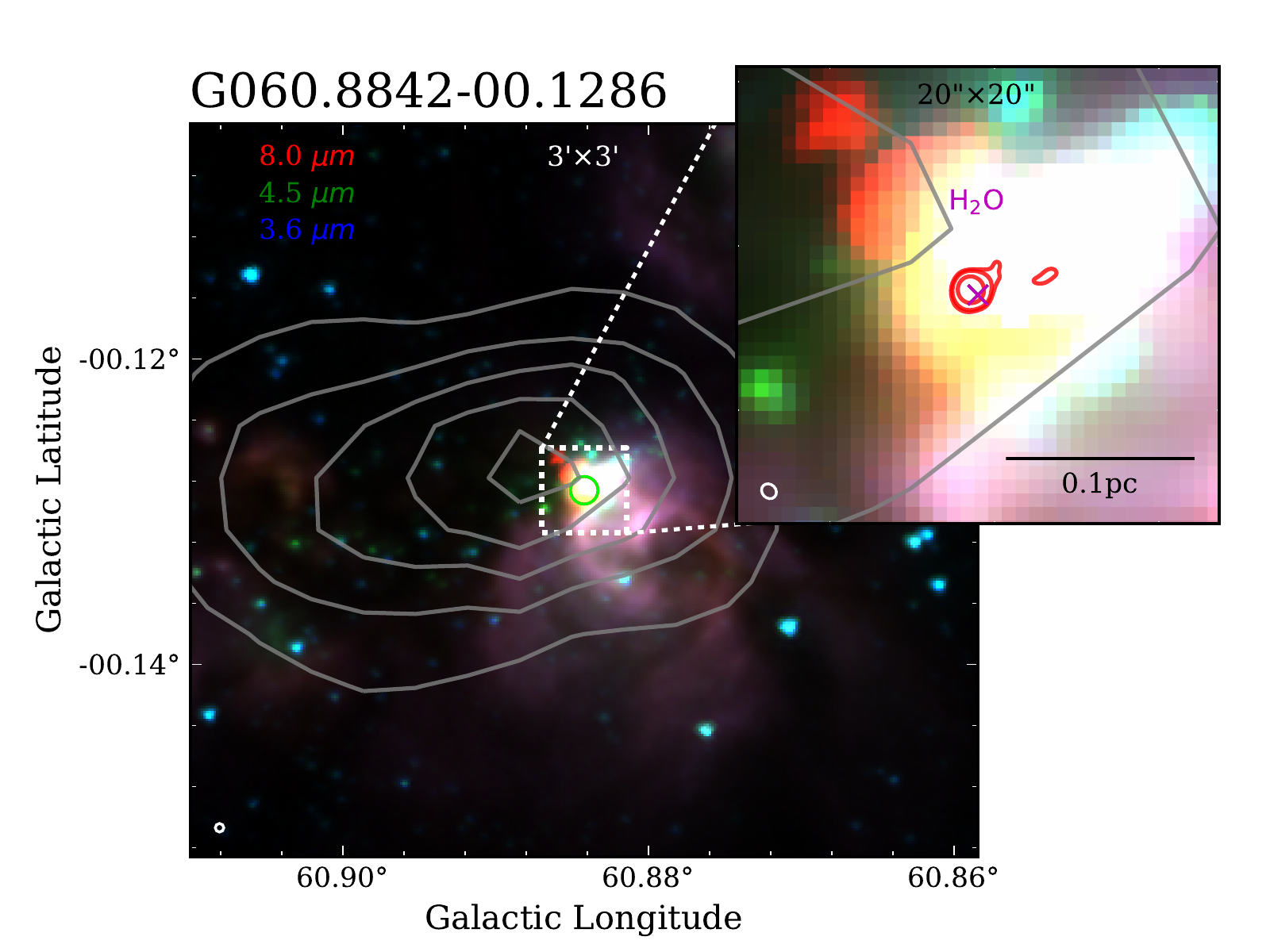}   &    
     \includegraphics[width = 0.42\textwidth]{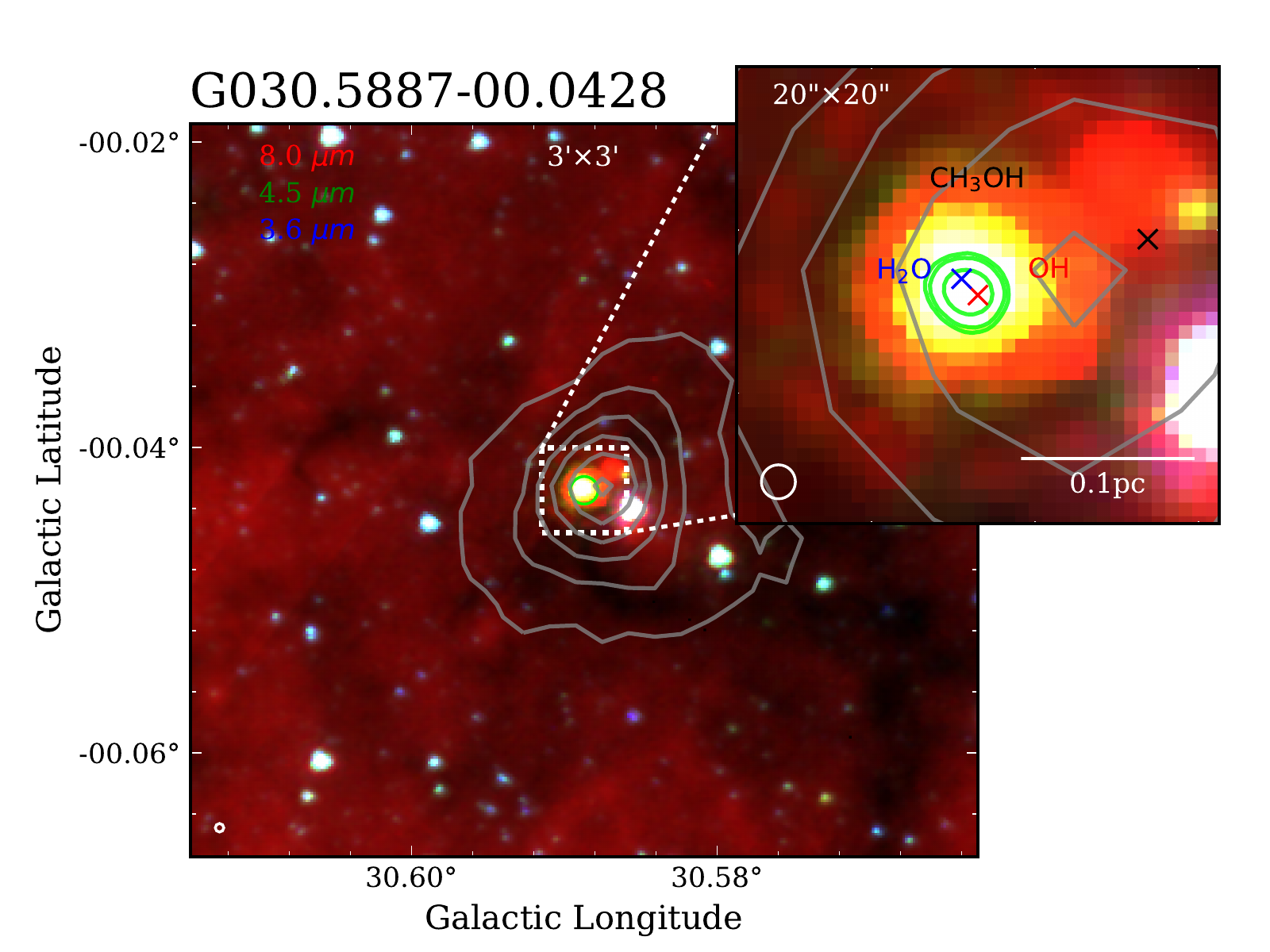} 
     \\
     \includegraphics[width = 0.42\textwidth]{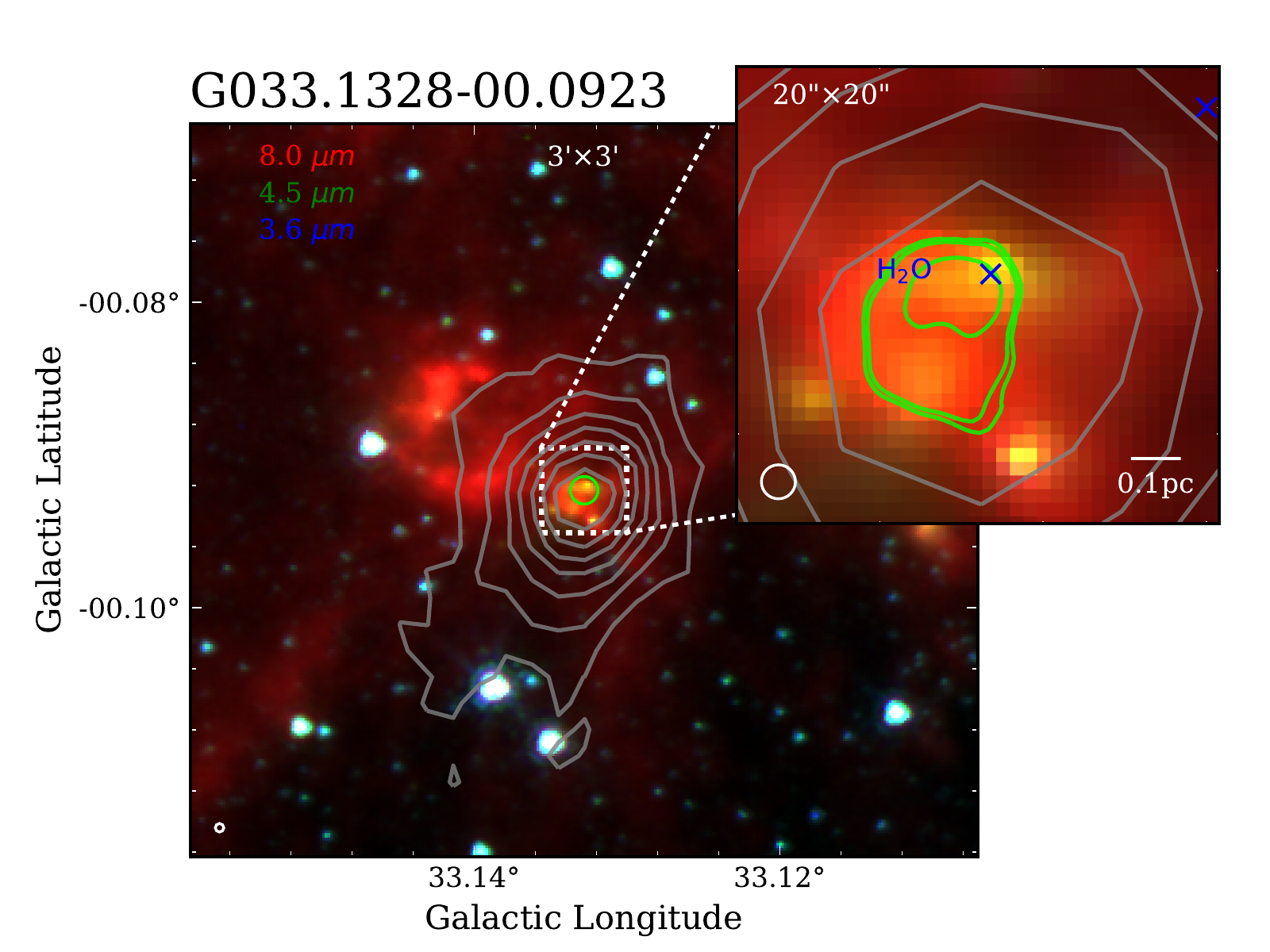}   &        
     \includegraphics[width = 0.42\textwidth]{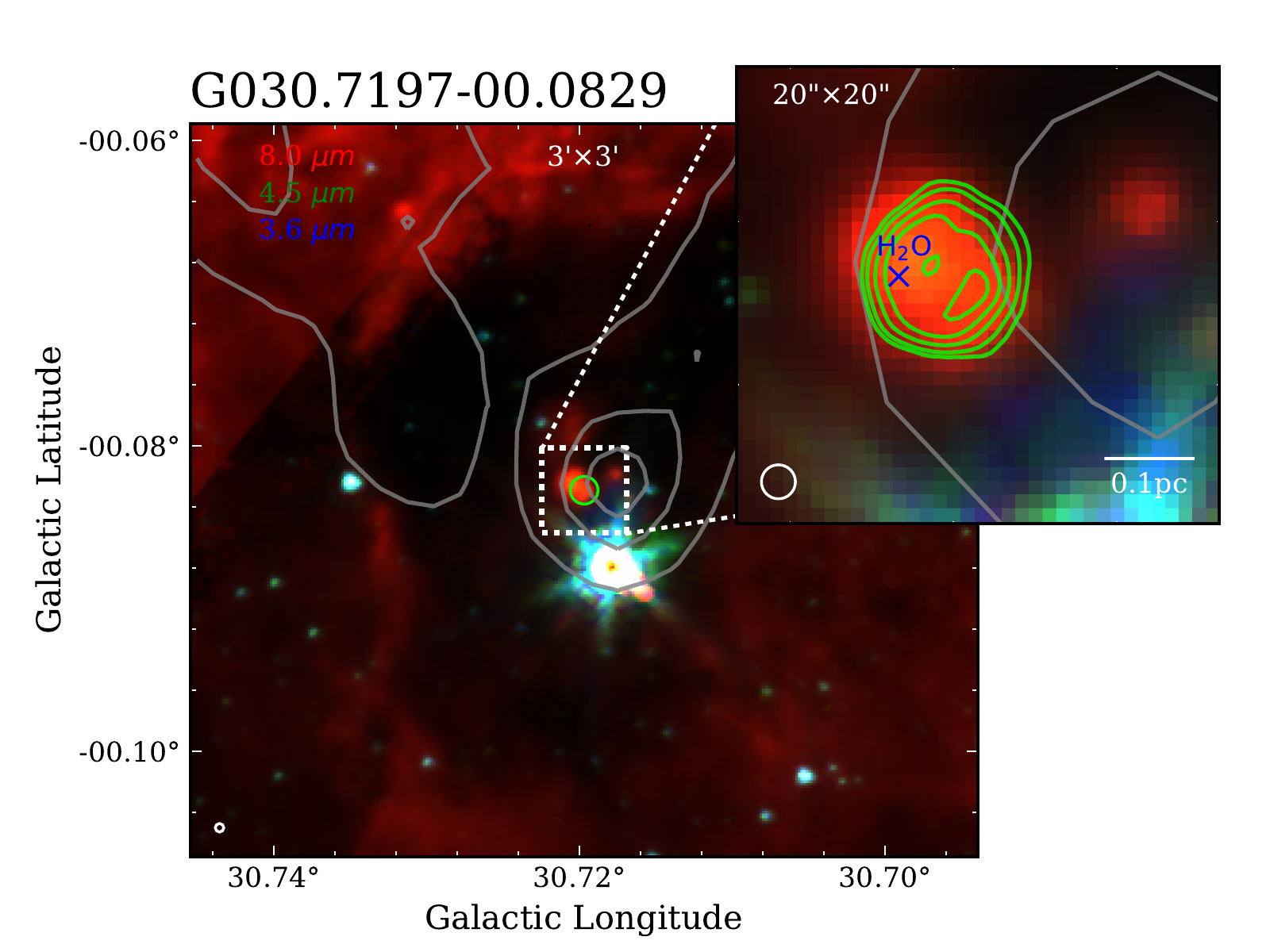}   \\
     \includegraphics[width = 0.42\textwidth]{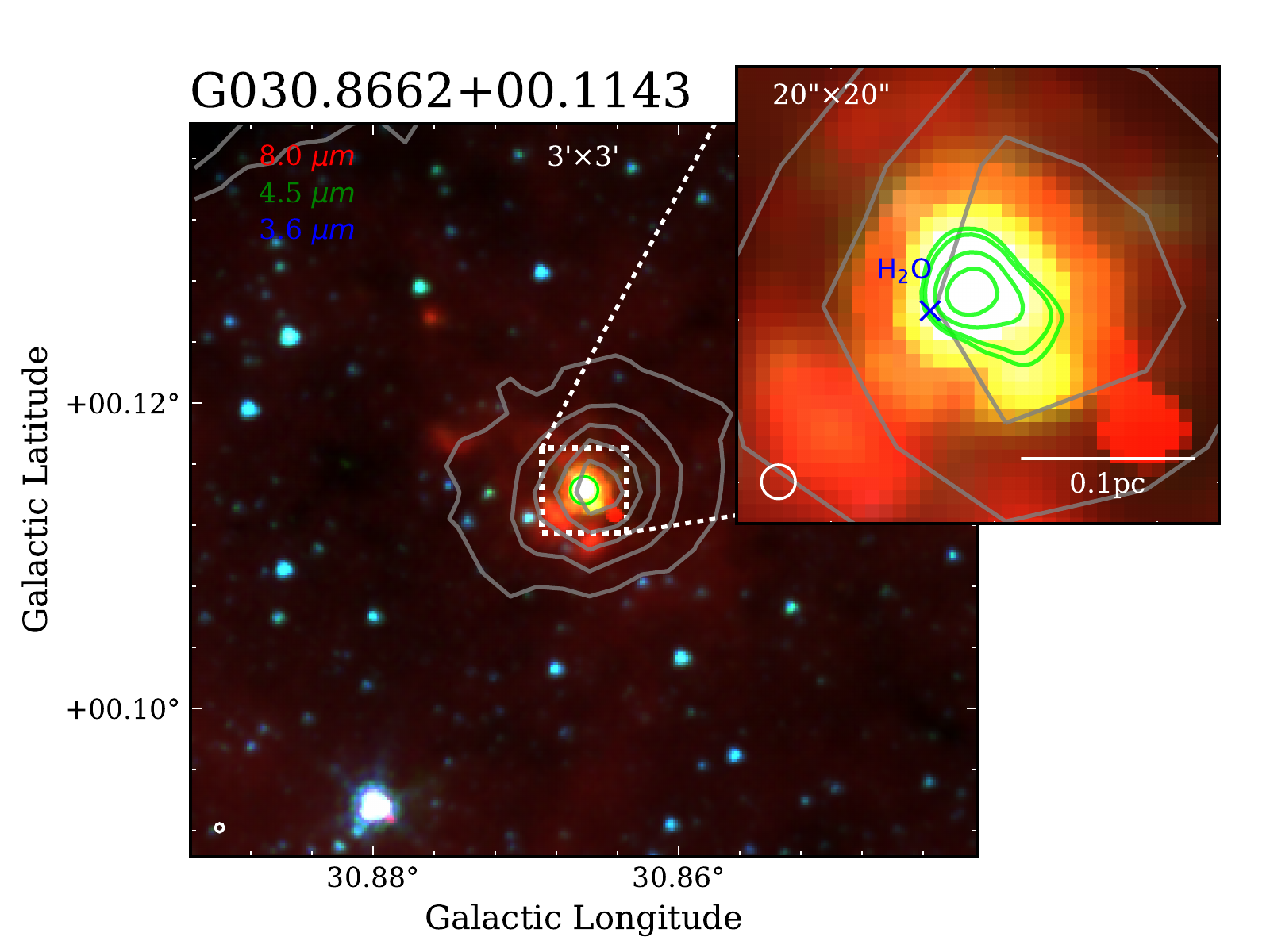}
     & \\
       \end{tabular} 
  \caption{As described in Fig.\,\ref{fig:optically_thick_hiis_1}, except in this case the sources are \hchii\ regions and intermediate objects between \hchii\ and \uchii\ regions. The gray contours in the image of G060.8842$-$00.1286 show the 500 $\mu$m emission from Hi$-$GAL. 
  The lime contours in the images of G030.7197$-$00.0829, G030.8662$+$00.1143, G030.5887$-$00.0428 and G33.1328$-$00.0923 show the 5 GHz emission from CORNISH survey.
  }
 \label{fig:optically_thick_hiis_3}
 \end{figure*}

This source shows compact radio emission at 5 GHz CORNISH, as shown in the middle-right panel of Fig. \ref{fig:optically_thick_hiis_3}. 
Its flux densities at high frequencies were obtained in project VLA18A-066, with 217.70\,mJy at 15.5\,GHz and 223.53\,mJy at 16.5\,GHz. 
With flux densities at low frequency of 1.4 GHz and 5 GHz  \citep[summarized in ][]{Yang2019MNRAS4822681Y}, 
its physical properties 
can be determined from the radio SED. 
Water, hydroxyl, and methanol maser sites \citep{Argon2000ApJS,Pestalozzi2005AA432737P,Urquhart2011MNRAS4181689U} are detected in its vicinity, and molecular outflows \citep{Yang2018ApJS2353Y} are found to be associated with its natal clump. 
Its natal clump AGAL030.588$-$00.042 has a mass of $758\,M_{\odot}$ and a bolometric luminosity of $1.12\times10^{4}\,L_{\odot}$  \citep{Urquhart2018MNRAS4731059U}, and shows a broad millimeter RRL $\rm H40\alpha$ with $\Delta V = \rm 56.2\,km\,s^{-1}$ \citep{Kim2017AA602A}.
It is the only radio source in the parent clump, and its spectral type B0.5, obtained from the bolometric luminosity, is consistent with that of a B0 star derived from the radio luminosity, indicating the absence of dust within this \hii\ region. 
The broad RRL line, compact size, and high electron density are consistent with this source being classified as an \hchii\ region.


\paragraph{G032.7441$-$00.0755: }
\label{G032p7441}

The radio emission associated with this source is weak and very compact and there is bright emission at 70$\mu m$ from the Hi$-$GAL survey \citep{Molinari2010PASP}, while no counterpart is seen at  mid-infrared wavelengths ($8\mu$m; see middle-right panel of Fig.\,\ref{fig:optically_thick_hiis_2}).
This source was found to host $\rm H_{2}O$ masers \citep{Caswell1983AuJPh36443C}, OH masers \citep{Caswell2013MNRAS4311180C}, and $\rm CH_{3}OH$ masers \citep{Bartkiewicz2016AA587A104B}, 
and is associated with CO outflows \citep{Yang2018ApJS2353Y}, 
broad molecular lines such as SiO (2-1) \citep{Csengeri2016AA586A149C}, N$_{2}$H$^{+}$, and $\rm HCO^{+}$ \citep{Shirley2013ApJS2092S} and millimeter RRLs ($\Delta V$ = 40.34\,\kms; \citealt{Kim2017AA602A}). 
The blueshifted and redshifted methanol masers spots mapped by \citet{Bartkiewicz2016AA587A104B} have a similar orientation to the blueshifted and redshifted outflows mapped by \citet{Yang2018ApJS2353Y}. 
Its physical parameters ($n_{\rm e} =\rm 2.79\times10^{5}\,cm^{-3}$, $diam=\rm 0.011\,pc$, $\rm EM = 8.28\times\,10^{8}\,pc\,cm^{-6}$, $\nu_{\rm t}=14.37\rm\,GHz$) are consistent with other \hchii\ regions and we therefore identify this as a new mid-infrared-dark \hchii\ region detection. 
Figure \ref{fig:optically_thick_hiis_2} shows that it is the only radio source in its natal clump. 
Its spectral type O7 derived from the bolometric luminosity is earlier than  O9.5 derived from the Lyman continuum flux, indicating that about 88\% of the Lyman continuum photons were absorbed by dust within this \hii\ region. It could be the best example to trace the dynamics associated with the final stages of accretion in massive star formation because it is still dark at $8\mu m$ and covers a significant broad component of ionized- (e.g., RRL), shocked- (e.g., SiO), and  molecular gas (e.g., CO).

\paragraph{G034.2572, G034.2573 and G034.2581: }
\label{G34p2572} 

These three \hii\ regions lie in G34.26$+$0.15, a well-studied complex region that contains three \uchii\ regions  \citep{Wood1989ApJS,Sewilo2004ApJ}: G34.26$+$0.15A (G034.2573$+$00.1523), G34.26$+$0.15B (G034.2581$+$00.1533) and G34.26$+$0.15C (G034.2572$+$00.1535); these are marked with red circles in the middle-left panel of Fig.\,\ref{fig:optically_thick_hiis_2}. 
This complex also hosts $\rm H_{2}O$ masers \citep{Hofner1996AAS120283H}, OH masers \citep{Forster1999AAS,RuizVelasco2016ApJ822101R}, $\rm CH_{3}OH$ masers \citep{Breen2015MNRAS4109B} and numerous molecules \citep{Fu2016RAA16182F,Kim2000ApJS131483K}, as well as infall/outflows traced by CO or water masers \citep[e.g.,][]{Wyrowski2016AA585A149W,vanderTak2019AA625A103V,Yang2018ApJS2353Y,Imai2011PASJ631293I}. Broad radio recombination lines (RRLs) are detected in G34.26$+$0.15B and G34.26$+$0.15C with a line-width of $\Delta V >\rm 40\,km\,s^{-1}$ \citep{Sewilo2004ApJ}.  
This is also found in their natal clump AGAL034.258$+$00.154 \citep{Kim2017AA602A,Kim2018AA616A107K}. G34.26$+$0.15B is considered to be a \hchii\ region candidate \citep[G034.2581$+$00.1533,][]{Sewilo2004ApJ,Yang2019MNRAS4822681Y}, which is blended with G34.26$+$0.15C in the C-band and X-band images, and is only resolved in the higher resolution K-band image. G034.2572$+$00.1535 is associated with G34.26$+$0.15C, which is an extended source, and can be resolved into three compact sources, all of which have RRL line widths of $\Delta V >\rm 40\,km\,s^{-1}$ \citep{Sewilo2004ApJ}. 

G034.2572$+$00.1535 is very likely to host candidates in an evolutionary stage between \hchii\ region and \uchii\ region. 
The nearby source G034.2573$+$00.1523 
is also likely to be associated with an \hchii\ region.


\paragraph{G043.1652, G043.1657 and G043.1665: }
\label{w49a}

These three sources are located in the well-known star-forming region W49A complex that is associated with CO outflows \citep{Scoville1986ApJ303416S}. 
As shown in the bottom-left panel of Fig.\,\ref{fig:optically_thick_hiis_1}, 
the three sources are associated with three \hchii\ regions W49A A (G043.1652$+$00.0129), W49A B (G043.1657$+$00.0116), and W49A G (G043.1665$+$00.0106) in the W49A complex  \citep{dePree1997ApJ,DePree2004ApJ,Sewilo2004ApJ}, which are found to be associated with many $\rm CH_{3}OH$ \citep{Bartkiewicz2014AA564A110B,Breen2015MNRAS4109B}, OH \citep{Argon2000ApJS}, and $\rm H_{2}O$ \citep{dePree2000ApJ308D,Urquhart2011MNRAS4181689U}  masers.
 G043.1657$+$00.0116 (W49A B) has $n_{\rm e} =\rm 1.57\times10^{5}\,cm^{-3}$, $diam=\rm 0.046\,pc$, $\rm EM=11.32\times10^{8}\,pc\,cm^{-6}$, and $\log N_{\rm Ly}\rm  = 48.69$, which is consistent with the typical value of \hchii\ regions, as reported by \citet{dePree2000ApJ308D}.

  G043.1652$+$00.0129 (W49A A) is resolved into two compact components at higher resolution $\sim0.05\arcsec$ \citep{dePree2000ApJ308D,DePree2004ApJ}. 
 Its physical properties such as $n_{\rm e}$ = 0.88$\times10^{5}\,cm^{-3}$, $diam$= 0.053\,pc,  EM= 4.15$\times10^{8}\,pc\,cm^{-6}$, and $\log N_{\rm Ly}$ = 48.91, are consistent with previous results in \citet{dePree1997ApJ} for W49 A at a similar resolution of $\sim 1\arcsec$. 
 However, the derived properties are slightly below the typical values of \hchii\ regions and also show smaller $n_{\rm e}$, smaller EM, and larger $diam$ compared to the results measured at higher resolution (0.05\arcsec) with $n_{\rm e} =\rm 6.1\times10^{5}\,cm^{-3}$, $diam=\rm 0.056\,pc$, and $\rm EM=83\times10^{8}\,pc\,cm^{-6}$ \citep{dePree2000ApJ308D}. 
 This might be due to the fact that our observation includes not only the two compact components but also a larger fraction of optically thin emission around them.

G043.1665$+$00.0106 (W49A G) is also multiply peaked at higher resolution $\sim0.05\arcsec$ \citep{dePree2000ApJ308D,DePree2004ApJ}. 
 Its physical properties, such as $n_{\rm e}$ = 0.24$\times10^{5}\,cm^{-3}$, $diam=$ 0.24\,pc, 
 EM= 1.22$\times10^{8}\,pc\,cm^{-6}$, and $\log N_{\rm Ly}\rm  = 49.55$, are consistent with the results in \citet{dePree1996ApJ} for W49A G. 
 The $n_{\rm e}$ is slightly smaller compared to the measurements at higher resolution with $n_{\rm e} >\rm 1.0\times10^{5}\,cm^{-3}$ for the two main compact components \citep{dePree2000ApJ308D}, which may result from the large amount of optically thin emission around these compact components.



\paragraph{G045.0712 and G045.0694: }
\label{G45p0712}

The radio emission consists of two distinct sources: the stronger source  G045.0712$+$00.1321 and the weaker source  G045.0694$+$00.1323, offset by $\sim$6\arcsec\, (as shown in bottom-right panel of Fig. \ref{fig:optically_thick_hiis_1}). 
G045.0712$+$00.1321 was identified as an \hchii\ region by \citet{Keto2008ApJ672} and \citet{Sewilo2011ApJS} (G45.07$+$0.13 NE). The physical properties of G045.0712$+$00.1321 
indicate that this \hchii\ region is associated with a O6.5 type massive star, which supports the previous results and classification by \citet{Sewilo2011ApJS}. 
The fainter of the two, G045.0694$+$00.1323, is likely to be transitioning into an \uchii\ region based on the distribution of radio properties shown in Fig.\,\ref{fig:classify_summary}. 
Their radio emission is coincident with a bright extended infrared source and a dense submillimeter clump, AGAL045.071$+$00.132, in \citet{Urquhart2018MNRAS4731059U}. 
The natal clump is associated with extended molecular outflows  aligned W to E  \citep{Yang2018ApJS2353Y}. 
This source is also host to $\rm H_{2}O$  \citep{Hofner1996AAS120283H}, OH  \citep{Argon2000ApJS}, and $\rm CH_{3}OH$ masers \citep{Kurtz2004ApJS155149K}. 
The presence of two very young \hii\ regions, molecular outflows, and three different species of masers would suggest that this clump hosts a young proto-cluster.


\paragraph{G045.4656$+$00.0452: }
\label{G045p4656}

This compact radio source is embedded in a dense molecular clump and is associated with an extended mid-infrared source, as well as water \citep{Forster1999AAS} and OH \citep{Argon2000ApJS} maser emissions (see bottom-left panel of  Fig.\,\ref{fig:optically_thick_hiis_2}). 
Its natal clump AGAL045.466+00.046 is also associated with bipolar outflows \citep{Yang2018ApJS2353Y} and  broad $\rm H39\alpha$ RRL ($\Delta v =  47.8$\,\kms; \citealt{Kim2017AA602A}).  \citet{Cyganowski2008AJ1362391C} identified this source as an extended green object associated with an infrared dark cloud. The physical parameters determined for this source ($n_{\rm e} =\rm 1.02\times10^{5}\,cm^{-3}$, $diam=\rm 0.023\,pc$, $\rm EM = 2.36\times\,10^{8}\,pc\,cm^{-6}$, $\nu_{\rm t}=7.89\rm\,GHz$) are consistent with this being classified as an \hchii\ region.

\paragraph{G061.4770$+$00.0892: }
\label{G061p4770}

This object is very compact with a deconvolved size similar to that of the beam ($\sim 0.7\arcsec$) at K-band, and its radio emission is blended with a nearby cometary \uchii\ region detected both in 5\,GHz CORNISH and X-band observations described in this work. 
However, the two sources are separated in the high-resolution observations \citep[$\sim0.4\arcsec$;][]{Wood1989ApJS} and our K-band observations ($\sim0.7\arcsec$). 
As shown in the bottom-right panel of Fig.\,\ref{fig:optically_thick_hiis_2}, 
the near-infrared RGB image of this source presents extended 4.5 $\mu$m emission and so it could be associated with an extended green object (EGO) as defined by \citet{Cyganowski2008AJ1362391C}.
A bipolar molecular outflow aligned NE to SW \citep{Phillips1991AAS88189P,White1992AA266452W} and water masers \citep{Henkel1986AA165197H,Svoboda2016ApJ82259S} are detected toward its parent cloud. 
Broad RRL components \citep{Garay1998ApJ501710G} and strong OH (1665/67 MHz) absorption \citep{Sarma2013ApJ76724S} are reported towards this source and the other physical properties derived from radio emission indicate that this source is likely to host an \hchii\ region.

\subsection{Intermediate objects between \hchii\ and \uchii\ regions}
\label{transient}

According to their physical properties, there are eight objects located in the evolutionary stages between \hchii\ regions and \uchii\ regions in Table \ref{tab:10hchii_candidates}. Two out of the eight sources (i.e., G034.2572$+$00.1535 and G045.0694$+$00.1323) are associated with clusters of \hii\ regions that have already been discussed in Sect.\,\ref{G34p2572}; in the following sections we provide brief notes on the other six intermediate objects.

\paragraph{G030.7197$-$00.0829: }
\label{G030p7197}
 
 This source was resolved at 5\,GHz by CORNISH.  
The physical properties ($n_{\rm e}=0.22\times\rm 10^{5}\,cm^{-3}$, $diam = 0.09\,\rm pc$, EM=$0.45\times10^{8}\rm\,pc\,cm^{-6}$, $\nu_{\rm t}=\rm 3.6\,GHz$) can be determined from the radio SED based on flux densities of 464.58 mJy at 1.4 GHz \citep{White2005AJ}, 969.33 mJy at 5 GHz \citep{Purcell2013ApJS}, and 570\,mJy at 43\,GHz \citep{Leto2009AA5071467L}.
These results are consistent with the measurements in \citet{Leto2009AA5071467L}. 
Its natal clump AGAL030.718$-$00.082 has a mass of $6.6\times10^{3}\,M_{\odot}$, a bolometric luminosity of  $5.5\times10^{4}\,L_{\odot}$  \citep{Urquhart2018MNRAS4731059U}, and a broad millimeter RRL $\rm H40\alpha$ with $\Delta V = \rm 43.0\,km\,s^{-1}$ \citep{Kim2017AA602A} , 
and is associated with CO outflows \citep{Yang2018ApJS2353Y}. 
Its Lyman continuum flux agrees with its bolometric luminosity, indicating a lack of dust within this \hii\ region. 
Therefore, this source appears to be an intermediate object between \hchii\ and \uchii\ regions.


\paragraph{G030.8662$+$00.1143: }
\label{G030p8662}
The SED of this resolved source was constructed from the flux densities of 137.17 mJy at 1.4 GHz and of 255.2\,mJy at 5 GHz from \citet{White2005AJ}, 306.0 mJy at 6.7\,GHz, and 356.0\,mJy at 8.4\,GHz from \citet{Walsh1998MNRAS301}, as well as 560\,mJy at 43\,GHz from \citet{Leto2009AA5071467L}. 
Its physical characteristics measured from the radio SED, such as $n_{\rm e}=$0.37$\times\rm 10^{5}\,cm^{-3}$, $diam = $0.03\,pc, EM=0.42$\times10^{8}\rm\,pc\,cm^{-6}$, and $\nu_{\rm t}=$3.5\,GHz, are consistent with previous measurements \citep{Leto2009AA5071467L}. 
Water maser sites \citep{Urquhart2009AA507795U,Urquhart2011MNRAS4181689U} are detected in its vicinity and molecular outflows \citep{Yang2018ApJS2353Y} are found to be associated with its natal clump. 
Its natal clump AGAL030.866$+$00.114 has a mass of $ 295\,M_{\odot}$, a bolometric luminosity of $ 1.30\times10^{4}\,L_{\odot}$  \citep{Urquhart2018MNRAS4731059U}, and a broad millimeter RRL $\rm H39\alpha$ with $\Delta V = \rm 44.9\,km\,s^{-1}$ \citep{Kim2017AA602A}.
Its spectral type B0.5 obtained from the bolometric luminosity is consistent with O9.5 derived from radio luminosity, indicating the absence of dust in this \hii\ region. 
Therefore, this source appears to be an intermediate object.


\paragraph{G033.1328$-$00.0923: }
\label{G033p1328}

This source shows extended emission at 5 GHz CORNISH, shown as lime contours in the bottom-left panel of Fig. \ref{fig:optically_thick_hiis_3}. 
With flux densities of 173.43 mJy at 1.4 GHz  and 378.59\,mJy at 5 GHz  summarized in \citet{Yang2019MNRAS4822681Y}, as well as 461.2\,mJy at 9\,GHz  and 675.3\,mJy at 15\,GHz  measured by \citet{Kurtz1994ApJS659K}, we construct its radio SED between 1 and 15 GHz.  
Its physical properties from the SED fitting are consistent with results in \citet{Kurtz1994ApJS659K}.
Water masers \citep{Pestalozzi2005AA432737P,Kurtz2005AJ130711K} are detected in its vicinity and molecular outflows \citep{Yang2018ApJS2353Y} are found to be associated with its natal clump. 
Its natal clump AGAL033.133$-$00.092 has a mass of $5.0\times10^{3}\,M_{\odot}$, a bolometric luminosity of  $1.1\times10^{5}\,L_{\odot}$  \citep{Urquhart2014AA568A41U,Urquhart2018MNRAS4731059U}, and a broad millimeter RRL $\rm H39\alpha$ with $\Delta V = \rm 43.0\,km\,s^{-1}$ \citep{Kim2017AA602A}. 
As it is only one radio source in the natal clump, its spectral type O7 obtained from the bolometric luminosity is consistent with O7.5 derived from the radio luminosity.
Therefore, this source is likely to be an intermediate object between \hchii\ and \uchii\ region.

\paragraph{G049.3666$-$00.3010:}
\label{G049p3666}

This object appears to have a nearby \uchii\ region to the east referenced as G049.3704$-$00.3012 (marked with a red circle in the upper-left panel of  Fig.\,\ref{fig:optically_thick_hiis_3}). 
Both of these \hii\ regions are embedded towards the center of the dense clump AGAL049.369$-$00.301, which has been associated with a broad H40$\alpha$ RRL with $\Delta V = 34.5\,\rm km\,s^{-1}$ \citep{Kim2017AA602A}. 
The optically thick radio source is coincident with an extended mid-infrared source, and two water masers have been detected in its vicinity \citep{Valdettaro2001AA368845V,Xi2015MNRAS4534203X}.


\paragraph{G051.6785$+$00.7193: }
\label{G051p6785}

This radio source is very compact at all radio bands presented in this work, while it can be resolved into two sources at high angular resolution $\sim 0.2\arcsec$ {at 1.3\,cm using the VLA} in \citet{Rodrguez2012ApJR}. 
The radio source is embedded in a very compact and centrally condensed ATLASGAL clump {AGAL051.678$+$00.719 with a mass of  $2.88\times10^{3}\,M_{\odot}$} and is associated with a very bright mid-infrared point source that has a luminosity of $1.0\times10^{5}\,L_{\odot}$. 
The natal clump is also associated with water and methanol masers  \citep{Sridharan2002ApJ566931S,Rodrguez2012ApJR}, and molecular outflows aligned with extended mid-infrared emission going from NE to SW \citep{Beuther2004ApJ608330B}, as presented in the upper-right panel of Fig.\,\ref{fig:optically_thick_hiis_3}.


\paragraph{G060.8842$-$00.1286: }
\label{G060p8842}

This object is southwest of the two \hii\ regions (see middle-left panel of Fig.\,\ref{fig:optically_thick_hiis_3}) in the massive star-forming region S87IRS1 \citep{Barsony1989ApJ345268B}, the other being a nearby  extended and weak \hii\ region \citep{Purcell2013ApJS} that has been resolved out at K-band in this work.  The S87IRS1 is associated with the clump JPSG060.886-00.129 in \citet{Eden2017MNRAS4692163E}, which is itself associated with a molecular outflow \citep{Barsony1989ApJ345268B,Xue2008ApJ680446X}. The radio source is associated with bright mid-infrared emission and coincident with a water maser \citep{Kurtz2005AJ130711K}. At high resolution $\sim$0.4\arcsec, the clump is found to be fragmented into multiple millimeter cores   \citep{Beuther2018AA617A100B}.
Its bolometric luminosity agrees with its radio luminosity, suggesting a lack of dust within this \hii\ region.


\subsection{\hchii\ regions not resolved in this work}
\label{hchii_not_resolved}

In addition to the optically thick radio sources identified in this work, we include notes on another four \hchii\ regions that have been identified in previous studies (e.g., \citealt{Wood1989ApJS}, \citealt{Sewilo2004ApJ} and \citealt{Zhang2014ApJ}) but are unresolved in our observations. 
Two of the four (G043.1652$+$00.0129 and G035.5781$-$00.0305) are unresolved mainly due to the fact that our observations include their nearby \uchii\ regions as the resolution is not sufficient to resolve the emission into individual sources.  The remaining two regions (G043.1665+00.0106 and G010.9584$+$0.0211) are not recovered by this work primarily because our observations include a large amount of surrounding ionized gas emission as this diffuse gas is optically thin. 
Therefore, the derived properties in this work represent average values for sources with co-existing emission from \hchii\ and  nearby \uchii\ regions or represent a complex weighted average over the compact sources plus the surrounding diffuse ionized gas, and thus do not satisfy the criteria for classification as \hchii\ regions. 
However, these sources have previously been identified as \hchii\ regions and we therefore include these sources in this section for completeness. 
The source names and derived properties are given towards the end of Table\,\ref{tab:10hchii_candidates}.  
Two sources (G043.1652$+$00.0129 and G043.1665$+$00.0106) in the W49A complex region have already been discussed together in Sect.\,\ref{w49a} and are therefore not described again here. Images of the remaining two \hchii\ regions are presented in Fig.\,\ref{fig:optically_thick_hiis_1} and brief notes are provided below. 


\paragraph{G010.9584$+$0.0221: }
\label{G010p9584}

This source is an \hchii\ region and is located in the western part of the G10.96$+$0.01 region and surrounded by more diffuse ionized gas, as suggested by \citet{Sewilo2004ApJ}.  
Its physical properties, such as $n_{\rm e} =\rm 0.36\times10^{5}\,cm^{-3}$, $diam=\rm 0.029\,pc$, $\rm EM=0.38\times10^{8}\,pc\,cm^{-6}$ and $\log N_{\rm Ly}\rm  = 47.35$, 
are all consistent with the results reported by \citet{Sewilo2004ApJ} and \citet{Sewilo2011ApJS}. 
In spite of the reported broad $\rm H92\alpha $ line with $ \Delta V = \rm 43.8\pm1.5 km\,s^{-1}$, the derived properties are slightly below the typical values of \hchii\ regions, which might be due to the previous VLA observations (\citealt{Sewilo2004ApJ}, \citealt{Sewilo2011ApJS}) 
and this work includes a significant amount of optically thin emission from the diffuse ionized gas around this source, and both results are likely to be underestimates by averaging over the compact source plus its surrounding ionized gas, as mentioned in \citet{Sewilo2004ApJ} and  \citet{Yang2019MNRAS4822681Y}.
Its natal clump has a mass of $398\,M_{\odot}$ and a bolometric luminosity of  $1.0\times10^{4}\,L_{\odot}$  \citep{Urquhart2018MNRAS4731059U}, 
and is associated with high velocity outflow wings identified in CO spectra from the SEDIGISM survey \citep{Schuller2017AA601A124S}. 
In this case, the luminosity and Lyman continuum {flux} are both contributed by the same source, meaning that the spectral type derived from the bolometric luminosity is consistent with that derived from the radio luminosity; B0.5 and B0, respectively.

\paragraph{G035.5781$-$00.0305: }
\label{G035p5781}

This radio emission can be resolved into two extremely close sources at 2\,cm and 3.6\,cm with a resolution of $<1\arcsec$ \citep{Kurtz1994ApJS659K}: the source to the west has been identified as an \hchii\ region G35.578$-$0.030 \citep{Zhang2014ApJ} and the source to the east as an \uchii\ , G35.578$-$0.031 \citep{Kurtz1994ApJS659K}. 
These are seen as a single blended source in our radio maps (see the middle-right panel of Fig.\,\ref{fig:optically_thick_hiis_1}). This source is associated with OH masers \citep{Argon2000ApJS} and $\rm H_{2}O$ masers \citep{Forster1999AAS,Urquhart2011MNRAS4181689U}.
The physical properties for the blended source G035.5781$-$00.0305 are $n_{\rm e} =\rm 0.22\times10^{5}\,cm^{-3}$, $diam=\rm 0.093\,pc$, $\rm EM=0.45\times10^{8}\,pc\,cm^{-6}$, and $\log N_{\rm Ly}\rm  = 48.36$. Thus, G035.5781$-$00.0305 in this work has smaller $n_{\rm e}$, smaller EM and larger $diam$ compared to the  \hchii\ region G35.578$-$0.030 in \citet{Zhang2014ApJ} with $n_{\rm e} =\rm 3.3\times10^{5}\,cm^{-3}$, $diam=\rm 0.018\,pc$, $\rm EM=1.9\times10^{9}\,pc\,cm^{-6}$. 
Its natal clump has a mass of   $6.8\times10^{3}\,M_{\odot}$ and a bolometric luminosity of $2.0\times10^{5}\,L_{\odot}$  \citep{Urquhart2018MNRAS4731059U}, which is associated with molecular outflows \citep{Yang2018ApJS2353Y}. 


\subsection{Summary}

In Table\,\ref{tab:10hchii_candidates} we summarize the physical properties of the sources of our sample and the associated discussion in the preceding text. Inspection of this table reveals that in addition to the physical properties ($n_{\rm e}$, $diam$, EM and RRL), which are typical for \hchii\,regions, all the sources of our sample are found to be embedded towards the centres of dense molecular clumps and are also commonly associated with various masers, molecular outflows, broad RRLs, and extended green objects, all of which are all signposts of active star formation.
The bolometric luminosities tend to be higher than the radio flux suggests, which is consistent with these being associated with a forming protocluster. 
These optically thick \hii\ regions are therefore the best examples to investigate the relation between \hchii\ regions and \uchii\ regions, to study the birth of \hii\ regions, and therefore to understand the final stages of accretion in massive star formation.

There are 13 \hchii\ regions, 3 \hchii\ region candidates, and 8 intermediate objects listed in Table \ref{tab:10hchii_candidates}.  
Among them, four \hchii\ regions and three \hchii\ region candidates are reported here for the first time. 
Based on the classification of \hchii\ regions in Table\,\ref{tab_classify_hii}, 
it is difficult to assess the completeness of  the sample of \hchii\ regions and intermediate \hii\ regions identified in this study because there are four \hchii\ regions, marked with an asterisk in Table\,\ref{tab:10hchii_candidates},  that are in very close proximity to other \uchii\ regions that we were not able to resolve.

\section{Discussion}
\label{sect:discussion}
\subsection{Implications of the evolution of young \hii\ regions}
\label{sec_implication_evolution}


\begin{figure*}
 \centering
 \begin{tabular}{cc}
  \includegraphics[width = 0.45\textwidth]
 {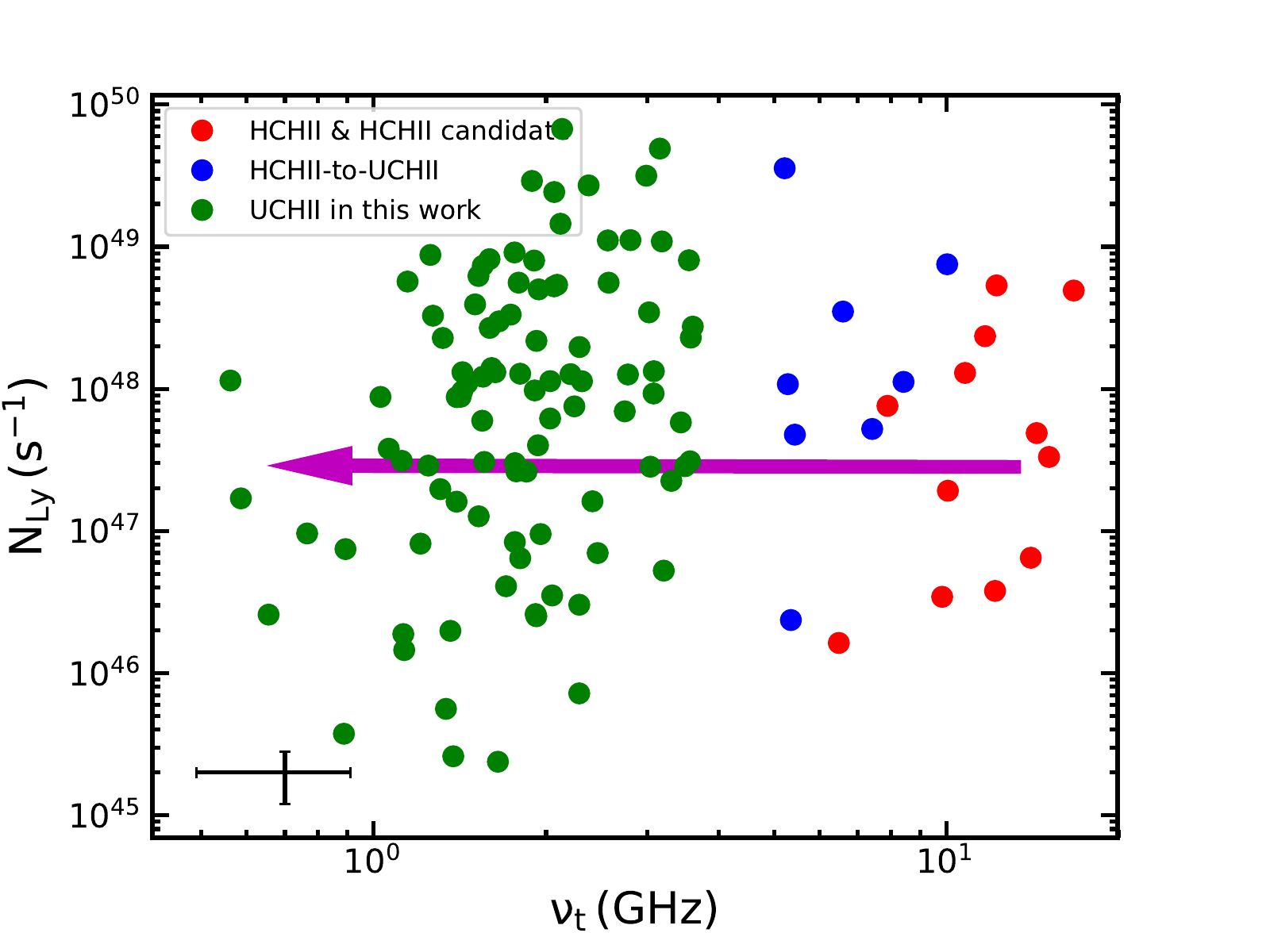} &
   \includegraphics[width = 0.45\textwidth]{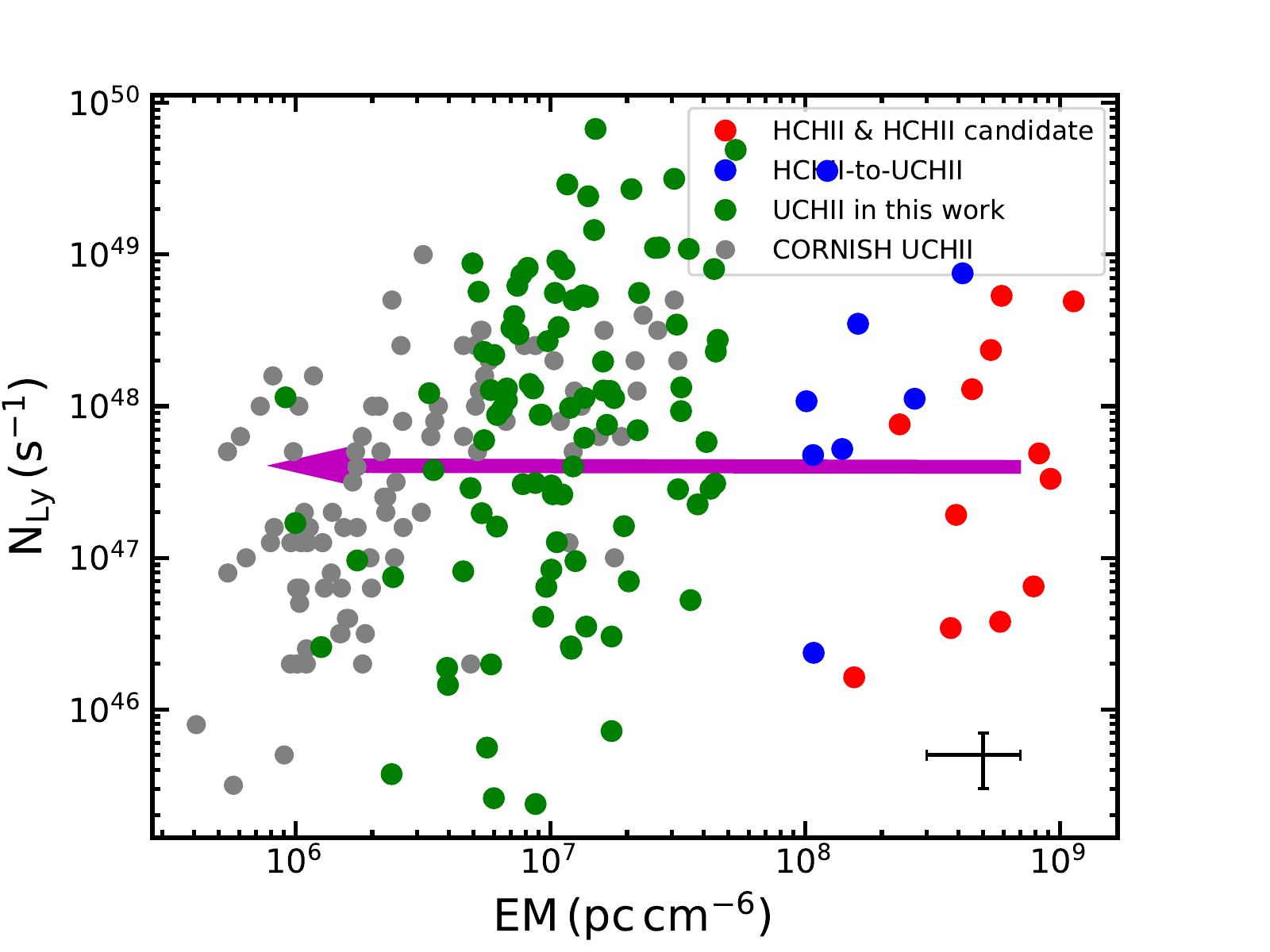} \\
  \includegraphics[width = 0.45\textwidth]{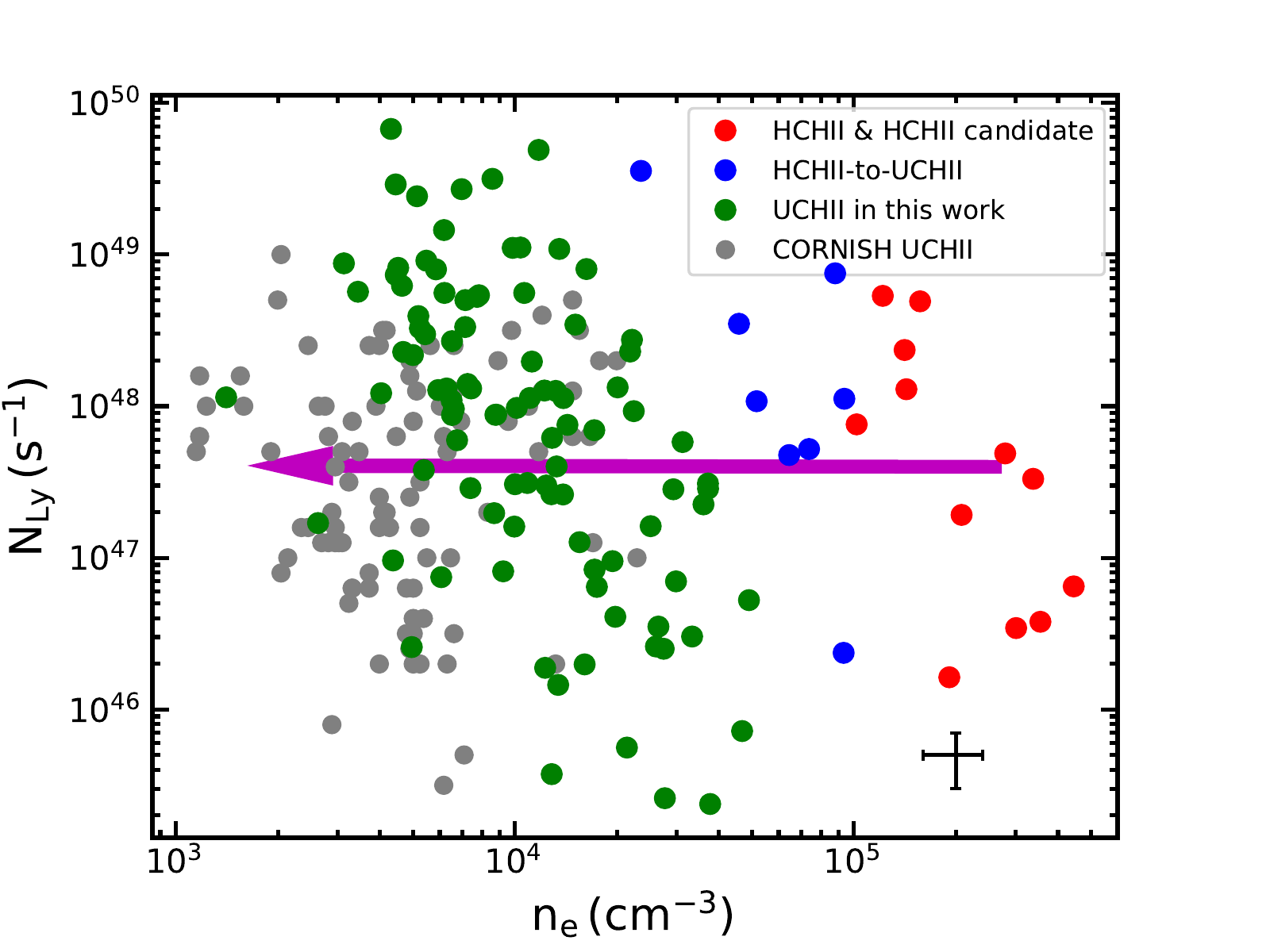} & 
 \includegraphics[width = 0.45\textwidth]{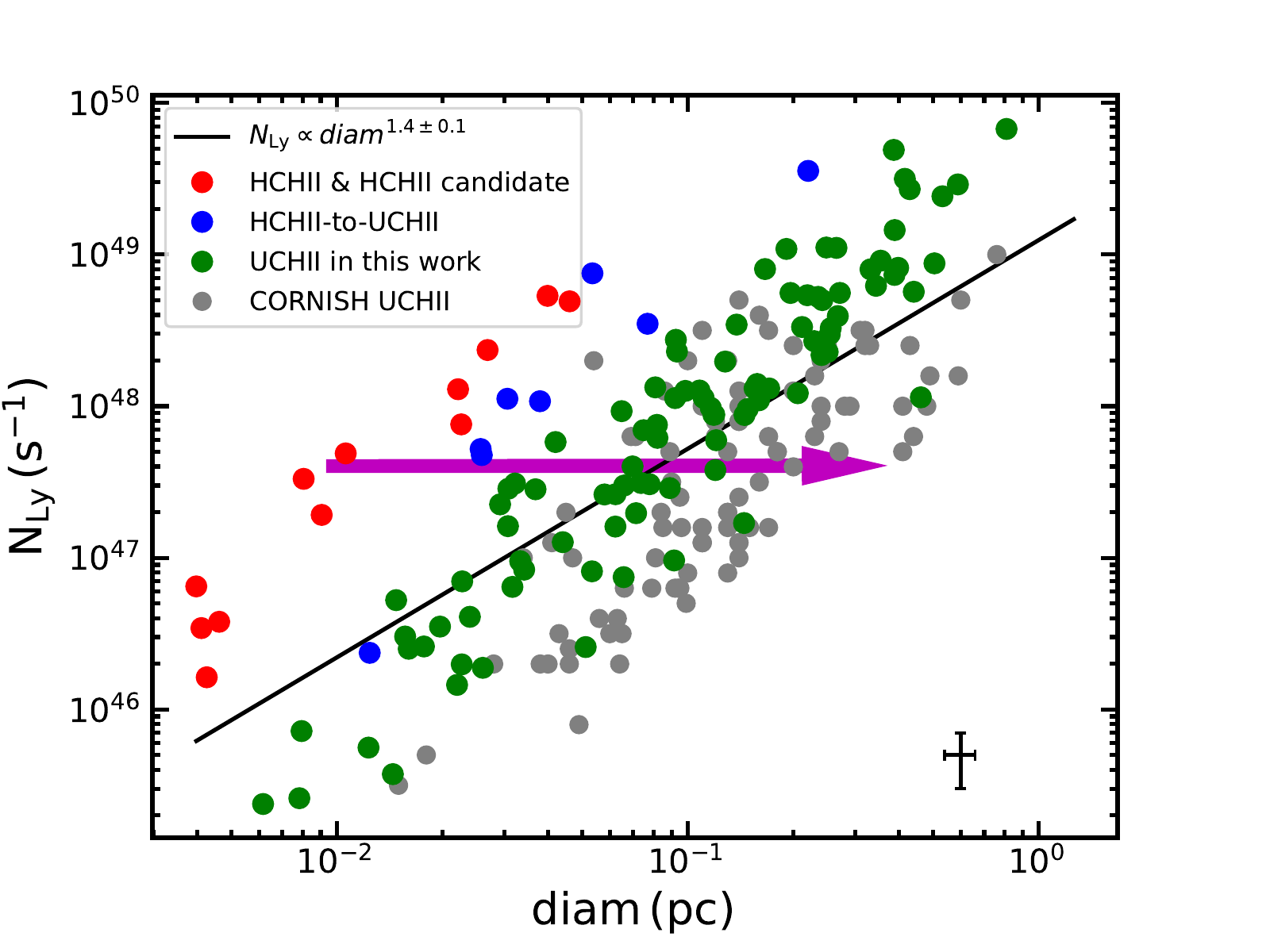}  \\ 
  \end{tabular}
 \caption{ The plots of the evolution and correlation of the derived physical parameters. 
 $\nu_{\rm t}$ vs.  $N_{\rm Ly}$ (upper-left), EM vs. $N_{\rm Ly}$ (upper-right), $n_{\rm e}$ vs. $N_{\rm Ly}$ (bottom-left),  and $diam$ vs. $N_{\rm Ly}$  (bottom-right) for \hchii\ regions (red dots), intermediate objects between \hchii\ region and \uchii\ regions (blue), \uchii\ regions in this work (green dots), and CORNISH \uchii\ regions (gray dots). 
 The CORNISH \uchii\ regions sample refers to the whole CORNISH \uchii\ regions sample from \citet{Kalcheva2018AA615A103K} by excluding \uchii\ regions in this work.
 The magenta arrow indicates the evolutionary trend of the physical properties. }
 \label{fig:evolution_phy2}
 \end{figure*}

As suggested by classical theoretical models  \citep{Dyson1995MNRAS277,Mezger1967ApJ}, \hii\, regions are expected to expand over time, which results in decreasing $n_{\rm e}$ and  EM and increasing $diam$, as seen in Fig.\,\ref{fig:classify_summary}. 
The plots shown in this figure display a clear evolutionary trend in $n_{\rm e}$, $diam$, and EM from \hchii\ regions to the intermediate objects between the \hchii\ and \uchii\ region stages. 
The mean values of physical properties range from $n_{\rm e}=\rm 2.5\times10^{5}\,cm^{-3}$, $diam=\rm\,0.012\,pc$, and $\rm EM=5.5\times10^{8}\,pc\,cm^{-6}$ for \hchii\ regions, to $n_{\rm e}=\rm 0.79\times\,10^{5}\,cm^{-3}$, $diam=\rm 0.03$, and $\rm EM=\rm 1.58\times10^{8}\,pc\,cm^{-6}$ for intermediate objects, and thus $n_{\rm e}$ tends to change quickly compared to the EM and $diam$ at the earliest times of \hii\ region stage.

To investigate the evolution of physical properties of \hii\ regions over a wide range of evolutionary stages, 
we add the CORNISH \uchii\ regions from  \citet{Kalcheva2018AA615A103K} that are presumably in a later stage compared to our sample. 
Evolution of the Lyman continuum flux $N_{\rm Ly}$, turnover frequency $\nu_{\rm t}$, and emission measure EM is presented in Fig. \ref{fig:evolution_phy2} for the three subsamples discussed here and for the four subsamples by adding the more evolved CORNISH \uchii\ regions. 
We see that $\nu_{\rm t}$ decreases as the \hii\ region evolves, from 11.5\,GHz for \hchii\ regions to 6.4\,GHz for intermediate objects, and to 1.8\,GHz for \uchii\ regions, as expected from the theoretical model in \citet{Mezger1967ApJ}. 
It is interesting to note that there is no obvious correlation between the Lyman continuum flux and the evolution of the \hii\ regions. Furthermore, we find no significant correlation between $N_{\rm Ly}$ and EM with $\rho = -0.01$ and $p$-value $= 0.85$, and between $N_{\rm Ly}$ and $n_{\rm e}$ with $\rho = -0.07$ and $p$-value $= 0.3$ in the four subsamples. In addition, the mean value of $N_{\rm Ly}\sim 10^{48}\,{\rm s}^{-1}$ is consistent throughout the four evolutionary phases, from the \hchii\ region and \hchii\ region candidates, to intermediate objects, to \uchii\ regions in this work, and to more evolved \uchii\ regions in CORNISH. 
These results suggest that there is effectively no evolution of the Lyman continuum photon flux with changes in the $\nu_{\rm t}$, $n_{\rm e}$, and EM, and by extension there is no increase in $N_{\rm Ly}$ with evolution of the \hii\ region.

As shown in the bottom-left panel of Fig.  \ref{fig:evolution_phy2}, the positive correlation between $N_{\rm Ly}$ and $diam$ is significant with $\rho = 0.5$ and $p$-value $\ll 0.001,$  using a partial correlation test to control the distance dependence, giving a power-law relation of $N_{\rm Ly}\propto diam^{1.4\pm0.1}$.
However, given the fact that there is little evidence of any sort of significant correlation between Lyman continuum flux and other parameters tracing the evolution of \hii\ regions, such as $\nu_{\rm t}$, $n_{\rm e}$, or EM as discussed above, this correlation is more likely to result from the fact that more luminous \hii\ regions expand more rapidly in their early stages but that the expansion speed will decrease over time, becoming similar to less luminous \hii\ regions. The evolution shown in bottom-left panel of Fig. 9 is therefore from left to right rather than diagonal from bottom-left to upper-right as suggested from the distribution. 
The flat evolution of $N_{\rm Ly}$ indicates that the value of $N_{\rm Ly}$ remains constant as the \hii\ region develops, and by extension that the ionizing flux from a young massive star remains constant during the evolutionary phases of \hii\ regions in this sample. 
This result is in agreement with the classical expansion model without gravity or the model with gravity in \citet{Keto2002ApJ980K} in which the $N_{\rm Ly}$ of the \hii\ region tends to stop increasing if it reaches the critical ratios where the accretion is quickly reduced.
Also, the constant $N_{\rm Ly}$ over time agrees with the results of \citet{Hosokawa2009ApJ691} and \citet{Hosokawa2010ApJ721} who showed that the luminosity and temperature of a bloated protostar remain almost unchanged in the last accretion phase.
Moreover, the almost unchanged $N_{\rm Ly}$ may also support the model of \citet{Peters2010ApJ711} who proposed that a shrinking \hii\ region has small fluctuations of 5\%--7\% in ionizing flux over time.


\begin{figure}
 \centering
    \includegraphics[width = 0.45\textwidth]{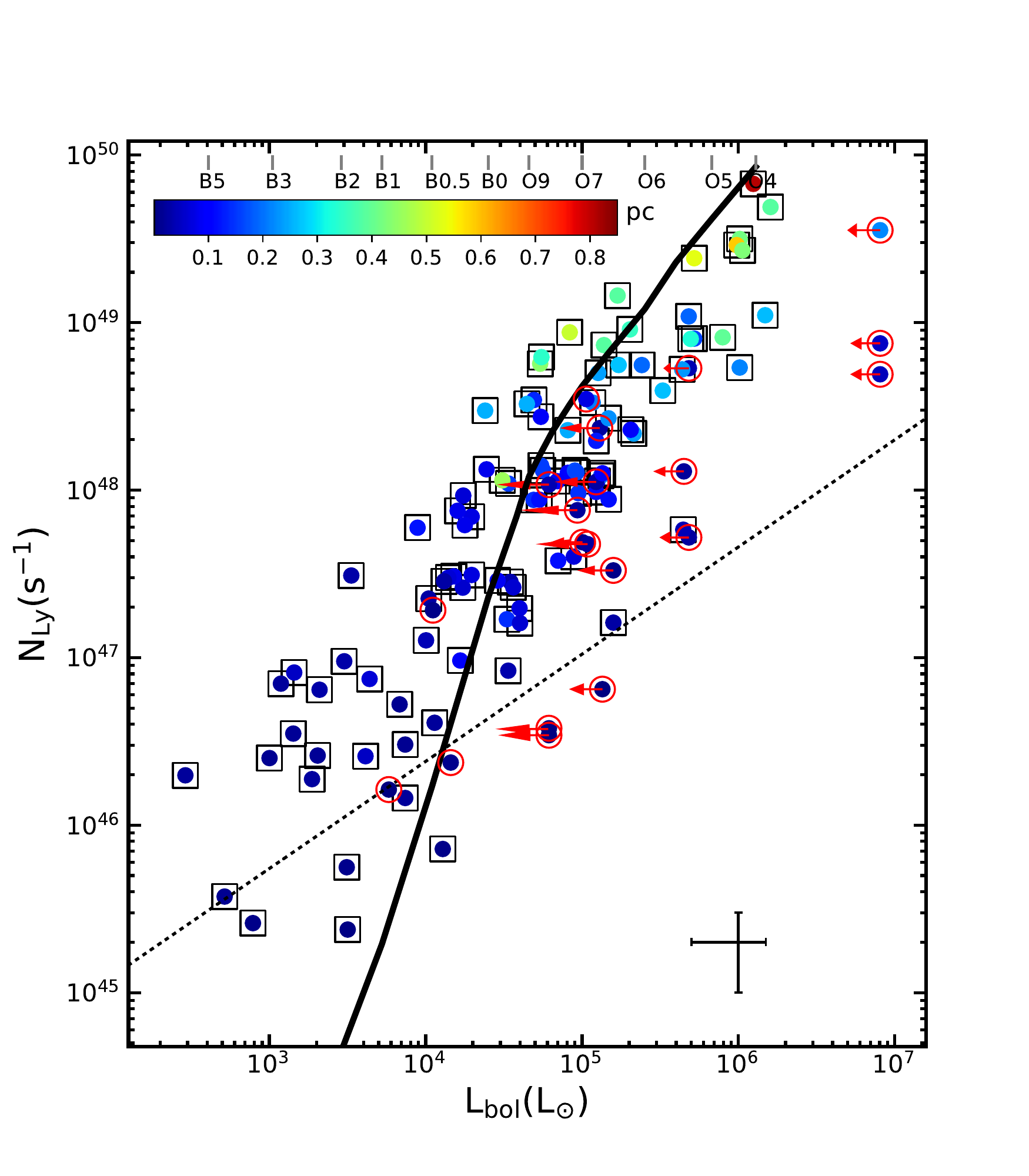}
 \caption{ Lyman continuum flux $N_{\rm Ly}$ vs. the bolometric luminosity $L_{\rm bol}$ for young \hii\ regions with rising spectra. 
 The black solid line refers to the expected Lyman continuum photon rate from a single ZAMS star of a given bolometric luminosity. The top axis lists the spectral type corresponding to a given bolometric luminosity taken from stellar models \citep{Thompson1984ApJ165T,Panagia1973AJ78,Davies2011MNRAS972D}.  
 The error bars in the bottom-right corner correspond to a $50\%$ uncertainty for $\rm L_{\rm bol}$ and $\rm N_{\rm Ly}$ \citep{Urquhart2018MNRAS4731059U}. 
 At the top, we show the color bar for the physical size of the sources, indicating the physical diameter in parsecs. The red circles and black squares refer to optically thick \hii\ regions ($\nu_{\rm t}>\rm 5\,GHz$) and optically thin \hii\ regions ($\nu_{\rm t}<\rm 5\,GHz$). 
 About $\rm 30\%$ of the sample is located in the forbidden region above the solid curve considering a $50\%$ uncertainty. 
 The dotted line represents the empirical relation between $\rm L_{\rm bol}$ and $\rm N_{Ly}$ for ionized jets from YSOs, with a power-law index of 0.64 derived by \citet{Purser2016MNRAS4601039P}.
 The red arrows for the optically thick \hii\ regions indicate that the bolometric luminosities are upper limits due to the presence of other \hii\ regions in the same clump. }
 
 \label{fig:lyman_vs_lum}%
 \end{figure}

\subsection{Lyman continuum$-$bolometric luminosity relationship}
\label{sec:disc_lym}

The measurements of Lyman continuum flux in the optically thin regime presented in {Sect.\,\ref{sec_lyc}} and the bolometric luminosity of the sample measured by previous studies (see Table \ref{tab_obser_list}) allow us to discuss the relation between Lyman continuum photons ($N_{Ly}$ ) and bolometric luminosity ($ L_{\rm bol}$), as well as Lyman continuum flux excess phenomenon in the  sample of young \hii\ regions.
There exists a significantly positive correlation between $ L_{\rm bol}$ and $N_{\rm Ly}$ with $\rho=0.54$ and $p$-value$\ll0.001$ when using the partial correlation test to remove the distance dependence, which is consistent with the correlation ($\rho=0.69$) calculated by \citet{Urquhart2013MNRAS435} for a sample of ultra-compact and compact \hii\ regions.

Figure\,\ref{fig:lyman_vs_lum} shows $N_{Ly}$ as a function of $ L_{\rm bol}$.
The color symbols indicate the physical size of the sample and the black solid line represents the upper limit of the expected Lyman continuum photon rates at specific given bolometric luminosities for ZAMS stars.
About $40\%$ of the sources in the sample are located in the forbidden region above this black line, suggesting a Lyman continuum excess. 
Considering a 50\% uncertainty on $N_{Ly}$ and $L_{bol}$, the fraction of Lyman excess sources in our sample is consistent with $\sim$30\% sources in previous work  \citep{Sanchez_Monge2013AA21S,Cesaroni2015AA579A}.  Those sources with Lyman excess are more likely to be associated with young B-type stars \citep[e.g.,][]{Sanchez_Monge2013AA21S,Lumsden2013ApJS208,Urquhart2013MNRAS435}.

Most of the optically thick \hii\ regions in the sample do not show a Lyman continuum excess; 
these are marked with red circles in Fig.\,\ref{fig:lyman_vs_lum} and located to the right of the black solid line representing the upper limit of the expected Ly continuum photons. 
The main reason for this is that many are embedded in clusters (as discussed in Sect.\,\ref{sect_optically_thick_hii}). 
Although it is possible that the Lyman flux has been underestimated because of filtering of some of the extended flux in the interferometric observations \citep[e.g.,][]{Urquhart2013MNRAS435}, 
and because of absorption by dust in the \hii\ region  \citep[e.g.,][]{Wood1989ApJS,Garay1993ApJ418368G}, 
it is unlikely these affects would be significant enough to result in these objects having a Lyman excess (in many cases the Lyman flux would need to have been underestimated by an order of magnitude or more).

It is possible that some of the optically thick objects we have detected are ionized jets whose radio emission also has positive spectral indices \citep{Moscadelli2016AA585A71M,Purser2016MNRAS4601039P}, and because there are very weak ($S_{\rm int}\sim$\,mJy) and compact ($diam\sim \rm 1000\,AU$) sources (see Sect.\,\ref{sect_optically_thick_hii}). 
We include the empirical relationship between bolometric luminosity ($L_{bol}$) and Lyman flux ($N_{\rm Ly}$) derived from young stellar objects (YSOs) in Fig.\,\ref{fig:lyman_vs_lum} (dotted diagonal line; \citealt{Purser2016MNRAS4601039P}).  
Given that it is likely that the Lyman continuum flux has been underestimated and the bolometric luminosity has been overestimated, only the optically thick sources located to the right of this relation are associated with radio jets; these are G030.0096, G060.8842, G034.2573, G034.2581, and G061.4770. 
The radio emission of the five sources are point-like as shown in Figs. \ref{fig:optically_thick_hiis_2} and \ref{fig:optically_thick_hiis_3}, and therefore no morphological evidence was found to indicate that they are radio jets, which implies that they are more likely to be \hchii\ regions as discussed in Sect.\,\ref{sect_optically_thick_hii}.  
Further observations are needed to reliably classify these objects.

In Fig.\,\ref{fig:lyman_vs_lum}, there are seven young \hii\ regions in Table\,\ref{tab:10hchii_candidates} located close to the black solid line, namely G010.9584, G030.0096, G030.5887, G030.8662, G060.8842, G030.7197, and G033.1328,
which means that their Lyman continuum fluxes agree well with their bolometric luminosities, and further indicates the absence of dust within these \hii\ regions to absorb the Lyman continuum photons. 
These seven objects are the only radio sources in the observed field of this work and in their parent clumps from  \citet{Urquhart2018MNRAS4731059U}. 
Three of the seven (G010.9584, G030.0096, and G030.5887) have been suggested to be in the \hchii\ region stage and the remaining four are expected to be in the intermediate stage between \hchii\  and \uchii\ regions. 
Except for three sources with no RRL information, the remaining five sources show broad RRL with line widths
$\rm \Delta V > 40\,km\,s^{-1}$, and all of them are associated with outflows and masers, as shown in Table \ref{tab:10hchii_candidates}.
These dust-free and young \hii\ regions are interesting cases to study the destruction of dust in the very young \hii\ regions because \hii\ regions are often expected to be dusty in the early stages, as discussed in Sect.\,\ref{sect_dust} and in \citet{Arthur2004ApJ608282A}.
Further investigations are needed to understand the absence of dust in these young \hii\ regions.

%


\section{Summary and conclusion}
\label{sec_conclusion}
In this work, we report the results of multi-band (8--12 GHz and 18--26 GHz), high angular-resolution ($\sim 1.7\arcsec$ and $\sim 0.7\arcsec$), VLA observations toward a sample of young \hii\ regions that are selected on the basis of rising spectra between 1 and 5 GHz in \citet{Yang2019MNRAS4822681Y}. 
We construct their radio SED between 1 GHz and 26 GHz and measure their physical properties for 116 young \hii\ regions by modeling each SED based on an ionization-bounded \hii\ region with standard uniform electron density. 
The sample has a mean electron density of $1.6\times10^{4}\,\rm cm^{-3}$, a mean diameter of $ 0.14\,\rm pc$, a mean emission measure of 1.9$\times10^{7}\rm pc\,cm^{-6}$, a mean turnover frequency of $3.29\,\rm GHz$, and a mean Lyman continuum flux of   6.5$\times10^{47}\,s^{-1}$. 
Based on these properties, there are a total of 20 \hchii\ regions and 3 candidates reported so far after combining our findings with the \hchii\ region catalog summarized in \citet{Yang2019MNRAS4822681Y}. 
This sample consists of a large number of \hchii\ regions and \uchii\ regions, which gives us a comprehensive picture of the physical condition and evolution of these young \hii\ regions.
The main results of our study can be summarized as follows:

\begin{enumerate}

    \item
         We identify 16 \hchii\ regions and 8 intermediate objects located between the class of \hchii\ and \uchii\ regions. 
     Four \hchii\ regions and three candidates are newly reported in this work, along with two new infrared-dark \hchii\ regions.  
     
     \item
          We discuss how the physical properties of \hii\ regions change as they evolve from \hchii\ regions to \uchii\ regions and then to compact \hii\ regions. While $n_{\rm e}$, $diam$, EM, and $\nu_{\rm t}$ all change during this evolution, the Lyman continuum flux stays relatively constant over time, suggesting that the accretion tends to be quickly reduced or could be halted at the earliest \hchii\ region stage in our sample.

    \item 
    These young and compact \hii\ regions are located in dusty clumps. 
    The mean fraction of ionizing flux absorbed by dust in \hii\ regions is 67\%, and the absorption fraction tends to be more significant for the more compact and younger \hii\ regions.
    Nevertheless, about 40\% of the sources show Lyman continuum excess and are preferentially associated with young B-type stars.

\end{enumerate}

In conclusion, 
young \hii\ regions are likely to
be located in dusty clumps. 
The youngest \hii\ regions, namely  \hchii\ regions and  intermediate objects between \hchii\ and \uchii, 
are found to be associated with star-forming activity such as that found in various masers, molecular outflows, broad RRLs, and extended green objects. Accretion at the two earliest stages of  \hii\ region evolution tends to be quickly reduced or stopped, and therefore these regions could be optimal tracers of the final stages of massive star formation.


%

\begin{acknowledgements}
We would like to thank the anonymous referee for the helpful comments. 
A. Y. Yang thanks Yan Gong for his helpful discussion. 
WWT acknowledges support from the National Key R$\&$D Programs of China (2018YFA0404203). 
 This work has made use of the SIMBAD database (CDS, Strasbourg, France).
The VLA is operated by the National Radio Astronomy Observatory, which is a facility of the National Science Foundation operated under cooperative agreement by Associated Universities, Inc.

\end{acknowledgements}

\bibliographystyle{aa}
\bibliography{ref}

\begin{thebibliography}{136}
\expandafter\ifx\csname natexlab\endcsname\relax\def\natexlab#1{#1}\fi

\bibitem[{{Afflerbach} {et~al.}(1996){Afflerbach}, {Churchwell}, {Acord},
  {Hofner}, {Kurtz}, \& {Depree}}]{Afflerbach1996ApJS106}
{Afflerbach}, A., {Churchwell}, E., {Acord}, J.~M., {et~al.} 1996, \apjs, 106,
  423

\bibitem[{{Argon} {et~al.}(2000){Argon}, {Reid}, \& {Menten}}]{Argon2000ApJS}
{Argon}, A.~L., {Reid}, M.~J., \& {Menten}, K.~M. 2000, \apjs, 129, 159

\bibitem[{{Arthur} {et~al.}(2004){Arthur}, {Kurtz}, {Franco}, \&
  {Albarr{\'a}n}}]{Arthur2004ApJ608282A}
{Arthur}, S.~J., {Kurtz}, S.~E., {Franco}, J., \& {Albarr{\'a}n}, M.~Y. 2004,
  \apj, 608, 282

\bibitem[{{Avalos} {et~al.}(2006){Avalos}, {Lizano}, {Rodr{\'{\i}}guez},
  {Franco-Hern{\'a}ndez}, \& {Moran}}]{Avalos2006ApJ641406A}
{Avalos}, M., {Lizano}, S., {Rodr{\'{\i}}guez}, L.~F., {Franco-Hern{\'a}ndez},
  R., \& {Moran}, J.~M. 2006, \apj, 641, 406

\bibitem[{{Barsony}(1989)}]{Barsony1989ApJ345268B}
{Barsony}, M. 1989, \apj, 345, 268

\bibitem[{{Bartkiewicz} {et~al.}(2014){Bartkiewicz}, {Szymczak}, \& {van
  Langevelde}}]{Bartkiewicz2014AA564A110B}
{Bartkiewicz}, A., {Szymczak}, M., \& {van Langevelde}, H.~J. 2014, \aap, 564,
  A110

\bibitem[{{Bartkiewicz} {et~al.}(2016){Bartkiewicz}, {Szymczak}, \& {van
  Langevelde}}]{Bartkiewicz2016AA587A104B}
{Bartkiewicz}, A., {Szymczak}, M., \& {van Langevelde}, H.~J. 2016, \aap, 587,
  A104

\bibitem[{{Battersby} {et~al.}(2011){Battersby}, {Bally}, {Ginsburg},
  {Bernard}, {Brunt}, {Fuller}, {Martin}, {Molinari}, {Mottram}, {Peretto},
  {Testi}, \& {Thompson}}]{Battersby2011AA535A128B}
{Battersby}, C., {Bally}, J., {Ginsburg}, A., {et~al.} 2011, \aap, 535, A128

\bibitem[{{Beltr{\'a}n} {et~al.}(2007){Beltr{\'a}n}, {Cesaroni}, {Moscadelli},
  \& {Codella}}]{Beltran2007AA}
{Beltr{\'a}n}, M.~T., {Cesaroni}, R., {Moscadelli}, L., \& {Codella}, C. 2007,
  \aap, 471, L13

\bibitem[{{Beltr{\'a}n} {et~al.}(2011){Beltr{\'a}n}, {Cesaroni}, {Zhang},
  {Galv{\'a}n-Madrid}, {Beuther}, {Fallscheer}, {Neri}, \&
  {Codella}}]{Beltr2011AA532A91B}
{Beltr{\'a}n}, M.~T., {Cesaroni}, R., {Zhang}, Q., {et~al.} 2011, \aap, 532,
  A91

\bibitem[{{Benjamin} {et~al.}(2003){Benjamin}, {Churchwell}, {Babler}, {Bania},
  {Clemens}, {Cohen}, {Dickey}, {Indebetouw}, {Jackson}, {Kobulnicky},
  {Lazarian}, {Marston}, {Mathis}, {Meade}, {Seager}, {Stolovy}, {Watson},
  {Whitney}, {Wolff}, \& {Wolfire}}]{Benjamin2003PASP}
{Benjamin}, R.~A., {Churchwell}, E., {Babler}, B.~L., {et~al.} 2003, \pasp,
  115, 953

\bibitem[{{Beuther} {et~al.}(2016){Beuther}, {Bihr}, {Rugel}, {Johnston},
  {Wang}, {Walter}, {Brunthaler}, {Walsh}, {Ott}, {Stil}, {Henning},
  {Schierhuber}, {Kainulainen}, {Heyer}, {Goldsmith}, {Anderson}, {Longmore},
  {Klessen}, {Glover}, {Urquhart}, {Plume}, {Ragan}, {Schneider},
  {McClure-Griffiths}, {Menten}, {Smith}, {Roy}, {Shanahan}, {Nguyen-Luong}, \&
  {Bigiel}}]{Beuther2016AA595A32B}
{Beuther}, H., {Bihr}, S., {Rugel}, M., {et~al.} 2016, \aap, 595, A32

\bibitem[{{Beuther} {et~al.}(2018){Beuther}, {Mottram}, {Ahmadi}, {Bosco},
  {Linz}, {Henning}, {Klaassen}, {Winters}, {Maud}, {Kuiper}, {Semenov},
  {Gieser}, {Peters}, {Urquhart}, {Pudritz}, {Ragan}, {Feng}, {Keto},
  {Leurini}, {Cesaroni}, {Beltran}, {Palau}, {S{\'a}nchez-Monge},
  {Galvan-Madrid}, {Zhang}, {Schilke}, {Wyrowski}, {Johnston}, {Longmore},
  {Lumsden}, {Hoare}, {Menten}, \& {Csengeri}}]{Beuther2018AA617A100B}
{Beuther}, H., {Mottram}, J.~C., {Ahmadi}, A., {et~al.} 2018, \aap, 617, A100

\bibitem[{{Beuther} {et~al.}(2004){Beuther}, {Schilke}, \&
  {Gueth}}]{Beuther2004ApJ608330B}
{Beuther}, H., {Schilke}, P., \& {Gueth}, F. 2004, \apj, 608, 330

\bibitem[{{Bihr} {et~al.}(2016){Bihr}, {Johnston}, {Beuther}, {Anderson},
  {Ott}, {Rugel}, {Bigiel}, {Brunthaler}, {Glover}, {Henning}, {Heyer},
  {Klessen}, {Linz}, {Longmore}, {McClure-Griffiths}, {Menten}, {Plume},
  {Schierhuber}, {Shanahan}, {Stil}, {Urquhart}, \& {Walsh}}]{Bihr2016AA}
{Bihr}, S., {Johnston}, K.~G., {Beuther}, H., {et~al.} 2016, \aap, 588, A97

\bibitem[{{Bisbas} {et~al.}(2015){Bisbas}, {Haworth}, {Williams}, {Mackey},
  {Tremblin}, {Raga}, {Arthur}, {Baczynski}, {Dale}, {Frostholm}, {Geen},
  {Haugb{\o}lle}, {Hubber}, {Iliev}, {Kuiper}, {Rosdahl}, {Sullivan}, {Walch},
  \& {W{\"u}nsch}}]{Bisbas2015MNRAS4531324B}
{Bisbas}, T.~G., {Haworth}, T.~J., {Williams}, R.~J.~R., {et~al.} 2015, \mnras,
  453, 1324

\bibitem[{{Blum} {et~al.}(1999){Blum}, {Damineli}, \&
  {Conti}}]{Blum1999AJ1171392B}
{Blum}, R.~D., {Damineli}, A., \& {Conti}, P.~S. 1999, \aj, 117, 1392

\bibitem[{{Breen} {et~al.}(2015){Breen}, {Fuller}, {Caswell}, {Green},
  {Avison}, {Ellingsen}, {Gray}, {Pestalozzi}, {Quinn}, {Richards}, {Thompson},
  \& {Voronkov}}]{Breen2015MNRAS4109B}
{Breen}, S.~L., {Fuller}, G.~A., {Caswell}, J.~L., {et~al.} 2015, \mnras, 450,
  4109

\bibitem[{{Carey} {et~al.}(2009){Carey}, {Noriega-Crespo}, {Mizuno}, {Shenoy},
  {Paladini}, {Kraemer}, {Price}, {Flagey}, {Ryan}, {Ingalls}, {Kuchar},
  {Pinheiro Gon{\c{c}}alves}, {Indebetouw}, {Billot}, {Marleau}, {Padgett},
  {Rebull}, {Bressert}, {Ali}, {Molinari}, {Martin}, {Berriman}, {Boulanger},
  {Latter}, {Miville-Deschenes}, {Shipman}, \& {Testi}}]{Carey2009PASP76C}
{Carey}, S.~J., {Noriega-Crespo}, A., {Mizuno}, D.~R., {et~al.} 2009, \pasp,
  121, 76

\bibitem[{{Caswell} {et~al.}(1983){Caswell}, {Batchelor}, {Forster}, \&
  {Wellington}}]{Caswell1983AuJPh36443C}
{Caswell}, J.~L., {Batchelor}, R.~A., {Forster}, J.~R., \& {Wellington}, K.~J.
  1983, Australian Journal of Physics, 36, 443

\bibitem[{{Caswell} {et~al.}(2013){Caswell}, {Green}, \&
  {Phillips}}]{Caswell2013MNRAS4311180C}
{Caswell}, J.~L., {Green}, J.~A., \& {Phillips}, C.~J. 2013, \mnras, 431, 1180

\bibitem[{{Cesaroni} {et~al.}(2019){Cesaroni}, {Beltr{\'a}n}, {Moscadelli},
  {S{\'a}nchez-Monge}, \& {Neri}}]{Cesaroni2019AA624A100C}
{Cesaroni}, R., {Beltr{\'a}n}, M.~T., {Moscadelli}, L., {S{\'a}nchez-Monge},
  {\'A}., \& {Neri}, R. 2019, \aap, 624, A100

\bibitem[{{Cesaroni} {et~al.}(2015){Cesaroni}, {Pestalozzi}, {Beltr{\'a}n},
  {Hoare}, {Molinari}, {Olmi}, {Smith}, {Stringfellow}, {Testi}, \&
  {Thompson}}]{Cesaroni2015AA579A}
{Cesaroni}, R., {Pestalozzi}, M., {Beltr{\'a}n}, M.~T., {et~al.} 2015, \aap,
  579, A71

\bibitem[{{Churchwell} {et~al.}(2009){Churchwell}, {Babler}, {Meade},
  {Whitney}, {Benjamin}, {Indebetouw}, {Cyganowski}, {Robitaille}, {Povich},
  {Watson}, \& {Bracker}}]{Churchwell2009PASP}
{Churchwell}, E., {Babler}, B.~L., {Meade}, M.~R., {et~al.} 2009, \pasp, 121,
  213

\bibitem[{{Codella} {et~al.}(2013){Codella}, {Beltr{\'a}n}, {Cesaroni},
  {Moscadelli}, {Neri}, {Vasta}, \& {Zhang}}]{Codella2013AA550A81C}
{Codella}, C., {Beltr{\'a}n}, M.~T., {Cesaroni}, R., {et~al.} 2013, \aap, 550,
  A81

\bibitem[{{Contreras} {et~al.}(2013){Contreras}, {Schuller}, {Urquhart},
  {Csengeri}, {Wyrowski}, {Beuther}, {Bontemps}, {Bronfman}, {Henning},
  {Menten}, {Schilke}, {Walmsley}, {Wienen}, {Tackenberg}, \&
  {Linz}}]{Contreras2013AA549A45C}
{Contreras}, Y., {Schuller}, F., {Urquhart}, J.~S., {et~al.} 2013, \aap, 549,
  A45

\bibitem[{{Csengeri} {et~al.}(2016){Csengeri}, {Leurini}, {Wyrowski},
  {Urquhart}, {Menten}, {Walmsley}, {Bontemps}, {Wienen}, {Beuther}, {Motte},
  {Nguyen-Luong}, {Schilke}, {Schuller}, {Zavagno}, \&
  {Sanna}}]{Csengeri2016AA586A149C}
{Csengeri}, T., {Leurini}, S., {Wyrowski}, F., {et~al.} 2016, \aap, 586, A149

\bibitem[{{Cyganowski} {et~al.}(2008){Cyganowski}, {Whitney}, {Holden},
  {Braden}, {Brogan}, {Churchwell}, {Indebetouw}, {Watson}, {Babler},
  {Benjamin}, {Gomez}, {Meade}, {Povich}, {Robitaille}, \&
  {Watson}}]{Cyganowski2008AJ1362391C}
{Cyganowski}, C.~J., {Whitney}, B.~A., {Holden}, E., {et~al.} 2008, \aj, 136,
  2391

\bibitem[{{Davies} {et~al.}(2011){Davies}, {Hoare}, {Lumsden}, {Hosokawa},
  {Oudmaijer}, {Urquhart}, {Mottram}, \& {Stead}}]{Davies2011MNRAS972D}
{Davies}, B., {Hoare}, M.~G., {Lumsden}, S.~L., {et~al.} 2011, \mnras, 416, 972

\bibitem[{{de Pree} {et~al.}(1996){de Pree}, {Gaume}, {Goss}, \&
  {Claussen}}]{dePree1996ApJ}
{de Pree}, C.~G., {Gaume}, R.~A., {Goss}, W.~M., \& {Claussen}, M.~J. 1996,
  \apj, 464, 788

\bibitem[{{De Pree} {et~al.}(1997){De Pree}, {Mehringer}, \&
  {Goss}}]{dePree1997ApJ}
{De Pree}, C.~G., {Mehringer}, D.~M., \& {Goss}, W.~M. 1997, \apj, 482, 307

\bibitem[{{De Pree} {et~al.}(2000){De Pree}, {Wilner}, {Goss}, {Welch}, \&
  {McGrath}}]{dePree2000ApJ308D}
{De Pree}, C.~G., {Wilner}, D.~J., {Goss}, W.~M., {Welch}, W.~J., \& {McGrath},
  E. 2000, \apj, 540, 308

\bibitem[{{De Pree} {et~al.}(2004){De Pree}, {Wilner}, {Mercer}, {Davis},
  {Goss}, \& {Kurtz}}]{DePree2004ApJ}
{De Pree}, C.~G., {Wilner}, D.~J., {Mercer}, A.~J., {et~al.} 2004, \apj, 600,
  286

\bibitem[{{Djordjevic} {et~al.}(2019){Djordjevic}, {Thompson}, {Urquhart}, \&
  {Forbrich}}]{Djordjevic2019MNRAS4871057D}
{Djordjevic}, J.~O., {Thompson}, M.~A., {Urquhart}, J.~S., \& {Forbrich}, J.
  2019, \mnras, 487, 1057

\bibitem[{{Dyson} \& {Williams}(1997)}]{Dyson1997pismbook}
{Dyson}, J.~E. \& {Williams}, D.~A. 1997, {The physics of the interstellar
  medium}, 2nd edn. (Bristol and Philadelphia: Institute of Physics Publishing)

\bibitem[{{Dyson} {et~al.}(1995){Dyson}, {Williams}, \&
  {Redman}}]{Dyson1995MNRAS277}
{Dyson}, J.~E., {Williams}, R.~J.~R., \& {Redman}, M.~P. 1995, \mnras, 277, 700

\bibitem[{{Eden} {et~al.}(2017){Eden}, {Moore}, {Plume}, {Urquhart},
  {Thompson}, {Parsons}, {Dempsey}, {Rigby}, {Morgan}, {Thomas}, {Berry},
  {Buckle}, {Brunt}, {Butner}, {Carretero}, {Chrysostomou}, {Currie},
  {deVilliers}, {Fich}, {Gibb}, {Hoare}, {Jenness}, {Manser}, {Mottram},
  {Natario}, {Olguin}, {Peretto}, {Pestalozzi}, {Polychroni}, {Redman},
  {Salji}, {Summers}, {Tahani}, {Traficante}, {diFrancesco}, {Evans}, {Fuller},
  {Johnstone}, {Joncas}, {Longmore}, {Martin}, {Richer}, {Weferling}, {White},
  \& {Zhu}}]{Eden2017MNRAS4692163E}
{Eden}, D.~J., {Moore}, T.~J.~T., {Plume}, R., {et~al.} 2017, \mnras, 469, 2163

\bibitem[{{Ellingsen} {et~al.}(2011){Ellingsen}, {Breen}, {Sobolev},
  {Voronkov}, {Caswell}, \& {Lo}}]{Ellingsen2011ApJ109E}
{Ellingsen}, S.~P., {Breen}, S.~L., {Sobolev}, A.~M., {et~al.} 2011, \apj, 742,
  109

\bibitem[{{Forster} \& {Caswell}(1999)}]{Forster1999AAS}
{Forster}, J.~R. \& {Caswell}, J.~L. 1999, \aaps, 137, 43

\bibitem[{{Forster} \& {Caswell}(2000)}]{Forster2000ApJ371F}
{Forster}, J.~R. \& {Caswell}, J.~L. 2000, \apj, 530, 371

\bibitem[{{Fu} \& {Lin}(2016)}]{Fu2016RAA16182F}
{Fu}, L. \& {Lin}, G.-M. 2016, Research in Astronomy and Astrophysics, 16, 182

\bibitem[{{Furuya} {et~al.}(2002){Furuya}, {Cesaroni}, {Codella}, {Testi},
  {Bachiller}, \& {Tafalla}}]{Furuya2002AA390L1F}
{Furuya}, R.~S., {Cesaroni}, R., {Codella}, C., {et~al.} 2002, \aap, 390, L1

\bibitem[{{Gao} {et~al.}(2019){Gao}, {Reich}, {Hou}, {Reich}, \&
  {Han}}]{Gao2019AA623A105G}
{Gao}, X.~Y., {Reich}, P., {Hou}, L.~G., {Reich}, W., \& {Han}, J.~L. 2019,
  \aap, 623, A105

\bibitem[{{Garay} \& {Lizano}(1999)}]{Garay1999PASP1049G}
{Garay}, G. \& {Lizano}, S. 1999, \pasp, 111, 1049

\bibitem[{{Garay} {et~al.}(1998){Garay}, {Lizano}, {G{\'o}mez}, \&
  {Brown}}]{Garay1998ApJ501710G}
{Garay}, G., {Lizano}, S., {G{\'o}mez}, Y., \& {Brown}, R.~L. 1998, \apj, 501,
  710

\bibitem[{{Garay} {et~al.}(1993){Garay}, {Rodriguez}, {Moran}, \&
  {Churchwell}}]{Garay1993ApJ418368G}
{Garay}, G., {Rodriguez}, L.~F., {Moran}, J.~M., \& {Churchwell}, E. 1993,
  \apj, 418, 368

\bibitem[{{Gibb} \& {Hoare}(2007)}]{Gibb2007MNRAS246G}
{Gibb}, A.~G. \& {Hoare}, M.~G. 2007, \mnras, 380, 246

\bibitem[{{Hatchell} {et~al.}(1998){Hatchell}, {Thompson}, {Millar}, \&
  {MacDonald}}]{Hatchell1998AAS13329H}
{Hatchell}, J., {Thompson}, M.~A., {Millar}, T.~J., \& {MacDonald}, G.~H. 1998,
  \aaps, 133, 29

\bibitem[{{Helfand} {et~al.}(2006){Helfand}, {Becker}, {White}, {Fallon}, \&
  {Tuttle}}]{Helfand2006AJ}
{Helfand}, D.~J., {Becker}, R.~H., {White}, R.~L., {Fallon}, A., \& {Tuttle},
  S. 2006, \aj, 131, 2525

\bibitem[{{Henkel} {et~al.}(1986){Henkel}, {Haschick}, \&
  {Guesten}}]{Henkel1986AA165197H}
{Henkel}, C., {Haschick}, A.~D., \& {Guesten}, R. 1986, \aap, 165, 197

\bibitem[{{Hoare} {et~al.}(2007){Hoare}, {Kurtz}, {Lizano}, {Keto}, \&
  {Hofner}}]{Hoare2007prplconfH}
{Hoare}, M.~G., {Kurtz}, S.~E., {Lizano}, S., {Keto}, E., \& {Hofner}, P. 2007,
  Protostars and Planets V, 181

\bibitem[{{Hoare} {et~al.}(2012){Hoare}, {Purcell}, {Churchwell}, {Diamond},
  {Cotton}, {Chandler}, {Smethurst}, {Kurtz}, {Mundy}, {Dougherty}, {Fender},
  {Fuller}, {Jackson}, {Garrington}, {Gledhill}, {Goldsmith}, {Lumsden},
  {Mart{\'{\i}}}, {Moore}, {Muxlow}, {Oudmaijer}, {Pandian}, {Paredes},
  {Shepherd}, {Spencer}, {Thompson}, {Umana}, {Urquhart}, \&
  {Zijlstra}}]{Hoare2012PASP}
{Hoare}, M.~G., {Purcell}, C.~R., {Churchwell}, E.~B., {et~al.} 2012, \pasp,
  124, 939

\bibitem[{{Hofner} \& {Churchwell}(1996)}]{Hofner1996AAS120283H}
{Hofner}, P. \& {Churchwell}, E. 1996, \aaps, 120, 283

\bibitem[{{Hollenbach} {et~al.}(1994){Hollenbach}, {Johnstone}, {Lizano}, \&
  {Shu}}]{Hollenbach1994ApJ654H}
{Hollenbach}, D., {Johnstone}, D., {Lizano}, S., \& {Shu}, F. 1994, \apj, 428,
  654

\bibitem[{{Hosokawa} \& {Omukai}(2009)}]{Hosokawa2009ApJ691}
{Hosokawa}, T. \& {Omukai}, K. 2009, \apj, 691, 823

\bibitem[{{Hosokawa} {et~al.}(2010){Hosokawa}, {Yorke}, \&
  {Omukai}}]{Hosokawa2010ApJ721}
{Hosokawa}, T., {Yorke}, H.~W., \& {Omukai}, K. 2010, \apj, 721, 478

\bibitem[{{Imai} {et~al.}(2011){Imai}, {Omi}, {Kurayama}, {Nagayama}, {Hirota},
  {Miyaji}, \& {Omodaka}}]{Imai2011PASJ631293I}
{Imai}, H., {Omi}, R., {Kurayama}, T., {et~al.} 2011, \pasj, 63, 1293

\bibitem[{{Kalcheva} {et~al.}(2018){Kalcheva}, {Hoare}, {Urquhart}, {Kurtz},
  {Lumsden}, {Purcell}, \& {Zijlstra}}]{Kalcheva2018AA615A103K}
{Kalcheva}, I.~E., {Hoare}, M.~G., {Urquhart}, J.~S., {et~al.} 2018, \aap, 615,
  A103

\bibitem[{{Keto}(2002)}]{Keto2002ApJ980K}
{Keto}, E. 2002, \apj, 580, 980

\bibitem[{{Keto}(2003)}]{Keto2003ApJ}
{Keto}, E. 2003, \apj, 599, 1196

\bibitem[{{Keto}(2007)}]{Keto2007ApJK}
{Keto}, E. 2007, \apj, 666, 976

\bibitem[{{Keto} \& {Wood}(2006)}]{Keto2006ApJ850K}
{Keto}, E. \& {Wood}, K. 2006, \apj, 637, 850

\bibitem[{{Keto} {et~al.}(2008){Keto}, {Zhang}, \& {Kurtz}}]{Keto2008ApJ672}
{Keto}, E., {Zhang}, Q., \& {Kurtz}, S. 2008, \apj, 672, 423

\bibitem[{{Kim} {et~al.}(2000){Kim}, {Cho}, {Chung}, {Kim}, {Roh}, {Kim},
  {Minh}, \& {Minn}}]{Kim2000ApJS131483K}
{Kim}, H.-D., {Cho}, S.-H., {Chung}, H.-S., {et~al.} 2000, \apjs, 131, 483

\bibitem[{{Kim} \& {Koo}(2001)}]{Kim2001ApJ549}
{Kim}, K.-T. \& {Koo}, B.-C. 2001, \apj, 549, 979

\bibitem[{{Kim} {et~al.}(2018){Kim}, {Urquhart}, {Wyrowski}, {Menten}, \&
  {Csengeri}}]{Kim2018AA616A107K}
{Kim}, W.~J., {Urquhart}, J.~S., {Wyrowski}, F., {Menten}, K.~M., \&
  {Csengeri}, T. 2018, \aap, 616, A107

\bibitem[{{Kim} {et~al.}(2017){Kim}, {Wyrowski}, {Urquhart}, {Menten}, \&
  {Csengeri}}]{Kim2017AA602A}
{Kim}, W.-J., {Wyrowski}, F., {Urquhart}, J.~S., {Menten}, K.~M., \&
  {Csengeri}, T. 2017, \aap, 602, A37

\bibitem[{{K{\"o}nig} {et~al.}(2017){K{\"o}nig}, {Urquhart}, {Csengeri},
  {Leurini}, {Wyrowski}, {Giannetti}, {Wienen}, {Pillai}, {Kauffmann},
  {Menten}, \& {Schuller}}]{Konig2017AA599A139K}
{K{\"o}nig}, C., {Urquhart}, J.~S., {Csengeri}, T., {et~al.} 2017, \aap, 599,
  A139

\bibitem[{{Kurtz}(2005)}]{Kurtz2005IAUS}
{Kurtz}, S. 2005, in IAU Symposium, Vol. 227, Massive Star Birth: A Crossroads
  of Astrophysics, ed. R.~{Cesaroni}, M.~{Felli}, E.~{Churchwell}, \&
  M.~{Walmsley}, 111--119

\bibitem[{{Kurtz} {et~al.}(2000){Kurtz}, {Cesaroni}, {Churchwell}, {Hofner}, \&
  {Walmsley}}]{Kurtz2000prplconfK}
{Kurtz}, S., {Cesaroni}, R., {Churchwell}, E., {Hofner}, P., \& {Walmsley},
  C.~M. 2000, Protostars and Planets IV, 299

\bibitem[{{Kurtz} {et~al.}(1994){Kurtz}, {Churchwell}, \&
  {Wood}}]{Kurtz1994ApJS659K}
{Kurtz}, S., {Churchwell}, E., \& {Wood}, D.~O.~S. 1994, \apjs, 91, 659

\bibitem[{{Kurtz} \& {Hofner}(2005)}]{Kurtz2005AJ130711K}
{Kurtz}, S. \& {Hofner}, P. 2005, \aj, 130, 711

\bibitem[{{Kurtz} {et~al.}(2004){Kurtz}, {Hofner}, \&
  {{\'A}lvarez}}]{Kurtz2004ApJS155149K}
{Kurtz}, S., {Hofner}, P., \& {{\'A}lvarez}, C.~V. 2004, \apjs, 155, 149

\bibitem[{{Leto} {et~al.}(2009){Leto}, {Umana}, {Trigilio}, {Buemi}, {Dolei},
  {Manzitto}, {Cerrigone}, \& {Siringo}}]{Leto2009AA5071467L}
{Leto}, P., {Umana}, G., {Trigilio}, C., {et~al.} 2009, \aap, 507, 1467

\bibitem[{{L{\'o}pez-Sepulcre} {et~al.}(2009){L{\'o}pez-Sepulcre}, {Codella},
  {Cesaroni}, {Marcelino}, \& {Walmsley}}]{Sepulcre2009AA499811L}
{L{\'o}pez-Sepulcre}, A., {Codella}, C., {Cesaroni}, R., {Marcelino}, N., \&
  {Walmsley}, C.~M. 2009, \aap, 499, 811

\bibitem[{{Lumsden} {et~al.}(2013){Lumsden}, {Hoare}, {Urquhart}, {Oudmaijer},
  {Davies}, {Mottram}, {Cooper}, \& {Moore}}]{Lumsden2013ApJS208}
{Lumsden}, S.~L., {Hoare}, M.~G., {Urquhart}, J.~S., {et~al.} 2013, \apjs, 208,
  11

\bibitem[{{Maud} {et~al.}(2015){Maud}, {Moore}, {Lumsden}, {Mottram},
  {Urquhart}, \& {Hoare}}]{Maud2015MNRAS453645M}
{Maud}, L.~T., {Moore}, T.~J.~T., {Lumsden}, S.~L., {et~al.} 2015, \mnras, 453,
  645

\bibitem[{{McKee} \& {Tan}(2003)}]{McKee2003ApJM}
{McKee}, C.~F. \& {Tan}, J.~C. 2003, \apj, 585, 850

\bibitem[{{McMullin} {et~al.}(2007){McMullin}, {Waters}, {Schiebel}, {Young},
  \& {Golap}}]{McMullin2007ASPC376127M}
{McMullin}, J.~P., {Waters}, B., {Schiebel}, D., {Young}, W., \& {Golap}, K.
  2007, Astronomical Society of the Pacific Conference Series, Vol. 376, {CASA
  Architecture and Applications}, ed. R.~A. {Shaw}, F.~{Hill}, \& D.~J. {Bell},
  127

\bibitem[{{Medina} {et~al.}(2019){Medina}, {Urquhart}, {Dzib}, {Brunthaler},
  {Cotton}, {Menten}, {Wyrowski}, {Beuther}, {Billington}, {Carrasco-Gonzalez},
  {Csengeri}, {Gong}, {Hofner}, {Nguyen}, {Ortiz-Le{\'o}n}, {Ott}, {Pandian},
  {Roy}, {Sarkar}, {Wang}, \& {Winkel}}]{Medina2019AA627A175M}
{Medina}, S. N.~X., {Urquhart}, J.~S., {Dzib}, S.~A., {et~al.} 2019, \aap, 627,
  A175

\bibitem[{{Mezger} \& {Henderson}(1967)}]{Mezger1967ApJ}
{Mezger}, P.~G. \& {Henderson}, A.~P. 1967, \apj, 147, 471

\bibitem[{{Molinari} {et~al.}(2010){Molinari}, {Swinyard}, {Bally}, {Barlow},
  {Bernard}, {Martin}, {Moore}, {Noriega-Crespo}, {Plume}, {Testi}, {Zavagno},
  {Abergel}, {Ali}, {Andr{\'e}}, {Baluteau}, {Benedettini}, {Bern{\'e}},
  {Billot}, {Blommaert}, {Bontemps}, {Boulanger}, {Brand}, {Brunt}, {Burton},
  {Campeggio}, {Carey}, {Caselli}, {Cesaroni}, {Cernicharo}, {Chakrabarti},
  {Chrysostomou}, {Codella}, {Cohen}, {Compiegne}, {Davis}, {de Bernardis}, {de
  Gasperis}, {Di Francesco}, {di Giorgio}, {Elia}, {Faustini}, {Fischera},
  {Fukui}, {Fuller}, {Ganga}, {Garcia-Lario}, {Giard}, {Giardino}, {Glenn},
  {Goldsmith}, {Griffin}, {Hoare}, {Huang}, {Jiang}, {Joblin}, {Joncas},
  {Juvela}, {Kirk}, {Lagache}, {Li}, {Lim}, {Lord}, {Lucas}, {Maiolo},
  {Marengo}, {Marshall}, {Masi}, {Massi}, {Matsuura}, {Meny}, {Minier},
  {Miville-Desch{\^e}nes}, {Montier}, {Motte}, {M{\"u}ller}, {Natoli}, {Neves},
  {Olmi}, {Paladini}, {Paradis}, {Pestalozzi}, {Pezzuto}, {Piacentini},
  {Pomar{\`e}s}, {Popescu}, {Reach}, {Richer}, {Ristorcelli}, {Roy}, {Royer},
  {Russeil}, {Saraceno}, {Sauvage}, {Schilke}, {Schneider-Bontemps},
  {Schuller}, {Schultz}, {Shepherd}, {Sibthorpe}, {Smith}, {Smith},
  {Spinoglio}, {Stamatellos}, {Strafella}, {Stringfellow}, {Sturm}, {Taylor},
  {Thompson}, {Tuffs}, {Umana}, {Valenziano}, {Vavrek}, {Viti}, {Waelkens},
  {Ward-Thompson}, {White}, {Wyrowski}, {Yorke}, \& {Zhang}}]{Molinari2010PASP}
{Molinari}, S., {Swinyard}, B., {Bally}, J., {et~al.} 2010, \pasp, 122, 314

\bibitem[{{Moscadelli} {et~al.}(2016){Moscadelli}, {S{\'a}nchez-Monge},
  {Goddi}, {Li}, {Sanna}, {Cesaroni}, {Pestalozzi}, {Molinari}, \&
  {Reid}}]{Moscadelli2016AA585A71M}
{Moscadelli}, L., {S{\'a}nchez-Monge}, {\'A}., {Goddi}, C., {et~al.} 2016,
  \aap, 585, A71

\bibitem[{{Motte} {et~al.}(2018){Motte}, {Bontemps}, \&
  {Louvet}}]{Motte2018ARAA41M}
{Motte}, F., {Bontemps}, S., \& {Louvet}, F. 2018, \araa, 56, 41

\bibitem[{{Murphy} {et~al.}(2010){Murphy}, {Cohen}, {Ekers}, {Green}, {Wark},
  \& {Moss}}]{Murphy2010MNRASa2}
{Murphy}, T., {Cohen}, M., {Ekers}, R.~D., {et~al.} 2010, \mnras, 405, 1560

\bibitem[{{Panagia}(1973)}]{Panagia1973AJ78}
{Panagia}, N. 1973, \aj, 78, 929

\bibitem[{{Panagia} \& {Felli}(1975)}]{Panagia1975AA391P}
{Panagia}, N. \& {Felli}, M. 1975, \aap, 39, 1

\bibitem[{{Pascucci} {et~al.}(2004){Pascucci}, {Apai}, {Henning}, {Stecklum},
  \& {Brandl}}]{Pascucci2004AA426523P}
{Pascucci}, I., {Apai}, D., {Henning}, T., {Stecklum}, B., \& {Brandl}, B.
  2004, \aap, 426, 523

\bibitem[{{Pestalozzi} {et~al.}(2005){Pestalozzi}, {Minier}, \&
  {Booth}}]{Pestalozzi2005AA432737P}
{Pestalozzi}, M.~R., {Minier}, V., \& {Booth}, R.~S. 2005, \aap, 432, 737

\bibitem[{{Peters} {et~al.}(2010){Peters}, {Banerjee}, {Klessen}, {Mac Low},
  {Galv{\'a}n-Madrid}, \& {Keto}}]{Peters2010ApJ711}
{Peters}, T., {Banerjee}, R., {Klessen}, R.~S., {et~al.} 2010, \apj, 711, 1017

\bibitem[{{Phillips} \& {Mampaso}(1991)}]{Phillips1991AAS88189P}
{Phillips}, J.~P. \& {Mampaso}, A. 1991, \aaps, 88, 189

\bibitem[{{Purcell} {et~al.}(2013){Purcell}, {Hoare}, {Cotton}, {Lumsden},
  {Urquhart}, {Chandler}, {Churchwell}, {Diamond}, {Dougherty}, {Fender},
  {Fuller}, {Garrington}, {Gledhill}, {Goldsmith}, {Hindson}, {Jackson},
  {Kurtz}, {Mart{\'{\i}}}, {Moore}, {Mundy}, {Muxlow}, {Oudmaijer}, {Pandian},
  {Paredes}, {Shepherd}, {Smethurst}, {Spencer}, {Thompson}, {Umana}, \&
  {Zijlstra}}]{Purcell2013ApJS}
{Purcell}, C.~R., {Hoare}, M.~G., {Cotton}, W.~D., {et~al.} 2013, \apjs, 205, 1

\bibitem[{{Purser} {et~al.}(2016){Purser}, {Lumsden}, {Hoare}, {Urquhart},
  {Cunningham}, {Purcell}, {Brooks}, {Garay}, {G{\'u}zman}, \&
  {Voronkov}}]{Purser2016MNRAS4601039P}
{Purser}, S.~J.~D., {Lumsden}, S.~L., {Hoare}, M.~G., {et~al.} 2016, \mnras,
  460, 1039

\bibitem[{{Rodr{\'{\i}}guez-Esnard} {et~al.}(2012){Rodr{\'{\i}}guez-Esnard},
  {Trinidad}, \& {Migenes}}]{Rodrguez2012ApJR}
{Rodr{\'{\i}}guez-Esnard}, T., {Trinidad}, M.~A., \& {Migenes}, V. 2012, \apj,
  761, 158

\bibitem[{{Ruiz-Velasco} {et~al.}(2016){Ruiz-Velasco}, {Felli}, {Migenes}, \&
  {Wiggins}}]{RuizVelasco2016ApJ822101R}
{Ruiz-Velasco}, A.~E., {Felli}, D., {Migenes}, V., \& {Wiggins}, B.~K. 2016,
  \apj, 822, 101

\bibitem[{{S{\'a}nchez-Monge} {et~al.}(2013){S{\'a}nchez-Monge}, {Beltr{\'a}n},
  {Cesaroni}, {Fontani}, {Brand}, {Molinari}, {Testi}, \&
  {Burton}}]{Sanchez_Monge2013AA21S}
{S{\'a}nchez-Monge}, {\'A}., {Beltr{\'a}n}, M.~T., {Cesaroni}, R., {et~al.}
  2013, \aap, 550, A21

\bibitem[{{Sarma} {et~al.}(2013){Sarma}, {Brogan}, {Bourke}, {Eftimova}, \&
  {Troland}}]{Sarma2013ApJ76724S}
{Sarma}, A.~P., {Brogan}, C.~L., {Bourke}, T.~L., {Eftimova}, M., \& {Troland},
  T.~H. 2013, \apj, 767, 24

\bibitem[{{Schlingman} {et~al.}(2011){Schlingman}, {Shirley}, {Schenk},
  {Rosolowsky}, {Bally}, {Battersby}, {Dunham}, {Ellsworth-Bowers}, {Evans},
  {Ginsburg}, \& {Stringfellow}}]{Schlingman2011ApJS19514S}
{Schlingman}, W.~M., {Shirley}, Y.~L., {Schenk}, D.~E., {et~al.} 2011, \apjs,
  195, 14

\bibitem[{{Schuller} {et~al.}(2017){Schuller}, {Csengeri}, {Urquhart},
  {Duarte-Cabral}, {Barnes}, {Giannetti}, {Hernand ez}, {Leurini}, {Mattern},
  {Medina}, {Agurto}, {Azagra}, {Anderson}, {Beltr{\'a}n}, {Beuther},
  {Bontemps}, {Bronfman}, {Dobbs}, {Dumke}, {Finger}, {Ginsburg}, {Gonzalez},
  {Henning}, {Kauffmann}, {Mac-Auliffe}, {Menten}, {Montenegro-Montes},
  {Moore}, {Muller}, {Parra}, {Perez-Beaupuits}, {Pettitt}, {Russeil},
  {S{\'a}nchez-Monge}, {Schilke}, {Schisano}, {Suri}, {Testi}, {Torstensson},
  {Venegas}, {Wang}, {Wienen}, {Wyrowski}, \&
  {Zavagno}}]{Schuller2017AA601A124S}
{Schuller}, F., {Csengeri}, T., {Urquhart}, J.~S., {et~al.} 2017, \aap, 601,
  A124

\bibitem[{{Schuller} {et~al.}(2009){Schuller}, {Menten}, {Contreras},
  {Wyrowski}, {Schilke}, {Bronfman}, {Henning}, {Walmsley}, {Beuther},
  {Bontemps}, {Cesaroni}, {Deharveng}, {Garay}, {Herpin}, {Lefloch}, {Linz},
  {Mardones}, {Minier}, {Molinari}, {Motte}, {Nyman}, {Reveret}, {Risacher},
  {Russeil}, {Schneider}, {Testi}, {Troost}, {Vasyunina}, {Wienen}, {Zavagno},
  {Kovacs}, {Kreysa}, {Siringo}, \& {Wei{\ss}}}]{Schuller2009AA}
{Schuller}, F., {Menten}, K.~M., {Contreras}, Y., {et~al.} 2009, \aap, 504, 415

\bibitem[{{Scoville} {et~al.}(1986){Scoville}, {Sargent}, {Sanders},
  {Claussen}, {Masson}, {Lo}, \& {Phillips}}]{Scoville1986ApJ303416S}
{Scoville}, N.~Z., {Sargent}, A.~I., {Sanders}, D.~B., {et~al.} 1986, \apj,
  303, 416

\bibitem[{{Sewilo} {et~al.}(2004){Sewilo}, {Churchwell}, {Kurtz}, {Goss}, \&
  {Hofner}}]{Sewilo2004ApJ}
{Sewilo}, M., {Churchwell}, E., {Kurtz}, S., {Goss}, W.~M., \& {Hofner}, P.
  2004, \apj, 605, 285

\bibitem[{{Sewi{\l}o} {et~al.}(2011){Sewi{\l}o}, {Churchwell}, {Kurtz}, {Goss},
  \& {Hofner}}]{Sewilo2011ApJS}
{Sewi{\l}o}, M., {Churchwell}, E., {Kurtz}, S., {Goss}, W.~M., \& {Hofner}, P.
  2011, \apjs, 194, 44

\bibitem[{{Shirley} {et~al.}(2013){Shirley}, {Ellsworth-Bowers}, {Svoboda},
  {Schlingman}, {Ginsburg}, {Rosolowsky}, {Gerner}, {Mairs}, {Battersby},
  {Stringfellow}, {Dunham}, {Glenn}, \& {Bally}}]{Shirley2013ApJS2092S}
{Shirley}, Y.~L., {Ellsworth-Bowers}, T.~P., {Svoboda}, B., {et~al.} 2013,
  \apjs, 209, 2

\bibitem[{{Spitzer}(1978)}]{Spitzer1978ppimbookS}
{Spitzer}, L. 1978, {Physical processes in the interstellar medium} (WILEY-VCH
  Verlag GmbH \& Co. KGaA)

\bibitem[{{Sridharan} {et~al.}(2002){Sridharan}, {Beuther}, {Schilke},
  {Menten}, \& {Wyrowski}}]{Sridharan2002ApJ566931S}
{Sridharan}, T.~K., {Beuther}, H., {Schilke}, P., {Menten}, K.~M., \&
  {Wyrowski}, F. 2002, \apj, 566, 931

\bibitem[{{Steggles}(2016)}]{Steggles2016PhDT374S}
{Steggles}, H.~G. 2016, PhD thesis, University of Leeds

\bibitem[{{Steggles} {et~al.}(2017){Steggles}, {Hoare}, \&
  {Pittard}}]{Steggles2017MNRAS4664573S}
{Steggles}, H.~G., {Hoare}, M.~G., \& {Pittard}, J.~M. 2017, \mnras, 466, 4573

\bibitem[{{Surcis} {et~al.}(2015){Surcis}, {Vlemmings}, {van Langevelde},
  {Hutawarakorn Kramer}, {Bartkiewicz}, \& {Blasi}}]{Surcis2015AA578A102S}
{Surcis}, G., {Vlemmings}, W.~H.~T., {van Langevelde}, H.~J., {et~al.} 2015,
  \aap, 578, A102

\bibitem[{{Svoboda} {et~al.}(2016){Svoboda}, {Shirley}, {Battersby},
  {Rosolowsky}, {Ginsburg}, {Ellsworth-Bowers}, {Pestalozzi}, {Dunham},
  {Evans}, {Bally}, \& {Glenn}}]{Svoboda2016ApJ82259S}
{Svoboda}, B.~E., {Shirley}, Y.~L., {Battersby}, C., {et~al.} 2016, \apj, 822,
  59

\bibitem[{{Thompson} {et~al.}(2015){Thompson}, {Beuther}, {Dickinson},
  {MOttram}, {Klaassen}, {Ginsburg}, {Longmore}, {Remijan}, \&
  {Menten}}]{thompson2015}
{Thompson}, M., {Beuther}, H., {Dickinson}, C., {et~al.} 2015, Advancing
  Astrophysics with the Square Kilometre Array (AASKA14), 126

\bibitem[{{Thompson} {et~al.}(2016){Thompson}, {Goedhart}, {Goedhart},
  {Benaglia}, {Beuther}, {Blomme}, {Chrysostomou}, {Clark}, {Dickinson},
  {Ellingsen}, {Fenech}, {Hindson}, {Longmore}, {van Langevelde}, {MacLeod},
  {Molinari}, {Prinja}, {Purcell}, {Stevens}, {Umana}, {Urquhart}, {Vlemmings},
  {Walsh}, {Yang}, \& {Zijlstra}}]{Thompson2016mksconfE15T}
{Thompson}, M., {Goedhart}, S., {Goedhart}, S., {et~al.} 2016, in MeerKAT
  Science: On the Pathway to the SKA, 15

\bibitem[{{Thompson} {et~al.}(2006){Thompson}, {Hatchell}, {Walsh},
  {MacDonald}, \& {Millar}}]{Thompson2006AA}
{Thompson}, M.~A., {Hatchell}, J., {Walsh}, A.~J., {MacDonald}, G.~H., \&
  {Millar}, T.~J. 2006, \aap, 453, 1003

\bibitem[{{Thompson}(1984)}]{Thompson1984ApJ165T}
{Thompson}, R.~I. 1984, \apj, 283, 165

\bibitem[{{Urquhart} {et~al.}(2007){Urquhart}, {Busfield}, {Hoare}, {Lumsden},
  {Clarke}, {Moore}, {Mottram}, \& {Oudmaijer}}]{Urquhart2007AA461}
{Urquhart}, J.~S., {Busfield}, A.~L., {Hoare}, M.~G., {et~al.} 2007, \aap, 461,
  11

\bibitem[{{Urquhart} {et~al.}(2014){Urquhart}, {Csengeri}, {Wyrowski},
  {Schuller}, {Bontemps}, {Bronfman}, {Menten}, {Walmsley}, {Contreras},
  {Beuther}, {Wienen}, \& {Linz}}]{Urquhart2014AA568A41U}
{Urquhart}, J.~S., {Csengeri}, T., {Wyrowski}, F., {et~al.} 2014, \aap, 568,
  A41

\bibitem[{{Urquhart} {et~al.}(2009{\natexlab{a}}){Urquhart}, {Hoare},
  {Lumsden}, {Oudmaijer}, {Moore}, {Brook}, {Mottram}, {Davies}, \&
  {Stead}}]{Urquhart2009AA507795U}
{Urquhart}, J.~S., {Hoare}, M.~G., {Lumsden}, S.~L., {et~al.}
  2009{\natexlab{a}}, \aap, 507, 795

\bibitem[{{Urquhart} {et~al.}(2009{\natexlab{b}}){Urquhart}, {Hoare},
  {Purcell}, {Lumsden}, {Oudmaijer}, {Moore}, {Busfield}, {Mottram}, \&
  {Davies}}]{Urquhart2009AA501}
{Urquhart}, J.~S., {Hoare}, M.~G., {Purcell}, C.~R., {et~al.}
  2009{\natexlab{b}}, \aap, 501, 539

\bibitem[{{Urquhart} {et~al.}(2018){Urquhart}, {K{\"o}nig}, {Giannetti},
  {Leurini}, {Moore}, {Eden}, {Pillai}, {Thompson}, {Braiding}, {Burton},
  {Csengeri}, {Dempsey}, {Figura}, {Froebrich}, {Menten}, {Schuller}, {Smith},
  \& {Wyrowski}}]{Urquhart2018MNRAS4731059U}
{Urquhart}, J.~S., {K{\"o}nig}, C., {Giannetti}, A., {et~al.} 2018, \mnras,
  473, 1059

\bibitem[{{Urquhart} {et~al.}(2011){Urquhart}, {Morgan}, {Figura}, {Moore},
  {Lumsden}, {Hoare}, {Oudmaijer}, {Mottram}, {Davies}, \&
  {Dunham}}]{Urquhart2011MNRAS4181689U}
{Urquhart}, J.~S., {Morgan}, L.~K., {Figura}, C.~C., {et~al.} 2011, \mnras,
  418, 1689

\bibitem[{{Urquhart} {et~al.}(2013){Urquhart}, {Thompson}, {Moore}, {Purcell},
  {Hoare}, {Schuller}, {Wyrowski}, {Csengeri}, {Menten}, {Lumsden}, {Kurtz},
  {Walmsley}, {Bronfman}, {Morgan}, {Eden}, \&
  {Russeil}}]{Urquhart2013MNRAS435}
{Urquhart}, J.~S., {Thompson}, M.~A., {Moore}, T.~J.~T., {et~al.} 2013, \mnras,
  435, 400

\bibitem[{{Valdettaro} {et~al.}(2001){Valdettaro}, {Palla}, {Brand},
  {Cesaroni}, {Comoretto}, {Di Franco}, {Felli}, {Natale}, {Palagi}, {Panella},
  \& {Tofani}}]{Valdettaro2001AA368845V}
{Valdettaro}, R., {Palla}, F., {Brand}, J., {et~al.} 2001, \aap, 368, 845

\bibitem[{{van der Tak} \& {Menten}(2005)}]{vandertak2005AA947V}
{van der Tak}, F.~F.~S. \& {Menten}, K.~M. 2005, \aap, 437, 947

\bibitem[{{van der Tak} {et~al.}(2019){van der Tak}, {Shipman}, {Jacq},
  {Herpin}, {Braine}, \& {Wyrowski}}]{vanderTak2019AA625A103V}
{van der Tak}, F.~F.~S., {Shipman}, R.~F., {Jacq}, T., {et~al.} 2019, \aap,
  625, A103

\bibitem[{{Walsh} {et~al.}(1998){Walsh}, {Burton}, {Hyland}, \&
  {Robinson}}]{Walsh1998MNRAS301}
{Walsh}, A.~J., {Burton}, M.~G., {Hyland}, A.~R., \& {Robinson}, G. 1998,
  \mnras, 301, 640

\bibitem[{{White} \& {Fridlund}(1992)}]{White1992AA266452W}
{White}, G.~J. \& {Fridlund}, C.~V.~M. 1992, \aap, 266, 452

\bibitem[{{White} {et~al.}(2005){White}, {Becker}, \& {Helfand}}]{White2005AJ}
{White}, R.~L., {Becker}, R.~H., \& {Helfand}, D.~J. 2005, \aj, 130, 586

\bibitem[{{Wood} \& {Churchwell}(1989)}]{Wood1989ApJS}
{Wood}, D.~O.~S. \& {Churchwell}, E. 1989, \apjs, 69, 831

\bibitem[{{Wyrowski} {et~al.}(2016){Wyrowski}, {G{\"u}sten}, {Menten},
  {Wiesemeyer}, {Csengeri}, {Heyminck}, {Klein}, {K{\"o}nig}, \&
  {Urquhart}}]{Wyrowski2016AA585A149W}
{Wyrowski}, F., {G{\"u}sten}, R., {Menten}, K.~M., {et~al.} 2016, \aap, 585,
  A149

\bibitem[{{Xi} {et~al.}(2015){Xi}, {Zhou}, {Esimbek}, {Wu}, {He}, {Ji}, \&
  {Tang}}]{Xi2015MNRAS4534203X}
{Xi}, H., {Zhou}, J., {Esimbek}, J., {et~al.} 2015, \mnras, 453, 4203

\bibitem[{{Xue} \& {Wu}(2008)}]{Xue2008ApJ680446X}
{Xue}, R. \& {Wu}, Y. 2008, \apj, 680, 446

\bibitem[{{Yang} {et~al.}(2019){Yang}, {Thompson}, {Tian}, {Bihr}, {Beuther},
  \& {Hindson}}]{Yang2019MNRAS4822681Y}
{Yang}, A.~Y., {Thompson}, M.~A., {Tian}, W.~W., {et~al.} 2019, \mnras, 482,
  2681

\bibitem[{{Yang} {et~al.}(2018){Yang}, {Thompson}, {Urquhart}, \&
  {Tian}}]{Yang2018ApJS2353Y}
{Yang}, A.~Y., {Thompson}, M.~A., {Urquhart}, J.~S., \& {Tian}, W.~W. 2018,
  \apjs, 235, 3

\bibitem[{{Yang} {et~al.}(2016){Yang}, {Tian}, {Zhu}, {Leahy}, \&
  {Wu}}]{Yang2016ApJS2236Y}
{Yang}, A.~Y., {Tian}, W.~W., {Zhu}, H., {Leahy}, D.~A., \& {Wu}, D. 2016,
  \apjs, 223, 6

\bibitem[{{Zhang} {et~al.}(2014){Zhang}, {Wang}, {Xu}, {Wyrowski}, \&
  {Menten}}]{Zhang2014ApJ}
{Zhang}, C.-P., {Wang}, J.-J., {Xu}, J.-L., {Wyrowski}, F., \& {Menten}, K.~M.
  2014, \apj, 784, 107

\bibitem[{{Zinnecker} \& {Yorke}(2007)}]{Zinnecker2007ARAA45}
{Zinnecker}, H. \& {Yorke}, H.~W. 2007, \araa, 45, 481

\end{thebibliography}

\begin{appendix}

\section{Additional Tables}

\begin{table*}
\caption[]{
{Sample of 118 rising-spectra \hii\ regions}
\label{tab_obser_list_appendix}
}
\centering
\footnotesize
\setlength{\tabcolsep}{2pt}
\begin{tabular}{l|c|c|c|c|c|l|c|c|c|c|c
}
\hline
\hline
Name  & $S_{1.4\,GHz}$ & $\rm S_{5\,GHz}$ & Dist & { $\log  L_{\rm bol}$} & [Ref.] & Name & $\rm S_{1.5\,GHz}$ & $S_{5\,GHz}$ & Dist & {$\log  L_{\rm bol}$ }& [Ref.] \\
    &  ($\rm mJy$) & ($\rm mJy $) & (kpc) & ($ L_{\odot}$) &  &  &  ($\rm mJy$) & ($\rm mJy$) & (kpc)  & ($ L_{\odot}$) &  \\
\hline
    & $\pm 10\%$ & $\pm 10\%$  & $\pm 10\%$ & $\pm 20\%$ &  &     & $\pm 10\%$ & $\pm 10\%$ & $\pm 10\%$ & $\pm 20\%$ &  \\
\hline
G010.3009$-$00.1477 & 426.2 & 631.4 & 3.5 & 5.2 & [1] & G027.3644$-$00.1657 & 45.0 & 60.1 & 8.0 & 4.8 & [1] \\
G010.4724$+$00.0275 & 31.3 & 38.4 & 8.5 & 5.7 & [1] & G027.9782$+$00.0789 & 89.3 & 124.0 & 4.8 & 4.2 & [2] \\
G010.6223$-$00.3788$\dagger$ & 327.6 & 483.3 & 2.4 &  5.7  & [1] & G028.2003$-$00.0494 &  $-$  & 161.0 & 6.1 & 5.1 & [1] \\
G010.6234$-$00.3837 & 571.3 & 1952.2 & 5.0 & 5.7 & [1] & G028.2879$-$00.3641 & 410.9 & 552.8 & 11.6 & 5.9 & [1] \\
G010.9584$+$00.0221 & 47.9 & 196.0 & 2.9 & 4.0 & [1] & G028.6082$+$00.0185 & 168.2 & 210.2 & 7.4 & 5.0 & [1] \\
G011.0328$+$00.0274 & 3.7 & 5.7 & 2.9 & 2.7 & [1] & G029.9559$-$00.0168 & 1610.8 & 3116.2 & 5.2 & 5.7 & [1] \\
G011.1104$-$00.3985 & 253.2 & 305.4 & 5.0 & 4.7 & [1] & G030.0096$-$00.2734 & 0.3 & 4.5 & 5.2 & 3.8 & [1] \\
G011.1712$-$00.0662 & 83.2 & 102.2 & 2.9 & 3.2 & [1] & G030.5353$+$00.0204 & 553.6 & 710.4 & 2.7 & 3.9 & [1] \\
G011.9368$-$00.6158 & 735.6 & 1155.9 & 3.4 & 4.8 & [1] & G031.0495$+$00.4697 & 10.7 & 13.6 & 2.0 & 3.5 & [1] \\
G011.9446$-$00.0369 & 251.1 & 943.6 & 3.1 & 4.3 & [1] & G031.1596$+$00.0448 & 20.7 & 23.8 & 2.7 & 3.3 & [1] \\
G012.1988$-$00.0345 & 47.6 & 62.7 & 11.9 & 5.0 & [1] & G031.2801$+$00.0632 & 144.5 & 268.9 & 5.2 & 4.8 & [1] \\
G012.2081$-$00.1019 & 127.9 & 207.9 & 13.4 & 5.5 & [1] & G032.4727$+$00.2036 & 56.0 & 97.4 & 3.0 & 3.5 & [1] \\
G012.4294$-$00.0479 & 24.4 & 45.2 & 2.6 & 3.2 & [1] & G032.7441$-$00.0755 & 0.3 & 7.9 & 11.7 & 5.0 & [1] \\
G012.8050$-$00.2007 & 4332.2 & 12616.4 & 2.6 &  $-$  & [1] & G032.7966$+$00.1909 & 1698.9 & 3123.4 & 13.0 & 6.1 & [1] \\
G012.8131$-$00.1976$\dagger$ & 907.7 & 1500.4 &  $-$  &  $-$  & [1] & G032.9273$+$00.6060 & 229.5 & 285.6 & 15.1 & 4.7 & [1] \\
G012.9995$-$00.3583 & 10.5 & 20.1 & 1.3 & 2.9 & [1] & G033.4163$-$00.0036 & 57.6 & 75.2 & 5.4 & 4.1 & [1] \\
G013.2099$-$00.1428 & 437.9 & 946.8 & 2.6 & 4.2 & [1] & G033.9145$+$00.1105 & 464.7 & 842.2 & 6.5 & 5.2 & [1] \\
G013.3850$+$00.0684 & 139.2 & 603.9 & 1.9 & 3.5 & [1] & G034.2572$+$00.1535 & 370.8 & 1762.6 & 1.6 & 4.8 & [1] \\
G014.7785$-$00.3328 & 15.4 & 18.2 & 3.1 & 3.0 & [1] & G034.2581$+$00.1533 &  $-$  & 35.9 & 1.6 & 4.8 & [1] \\
G016.1448$+$00.0088 & 8.8 & 14.8 & 12.3 & 4.2 & [1] & G034.4032$+$00.2277 & 5.3 & 8.9 & 1.6 & 3.5 & [1] \\
G016.3913$-$00.1383 & 40.8 & 124.3 & 1.9 & 2.5 & [1] & G035.0242$+$00.3502 & 5.3 & 11.4 & 2.3 & 4.1 & [1] \\
G016.9445$-$00.0738 & 258.5 & 519.3 & 15.9 & 5.2 & [1] & G035.4669$+$00.1394 & 235.1 & 317.6 & 8.5 & 5.3 & [1] \\
G017.0299$-$00.0696 & 2.0 & 5.4 & 10.1 & 4.1 & [1] & G035.5781$-$00.0305 & 38.0 & 187.8 & 10.4 & 5.3 & [1] \\
G017.1141$-$00.1124 & 14.2 & 17.2 & 10.1 & 4.5 & [1] & G036.4057$+$00.0226 & 17.3 & 22.3 & 3.5 & 3.9 & [1] \\
G018.1460$-$00.2839$\dagger$ & 151.2 & 856.2 &  $-$  &  $-$  & [1] & G037.5457$-$00.1120 & 252.3 & 406.5 & 9.7 & 5.1 & [1] \\
G018.3024$-$00.3910 & 846.8 & 1277.9 & 3.2 & 4.7 & [1] & G037.7347$-$00.1128 & 12.3 & 16.0 & 9.7 & 4.6 & [1] \\
G018.4433$-$00.0056 & 56.2 & 81.3 & 11.9 & 4.5 & [2] & G037.7633$-$00.2167$\dagger$ & 295.3 & 337.6 &  $-$  &  $-$  & [1] \\
G018.4614$-$00.0038 & 128.2 & 342.1 & 11.8 & 5.4 & [1] & G037.8731$-$00.3996 & 1279.3 & 2561.2 & 9.7 & 5.7 & [1] \\
G018.6654$+$00.0294 & 3.7 & 5.7 & 10.1 & 4.5 & [1] & G037.9723$-$00.0965 & 10.2 & 20.9 & 16.6 & 4.3 & [2] \\
G018.7106$+$00.0002 & 41.0 & 107.5 & 2.4 & 3.1 & [1] & G038.8756$+$00.3080 & 191.2 & 311.3 & 14.2 & 4.7 & [1] \\
G018.7612$+$00.2630 & 26.4 & 51.4 & 14.1 & 5.1 & [1] & G039.1956$+$00.2255 & 14.2 & 62.3 & 14.5 & 4.4 & [1] \\
G018.8250$-$00.4675 & 9.1 & 11.4 & 5.0 & 3.6 & [1] & G039.7277$-$00.3973$\dagger$ & 112.3 & 133.3 &  $-$  &  $-$  & [1] \\
G018.8338$-$00.3002 & 108.4 & 131.4 & 12.7 & 4.9 & [1] & G039.8824$-$00.3460 & 247.0 & 276.9 & 9.3 & 4.6 & [1] \\
G019.0754$-$00.2874 & 333.7 & 380.7 & 5.0 & 5.1 & [1] & G042.4345$-$00.2605 & 38.5 & 83.7 & 4.4 & 4.0 & [1] \\
G019.4912$+$00.1352 & 269.3 & 415.1 & 13.7 & 5.1 & [1] & G043.1651$-$00.0283 & 564.3 & 2714.3 & 11.1 & 6.2 & [1] \\
G019.6087$-$00.2351 & 855.6 & 2900.9 & 12.6 & 6.0 & [1] & G043.1652$+$00.0129 &  $-$  & 160.1 & 11.1 & 6.9 & [1] \\
G019.6090$-$00.2313 & 126.5 & 259.9 & 12.6 & 6.0 & [1] & G043.1657$+$00.0116 &  $-$  & 98.2 & 11.1 & 6.9 & [1] \\
G019.7407$+$00.2821 & 44.4 & 239.0 & 14.0 & 4.7 & [1] & G043.1665$+$00.0106 & 237.8 & 1365.7 & 11.1 & 6.9 & [1] \\
G019.7549$-$00.1282 & 10.6 & 36.5 & 7.8 & 4.5 & [1] & G043.1778$-$00.5181 & 122.9 & 181.7 & 8.0 & 5.0 & [1] \\
G020.0720$-$00.1421$\dagger$ & 138.1 & 210.1 &  $-$  &  $-$  & [1] & G045.0694$+$00.1323 & 17.9 & 46.2 & 8.0 & 5.7 & [1] \\
G020.0809$-$00.1362 & 104.4 & 498.2 & 12.6 & 5.7 & [1] & G045.0712$+$00.1321 & 61.6 & 146.7 & 8.0 & 5.7 & [1] \\
G020.3633$-$00.0136 & 32.2 & 55.1 & 3.4 & 3.3 & [1] & G045.1223$+$00.1321 & 1346.0 & 2984.3 & 8.0 & 6.0 & [1] \\
G021.3571$-$00.1766 & 18.5 & 24.9 & 10.0 & 4.6 & [1] & G045.4545$+$00.0591 & 492.5 & 1029.5 & 6.9 & 5.6 & [2] \\
G021.3855$-$00.2541 & 51.1 & 113.9 & 10.0 & 4.9 & [1] & G045.4656$+$00.0452 & 28.9 & 62.3 & 6.6 & 5.0 & [2] \\
G021.4257$-$00.5417 & 78.9 & 94.8 & 3.5 & 4.2 & [1] & G045.4790$+$00.1294 & 380.2 & 504.2 & 6.0 & 4.5 & [2] \\
G023.2654$+$00.0765 & 55.6 & 88.6 & 5.9 & 4.3 & [1] & G048.6057$+$00.0228 & 6.6 & 36.2 & 10.8 & 5.6 & [1] \\
G023.4553$-$00.2010 & 2.9 & 14.4 & 5.9 & 3.8 & [1] & G048.6099$+$00.0270 & 56.5 & 131.2 & 9.8 & 5.1 & [4] \\
G024.5065$-$00.2224 & 153.7 & 205.6 & 5.8 & 4.7 & [1] & G048.9296$-$00.2793 & 66.9 & 185.4 & 5.6 & 4.2 & [2] \\
G024.7898$+$00.0833 &  $-$  & 12.5 & 6.4 & 5.2 & [1] & G049.2679$-$00.3374 & 64.4 & 102.6 & 5.6 & 4.5 & [2] \\
G024.9237$+$00.0777$\dagger$ & 57.1 & 172.5 & 12.2 &  $-$  & [1] & G049.3704$-$00.3012 & 252.9 & 414.4 & 5.4 & 5.1 & [3] \\
G025.3948$+$00.0332 & 203.7 & 296.9 & 15.6 & 5.3 & [1] & G049.4905$-$00.3688 & 1165.6 & 3821.7 & 5.3 & 6.2 & [1] \\
G025.3970$+$00.5614 & 93.7 & 121.2 & 13.9 & 5.1 & [1] & G050.3152$+$00.6762 & 38.7 & 81.3 & 1.8 & 3.3 & [1] \\
G025.3981$-$00.1411 & 1351.5 & 2132.2 & 10.2 & 6.0 & [1] & G051.6785$+$00.7193 & 2.8 & 22.6 & 10.9 & 5.0 & [1] \\
G025.7157$+$00.0487 & 15.7 & 20.8 & 9.4 & 4.6 & [1] & G052.7533$+$00.3340 & 264.0 & 386.0 & 9.0 & 4.4 & [2] \\
G025.8011$-$00.1568 & 19.2 & 31.9 & 10.2 & 4.9 & [1] & G053.9589$+$00.0320 & 40.8 & 46.0 & 4.0 & 3.6 & [1] \\
G026.5444$+$00.4169 & 301.0 & 413.4 & 9.8 & 5.2 & [1] & G060.8842$-$00.1286 &  $-$  & 18.7 & 2.5 & 4.2 & [2] \\
G027.2800$+$00.1447 & 370.4 & 428.0 & 13.4 & 4.9 & [1] & G061.4763$+$00.0892 & 252.7 & 718.7 & 4.1 & 5.1 & [2] \\
G030.5887$-$00.0428$\star$ & 7.9 & 92.37 & 2.7 & 4.0 & [1] & G030.8662$+$00.1143$\star$  &  137.17  & 255.2 & 2.7 & 4.1 & [1] \\
G030.7197$-$00.0829$\star$ & 464.58 & 969.33 & 5.2 & 4.7 & [1] & G033.1328$-$00.0923$\star$ & 173.43 & 378.59 & 9.4 & 5.0 & [1] \\
\hline
\hline
\end{tabular}
\tablefoot{
 This table will be available in electronic form at the CDS.
$\dagger$ refers to the sources in our observation with poor-quality images.
$\star$ refers to the 4 sources with data from archives and the literature mentioned in Sect\,\ref{sect_optically_thick_hii}.
Columns: 
(1) source name; 
(2) and (3) flux density at 1.4 GHz and 5 GHz taken from \citet{Yang2019MNRAS4822681Y}; 
(4) heliocentric distance; 
(5) bolometric luminosity;
(6) reference for heliocentric distance and bolometric luminosity: 
[1] \citet{Urquhart2018MNRAS4731059U}, 
[2] \citet{Cesaroni2015AA579A},
[3] \citet{Urquhart2013MNRAS435},
[4] \citet{Kalcheva2018AA615A103K}. 
The symbol $\pm 10\%$ refers to an overall estimation of the percentage error at 1.4 GHz and 5 GHz.
}
\end{table*}


\begin{table*}
\setlength{\tabcolsep}{0.6pt}
\caption[]{ 
Observation results of 112 young \hii\ regions at X-band (8--12\,GHz) and K-band (18--26\,GHz). 
\label{summary_hii_obsparam_appendix}}
\centering
\footnotesize
\begin{tabular}{p{3.1cm}|C{1.75cm}|C{1.0cm}|C{1.1cm}|C{1.1cm}|C{1.1cm}|C{1.4cm}|C{1.75cm}|C{1.0cm}|C{1.1cm}|C{1.1cm}|C{1.1cm}|C{1.3cm}}

\hline
\hline
Name & $ S_{\rm Peak}(X)$ & $\sigma$(X) & $S_{\rm 9\,GHz}$ &$S_{\rm 10\,GHz}$ & $ S_{\rm 11\,GHz}$ & $ {\theta}_{\rm s}$(X) & $S_{\rm Peak}$(K)  & $\sigma$(K) & $ S_{\rm 20\,GHz}$ & $S_{\rm 22\,GHz}$ & $ S_{\rm 24\,GHz}$    & ${\theta}_{\rm s}$(K)  \\ 
  & $\rm (mJy/beam)$& $\rm (mJy)$  & $\rm (mJy)$ & $\rm (mJy)$ & $\rm (mJy)$   
 & $\rm ({\arcsec\times \arcsec})$  
 & $\rm (mJy/beam)$ & $\rm (mJy)$ & $\rm (mJy)$ & $\rm (mJy)$ & $\rm (mJy)$ 
 & $\rm {(\arcsec\times \arcsec)}$  \\ 
  &   &   & $ \pm10\%$ & $\pm10\%$ & $\pm10\%$   
 &   
 & $\pm10\%$ &  & $\pm10\%$ & $\pm10\%$ & $\pm10\%$ 
 &   \\ 
\hline
(1)   & (2)  & (3) & (4)  & (5)  & (6) & (7) & (8) & (9) & (10) & (11) & (12) & (13) \\
\hline
\hline
  G010.3009$-$00.1477$\oplus$ &    88.1   &    3.1  &    700.3   &    686.7   &     661.4   &      6.8$\times$6.6  &     15.9  &     0.7   & 433.4  &     419.6  &  392.6  &  6.5$\times$6.4   \\ 
  G010.4724$+$00.0275 &    82.5   &    1.6  &    100.7   &    105.9   &     115.4   &      1.4$\times$0.4  &     66.9  &     0.8   &     156.8  &     159.9  &     172.0  &      1.6$\times$0.5   \\ 
  G010.6234$-$00.3837$\oplus$ &    1099.9 &    7.9  &    3071.4  &    3072.1  &     3314.8  &      4.2$\times$3.8  &     572.3 &     14.6  &     2884.8 &     2857.2 &     2851.5 &      3.1$\times$3.0   \\ 
  G010.9584$+$00.0221          &    186.3  &    1.5  &    258.3   &    256.2   &     265.0   &      1.2$\times$0.9  &     91.3  &     1.2   &     210.7  &     202.6  &     200.6  &      1.0$\times$0.7   \\ 
   G011.0328$+$00.0274    &    3.9    &    0.2  &    4.8     &    4.3     &     4.1     &      1.3$\times$0.9  &     1.7   &     0.1   &     3.4    &     2.9    &     2.8    &      0.9$\times$0.4   \\ 
  G011.1104$-$00.3985$\oplus$ &    70.5   &    0.7  &    350.4   &    334.8   &     327.7   &      9.5$\times$9.4  &     15.6  &     0.4   &     136.1  &     123.3  &     126.1  &      2.2$\times$1.7   \\ 
  G011.1712$-$00.0662$\oplus$          &    4.1    &    0.1  &    95.1    &    92.7    &     100.1   &      11.9$\times$8.6  &     0.6   &     0.1   &     $-$    &     $-$    &     $-$    &     $-$              \\ 
    G011.9368$-$00.6158$\oplus$ &    306.7  &    2.0  &    1116.4  &    1083.5  &     1098.3  &      3.4$\times$3.2  &     76.4  &     2.0   &     656.1  &     652.4  & 629.6 & 2.8$\times$1.8   \\ 
    G011.9446$-$00.0369$\oplus$ &    85.7   &    2.0  &    709.6   &    691.4   &    724.6   &      6.3$\times$4.7  &     20.0  &     0.6   & 307.2  &     291.6  &  289.9  &   4.3$\times$2.1   \\ 
   G012.1988$-$00.0345  &    29.6   &    0.4  &    66.0    &    64.8    &     63.9    &      2.0$\times$1.9  &     6.5   &     0.2   &     59.6   &     54.7   &     55.9   &      2.0$\times$2.0   \\ 
  G012.2081$-$00.1019          &    88.0   &    0.8  &    212.5   &    209.3   &     206.5   &      2.3$\times$1.9  &     27.2  &     0.6   &     159.0  &     142.1  &     140.8  &      2.0$\times$1.2   \\ 
   G012.4294$-$00.0479  &    19.9   &    0.4  &    51.7    &    49.8    &     49.8    &      2.8$\times$2.4  &     6.6   &     0.2   &     28.0   &     22.9   &     26.9   &      1.7$\times$0.9   \\ 
    G012.8050$-$00.2007$\oplus$          &    858.6  &    11.0 &    16097.6 &    15749.0 &     16931.9 &      17.1$\times$17.0  &     302.3 &     4.9   &     6400.9 &  6240.9 &  6252.3 &      8.1$\times$1.1   \\ 
   G012.9995$-$00.3583    &    8.4    &    0.1  &    14.6    &    14.7    &     15.0    &      1.9$\times$1.1  &     3.3   &     0.1   &     12.4   &     11.1   &     11.2   &      1.8$\times$0.8   \\ 
 G013.2099$-$00.1428$\oplus$ &    57.4   &    0.4  &    1080.7  &    1065.6  &     965.8   &      11.7$\times$11.6  &     10.7  &     0.3   &  385.9  &     359.6  &     354.2  &      9.0$\times$9.0   \\ 
    G013.3850$+$00.0684$\oplus$ &    15.0   &    0.6  &    738.1   &    819.3   &     812.9   &      21.3$\times$21.2  &     1.6   &     0.1   &     $-$  &  $-$  &   $-$  &  $-$   \\ 
  G014.7785$-$00.3328          &    18.9   &    0.2  &    25.3    &    25.0    &     25.4    &      1.1$\times$0.8  &     9.7   &     0.1   &     17.7   &     16.3   &     15.9   &      0.7$\times$0.5   \\ 
   G016.1448$+$00.0088    &    13.7   &    0.1  &    16.6    &    16.6    &     16.6    &      0.9$\times$0.8  &     7.7   &     0.1   &     14.3   & 13.4   & 13.0   & 0.7$\times$0.5   \\ 
  G016.3913$-$00.1383$\oplus$ &    2.5    &    0.1  &    50.0    &    52.5    &     46.9    &      9.9$\times$8.1  &     0.4   &     0.1   &     31.9   &  21.0   & 22.9   & 7.0$\times$4.6   \\ 
  G016.9445$-$00.0738$\oplus$          &    138.3  &    0.4  &    545.2   &    548.9   &     546.6   &      3.5$\times$2.5  &     30.6  &     0.3   &     529.2  &     513.4  &     515.6  &      3.5$\times$2.7   \\ 
   G017.0299$-$00.0696    &    2.7    &    0.1  &    3.9     &    3.8     &     3.9     &      1.6$\times$0.5  &     1.1   &     0.1   &     1.9    &     2.1    &     3.1    &      0.8$\times$0.5   \\ 
  G017.1141$-$00.1124          &    4.9    &    0.1  &    16.5    &    15.9    &     16.3    &      3.1$\times$2.5  &     1.0   &     0.1   &     10.3   &     8.2    &     11.5   &      2.5$\times$1.6   \\ 
 G018.3024$-$00.3910$\oplus$ &    56.9   &    0.7  &    1281.9  &    1308.6  &     1180.6  &      15.5$\times$13.3  &     12.7  &     1.0   &    $-$  & $-$  &    $-$  &     $-$   \\ 
  G018.4433$-$00.0056          &    38.2   &    2.6  &    73.0    &    73.9    &     95.3    &      1.8$\times$1.5  &     6.3   &     0.7   &     72.4   &     57.4   &     76.1   &      2.2$\times$1.8   \\ 
  G018.4614$-$00.0038          &    178.6  &    0.8  &    387.2   &    383.5   &     394.4   &      2.1$\times$1.8  &     54.0  &     0.9   &     355.7  &     322.1  &     388.6  &      2.1$\times$1.6   \\ 
   G018.6654$+$00.0294    &    4.9    &    0.8  &    6.9     &    6.7     &     6.6     &      1.7$\times$1.4  &     2.5   &     0.1   &     5.3    &     5.4    &     5.5    &      0.8$\times$0.7   \\ 
  G018.7106$+$00.0002          &    88.5   &    0.5  &    116.9   &    116.2   &     116.0   &      1.1$\times$0.9  &     42.5  &     0.4   &     104.2  &     101.2  &     100.8  &      0.9$\times$0.8   \\ 
   G018.7612$+$00.2630    &    42.6   &    0.2  &    55.7    &    55.3    &     55.4    &      1.1$\times$0.9  &     18.9  &     0.2   &     45.6   &     47.6   &     50.4   &      0.9$\times$0.8   \\ 
   G018.8250$-$00.4675    &    4.4    &    0.1  &    10.2    &    9.9     &     10.5    &      2.5$\times$2.2  &     1.2   &     0.1   &     8.3    &     6.3    &     7.7    &      1.6$\times$1.5   \\ 
  G018.8338$-$00.3002$\oplus$          &    56.0   &    0.5  &    132.6   &    135.2   &     139.0   &      5.8$\times$5.6  &     52.8  &     0.3   &     65.9   &     66.4   &     65.1   &      0.4$\times$0.3   \\ 
 G019.0754$-$00.2874$\oplus$          &    19.5   &    0.7  &  369.2   &  326.8   &  355.5   &      15.2$\times$15.2  &   6.5   &     0.5   &   $-$  &     $-$  &    $-$  &     $-$   \\ 
  G019.4912$+$00.1352$\oplus$ &    23.1   &    0.4  &    355.3   &    374.2   &     417.1   &      9.6$\times$8.1  &     5.0   &     0.3   &     176.0  &     151.5  &     196.4  &      5.9$\times$4.4   \\ 
 G019.6087$-$00.2351$\oplus$         &    249.2  &    3.5  &    3535.0  &    3145.6  &     3497.1  &      14.0$\times$13.9  &     45.6  &     1.0   & $-$  &     $-$  & $-$  &  $-$  \\ 
   G019.6090$-$00.2313          &    255.4  &    3.0  &    247.0   &    259.8   &     241.0   &      3.1$\times$1.1  &     50.1  &     1.5   &     130.0  &     133.0  & 145.9  &      1.4$\times$0.6   \\ 
  G019.7407$+$00.2821$\oplus$         &    3.6    &    0.1  &    166.4   &    168.5   &     167.7   &      21.3$\times$21.2  &     0.6   &     0.1   &     $-$    &     $-$    &     $-$    &     $-$              \\ 
   G019.7549$-$00.1282    &    40.3   &    0.4  &    44.6    &    44.5    &     44.8    &      0.6$\times$0.5  &     29.8  &     0.5   &     40.8   &     40.1   &     39.1   &      0.5$\times$0.4   \\ 
  G020.0809$-$00.1362          &    205.7  &    2.3  &    645.7   &    656.2   &     696.5   &      3.8$\times$1.6  &     188.7 &     1.6   &     387.0  &     381.0  &     400.0  &      1.0$\times$0.5   \\ 
    G020.3633$-$00.0136          &    17.6   &    0.3  &    53.8    &    53.2    &     52.4    &      2.8$\times$2.3  &     3.9   &     0.1   & 45.9   &     46.6   &     47.5 &      2.8$\times$2.3   \\ 
   G021.3571$-$00.1766    &    16.7   &    0.3  &    25.4    &    25.1    &     24.7    &      1.4$\times$1.1  &     6.9   &     0.1   &     22.7   &     21.1   &     23.8   &      1.2$\times$1.1   \\ 
  G021.3855$-$00.2541          &    81.2   &    0.4  &    122.8   &    121.4   &     120.8   &      1.4$\times$1.0  &     39.5  &     0.1   &     103.9  &     101.7  &     108.8  &      1.1$\times$0.9   \\ 
   G021.4257$-$00.5417$\oplus$          &    2.3    &    0.2  &    72.7    &    75.0    &     71.1    &      11.8$\times$10.6  &     0.4   &     0.1   & $-$  &     $-$   &  $-$   &  $-$   \\ 
    G023.2654$+$00.0765$\oplus$          &    20.1   &    0.2  &    83.3    &    85.6    &     82.8    &      5.3$\times$5.0  &     5.5   &     0.2   &     47.7   &     43.1   &   46.7   &      2.7$\times$1.5   \\ 
   G023.4553$-$00.2010    &    13.7   &    0.5  &    13.8    &    12.9    &     10.9    &      1.2$\times$0.8  &     11.0  &     0.1   &     13.9   &     13.5   &     13.2   &      0.3$\times$0.3   \\ 
  G024.5065$-$00.2224$\oplus$          &    19.9   &    0.7  &    257.0   &    249.5   &     258.1   &      7.7$\times$7.5  &     4.6   &     0.2   &     127.3  &     119.0  &  117.5  &      5.2$\times$5.1   \\ 
   G024.7889$+$00.0824$\dagger$  &    25.6   &    1.1  &    33.1    &    37.8    &     35.7    &      1.9$\times$1.1  &     6.9   &     0.8   &     23.7   &     22.9   &     20.1   &      1.4$\times$1.2   \\ 
   G024.7898$+$00.0833    &    32.4   &    1.1  &    31.6    &    32.9    &     30.8    &      1.4$\times$0.9  &     65.0  &     0.8   &     65.8   &     72.3   &     71.2   &      0.3$\times$0.2   \\ 
  G025.3948$+$00.0332          &    35.9   &    0.9  &    367.9   &    358.2   &     369.9   &      5.1$\times$4.9  &     7.4   &     0.4   &     318.9  &     278.8  &     303.5  &      4.7$\times$4.4   \\ 
 G025.3970$+$00.5614          &    117.8  &    0.7  &    158.5   &    164.8   &     163.1   &      1.3$\times$0.7  &     71.4  &     0.6   &  158.5  &     153.5  & 161.3 &      0.8$\times$0.8   \\ 
  G025.3981$-$00.1411$\oplus$          &    180.3  &    3.8  &    2444.6  &    2668.5  &     2510.9  &      7.5$\times$7.5  &     36.0  &     2.7   &     711.0  &     734.0  &     725.0  &      4.7$\times$4.7   \\ 
  G025.7157$+$00.0487          &    12.4   &    0.2  &    20.8    &    21.3    &     22.5    &      1.5$\times$1.4  &     3.9   &     0.2   &     15.8   &     15.8   &     18.0   &      1.4$\times$1.1   \\ 
   G025.8011$-$00.1568    &    26.6   &    0.2  &    37.1    &    36.8    &     36.3    &      1.1$\times$1.0  &     10.9  &     0.1   &     33.9   &     32.2   &     33.3   &      1.1$\times$1.0   \\ 
  G026.5444$+$00.4169$\oplus$          &    15.3   &    1.0  &    532.2   &    556.4   &     598.7   &      17.4$\times$17.3  &     2.5   &     0.3   &     $-$    &     $-$    &     $-$    &     $-$              \\
  G027.2800$+$00.1447$\oplus$          &    46.0   &    0.6  &    433.7   &    466.2   &     446.7   &      6.0$\times$5.1  &     9.6   &     0.6   &     357.2  &     289.2  &     302.8  &      5.6$\times$3.7   \\ 
  G027.3644$-$00.1657          &    37.0   &    0.2  &    58.5    &    56.7    &     55.8    &      1.4$\times$1.1  &     14.9  &     0.2   &     40.1   &     38.4   &     40.7   &      0.9$\times$0.8   \\ 
  G027.9782$+$00.0789$\oplus$          &    5.5    &    0.2  &    125.3   &    127.0   &     144.8   &      12.9$\times$12.8  &     0.7   &     0.1   &     $-$    &     $-$    &     $-$    &     $-$              \\
  G028.2003$-$00.0494          &    259.6  &    1.8  &    298.0   &    320.0   &     339.0   &      0.9$\times$0.6  &     339.1 &     2.1   &     558.0  &     582.0  &     627.0  &      0.7$\times$0.5   \\ 
  G028.2879$-$00.3641          &    123.9  &    1.2  &    592.7   &    581.1   &     601.7   &      4.0$\times$3.8  &     42.5  &     1.3   &     614.6  &     577.6  &     549.7  &      2.1$\times$1.0   \\ 
   G028.6082$+$00.0185$\oplus$          &    52.2   &    1.1  &    226.4   &    228.7   &     242.5   &      3.4$\times$3.1  &     14.7  &     0.3   &     188.0  &     173.0  &    190.8  &      3.1$\times$2.4   \\ 
   G029.9559$-$00.0168$\oplus$          &    357.2  &    7.5  &    2667.2  &    2826.8  &    2951.5  &      5.9$\times$4.2  &     103.2 &     2.4   &     1519.0 &     1353.6 &   1502.1  &      4.4$\times$1.8   \\ 
   G030.0096$-$00.2734    &    5.6    &    0.1  &    5.7     &    5.8     &     5.8     &      0.3$\times$0.2  &     4.9   &     0.1   &     5.8    &     5.5    &     5.3    &      0.2$\times$0.2   \\ 
 G030.5353$+$00.0204$\oplus$          &    103.9  &    1.3  &   684.7  &  676.5   & 682.7 &      6.7$\times$6.7  &     25.1  &     0.8   &     379.7  &     380.5  &     391.2  &     5.0$\times$5.0   \\ 
   G031.0495$+$00.4697    &    10.4   &    0.1  &    13.5    &    13.4    &     13.1    &      1.2$\times$0.7  &     3.4   &     0.1   &     12.5   &     11.7   &     13.1   &      1.3$\times$0.8   \\ 
   G031.1596$+$00.0448    &    18.6   &    0.1  &    24.9    &    24.7    &     24.4    &      1.1$\times$0.9  &     7.1   &     0.1   &     20.6   &     19.1   &     20.4   &      0.9$\times$0.8   \\ 
 G031.2801$+$00.0632$\oplus$          &    14.3   &    0.6  &    353.0   &    400.1   &     416.5   &      10.9$\times$10.2  &     2.1   &     0.3   &  $-$ &  $-$  & $-$  & $-$   \\ 
\hline
\end{tabular}
\end{table*}
\addtocounter{table}{-1}
\begin{table*}
\setlength{\tabcolsep}{0.6pt}
\caption[]{ 
--continued\, Observation results of 112 young \hii\ regions at X-band (8--12\,GHz) and K-band (18--26\,GHz).}
\centering
\footnotesize
\begin{tabular}{p{3.1cm}|C{1.75cm}|C{1.0cm}|C{1.1cm}|C{1.1cm}|C{1.1cm}|C{1.4cm}|C{1.75cm}|C{1.0cm}|C{1.1cm}|C{1.1cm}|C{1.1cm}|C{1.3cm}}
\hline
\hline
Name & $ S_{\rm Peak}(X)$ & $\sigma$(X) & $S_{\rm 9\,GHz}$ &$S_{\rm 10\,GHz}$ & $ S_{\rm 11\,GHz}$ & $ {\theta}_{\rm s}$(X) & $S_{\rm Peak}$(K)  & $\sigma$(K) & $ S_{\rm 20\,GHz}$ & $S_{\rm 22\,GHz}$ & $ S_{\rm 24\,GHz}$    & ${\theta}_{\rm s}$(K)  \\ 
  & $\rm (mJy/beam)$& $\rm (mJy)$  & $\rm (mJy)$ & $\rm (mJy)$ & $\rm (mJy)$   
 & $\rm ({\arcsec\times \arcsec})$  
 & $\rm (mJy/beam)$ & $\rm (mJy)$ & $\rm (mJy)$ & $\rm (mJy)$ & $\rm (mJy)$ 
 & $\rm {(\arcsec\times \arcsec)}$  \\ 
  &   &   & $ \pm10\%$ & $\pm10\%$ & $\pm10\%$   
 &   
 & $\pm10\%$ &  & $\pm10\%$ & $\pm10\%$ & $\pm10\%$ 
 &   \\ 
\hline
(1)   & (2)  & (3) &  (4)  & (5)  & (6) & (7) & (8) & (9) & (10) & (11) & (12) & (13) \\
\hline
\hline
  G032.4727$+$00.2036          &    60.1   &    0.2  &    102.4   &    101.1   &     100.1   &      1.5$\times$1.2  &     24.2  &     0.2   &     86.7   &     82.2   &     85.7   &      1.4$\times$0.9   \\ 
   G032.7441$-$00.0755    &    15.1   &    0.1  &    16.2    &    17.3    &     18.8    &      1.0$\times$0.7  &     31.4  &     0.1   &     29.8   &     31.8   &     34.7   &      0.1$\times$0.1   \\ 
 G032.7966$+$00.1909$\oplus$          &    623.5  &    6.2  &    3768.0  &    3813.0  &     3925.0  &      11.5$\times$11.4  &     238.8 &     6.1   &    $-$ &     $-$ &    $-$ &  $-$   \\ 
   G032.9273$+$00.6060$\oplus$          &    54.3   &    0.5  &    278.5   &  272.8   &     284.5   &      7.6$\times$7.6  &     25.1  &     0.2   &    $-$   &     $-$  &     $-$  &      $-$   \\ 
  G033.4163$-$00.0036$\oplus$          &    7.0    &    0.3  &    111.5   &    98.4    &     101.8   &      10.1$\times$9.9  &     1.1   &     0.1   &     30.5   &     18.5   &     20.9   &      3.8$\times$2.2   \\
  G033.9145$+$00.1105$\oplus$          &    121.2  &    1.9  &    732.4   &    737.6   &     740.7   &      6.0$\times$6.0  &     31.5  &     0.7   &  270.9  &     263.9  &  288.9  &      2.8$\times$1.4   \\ 
  G034.2572$+$00.1535${ \oplus}$          &    956.0  &    11.0 &    2710.0  &    2970.0  &     3050.0  &      2.9$\times$1.9  &     648.0 &     5.5   &     3750.0 &     3760.0 &     4200.0 &      1.8$\times$1.2   \\ 
 G034.2573$+$00.1523$\dagger$  &    117.0  &    10.8 &    145.6   &    141.5   &     180.0   &      1.7$\times$0.6  &     60.7  &     7.5   &     57.0   &     45.4   &     33.1   &      0.6$\times$0.6   \\ 
   G034.2581$+$00.1533          &   182.0 &    11.4 &    101.7   &    110.5   &     116.9   &      1.8$\times$1.4  &    88.0 &     10.0  &     120.0  &     120.0  &     169.0  &      0.4$\times$0.2   \\ 
   G034.4032$+$00.2277    &    7.7    &    0.1  &    9.0     &    8.9     &     8.7     &      0.8$\times$0.5  &     5.1   &     0.1   &     6.9    &     7.0    &     6.9    &      0.4$\times$0.4   \\ 
   G035.0242$+$00.3502    &    12.0   &    0.1  &    13.1    &    13.0    &     13.0    &      0.8$\times$0.6  &     9.6   &     0.1   &     11.8   &     11.9   &     12.2   &      0.3$\times$0.3   \\ 
  G035.4669$+$00.1394$\oplus$          &    34.3   &    0.3  &    309.0   &    288.0   &     301.0   &      6.3$\times$6.1  &     6.9   &     0.4   &   163.0  &     148.0  &     173.5  &      4.6$\times$3.5   \\ 
  G035.5781$-$00.0305          &    141.1  &    0.7  &    218.2   &    202.8   &     212.1   &      1.3$\times$0.9  &     147.7 &     1.1   &     209.2  &     202.0  &     207.6  &      0.6$\times$0.2   \\ 
  G036.4057$+$00.0226          &    15.4   &    0.45 &    23.5    &    23.6    &     24.1    &      2.3$\times$1.6  &     4.4   &     0.32  &     18.9   &     18.9   &     19.6   &      1.4$\times$1.3   \\ 
   G036.4062$+$00.0221$\dagger$  &    10.4   &    0.45 &    11.4    &    11.3    &     10.5    &      2.3$\times$1.7  &     7.8   &     0.32  &     8.1    &     7.7    &     7.9    &      1.9$\times$1.7   \\ 
  G037.5457$-$00.1120$\oplus$          &    38.7   &    1.0  &    450.0   &    499.0   &     550.0   &      7.0$\times$5.5  &     8.4   &     0.4   &     167.8  &     142.3  &     168.8  &      3.5$\times$2.8   \\ 
   G037.7347$-$00.1128    &    12.9   &    0.2  &    16.3    &    16.3    &     16.7    &      1.1$\times$0.9  &     6.1   &     0.1   &     14.6   &     13.9   &     14.4   &      0.8$\times$0.7   \\ 
  G037.8731$-$00.3996$\oplus$          &    552.8  &    7.0  &    2588.0  &    2465.0  &     2720.0  &      4.1$\times$3.5  &     264.1 &     3.5   &     1320.0 &     1201.0 &     1153.0 &      1.2$\times$1.1   \\ 
   G037.9723$-$00.0965    &    9.2    &    0.1  &    21.0    &    21.5    &     20.5    &      1.9$\times$1.9  &     2.5   &     0.1   &     18.9   &     17.0   &     20.6   &      1.9$\times$1.7   \\ 
  G038.8756$+$00.3080          &    89.4   &    0.7  &    305.2   &    295.2   &     318.1   &      2.8$\times$2.5  &     22.8  &     0.8   &     231.9  &     210.5  &     238.2  &      2.2$\times$1.9   \\ 
  G039.1956$+$00.2255          &    48.4   &    0.2  &    62.0    &    60.6    &     60.6    &      0.8$\times$0.8  &     22.6  &     0.3   &     58.3   &     56.7   &     55.8   &      0.8$\times$0.8   \\ 
    G039.8824$-$00.3460$\oplus$          &    65.8   &    0.4  &    367.4   &    361.7   &     373.8   &      3.8$\times$3.5  &     14.7  &     0.6   &  214.5  &     212.0  &     226.4  &       3.2$\times$3.2   \\ 
   G042.4345$-$00.2605$\oplus$          &    23.4   &    0.4  &    66.2    &    62.7    &     61.3    &      3.1$\times$2.7  &     6.0   &     1.2   &      42.3   &     40.0   &   41.4   &  2.7$\times$1.4   \\ 
    G043.1651$-$00.0283$\oplus$          &    544.9  &    13.5 &    3809.0  &    3607.1  &     4020.7  &      7.6$\times$7.5  &     215.6 &     4.5   &     2023.9 &     1941.5 &    $-$    &      3.4$\times$3.3   \\ 
  G043.1652$+$00.0129          &    286.0  &    17.1 &    310.0   &    326.0   &  338.2   &      1.7$\times$1.4  &     345.0 &     10.5  &     529.0  &     545.0  &     516.0  &      0.6$\times$0.4   \\ 
  G043.1657$+$00.0116          &    271.0  &    16.4 &    370.6   &    381.7   &     416.0   &      2.4$\times$1.8  &     132.0 &     13.1  &     $-$    &     $-$    &     $-$    &     $-$              \\ 
 G043.1665$+$00.0106$\oplus$          &    524.4  &    18.0 &    2740.0  &    2766.6  &     2900.0  &      3.6$\times$3.2  &     292.9 &     11.4  & 2257.4 &     2068.0 & 2293.9 &      2.9$\times$2.8   \\ 
 G043.1778$-$00.5181          &    12.9   &    0.5  &    150.2   &    146.4   &   160.0   &  6.9$\times$6.9  &     1.8   &     0.3   &     39.5  &     40.3   &     41.0   &      4.3$\times$2.8   \\ 
 G045.0694$+$00.1323          &    35.4  &    1.7  &    71.0    &    78.0    &     80.0    &      1.8$\times$1.8  &     448.1 &     4.2   &     $-$    &     $-$    &     $-$    &     $-$              \\ 
  G045.0712$+$00.1321          &  $-$  &    1.3  &    388.0   &    424.0   &     461.2   &      0.8$\times$0.6  &     436.5 &     4.2   &     750.0  &     745.0  &     806.5  &      0.6$\times$0.5   \\ 
 G045.1223$+$00.1321$\oplus$          &    831.8  &    5.4  &    4191.0  &    4036.4  &     4197.0  &      12.8$\times$12.7  &     517.5 &     5.7   &     2860.1 &     2847.0 &  2888.9 &      2.7$\times$2.6   \\ 
  G045.4545$+$00.0591$\oplus$          &    79.2   &    6.6  &    940.0   &    1057.0  &     1020.0  &      8.8$\times$8.7  &     12.5  &     1.0   &     194.0  &     177.0  &     162.0  &      6.2$\times$6.1   \\ 
   G045.4656$+$00.0452    &    109.7  &    2.5  &    135.6   &    139.2   &     134.0   &      1.2$\times$0.5  &     93.2  &     0.1   &     139.4  &     155.5  &     143.0  &      0.8$\times$0.4   \\ 
  G045.4790$+$00.1294$\oplus$          &    17.0   &    2.6  &    326.0   &    304.0   &     349.3   &      8.6$\times$8.4  &     0.4   &     0.1   &     $-$    &     $-$    &     $-$    &     $-$              \\ 
   G048.6057$+$00.0228    &    28.1   &    0.3  &    44.5    &    47.6    &     44.3    &      1.8$\times$1.3  &     9.6   &     0.3   &     31.6   &     29.8   &     32.0   &      0.9$\times$0.9   \\ 
  G048.6099$+$00.0270$\oplus$          &    27.8   &    0.6  &    98.2    &    97.1    &     102.0   &      6.9$\times$6.8  &     0.4   &     0.6   &     $-$    &     $-$    &     $-$    &     $-$              \\ 
  G048.9296$-$00.2793$\oplus$          &    15.0   &    1.6  &    316.0   &    283.0   &     311.2   &      9.2$\times$8.1  &     1.2   &     1.6   &     32.1   &     29.9   &     30.2   &      3.6$\times$3.6   \\ 
  G049.2679$-$00.3374$\oplus$          &    6.5    &    0.4  &    72.2    &    71.0    &     81.3    &      5.1$\times$5.0  &     1.1   &     0.2   &     32.7   &     28.6  &     35.6  &      3.2$\times$3.2   \\ 
   G049.3666$-$00.3010$\dagger$  &    54.0   &    4.9  &    387.0   &    367.3   &     406.7   &      6.4$\times$6.2  &     1.0   &     1.3   &     $-$    &     $-$    &     $-$    &     $-$              \\
   G049.3704$-$00.3012          &    54.8   &    4.9  &    652.4   &    646.0   &     667.0   &      9.1$\times$9.0  &     5.5   &     1.3   &     269.0  &     177.0  &     152.5  &      4.8$\times$3.8   \\ 
  G049.4905$-$00.3688$\oplus$          &    667.2  &    9.9  &    3900.0  &    3770.0  &     4020.0  &      3.9$\times$3.7  &     186.7 &     8.2   &    2591.4 &     2336.9 &     2470.8 &     2.8$\times$2.8   \\ 
  G050.3152$+$00.6762          &    47.1   &    0.4  &    79.6    &    76.8    &     76.1    &      1.3$\times$1.2  &     15.9  &     0.2   &     62.7   &     59.4   &     63.9   &      1.3$\times$1.0   \\ 
   G051.6785$+$00.7193    &    27.4   &    0.1  &    30.7    &    31.2    &     32.4    &      1.1$\times$0.5  &     33.9  &     0.2   &     35.0   &     35.8   &     38.2   &      0.2$\times$0.1   \\ 
    G052.7533$+$00.3340$\oplus$          &    17.1   &    0.2  &    370.2   &    368.6   &    371.6   &      8.9$\times$8.0  &     2.1   &     0.3   &     $-$    &     $-$    &     $-$    &     $-$              \\ 
  G053.9589$+$00.0320          &    20.5   &    0.1  &    45.3    &    44.6    &     43.5    &      1.8$\times$1.6  &     5.7   &     0.2   &     35.0   &     31.8   &     38.2   &      1.7$\times$1.2   \\ 
   G060.8842$-$00.1286    &    24.1   &    0.8  &    34.1    &    36.2    &     32.6    &      1.7$\times$1.5  &     11.1  &     0.2   &     24.9   &     23.3   &     24.2   &      0.7$\times$0.7   \\ 
    G061.4763$+$00.0892$\oplus$          &    129.2  &    3.5  &  580.4   &    550.0   &     592.1   &      4.5$\times$4.3  &     56.8  &     1.5   &     334.0  &     324.0  &     400.0  &      3.5$\times$2.0   \\ 
   G061.4770$+$00.0892$\dagger$  &    125.0  &    3.5  &    153.0   &    155.0   &     161.0   &      1.61$\times$1.0  &     29.0  &     1.5   &     137.5  &     127.9  &     157.0  &      0.9$\times$0.6   \\ 
\hline
\hline
\end{tabular}
\tablefoot{
 This table will be available in electronic form at the CDS.
 The  `-' symbol means no measurement is available.
$\dagger$ refers to the  five added \uchii\ regions in the observed fields with rising spectra between C and X band; see Sect\,\ref{sect_obs_results}. 
$\oplus$ indicates that the sources are extended and their K-band flux densities should be considered as lower limits. 
Columns: 
(1) Source name;
(2) and (3) peak flux density and local RMS at X-band; 
(4-6) flux density at 9\,GHz, 10\,GHz and 11\,GHz, respectively;
(7) deconvolved source size at X-band;
(8) and (9) peak flux density and RMS at K-band;
(10-12) flux density at 20\,GHz, 22\,GHz and 24\,GHz, respectively;
(13) deconvolved source size at K-band.
The symbol of $\pm 10\%$ refers to an overall estimation of the percentage error at each frequency. 
}
\end{table*}

\begin{table*}
\setlength{\tabcolsep}{0.6pt}
\centering
\caption[]{{ \it \rm \large
Derived physical properties of 116 young \hii\ regions. 
}\label{112hii_result_phyparam_append}}
\begin{tabular}{p{3.2cm}|C{1.6cm}|C{1.2cm}|C{2.0cm}|C{1.2cm}|C{1.3cm}|C{1.9cm}|C{1.2cm}}
\hline
\hline
Name &   $ n_{\rm e}$ & $ diam$ & EM & $ \nu_{\rm t}$ & $\log N_{\rm Ly}$ & Spectral & $f_{\rm d}$ \\
\hline
 &   ($\rm 10^{5}\,cm^{-3}$) & ($\rm pc$) & ($\rm 10^{7}\,pc\,cm^{-6}$) & ($\rm GHz$) &  ($\rm s^{-1}$) & Type &   \\
\hline
(1)   &  (2)  & (3) & (4) & (5) & (6) & (7) & (8)  \\
\hline
\hline
  G010.3009$-$00.1477 & 0.09 & 0.119 & 0.92   & 1.69  & 47.94   & O9.5          & 0.86           \\ 
  G010.4724$+$00.0275 & 1.43 & 0.022 & 45.2   & 10.77 & 48.11   & O9            & 0.94           \\ 
  G010.6234$-$00.3837 & 0.16 & 0.166 & 4.39   & 3.55  & 48.9    & O6.5          & 0.81           \\ 
  G010.9584$+$00.0221 & 0.36 & 0.029 & 3.78   & 3.31  & 47.35   & B0            & $-$            \\ 
  G011.0328$+$00.0274 & 0.13 & 0.014 & 0.24   & 0.89  & 45.57   & B1            & $-$            \\ 
  G011.1104$-$00.3985 & 0.07 & 0.145 & 0.62   & 1.4   & 47.94   & O9.5          & 0.27           \\ 
  G011.1712$-$00.0662 & 0.09 & 0.053 & 0.45   & 1.21  & 46.91   & B0            & $-$            \\ 
  G011.9368$-$00.6158 & 0.07 & 0.155 & 0.86   & 1.63  & 48.12   & O9            & 0.19           \\ 
  G011.9446$-$00.0369 & 0.17 & 0.075 & 2.2    & 2.56  & 47.84   & O9.5          & $-$            \\ 
  G012.1988$-$00.0345 & 0.07 & 0.148 & 0.65   & 1.43  & 47.98   & O9            & 0.77           \\ 
  G012.2081$-$00.1019 & 0.05 & 0.268 & 0.72   & 1.5   & 48.59   & O7            & 0.83           \\ 
  G012.4294$-$00.0479 & 0.26 & 0.02  & 1.38   & 2.05  & 46.55   & B0.5          & $-$            \\ 
  G012.8050$-$00.2007 & 0.1  & 0.248 & 2.68   & 2.81  & 49.05   & O6            & $-$            \\ 
  G012.9995$-$00.3583 & 0.28 & 0.008 & 0.6    & 1.38  & 45.41   & B1            & $-$            \\ 
  G013.2099$-$00.1428 & 0.14 & 0.082 & 1.67   & 2.24  & 47.88   & O9.5          & $-$            \\ 
  G013.3850$+$00.0684 & 0.37 & 0.032 & 4.43   & 3.57  & 47.49   & B0            & $-$            \\ 
  G014.7785$-$00.3328 & 0.27 & 0.016 & 1.21   & 1.92  & 46.4    & B0.5          & $-$            \\ 
  G016.1448$+$00.0088 & 0.14 & 0.058 & 1.11   & 1.85  & 47.42   & B0            & $-$            \\ 
  G016.3913$-$00.1383 & 0.16 & 0.023 & 0.59   & 1.36  & 46.3    & B0.5          & $-$            \\ 
  G016.9445$-$00.0738 & 0.06 & 0.389 & 1.48   & 2.12  & 49.16   & O6            & $-$            \\ 
  G017.0299$-$00.0696 & 0.2  & 0.024 & 0.94   & 1.7   & 46.61   & B0.5          & $-$            \\ 
  G017.1141$-$00.1124 & 0.03 & 0.145 & 0.1    & 0.59  & 47.23   & B0            & 0.75           \\ 
  G018.3024$-$00.3910 & 0.07 & 0.158 & 0.83   & 1.61  & 48.15   & O8.5          & 0.14           \\ 
  G018.4433$-$00.0056 & 0.07 & 0.16  & 0.68   & 1.46  & 48.04   & O9            & $-$            \\ 
  G018.4614$-$00.0038 & 0.11 & 0.196 & 2.23   & 2.57  & 48.75   & O6.5          & 0.54           \\ 
  G018.6654$+$00.0294 & 0.17 & 0.034 & 1.01   & 1.76  & 46.92   & B0            & 0.88           \\ 
  G018.7106$+$00.0002 & 0.3  & 0.023 & 2.03   & 2.46  & 46.84   & B0            & $-$            \\ 
  G018.7612$+$00.2630 & 0.11 & 0.111 & 1.36   & 2.03  & 48.05   & O9            & 0.83           \\ 
  G018.8250$-$00.4675 & 0.05 & 0.051 & 0.13   & 0.66  & 46.41   & B0.5          & $-$            \\ 
  G018.8338$-$00.3002 & 0.05 & 0.251 & 0.55   & 1.32  & 48.36   & O8            & 0.29           \\ 
  G019.0754$-$00.2874 & 0.06 & 0.205 & 0.33   & 1.04  & 47.99   & O9            & 0.85           \\ 
  G019.4912$+$00.1352 & 0.04 & 0.388 & 0.77   & 1.55  & 48.87   & O6.5          & $-$            \\ 
  G019.6087$-$00.2351 & 0.09 & 0.416 & 3.06   & 2.99  & 49.5    & O5            & 0.63           \\ 
  G019.6090$-$00.2313 & 0.08 & 0.219 & 1.34   & 2.02  & 48.63   & O6.5          & 0.95           \\ 
  G019.7407$+$00.2821 & 0.15 & 0.138 & 3.13   & 3.02  & 48.54   & O7.5          & $-$            \\ 
  G019.7549$-$00.1282 & 0.29 & 0.037 & 3.17   & 3.04  & 47.45   & B0            & 0.59           \\ 
  G020.0809$-$00.1362 & 0.14 & 0.191 & 3.49   & 3.18  & 49.04   & O6            & 0.52           \\ 
  G020.3633$-$00.0136 & 0.17 & 0.032 & 0.96   & 1.73  & 46.81   & B0            & $-$            \\ 
  G021.3571$-$00.1766 & 0.13 & 0.062 & 1.02   & 1.77  & 47.42   & B0            & 0.62           \\ 
  G021.3855$-$00.2541 & 0.12 & 0.108 & 1.61   & 2.21  & 48.1    & O9            & 0.61           \\ 
  G021.4257$-$00.5417 & 0.04 & 0.091 & 0.17   & 0.77  & 46.98   & B0            & 0.58           \\ 
  G023.2654$+$00.0765 & 0.11 & 0.073 & 0.87   & 1.65  & 47.49   & B0            & $-$            \\ 
  G023.4553$-$00.2010 & 0.49 & 0.015 & 3.55   & 3.21  & 46.72   & B0.5          & $-$            \\ 
  G024.5065$-$00.2224 & 0.08 & 0.118 & 0.91   & 1.68  & 47.94   & O9.5          & 0.46           \\ 
  G024.7889$+$00.0824 & 0.25 & 0.031 & 1.94   & 2.41  & 47.21   & B0            & 0.92           \\ 
  G024.7898$+$00.0833 & 3.38 & 0.008 & 91.84  & 15.1  & 47.52   & B0            & 0.92           \\ 
  G025.3948$+$00.0332 & 0.05 & 0.36  & 1.01   & 1.76  & 48.96   & O6            & 0.24           \\ 
  G025.3970$+$00.5614 & 0.07 & 0.211 & 1.08   & 1.82  & 48.52   & O7.5          & 0.20           \\ 
  G025.3981$-$00.1411 & 0.04 & 0.589 & 1.17   & 1.89  & 49.46   & O5.5          & 0.66           \\ 
  G025.7157$+$00.0487 & 0.09 & 0.071 & 0.54   & 1.31  & 47.29   & B0            & 0.72           \\ 
  G025.8011$-$00.1568 & 0.13 & 0.07  & 1.23   & 1.94  & 47.6    & O9.5          & 0.87           \\ 
  G026.5444$+$00.4169 & 0.06 & 0.271 & 1.04   & 1.79  & 48.75   & O6.5          & 0.15           \\ 
  G027.2800$+$00.1447 & 0.03 & 0.505 & 0.49   & 1.26  & 48.94   & O6.5          & $-$            \\ 
  G027.3644$-$00.1657 & 0.05 & 0.12  & 0.35   & 1.06  & 47.58   & B0            & 0.83           \\ 
  G027.9782$+$00.0789 & 0.1  & 0.078 & 0.78   & 1.56  & 47.49   & B0            & $-$            \\ 
  G028.2003$-$00.0494 & 1.41 & 0.027 & 53.54  & 11.68 & 48.37   & O8            & 0.43           \\ 
  G028.2879$-$00.3641 & 0.05 & 0.398 & 0.81   & 1.59  & 48.91   & O6.5          & 0.80           \\ 
  G028.6082$+$00.0185 & 0.06 & 0.17  & 0.68   & 1.46  & 48.12   & O9            & 0.60           \\ 
  G029.9559$-$00.0168 & 0.06 & 0.331 & 1.14   & 1.87  & 48.9    & O6.5          & 0.65           \\ 
  G030.0096$-$00.2734 & 1.91 & 0.004 & 15.57  & 6.49  & 46.21   & B0.5          & $-$            \\ 

  \hline
\end{tabular}
\end{table*}    
\addtocounter{table}{-1}
\begin{table*}
\setlength{\tabcolsep}{0.6pt}
\centering
\caption[]{{ \it \rm \large --continuue\, Derived physical properties of 116 young \hii\ regions.
}}
\begin{tabular}{p{3.2cm}|C{1.6cm}|C{1.2cm}|C{2.0cm}|C{1.2cm}|C{1.3cm}|C{1.9cm}|C{1.2cm}}
\hline
\hline
Name &   $ n_{\rm e}$ & $  diam$ & EM & $ \nu_{\rm t}$  & $\log N_{\rm Ly}$ & Spectral & $f_{\rm d}$ \\
\hline
 &   ($\rm 10^{5}\,cm^{-3}$) & ($\rm pc$) & ($\rm 10^{7}\,pc\,cm^{-6}$) & ($\rm GHz$) & ($\rm s^{-1}$) & Type &   \\
\hline
(1)   &  (2)  & (3) & (4) & (5) & (6) & (7) & (8)  \\
\hline
\hline
  G030.5353$+$00.0204 & 0.07 & 0.12  & 0.55   & 1.32  & 47.73   & O9.5          & $-$            \\ 
  G030.5887$-$00.0428 & 2.08 & 0.009 & 39.1   & 10.06 & 47.28   & B0            & $-$            \\ 
  G030.7197$-$00.0829 & 0.22 & 0.093 & 4.54   & 3.61  & 48.44   & O7.5          & $-$            \\ 
  G030.8662$+$00.1143 & 0.37 & 0.031 & 4.25   & 3.5   & 47.46   & O9.5          & $-$            \\ 
  G031.0495$+$00.4697 & 0.21 & 0.012 & 0.56   & 1.34  & 45.75   & B1            & $-$            \\ 
  G031.1596$+$00.0448 & 0.12 & 0.026 & 0.39   & 1.13  & 46.27   & B0.5          & $-$            \\ 
  G031.2801$+$00.0632 & 0.14 & 0.092 & 1.78   & 2.31  & 48.05   & O9            & 0.49           \\ 
  G032.4727$+$00.2036 & 0.19 & 0.033 & 1.25   & 1.96  & 46.98   & B0            & $-$            \\ 
  G032.7441$-$00.0755 & 2.79 & 0.011 & 82.77  & 14.37 & 47.69   & O9.5          & 0.88           \\ 
  G032.7966$+$00.1909 & 0.04 & 0.81  & 1.5    & 2.13  & 49.83   & O4            & 0.21           \\ 
  G032.9273$+$00.6060 & 0.03 & 0.44  & 0.52   & 1.29  & 48.83   & O6.5          & $-$            \\ 
  G033.1328$-$00.0923 & 0.21 & 0.1   & 4.56   & 3.61  & 48.54   & O7.5          & 0.16           \\ 
  G033.4163$-$00.0036 & 0.12 & 0.066 & 1.01   & 1.76  & 47.48   & B0            & $-$            \\ 
  G033.9145$+$00.1105 & 0.07 & 0.229 & 0.98   & 1.74  & 48.57   & O8            & 0.44           \\ 
  G034.2572$+$00.1535 & 0.52 & 0.038 & 10.12  & 5.28  & 48.03   & O9            & 0.51           \\ 
  G034.2573$+$00.1523 & 3.55 & 0.005 & 58.23  & 12.16 & 46.58   & B0.5          & 0.98           \\ 
  G034.2581$+$00.1533 & 3.01 & 0.004 & 37.26  & 9.83  & 46.54   & B0.5          & 0.98           \\ 
  G034.4032$+$00.2277 & 0.38 & 0.006 & 0.87   & 1.65  & 45.38   & B1            & $-$            \\ 
  G035.0242$+$00.3502 & 0.47 & 0.008 & 1.74   & 2.28  & 45.86   & B0.5          & 0.58           \\ 
  G035.4669$+$00.1394 & 0.05 & 0.24  & 0.6    & 1.37  & 48.33   & O8            & 0.82           \\ 
  G035.5781$-$00.0305 & 0.22 & 0.093 & 4.46   & 3.58  & 48.36   & O8            & 0.81           \\ 
  G036.4057$+$00.0226 & 0.33 & 0.016 & 1.74   & 2.29  & 46.48   & B0.5          & $-$            \\ 
  G036.4062$+$00.0221 & 0.13 & 0.022 & 0.4    & 1.13  & 46.16   & B0.5          & $-$            \\ 
  G037.5457$-$00.1120 & 0.07 & 0.242 & 1.23   & 1.94  & 48.7    & O7            & 0.24           \\ 
  G037.7347$-$00.1128 & 0.1  & 0.062 & 0.62   & 1.4   & 47.21   & B0            & 0.77           \\ 
  G037.8731$-$00.3996 & 0.05 & 0.532 & 1.41   & 2.07  & 49.38   & O5.5          & 0.42           \\ 
  G037.9723$-$00.0965 & 0.13 & 0.082 & 1.36   & 2.03  & 47.79   & O9.5          & $-$            \\ 
  G038.8756$+$00.3080 & 0.05 & 0.344 & 0.74   & 1.52  & 48.79   & O6.5          & $-$            \\ 
  G039.1956$+$00.2255 & 0.2  & 0.081 & 3.26   & 3.08  & 48.12   & O9            & $-$            \\ 
  G039.8824$-$00.3460 & 0.05 & 0.256 & 0.7    & 1.48  & 48.51   & O7.5          & $-$            \\ 
  G042.4345$-$00.2605 & 0.15 & 0.044 & 1.06   & 1.8   & 47.1    & B0            & $-$            \\ 
  G043.1651$-$00.0283 & 0.12 & 0.387 & 5.33   & 3.9   & 49.69   & O5            & 0.42           \\ 
  G043.1652$+$00.0129 & 0.88 & 0.053 & 41.47  & 10.34 & 48.91   & O6.5          & 0.90           \\ 
  G043.1657$+$00.0116 & 1.57 & 0.046 & 113.18 & 16.68 & 48.69   & O7            & 0.94           \\ 
  G043.1665$+$00.0106 & 0.24 & 0.22  & 12.21  & 5.78  & 49.55   & O5            & 0.58           \\ 
  G043.1778$-$00.5181 & 0.06 & 0.165 & 0.58   & 1.36  & 48.11   & O9            & 0.71           \\ 
  G045.0694$+$00.1323 & 0.74 & 0.026 & 13.97  & 6.16  & 47.76   & O9.5          & 0.97           \\ 
  G045.0712$+$00.1321 & 1.22 & 0.04  & 58.98  & 12.23 & 48.73   & O6.5          & 0.76           \\ 
  G045.1223$+$00.1321 & 0.07 & 0.429 & 2.08   & 2.49  & 49.43   & O5.5          & 0.68           \\ 
  G045.4545$+$00.0591 & 0.08 & 0.235 & 1.4    & 2.06  & 48.72   & O6.5          & 0.77           \\ 
  G045.4656$+$00.0452 & 1.02 & 0.023 & 23.56  & 7.89  & 47.88   & O9.5          & 0.81           \\ 
  G045.4790$+$00.1294 & 0.01 & 0.462 & 0.09   & 0.56  & 48.06   & O9            & $-$            \\ 
  G048.6057$+$00.0228 & 0.31 & 0.042 & 4.1    & 3.44  & 47.76   & O9.5          & 0.97           \\ 
  G048.6099$+$00.0270 & 0.1  & 0.116 & 1.19   & 1.91  & 47.99   & O9            & 0.85           \\ 
  G048.9296$-$00.2793 & 0.22 & 0.065 & 3.26   & 3.08  & 47.97   & O9            & $-$            \\ 
  G049.2679$-$00.3374 & 0.07 & 0.089 & 0.49   & 1.25  & 47.46   & B0            & $-$            \\ 
  G049.3666$-$00.3010 & 0.94 & 0.031 & 26.89  & 8.41  & 48.05   & O9            & 0.83           \\ 
  G049.3704$-$00.3012 & 0.11 & 0.128 & 1.61   & 2.2   & 48.34   & O8            & 0.67           \\ 
  G049.4905$-$00.3688 & 0.1  & 0.27  & 2.57   & 2.75  & 49.04   & O6            & 0.85           \\ 
  G050.3152$+$00.6762 & 0.26 & 0.018 & 1.2    & 1.92  & 46.42   & B0.5          & $-$            \\ 
  G051.6785$+$00.7193 & 0.64 & 0.026 & 10.74  & 5.44  & 47.68   & O9.5          & 0.88           \\ 
  G052.7533$+$00.3340 & 0.05 & 0.254 & 0.75   & 1.53  & 48.51   & O7.5          & $-$            \\ 
  G053.9589$+$00.0320 & 0.06 & 0.066 & 0.24   & 0.89  & 46.87   & B0            & $-$            \\ 
  G060.8842$-$00.1286 & 0.93 & 0.012 & 10.8   & 5.35  & 46.4    & B0.5          & $-$            \\ 
  G061.4763$+$00.0892 & 0.13 & 0.09  & 1.72   & 2.27  & 48.1    & O9            & 0.81           \\ 
  G061.4770$+$00.0892 & 4.45 & 0.004 & 78.76  & 14.04 & 46.81   & B0            & 0.99           \\ 

\hline
\hline
\end{tabular}
\tablefoot{
 This table will be available in electronic form at the CDS.
The fraction of Lyman
continuum photons absorbed by dust within \hii\ regions $f_{\rm d}$ should be taken as upper limits. 
The  `-' symbol means no measurement is available.  
Columns: 
(1) source name;
(2) electron density;
(3) physical diameter;
(4) emission measure;
(5) turnover frequency;
(6) Lyman continuum flux;
(7) spectral type; 
(8) dust absorption fraction.
}
\end{table*}

\clearpage
\section{Additional images}

\begin{figure*}
 \centering
 \caption{Examples of the best-fitting SEDs and the radio images in C-band, X-band, and K-band observations.}
 \begin{tabular}{cccc}
\includegraphics[width = 0.24\textwidth]{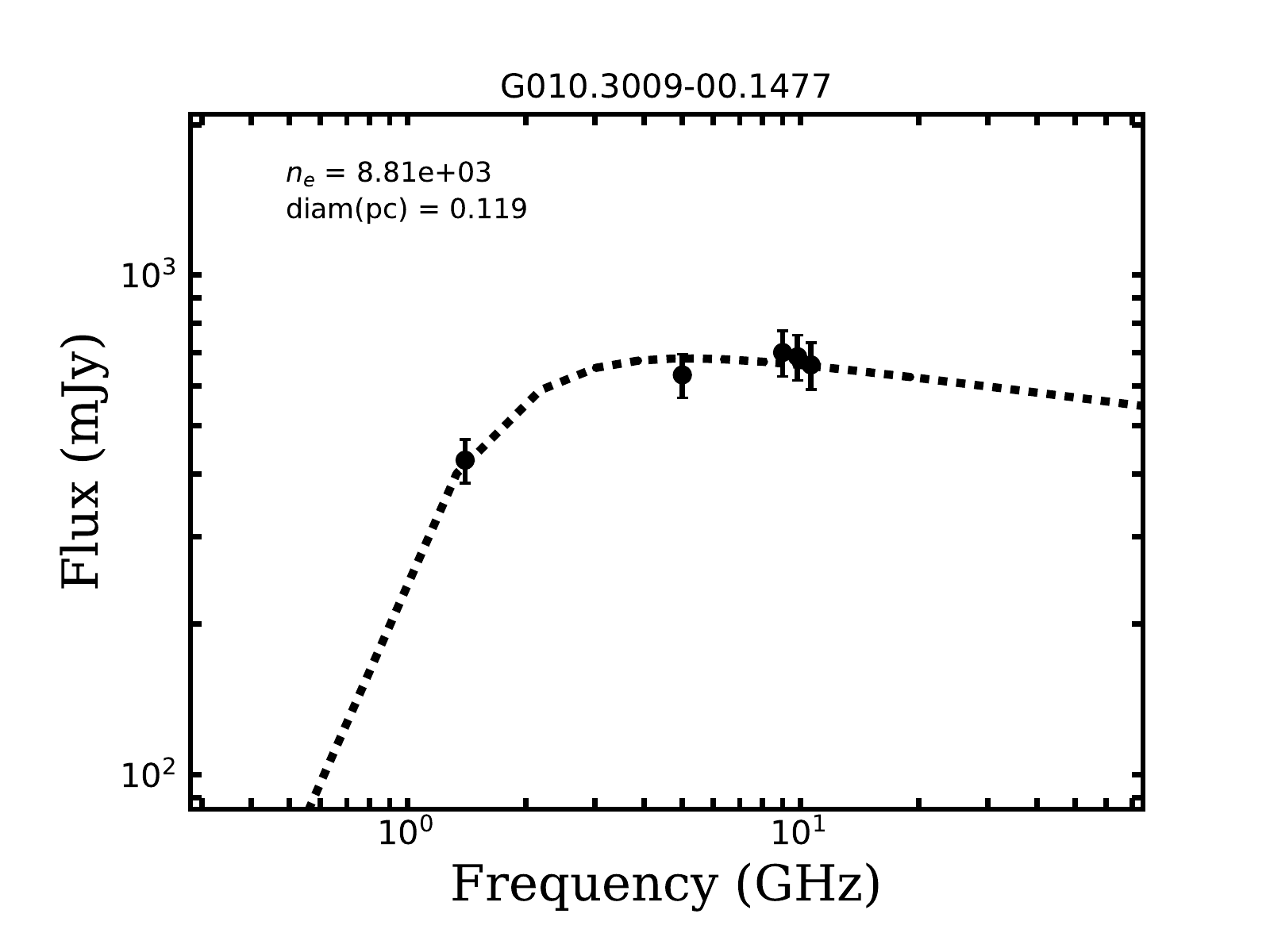} &
 \hspace{-5mm} 
 \includegraphics[width = 0.24\textwidth]{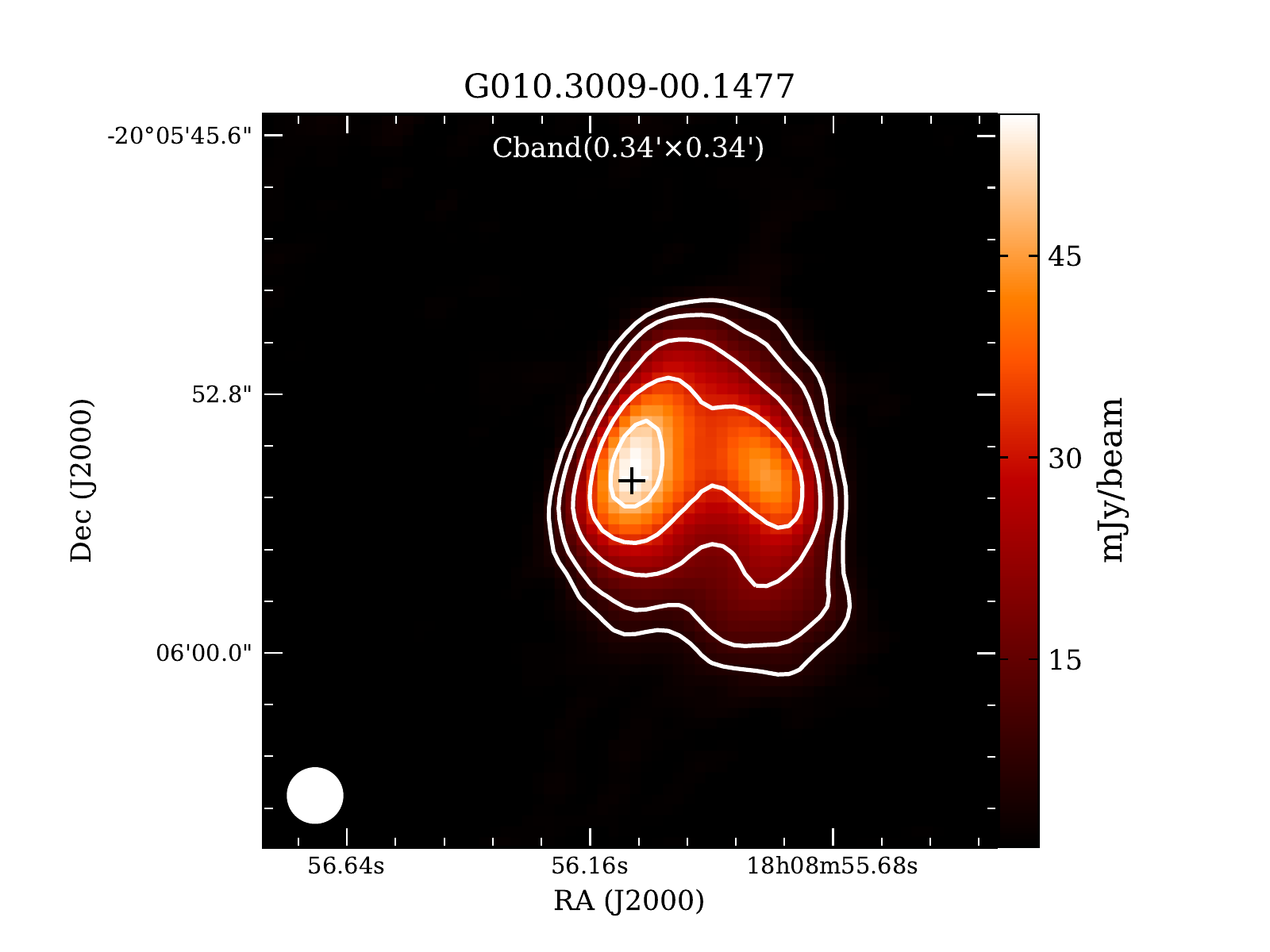}&
  \hspace{-5mm} 
 \includegraphics[width = 0.24\textwidth]{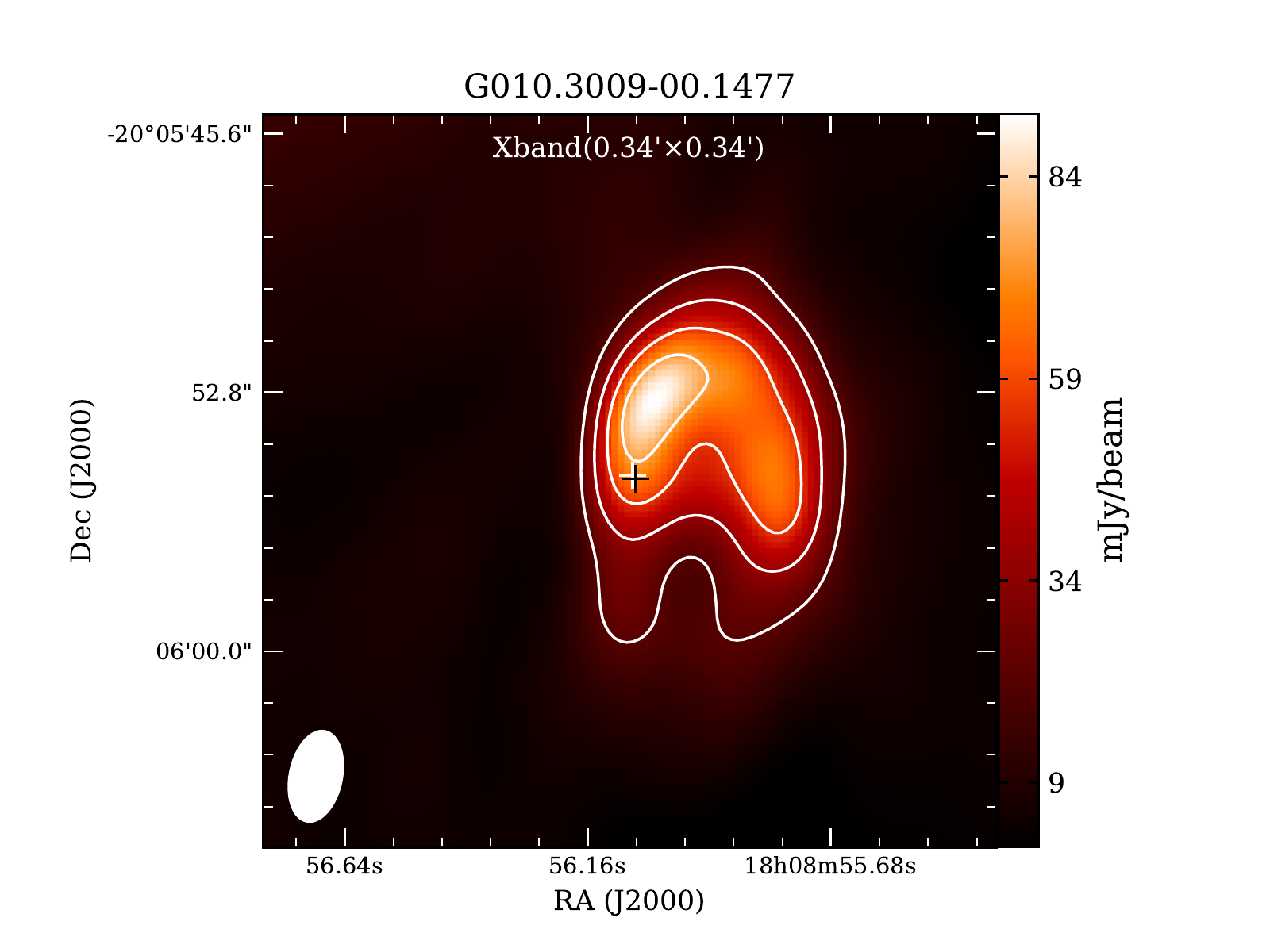}&
  \hspace{-5mm} 
  \includegraphics[width = 0.24\textwidth]{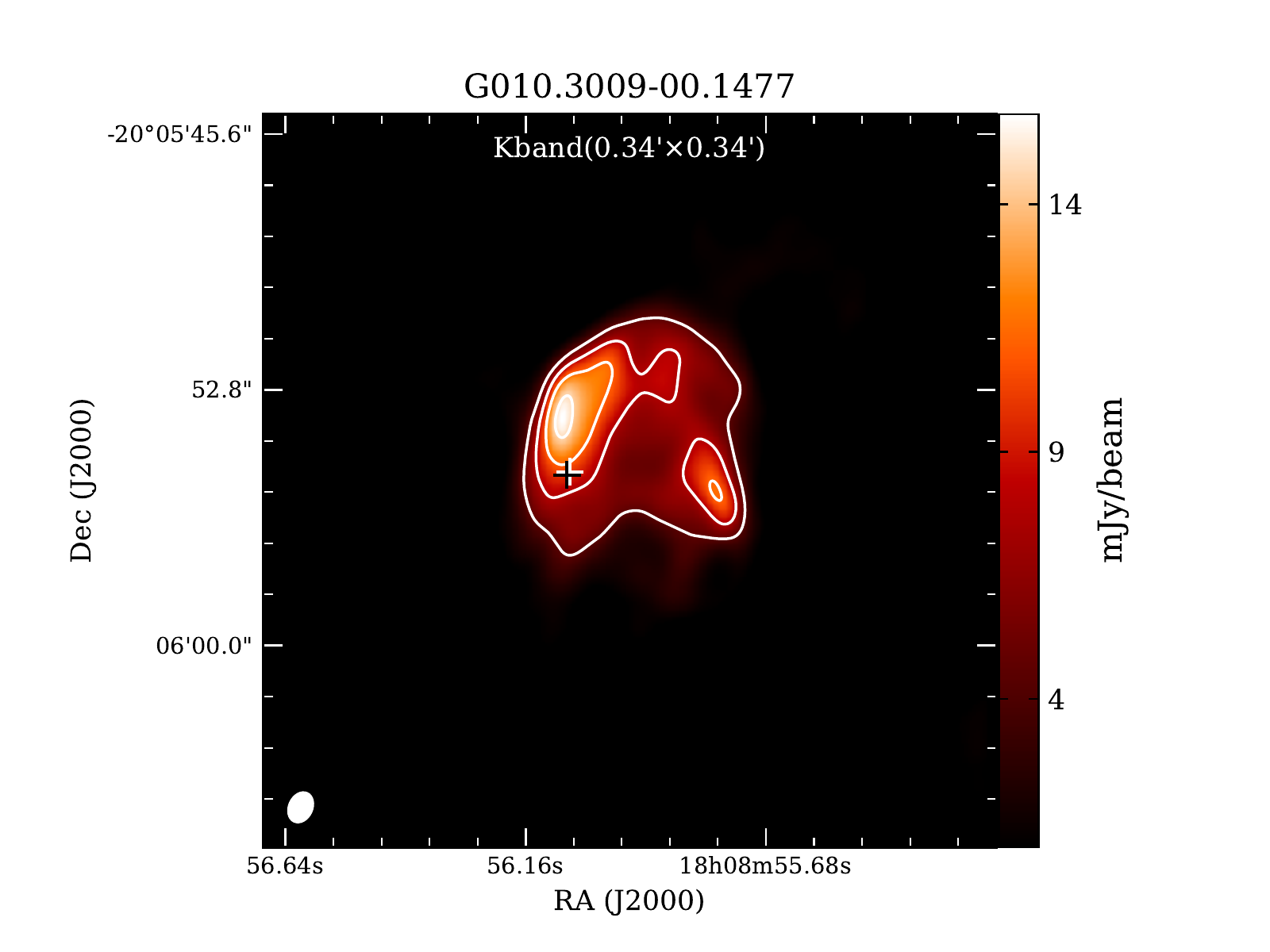} \\
  \includegraphics[width = 0.24\textwidth]{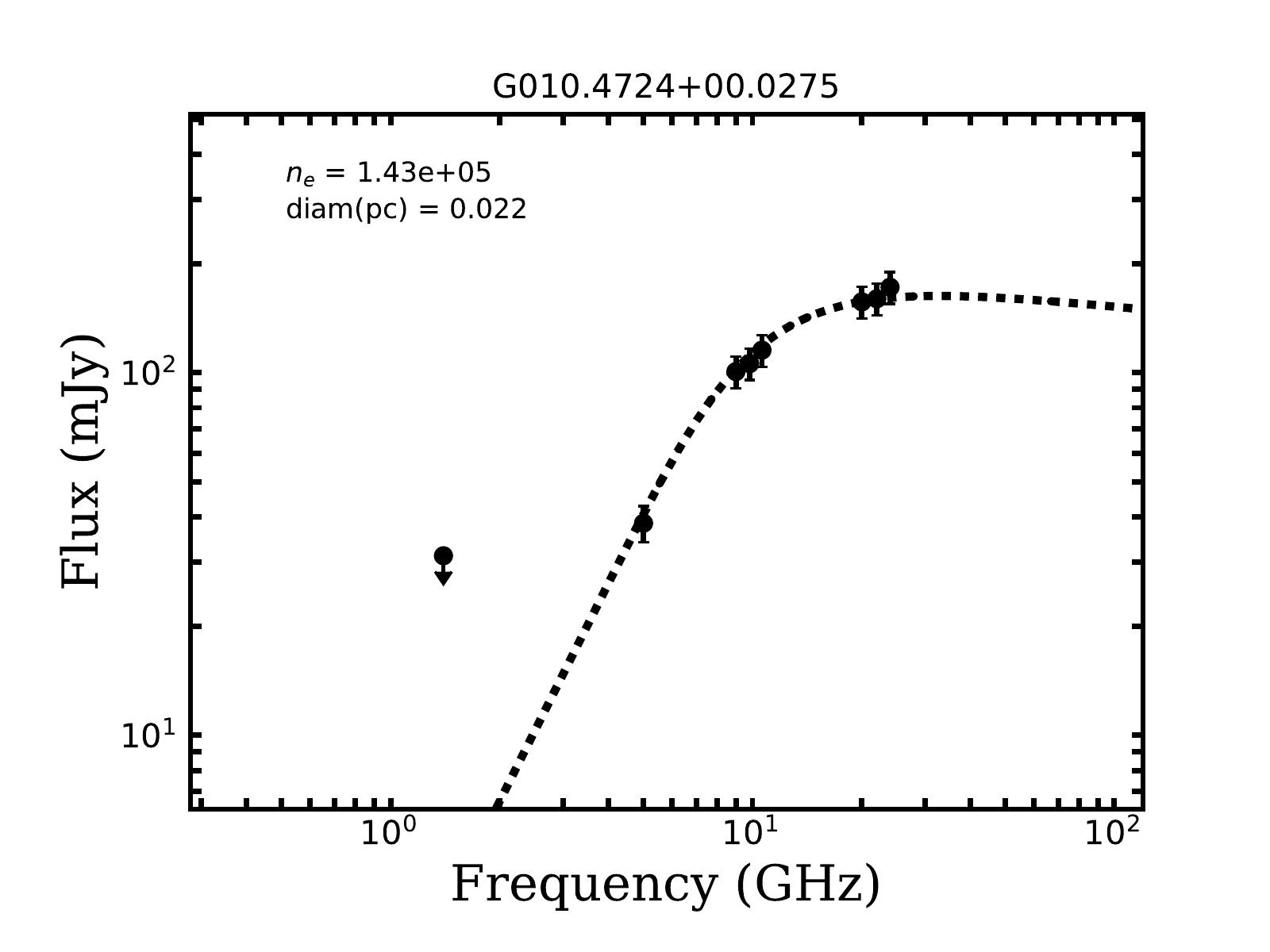} &
 \hspace{-5mm} 
 \includegraphics[width = 0.24\textwidth]{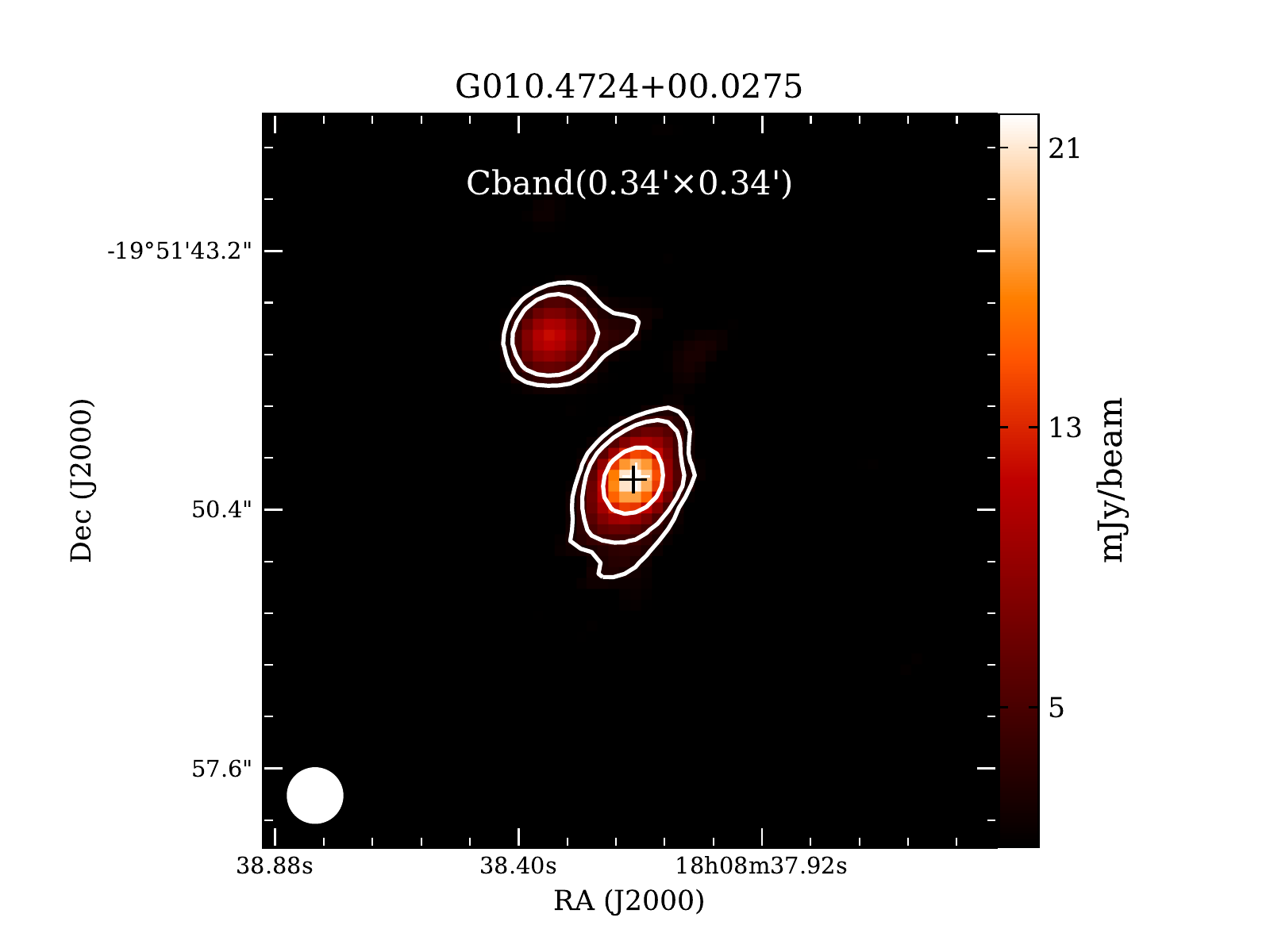}&
  \hspace{-5mm} 
 \includegraphics[width = 0.24\textwidth]{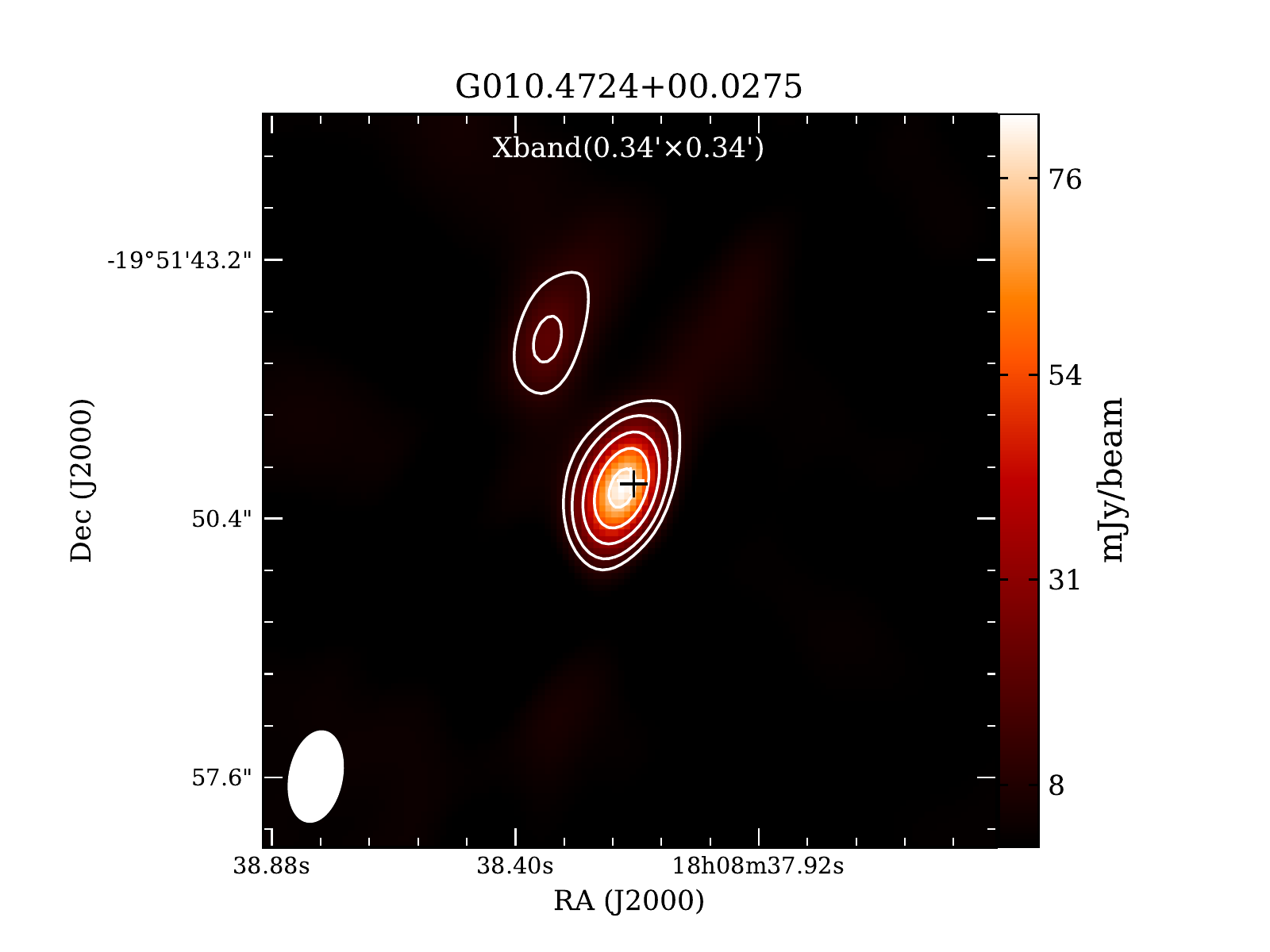}&
  \hspace{-5mm} 
  \includegraphics[width = 0.24\textwidth]{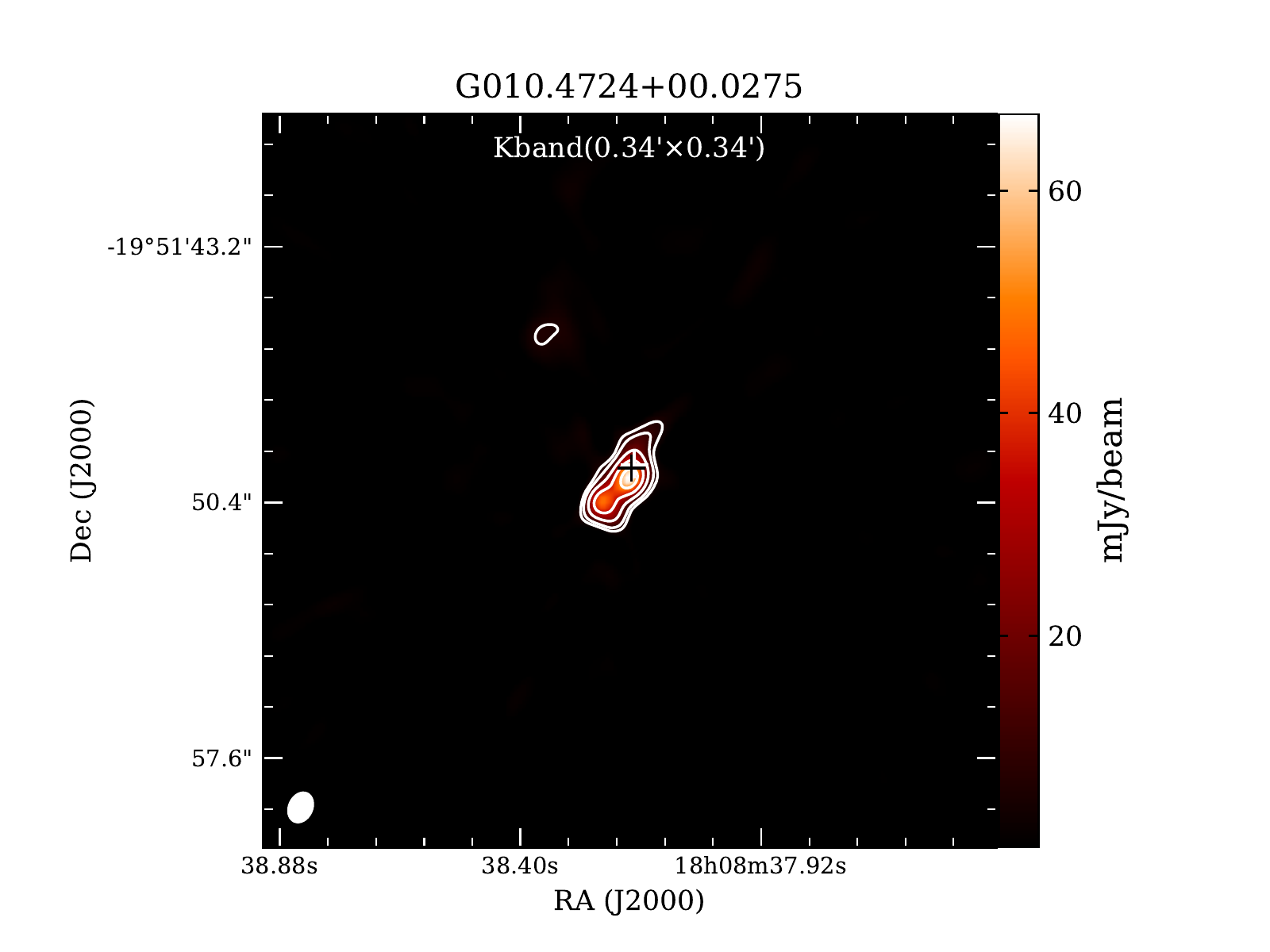} \\
    \includegraphics[width = 0.24\textwidth]{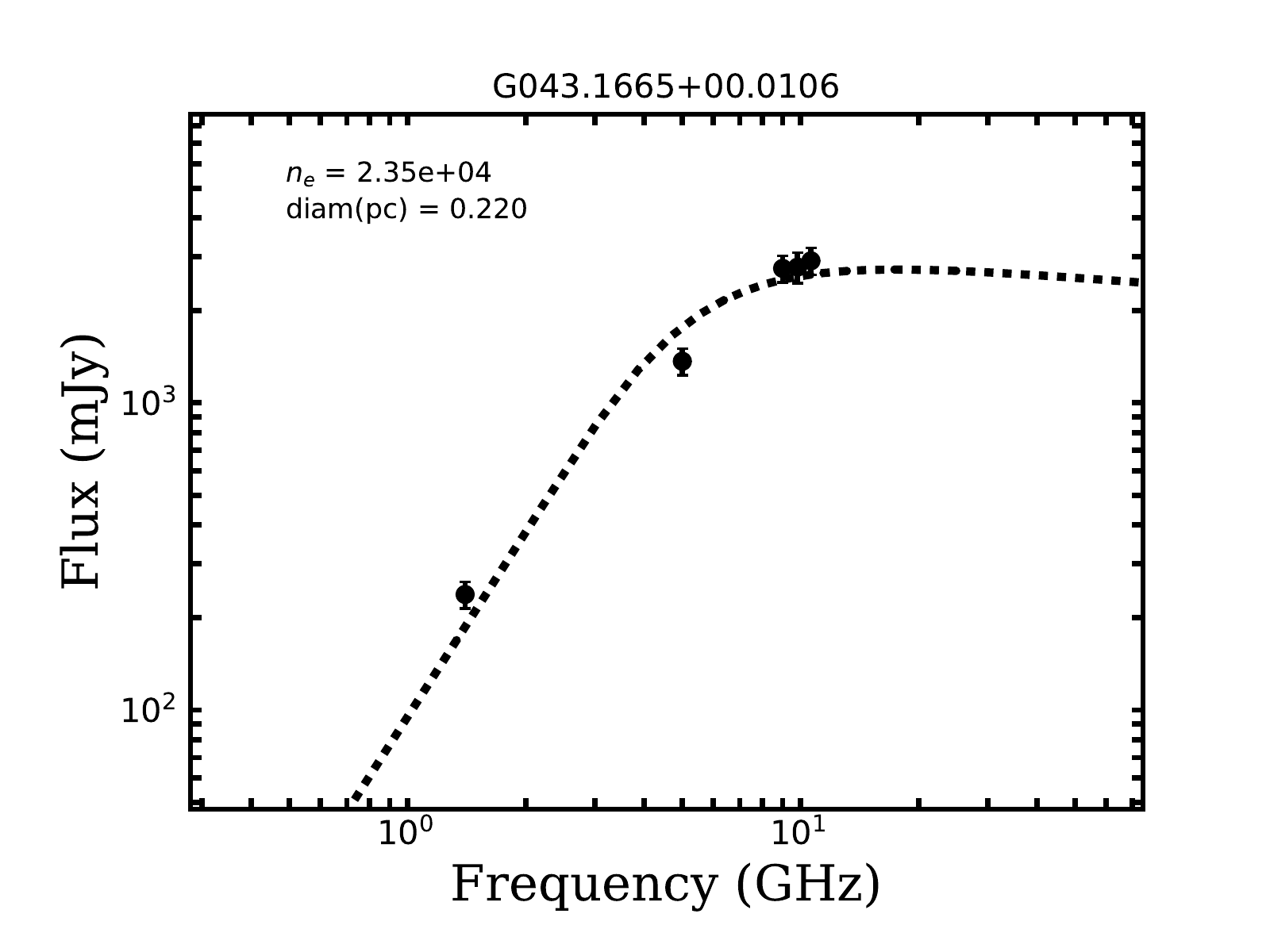} &
 \hspace{-5mm} 
 \includegraphics[width = 0.24\textwidth]{./G043p1665+00p0106_ff_radio_imgC1.pdf}&
  \hspace{-5mm} 
 \includegraphics[width = 0.24\textwidth]{./G043p1665+00p0106_ff_radio_imgX.pdf}&
  \hspace{-5mm} 
  \includegraphics[width = 0.24\textwidth]{./G043p1665+00p0106_ff_radio_imgK1.pdf} \\
  (a)~Radio~~SED   & (b)~C-band image &  (c)~X-band~image & (d)~K-band~image \\
\end{tabular}
 \begin{tablenotes}
 \item
 Panel (a): Radio SED and best-fitting model for each source in the sample. 
The SED shows the free-free emission fit to flux density points between 1 and 26 GHz for a single compact source, 
while the extended source has the best fit to flux density points between 1 and 11 GHz as their K-band flux measurements are not reliable owing to the shortage of short baseline spacings. 
The uncertainties on flux measurements of these points are used to constrain the fitting process and to obtain the best estimate. 
The best-fitting results of electron density $n_{\rm e}(\,\rm cm^{-3})$ and physical diameter $diam\,(\rm pc)$ for each source are shown in the upper-left corner of each figure. 
Panels (b), (c) and (d): Radio images in C-band, X-band, and K-band marked with the positions of the young \uchii\ regions in each image, including single-component compact sources, extended sources, and cluster sources. 
The C-band images are taken from the CORNISH survey and are used to compare with the images at X-band in this work as the X-band observations have comparable beam sizes to those of the CORNISH survey. 
For some sources, the K-band images are not shown because of the poor quality of observational data at K-band. 
The lime polygons in the C-band images shown for some sources are similar to the defined region in the CORNISH survey. 
The lime polygons in the X-band images shown for some sources refer to the manually drawn emission regions used to measure the observational results following the same strategy in CORNISH survey. 
The white contour levels in the images are equally spaced by 5$\sigma$ and start at a level of 5$\sigma$. 
The image size of each target is shown in the upper-middle part of each image. 
The beam sizes for C-band ($1.5\arcsec$), X-band ($\sim 1.7\arcsec$), and K-band ($0.7\arcsec$) are shown in the lower-left corner of each image. 
Note: Figures for the full sample are available in electronic form at the Zenodo via \url{https://doi.org/10.5281/zenodo.4293684}.
   \end{tablenotes}
\label{summary_sed_multiband_images}
\end{figure*}
\end{appendix}

\end{document}